\documentclass[apj,twocolappendix, numberedappendix]{emulateapj}
\usepackage[colorlinks,citecolor=blue]{hyperref}
\usepackage{textcomp}
\usepackage{subfigure}
\usepackage{verbatim}
\usepackage{bm}
\usepackage{hyperref}
\usepackage{amsmath}
\usepackage{graphicx}
\usepackage{epstopdf}
\usepackage{amssymb}
\usepackage{extarrows}
\usepackage{color}
\usepackage{CJK}

\newcommand{\ba}{\begin{eqnarray}}
\newcommand{\ea}{\end{eqnarray}}
\newcommand{\be}{\begin{equation}}
\newcommand{\ee}{\end{equation}}
\newcommand{\gr}{\mathrm{GR}}
\newcommand{\gw}{\mathrm{GW}}

\newcommand{\m}{\mathrm{max}}
\newcommand{\mi}{\mathrm{min}}
\newcommand{\qu}{\mathrm{quad}}
\newcommand{\oct}{\mathrm{oct}}

\newcommand{\au}{\mathrm{AU}}

\newcommand{\IN}{\mathrm{in}}
\newcommand{\OUT}{\mathrm{out}}
\newcommand{\h}{\mathrm{Hubble}}

\newcommand{\lk}{\mathrm{LK}}
\newcommand{\merger}{\mathrm{merger}}
\newcommand{\eff}{\mathrm{eff}}
\newcommand{\f}{\mathrm{f}}
\newcommand{\SL}{\mathrm{sl}}
\newcommand{\li}{\mathrm{lim}}
\newcommand{\tot}{\mathrm{tot}}

\newcommand{\SB}{\mathrm{sb}}
\def\e1{e_1^2}

\begin{document}

\title{Black Hole and Neutron Star Binary Mergers in Triple Systems: \\Merger Fraction and Spin-Orbit Misalignment}
\author{Bin Liu$^{1,2,3}$, Dong Lai$^{2,1,4}$}
\affil{$^{1}$ Shanghai Astronomical Observatory, Chinese Academy of Sciences, 80 Nandan Road, Shanghai 200030, China\\
$^{2}$ Cornell Center for Astrophysics and Planetary Science, Cornell University, Ithaca, NY 14853, USA\\
$^{3}$ Key Laboratory for the Structure and Evolution of Celestial Objects, Chinese Academy of Sciences, Kunming 650011, China\\
$^{4}$ Tsung-Dao Lee Institute, Shanghai 200240, China
}

\begin{abstract}
Black hole (BH) mergers driven by gravitational perturbations of external companions constitute an
important class of formation channels for merging BH binaries
detected by LIGO. We have studied the orbital and spin evolution of
binary BHs in triple systems, where the tertiary companion excites
large eccentricity in the inner binary through Lidov-Kozai oscillations,
causing the binary to merge via gravitational radiation. Using the
single-averaged and double-averaged secular dynamics
of triples (where the equations of motion are averaged over the inner
orbit and both orbits, respectively), we perform a large set of numerical
integrations to determine the merger window (the range of companion
inclinations that allows the inner binary to merge within $\sim$10~Gyrs) and
the merger fraction as a function of various system parameters (e.g.,
the binary masses $m_1,~m_2$ and initial semi-major axis $a_0$, the
mass, semi-major axis and eccentricity $e_{\rm
  out}$ of the outer companion).  For typical BH binaries ($m_{1,2}\simeq
20M_\odot-30M_\odot$ and $a_0\gtrsim 10$~AU), the merger
fraction increases rapidly with $e_{\rm out}$ because
of the octupole perturbation, ranging from $\sim 1\%$ at $e_{\rm
  out}=0$ to $10-20\%$ at $e_{\rm out}=0.9$.  We derive the analytical
expressions and approximate scaling relations for the merger window
and merger fraction for systems with negligible octupole effect, and
apply them to neutron star binary mergers in triples.  We also
follow the spin evolution of the BHs during the companion-induced
orbital decay, where de-Sitter spin precession competes with
Lidov-Kozai orbital precession/nutation.  Starting from aligned spin
axes (relative to the orbital angular momentum axis), a wide range of
final spin-orbit misalignment angle $\theta_{\rm sl}^{\rm f}$ can be
generated when the binary enters the LIGO sensitivity band.  For
systems where the octupole effect is small (such as those with
$m_1\simeq m_2$ or $e_{\rm out}\sim 0$), the distribution of
$\theta_{\rm sl}^\f$ peaks around $90^\circ$. As the octuple effect
increases, a more isotropic distribution of final spin axis is
produced.  Overall, merging BH binaries produced by Lidov-Kozai
oscillations in triples exhibit a unique distribution of the effective
(mass-weighted) spin parameter $\chi_{\rm eff}$; this may be used to
distinguish this formation channel from other dynamical channels.
\end{abstract}
\keywords{binaries: general - black hole physics - gravitational waves
  - stars: black holes - stars: kinematics and dynamics}

\section{Introduction}
\label{sec 1}

Over the last two years, several mergers of black hole (BH) and
neutron star (NS) binaries have been observed in gravitational
waves by aLIGO/VIRGO \citep[e.g.,][]{Abbott 2016a,Abbott
  2016b,Abbott 2017a,Abbott 2017b,Abbott 2017c,Abbott 2017d}.  With
the estimated binary BH merger rate of 10-200 Gpc$^{-3}$yr$^{-1}$,
many hundreds of BH mergers are expected to be detected in
the coming years. It is therefore important to systematically study
various formation mechanisms of such compact binaries and their
observable signatures.

The formation channels of merging BH binaries can be broadly divided
into two categories: isolated binary evolution and dynamical
formation, corresponding to different ways of bringing widely
separated BHs into sufficiently close orbits to allow
gravitational-radiation driven binary coalescence. In the isolated binary
evolution scenario, massive stellar binaries formed with relatively
small separations ($\lesssim 10$~AU)
are tighten in orbit by the drag forces through common-envelop phases
\citep[e.g.,][]{Lipunov 1997,Lipunov 2017,Podsiadlowski
  2003,Belczynski 2010,Belczynski 2016,Dominik 2012, Dominik
  2013,Dominik 2015} or through chemically homogeneous evolution
associated with rapid stellar rotations \citep[e.g.,][]{Mandel and de
  Mink 2016,Marchant 2016}. The dynamical formation mechanism
includes various ``flavors'', all involving gravitational interactions
between multiple stars/BHs. In one class of scenarios, binary BHs become bound
and tighten through three-body encounters and/or secular interactions
in dense star clusters \citep[e.g.,][]{Portegies
  2000,Miller 2002,Miller 2009,O'Leary 2006, Banerjee 2010,Downing
  2010,Rodriguez 2015,Chatterjee 2017,Samsing 2018} or galactic nuclei
\citep[e.g.,][]{O'Leary 2009,Antonini 2012,Antonini and Rasio
  2016,VanLandingham 2016,Petrovich 2017,Hoang 2017,Leigh 2018};
alternatively, binary BH mergers can be induced in triples in the galactic fields
\citep[e.g.,][]{Silsbee and Tremaine 2017,Antonini 2017}.

Despite many studies, there are large uncertainties in the predicted
event rates and binary BH properties in various formation
scenarios. Some of these involve uncertainties in the physical
processes (e.g. common-envelop evolution), while others are
``environmental'' uncertainties (e.g. BH population in clusters,
orbital parameter distributions in triples).  In particular, it is
difficult to distinguish different formation mechanisms on the basis
of event rates and mass measurements of merging binaries.
Other discriminant observables would be desirable.
In the dynamical channel, a BH binary
could acquire substantial eccentricity through close encounters, so the
detection of eccentric merging binaries
would indicate certain dynamical processes at work
\citep[e.g.,][]{Gultekin 2006,O'Leary 2009,Antonini 2012,Cholis
  2016,Samsing 2017,Silsbee and Tremaine 2017,Chen xian,Antonini
  2017}. However, due to the efficient eccentricity damping by gravitational wave
emission, the majority of the merging binaries will be fully circularized as they enter
the aLIGO/VIRGO frequency band ($\gtrsim 10$~Hz) regardless of the formation channels.
Another potentially valuable observable is the
BH spin, which is expected to carry information on the binary formation history.
In particular, through the phase shift in the binary inspiral waveform,
one can directly measure the mass-weighted average of dimensionless spin parameter,
\be\label{eq:effective spin parameter}
\chi_{\eff}\equiv\frac{m_1 \boldsymbol{\chi}_1+m_2 \boldsymbol{\chi}_2}{m_1+m_2}\cdot\hat{\mathbf{L}},
\ee
where $m_{1,2}$ are the masses of BHs, $\boldsymbol{\chi}_{1,2}=c\mathbf{S}_{1,2}/(Gm_{1,2}^2)$ are the
dimensionless BH spins, and $\hat{\mathbf{L}}$ is the unit orbital angular momentum vector.
In the isolated binary evolution channel, because of mass transfer
and accretion in the common envelope phase,
the BH spin tends to be aligned with the orbital angular
momentum, although velocity kick during BH formation may introduce small misalignment
\citep[e.g.,][]{Postnov 2017,Belczynski 2017}.  On the other hand, in the
dynamical formation channel, the BH spin axis has a propensity to
point in any direction. Therefore, the distribution of spin tilts is
of great importance and could be used as a probe to understand merging
binary formation channels \citep[e.g.,][]{Rodriguez ApJL,Will Nature}.
The five BH binaries detected by aLIGO so far have relatively small $\chi_{\rm eff}$
($-0.06^{+0.14}_{-0.14}$ for GW150914, $0.21^{+0.2}_{-0.1}$ for GW151226, $-0.12^{+0.21}_{-0.3}$ for GW170104,
$0.07^{+0.23}_{-0.09}$ for GW170608 and $0.06^{+0.12}_{-0.12}$ for GW170814).
This could be either the result of slowly-spinning
BHs \citep[e.g.,][]{Zaldarriaga} or large spin-orbit misalignments.
Of particular interest is that
GW170104 \citep[e.g.,][]{Abbott 2017a} has a negative $\chi_{\eff}$ value (with appreciable error bars),
implying that the configurations with both component spins positively aligned with the orbital
angular momentum are disfavored.
Such a negative $\chi_{\eff}$ value may not be produced in the standard binary evolution channel,
but would be natural if the binary is dynamically formed.

In this work, we study the orbital and spin evolution of merging BH binaries and NS binaries
in the presence of an external companion.
It is well known that a tertiary body on an inclined
orbit can accelerate the orbital decay of an inner binary by inducing
Lidov-Kozai (LK) eccentrcity/inclination oscillations
\citep[e.g.,][]{Lidov,Kozai}. This effect was first studied
in the context of supermassive BH binary mergers \citep[e.g.,][]{Blaes 2002}.
There have been a number of previous studies of LK-induced mergers of stellar-mass BH binaries
in globular clusters or active galactic nuclei \citep[e.g.,][]{Miller 2002,Wen 2003,Thompson 2011,Antonini 2012,Antonini 2014,Hoang 2017}
and in the galactic fields \citep[e.g.,][]{Antonini 2017,Silsbee and Tremaine 2017}.
Many of these works involved population synthesis calculations, adopting
various assumptions on the BH binary/triple parameters and distributions and accounting
for the effects of cluster dynamics. Such approaches are important, but
it can be difficult to know how the numerical results (such as the predicted binary
merger rates) depend on the input parameters and assumptions.
In this paper we focus on the ``clean'' problem of isolated triples. Using the secular
equations of motion of hierarchical triples (both the octupole-level ``double-averaged'' equations
and ``single-averaged'' equations that we develop in this paper),
we systematically examine the ``merger window'' (i.e., the range of inclination angles between the
inner binary and the outer companion that induces binary merger) and merger fraction
as a function of BH and companion masses and orbital parameters.
Guided by numerical integrations and analytic estimates, we
identify the key parameters and scaling relations for understanding LK-induced mergers.

Another important goal of our work is to examine how misalignments between
the BH spins and the orbital angular momentum in the BH binaries
can be produced in LK-induced mergers. This problem was first studied in
our recent paper \citep[]{Liu-ApJL}, where we
focused on BH binaries with small initial orbital separations ($\lesssim 1\au$) such that
the external companion induces zero or only modest ($e \lesssim 0.9$)
eccentricity excitation in the inner binary. We found that starting from aligned
BH spins, a wide range of spin-orbit misalignments (including retrograde spins)
can be generated. In this paper, we consider more general, wide BH binaries (such that
the binaries have no chance of merging by themselves within $\sim10^{10}$ yrs) where
an external companion induces extreme eccentricity excitation and merger of the binary.
As we show in this paper, the BH spin exhibits a wide range of evolutionary paths, and
different distributions of final spin-orbit misalignments
can be produced depending on the system parameters.

Our paper is organized as follows. In Section \ref{sec 2},
we present the equations for calculating the evolution of triples
including gravitational radiation. These equations are
based on the single averaging (for the inner orbit) and double averaging (for both inner and outer orbits) approximations
for the orbital evolution of hierarchical triples. We also present the basic properties of LK oscillations
for general triple systems; these are useful for determining analytical expressions of
the merger windows and merger fractions for ``quadrupole" systems.
In Section \ref{sec 3}, we perform a large set of numerical integrations to determine
the merger windows for LK-induced binary mergers,
assuming isotropic distribution of the orientations of tertiary companions.
The associated merger fractions of BH binaries and NS binaries are obtained,
including various analytical/scaling relations and fitting formulae.
In Section \ref{sec 4}, we study the BH spin evolution during LK-induced binary mergers.
We identify various dynamical behaviours for the spin evolution and calculate the distributions
of the spin-orbit misalignment angle and the effective spin parameter $\chi_\eff$ when
the binary enters the LIGO/VIRGO band.
We summarize our main results in Section \ref{sec 5}.

\section{Lidov-Kozai Oscillations in triples with gravitational radiation}
\label{sec 2}
We consider a hierarchical triple system, composed of
an inner BH binary of masses $m_1$, $m_2$
and a distant companion of mass $m_3$
that moves around the center of mass of the inner bodies.
The reduced mass for the inner binary is $\mu_\IN\equiv m_1m_2/m_{12}$, with $m_{12}\equiv m_1+m_2$.
Similarly, the outer binary has $\mu_\OUT\equiv(m_{12}m_3)/m_{123}$ with $m_{123}\equiv m_{12}+m_3$.
The semi-major axes and eccentricities are denoted by $a_\IN$, $a_\OUT$ and $e_\IN$, $e_\OUT$, respectively.
The orbital angular momenta of two orbits are
\begin{eqnarray}
&&\textbf{L}_\IN=\mathrm{L}_\IN\hat{\textbf{L}}_\IN=\mu_\IN\sqrt{G m_{12}a_\IN(1-e_\IN^2)}\,\hat{\textbf{L}}_\IN,\\
&&\textbf{L}_\OUT=\mathrm{L}_\OUT\hat{\textbf{L}}_\OUT=\mu_\OUT\sqrt{G m_{123}a_\OUT(1-e_\OUT^2)}\,\hat{\textbf{L}}_\OUT,
\end{eqnarray}
where $\hat{\bf L}_\IN$ and $\hat{\bf L}_\OUT$ are unit vectors.
Similarly, we define the eccentricity vectors as $\textbf{e}_\IN=e_\IN\hat{\textbf{e}}_\IN$ and $\textbf{e}_\OUT=e_\IN\hat{\textbf{e}}_\OUT$.
Throughout the paper, for convenience of notation, we will frequently omit the subscript ``$\IN$"
for the inner orbit.

To study the evolution of the merging inner BH binary under the influence of the tertiary companion,
we first develop the secular equations of motion in terms of the angular momentum $\textbf{L}$ and eccentricity $\mathbf{e}$ vectors:
\begin{eqnarray}
&&\frac{d \textbf{L}}{dt}=\frac{d \textbf{L}}{dt}\bigg|_{\mathrm{LK}}+\frac{d \textbf{L}}{dt}\bigg|_{\mathrm{GW}}~,\label{eq:Full Kozai 1}\\
&&\frac{d \mathbf{e}}{dt}=\frac{d \mathbf{e}}{dt}\bigg|_{\mathrm{LK}}+\frac{d \mathbf{e}}{dt}\bigg|_{\mathrm{GR}}+\frac{d \mathbf{e}}{dt}\bigg|_{\mathrm{GW}}~,\label{eq:Full Kozai 2}
\end{eqnarray}
where we include the contributions from the external companion that generate LK oscillations
(to be discussed in Section \ref{sec 2 1}),
the general relatively (GR) post-Newtonian correction,
and the dissipation due to gravitational waves (GW) emission.

General Relativity (1-PN correction) introduces pericenter precession as
\be\label{eq:e GR}
\frac{d \mathbf{e}}{dt}\bigg|_{\mathrm{GR}}=\Omega_\mathrm{GR}\hat{\textbf{L}}\times\mathbf{e},
\ee
with the precession rate given by
\be\label{eq:GR}
\Omega_\mathrm{GR}=\frac{3Gn m_{12}}{c^2a(1-e^2)},
\ee
where $n=(G m_{12}/a^3)^{1/2}$ is the mean motion of the inner binary.
Gravitational radiation draws energy and angular momentum from the BH orbit .
The rates of change of $\textbf{L}$ and $\mathbf{e}$ are given by \citep[]{Peters 1964}
\begin{eqnarray}
&&\frac{d \textbf{L}}{dt}\bigg|_{\mathrm{GW}}=-\frac{32}{5}\frac{G^{7/2}}{c^5}\frac{\mu^2 m_{12}^{5/2}}{a^{7/2}}
\frac{1+7e^2/8}{(1-e^2)^2}\hat{\textbf{L}},\label{eq:GW 1}\\
&&\frac{d \mathbf{e}}{dt}\bigg|_{\mathrm{GW}}=-\frac{304}{15}\frac{G^3}{c^5}\frac{\mu m_{12}^2}{a^4(1-e^2)^{5/2}}
\bigg(1+\frac{121}{304}e^2\bigg)\mathbf{e}.\label{eq:GW 2}
\end{eqnarray}
The associated orbital decay rate is
\be\label{eq:decay rate}
\begin{split}
\bigg(\frac{\dot{a}}{a}\bigg)_{\mathrm{GW}}&\equiv-\frac{1}{T_\gw}\\
&=-\frac{64}{5}\frac{G^3\mu m_{12}^2}{c^5a^4}\frac{1}{(1-e^2)^{7/2}}
\bigg(1+\frac{73}{24}e^2+\frac{37}{96}e^4\bigg).
\end{split}
\ee
The merger time due to GW radiation of an isolated binary with the
initial semi-major axis $a_0$ and eccentricity $e_0=0$ is given by
\ba\label{eq:Tmerger}
T_\mathrm{m,0}&&=\frac{5c^5 a_0^4}{256 G^3 m_{12}^2 \mu}\\
&&\simeq10^{10}\bigg(\frac{60M_\odot}{m_{12}}\bigg)^2\bigg(\frac{15M_\odot}{\mu}\bigg)\bigg(\frac{a_0}{0.202\au}\bigg)^4\mathrm{yrs}\nonumber.
\ea
Thus, for typical BH binaries ($m_1\sim m_2\sim30M_\odot$),
only for separations less than about $0.2$AU can the isolated binary
be allowed to merge within a Hubble time ($T_\h\equiv10^{10}$yrs).
In this paper, we will consider much larger initial binary separations ($a_0\gtrsim10\au$),
so that merger is possible only when the tertiary companion induces extreme eccentricity excitations
in the inner binary.

\subsection{Orbital Evolution in the Secular Approximation}
\label{sec 2 1}
If we introduce the instantaneous separation between the inner bodies as $\mathbf{r}\equiv\mathrm{r}\hat{\mathbf{r}}$,
and the separation between the external perturber and the center of mass of the
inner bodies as $\mathbf{r}_\OUT\equiv\mathrm{r}_\OUT\hat{\mathbf{r}}_\OUT$,
then the complete Hamiltonian of the system can then be written as \citep[e.g.,][]{Harrington}
\be
\mathcal{H}=\frac{1}{2}\mu|\dot{\mathbf{r}}|^2+\frac{1}{2}\mu_\OUT|\dot{\mathbf{r}}_\OUT|^2-\frac{Gm_1 m_2}{r}-\frac{Gm_{12}m_3}{r_\OUT}+\Phi,
\ee
where
\be\label{eq:total potential}
\Phi=-Gm_1m_2m_3\sum\limits_{l = 2}^\infty\bigg[\frac{m_1^{l-1}+(-1)^lm_2^{l-1}}{m_{12}^l}\bigg]\frac{r^l}{r_\OUT^{l+1}}P_l(\cos \theta).
\ee
Here $P_l(x)$ is the Legendre polynomial of degree $l$ and $\theta$ is the
angle between $\mathbf{r}$ and $\mathbf{r}_\OUT$.

\subsubsection{Double-Averaged Secular Equations}
\label{sec 2 1 1}
For the sufficiently hierarchical systems, the angular momenta of the inner and outer binaries exchange periodically over a long timescale
(longer than the companion's orbital period),
while the exchange of energy is negligible.
The orbital evolution of the triple system can be studied by expanding the Hamiltonian to the octupole order
and averaging over both the inner and outer orbits (double averaging), i.e., $\Phi=\Phi_\qu+\Phi_\oct$.
The quadrupole ($l=2$) piece is given by
\be\label{eq:DA potential quad}
\begin{split}
\langle\langle\Phi_\qu\rangle\rangle=&
\frac{\mu\Phi_0}{8}\bigg[1-6e^2-3(1-e^2)(\hat{\bf L}\cdot\hat{\bf L}_\OUT)^2\\
&+15e^2(\hat{\bf e}\cdot\hat{\bf L}_\OUT)^2\bigg],
\end{split}
\ee
and the octupole ($l=3$) potential is
\ba
\langle\langle\Phi_\oct\rangle\rangle=&&
\frac{15\mu\Phi_0\varepsilon_{\oct}}{64}
\Bigg\{e(\hat{\bf e}\cdot\hat{\bf e}_\OUT)\bigg[8e^2-1\\\nonumber
&&-35e^2(\hat{\bf e}\cdot\hat{\bf L}_\OUT)^2
+5(1-e^2)(\hat{\bf L}\cdot\hat{\bf L}_\OUT)^2\bigg]\\\nonumber
&&+10e(1-e^2)(\hat{\bf e}\cdot\hat{\bf L}_\OUT)(\hat{\bf L}\cdot\hat{\bf e}_\OUT)(\hat{\bf L}\cdot\hat{\bf L}_\OUT)\Bigg\},
\ea
where
\be
\Phi_0\equiv\frac{Gm_3a^2}{a_\OUT^3(1-e_\OUT^2)^{3/2}},
\ee
and
\be\label{eq:varepsilon oct}
\varepsilon_{\rm oct}\equiv {m_1-m_2\over m_{12}}\left({a\over a_\OUT}\right)
{e_\OUT\over 1-e_\OUT^2}.
\ee
The explicit expressions for $(d \textbf{L}/dt)_\lk$, $(d \textbf{e}/dt)_\lk$
and for $(d \textbf{L}_\OUT/dt)_\lk$, $(d \textbf{e}_\OUT/dt)_\lk$ are provided in \citet{Liu et al 2015}.
In general, $\dot{\textbf{L}}_{\IN,\OUT}$ and $\dot{\mathbf{e}}_{\IN,\OUT}$
consist of two pieces: a quadrupole term and an octupole term.
The quadrupole term induces the oscillations in the eccentricity
and mutual orbital inclination on the timescale of
\be
t_\lk=\frac{1}{n}\frac{m_{12}}{m_3}\bigg(\frac{a_{\OUT,\eff}}{a}\bigg)^3,
\ee
where the effective outer binary separation is defined as
\be\label{eq:aout eff}
a_{\OUT,\eff}\equiv a_\OUT\sqrt{1-e^2_\OUT}.
\ee
The octupole piece is quantified by terms proportional to $\varepsilon_\oct$,
which measures the relative strength of the octupole potential compared to the quadrupole one.

For systems that can be correctly described by the double-averaged equations,
the eccentricity variation timescale of the inner binary must be longer than the period of companion's orbit.
Otherwise, the secular equations may break down \citep[e.g.,][]{Seto PRL,Antonini 2014}.
Note that when the eccentricity of the inner binary is excited to the maximum value $e_\m$,
the eccentricity vector $\textbf{e}$ evolves on the timescale of $t_\lk\sqrt{1-e_\m^2}$
\citep[e.g.,][]{Anderson et al HJ}, much shorter than the quadrupole LK period ($\sim t_\lk$).
Thus, for the double-averaged secular equations to be valid, we require
\be\label{eq:DA}
t_\lk\sqrt{1-e_\m^2}\gtrsim P_\OUT,
\ee
where $P_\OUT$ is the period of the outer binary.

\subsubsection{Single-Averaged Secular Equations}
\label{sec 2 1 2}
For moderately hierarchical systems, the change in the angular momentum of the inner binary may be
significant within one period of the outer orbit,
and the short-term ($\lesssim P_\OUT$) oscillations of the system cannot be ignored.
In this case, the double-averaged secular equations break down, and we can use
the single-averaged secular equations (only averaging over the inner orbital period).

Averaging over the inner orbit, the quadrupole term in Equation (\ref{eq:total potential}) becomes
\be\label{eq:potential quad}
\langle\Phi_\qu\rangle=\frac{\mu\Phi'_0}{4}\bigg[-1+6e^2
+3(\mathbf{j}\cdot\hat{\mathbf{r}}_\OUT)^2
-15(\mathbf{e}\cdot\hat{\mathbf{r}}_\OUT)^2\bigg],
\ee
and the octupole term is
\ba\label{eq:potential oct}
&&\langle\Phi_\oct\rangle=\frac{5 \mu \Phi'_0\varepsilon'_\oct}{16}\bigg[
(3-24e^2)(\mathbf{e}\cdot\hat{\mathbf{r}}_\OUT)\nonumber\\
&&~~~~~~~~~~~-15(\mathbf{j}\cdot\hat{\mathbf{r}}_\OUT)^2(\mathbf{e}\cdot\hat{\mathbf{r}}_\OUT)
+35(\mathbf{e}\cdot\hat{\mathbf{r}}_\OUT)^3\bigg],
\ea
where
\be
\mathbf{j}\equiv j\hat{\textbf{L}}=\sqrt{1-e^2}\hat{\textbf{L}}
\ee
is the dimensionless angular momentum vector,
and the coefficients $\Phi'_0$ and $\varepsilon'_\oct$ are given by
\be
\Phi'_0=\frac{Gm_3a^2}{r_\OUT^3},
\ee
and
\be
\varepsilon'_\oct=\frac{m_1-m_2}{m_{12}}\frac{a}{r_\OUT}.
\ee

In terms of the averaged potentials, the equations of motion for the
inner orbital vectors $\mathbf{j}$ and $\mathbf{e}$ are \citep[e.g.,][]{Tremaine 2009}
\ba
\frac{d\mathbf{j}}{dt}&&=-\frac{1}{\mu \sqrt{Gm_{12}a}}
\big(\mathbf{j}\times\nabla_\mathbf{j}\langle\Phi\rangle+\mathbf{e}\times\nabla_\mathbf{e}\langle\Phi\rangle\big)
\label{eq:equations of motion 1},\\
\frac{d\mathbf{e}}{dt}&&=-\frac{1}{\mu \sqrt{Gm_{12}a}}
\big(\mathbf{j}\times\nabla_\mathbf{e}\langle\Phi\rangle+\mathbf{e}\times\nabla_\mathbf{j}\langle\Phi\rangle\big)
\label{eq:equations of motion 2}.
\ea
Substituting Equation (\ref{eq:potential quad}) into Equations (\ref{eq:equations of motion 1}) and (\ref{eq:equations of motion 2}),
the quadrupole level equations can be obtained:
\be\label{eq:SA quad j}
\frac{d\mathbf{j}}{dt}\Bigg|_\qu=\frac{3}{2t'_\lk}\bigg[5(\mathbf{e}\cdot\hat{\mathbf{r}}_\OUT)\mathbf{e}\times\hat{\mathbf{r}}_\OUT-
(\mathbf{j}\cdot\hat{\mathbf{r}}_\OUT)\mathbf{j}\times\hat{\mathbf{r}}_\OUT\bigg],
\ee
\ba\label{eq:SA quad e}
&&\frac{d\mathbf{e}}{dt}\Bigg|_\qu=\frac{3}{2t'_\lk}\bigg[5(\mathbf{e}\cdot\hat{\mathbf{r}}_\OUT)\mathbf{j}\times\hat{\mathbf{r}}_\OUT\nonumber\\
&&~~~~~~~~~~~~~-(\mathbf{j}\cdot\hat{\mathbf{r}}_\OUT)\mathbf{e}\times\hat{\mathbf{r}}_\OUT
-2\mathbf{j}\times\mathbf{e}\bigg].
\ea
Similarly, the octupole contributions are
\ba\label{eq:SA oct j}
&&\frac{d\mathbf{j}}{dt}\Bigg|_\oct=\frac{15\varepsilon'_\oct}{16t'_\lk}
\bigg[10(\mathbf{j}\cdot\hat{\mathbf{r}}_\OUT)(\mathbf{e}\cdot\hat{\mathbf{r}}_\OUT)\mathbf{j}\times\hat{\mathbf{r}}_\OUT\nonumber\\
&&~~~~~~~~~~~-(1-8e^2)\mathbf{e}\times\hat{\mathbf{r}}_\OUT
+5(\mathbf{j}\cdot\hat{\mathbf{r}}_\OUT)^2\mathbf{e}\times\hat{\mathbf{r}}_\OUT\nonumber\\
&&~~~~~~~~~~~-35(\mathbf{e}\cdot\hat{\mathbf{r}}_\OUT)^2\mathbf{e}\times\hat{\mathbf{r}}_\OUT\bigg],
\ea
\ba\label{eq:SA oct e}
&&\frac{d\mathbf{e}}{dt}\Bigg|_\oct=\frac{15\varepsilon'_\oct}{16t'_\lk}
\bigg[16(\mathbf{e}\cdot\hat{\mathbf{r}}_\OUT)\mathbf{j}\times\hat{\mathbf{e}}-
(1-8e^2)\mathbf{j}\times\hat{\mathbf{r}}_\OUT\nonumber\\
&&~~~~~~~~~~~+5(\mathbf{j}\cdot\hat{\mathbf{r}}_\OUT)^2\mathbf{j}\times\hat{\mathbf{r}}_\OUT
-35(\mathbf{e}\cdot\hat{\mathbf{r}}_\OUT)^2\mathbf{j}\times\hat{\mathbf{r}}_\OUT\nonumber\\
&&~~~~~~~~~~~+10(\mathbf{j}\cdot\hat{\mathbf{r}}_\OUT)(\mathbf{e}\cdot\hat{\mathbf{r}}_\OUT)\mathbf{e}\times\hat{\mathbf{r}}_\OUT
\bigg].
\ea
In the above, we have defined the single-averaged (quadrupole) LK timescale as
\be
t'_\lk=\frac{1}{n}\frac{m_{12}}{m_3}\bigg(\frac{r_\OUT}{a}\bigg)^3.
\ee
The evolution equations of $\mathbf{L}$ and $\mathbf{e}$ are
\ba
&&\frac{d \textbf{L}}{dt}\bigg|_{\mathrm{LK}}=\mu\sqrt{Gm_{12}a}\bigg(\frac{d \textbf{j}}{dt}\bigg|_\qu+\frac{d \textbf{j}}{dt}\bigg|_\oct\bigg)
\label{eq:SA L},\\
&&\frac{d \textbf{e}}{dt}\bigg|_{\mathrm{LK}}=\frac{d \textbf{e}}{dt}\bigg|_\qu+\frac{d \textbf{e}}{dt}\bigg|_\oct.
\ea
On the other hand, for the external companion, the dynamics is governed by
\be\label{eq:third body}
\mu_\OUT \frac{d^2\mathbf{r}_\OUT}{dt^2}=\nabla_{\mathbf{r}_\OUT}\bigg(\frac{Gm_{12}m_3}{r_\OUT}\bigg)-
\nabla_{\mathbf{r}_\OUT}\Big(\langle\Phi_\qu\rangle+\langle\Phi_\oct\rangle\Big).
\ee
The explicit form can be obtained by substituting Equations (\ref{eq:potential quad}) and (\ref{eq:potential oct})
into Equation (\ref{eq:third body}).
Equations (\ref{eq:SA quad j})-(\ref{eq:SA oct e}) and (\ref{eq:SA L})-(\ref{eq:third body}),
together with Equations (\ref{eq:e GR})-(\ref{eq:GW 2}), completely determine the evolution of
the triple system in the single averaging approximation.

The single-averaged equations are applicable to a wider range of system parameters than the double-averaged equations.
Nevertheless, their validity still requires that the eccentricity evolution timescale at $e\sim e_\m$ be longer than
the orbital period of the inner binary, i.e.,
\be\label{eq:SA}
t_\lk\sqrt{1-e_\m^2}\gtrsim P_\IN.
\ee

\subsection{Lidov-Kozai Eccentricity Excitation: Analytical Results}
\label{sec 2 2}
Before exploring the LK-induced mergers systematically (Section \ref{sec 3}), we
summarize some key analytical results for LK eccentricity excitations.
It is well known that short-range force effects (such as GR-induced apsidal precession; see Equation \ref{eq:e GR})
play an important role in determining the maximum eccentricity $e_\m$ in LK oscillations
\citep[e.g.,][]{Holman,Fabrycky and Tremaine 2007}.
Analytical expression for $e_\m$ for general hierarchical triples (arbitrary masses) can be
obtained in the double-averaged secular approximation when the disturbing potential
is truncated to the quadrupole order \citep[]{Liu et al 2015,Anderson et al HJ,Anderson et al 2017}.

In the absence of energy dissipation, the evolution of the
triple is governed by two conservation laws. The first is
the total orbital angular momentum of the system, $\mathbf{L}_\tot=\mathbf{L}+\mathbf{L}_\OUT$.
In the quadrupole approximation, $e_\OUT$
is constant, and the conservation of $|\mathbf{L}_\tot|$ implies
\be\label{eq:Kozai constant}
K\equiv j\cos I-\frac{\eta}{2}e^2=\mathrm{constant},
\ee
where $j=\sqrt{1-e^2}$, $I$ is the angle between $\hat{\bf L}$ and $\hat{\bf L}_\OUT$, and we have defined
\be\label{eq:ratio of L}
\eta\equiv\bigg(\frac{L}{L_\OUT}\bigg)_{e=0}=\frac{\mu}{\mu_\OUT}\bigg[\frac{m_{12}a}{m_{123}a_\OUT(1-e_\OUT^2)}\bigg]^{1/2}.
\ee
In the limit of $L\ll L_\OUT$, Equation (\ref{eq:Kozai constant}) reduces to the usual ``Kozai constant,"
$\sqrt{1-e^2}\cos I=$ constant.

The second conserved quantity is the total energy.
In the double averaging approximation, it is given by (to the quadrupole order)
\be\label{eq:energy}
\Phi=\langle\langle\Phi_\qu\rangle\rangle+\langle\Phi_\gr\rangle,
\ee
where $\langle\langle\Phi_\qu\rangle\rangle$ is given by Equation (\ref{eq:DA potential quad}), and
$\langle\Phi_\gr\rangle$ is given by
\be
\langle\Phi_\gr\rangle=-\frac{3G^2\mu m_{12}^2}{a^2c^2 j}=-\varepsilon_\gr\frac{\mu\Phi_0}{j},
\ee
with
\ba\label{eq:epsilonGR}
&&\varepsilon_\gr= \frac{3Gm_{12}^2a_{\OUT,\eff}^3}{c^2a^4m_3}\\
&&\simeq3.6\times10^{-5}\bigg(\!\frac{m_{12}}{60M_\odot}\!\bigg)^{\!\!2}\!\bigg(\!\frac{m_3}{30M_\odot}\!\bigg)^{\!\!-1}
\!\bigg(\!\frac{a_{\OUT,\eff}}{10^3\au}\!\bigg)^{\!\!3}\!\bigg(\!\frac{a}{10^2\au}\!\bigg)^{\!\!-4}\nonumber.
\ea
Using Equations (\ref{eq:Kozai constant}) and (\ref{eq:energy}), the maximum eccentricity
$e_\m$ attained in the LK oscillation (starting from an initial $I_0$ and $e_0\simeq 0$)
can be calculated analytically \citep[]{Liu et al 2015,Anderson et al 2017}:
\ba
&&\!\!\!\frac{3}{8}\frac{j^2_\mi-1}{j^2_\mi}\bigg[5\left(\cos I_0+\frac{\eta}{2}\right)^2-
\Bigl(3+4\eta\cos I_0+\frac{9}{4}\eta^2\Bigr)j^2_\mi \nonumber\\
&&\quad +\eta^2j^4_\mi\bigg]+\varepsilon_\gr \left(1-j_\mi^{-1}\right)=0,
\label{eq:EMAX}
\ea
where $j_\mi\equiv\sqrt{1-e_\m^2}$.
Note that in the limit of $\eta\rightarrow0$ and $\varepsilon_\gr\rightarrow 0$,
Equation~(\ref{eq:EMAX}) yields the well-known relation $e_\m=\sqrt{1-(5/3)\cos^2 I_0}$.
For general $\eta$,
the maximum possible $e_\m$ for all values of $I_0$, called
$e_\li$, is achieved at $I_{0,\li}$ that satisfies $de_\m/dI_0=0$, i.e.
\be\label{eq:I0lim}
\cos I_{0,\li}=\frac{\eta}{2}\bigg(\frac{4}{5}j_\li^2-1\bigg).
\ee
Substituting Equation (\ref{eq:I0lim}) into Equation (\ref{eq:EMAX}), we find that the
limiting eccentricity $e_\li$, the maximum of the $e_\m(I_0)$ curve, is determined by
\be\label{eq:ELIM}
\frac{3}{8}(j_\li^2-1)\left[-3+\frac{\eta^2}{4}\left(\frac{4}{5}j_\li^2-1\right)\right]+
\varepsilon_\gr \left(1-j_\li^{-1}\right)=0.
\ee
On the other hand, eccentricity excitation ($e_\m\ge 0$) occurs only within a window of inclinations
$(\cos I_0)_-\leqslant\cos I_0\leqslant(\cos I_0)_+$, where \citep{Anderson et al 2017}
\be\label{eq:Kozai window I}
(\cos I_0)_\pm=\frac{1}{10}\Big(-\eta\pm\sqrt{\eta^2+60-\frac{80}{3}\varepsilon_\gr}\Big).
\ee
This window exists only when
\be
\varepsilon_\gr\leq \frac{9}{4}+\frac{3}{80}\eta^2.
\label{eq:grlim}\ee
In another word, no eccentricity excitation is possible when Equation (\ref{eq:grlim}) is not satisfied.

\begin{figure}
\begin{centering}
\begin{tabular}{cc}
\includegraphics[width=8cm]{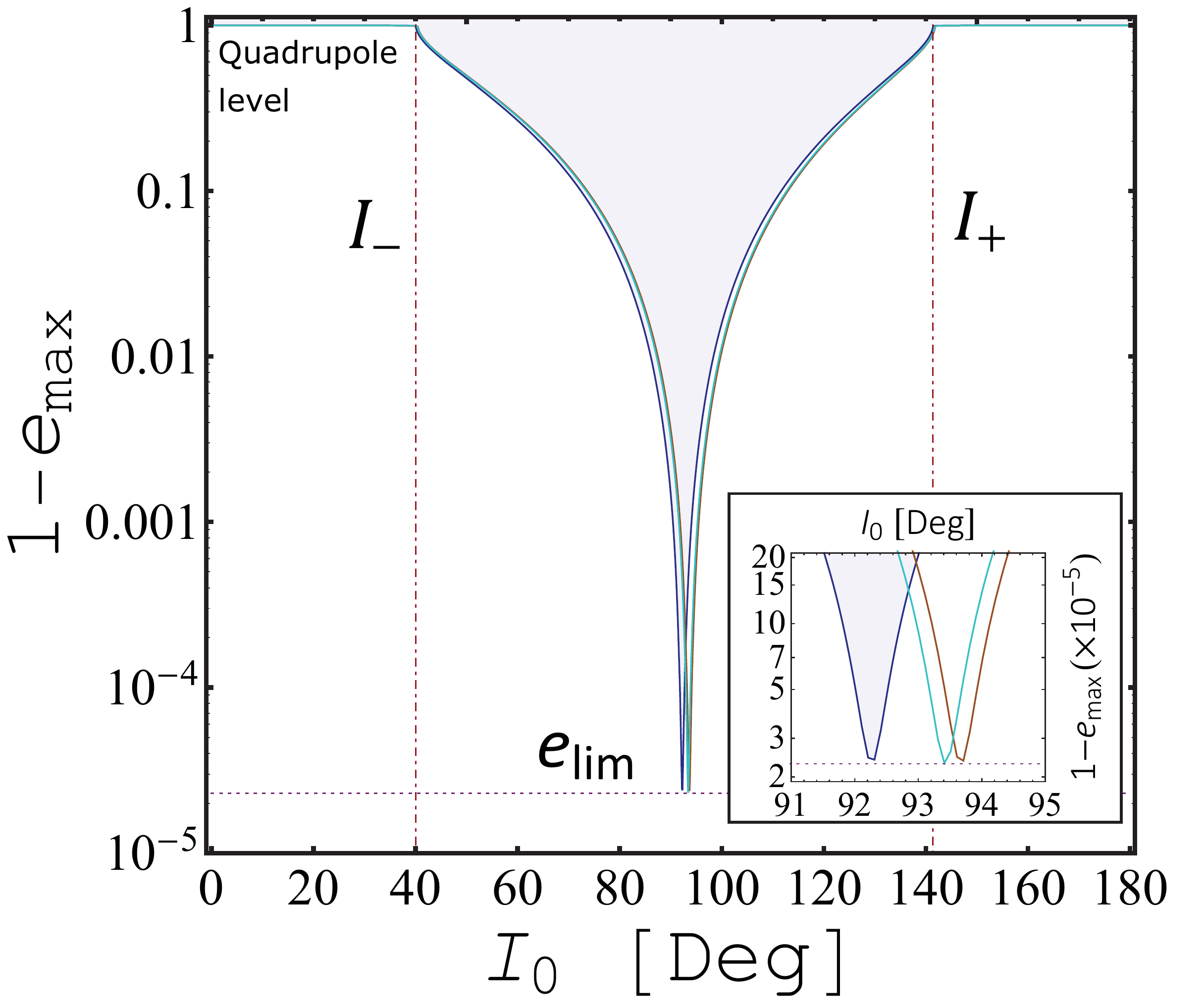}
\end{tabular}
\caption{The maximum eccentricity of the inner BH binary versus the initial inclination $I_0$ of the
tertiary companion, calculated using Equation~(\ref{eq:EMAX}).
The inner binary has $m_1=m_2=30M_\odot$, $a=100\au$, and initial $e_0=0$.
The parameters of the companion are $m_3=30M_\odot$, $a_\OUT=6000\au$ and $e_\OUT=0.001$ (blue);
$m_3=20M_\odot$, $a_\OUT=5241\au$ and $e_\OUT=0.001$ (cyan);
$m_3=20M_\odot$, $a_\OUT=6000\au$ and $e_\OUT=0.487$ (brown).
The $e_{\rm max}(I_0)$ curve depends mainly on $m_3/a_{\OUT,\eff}^3$.
The horizontal ($e_\li$) and vertical ($I_\pm$) lines are given by Equations~(\ref{eq:ELIM})
and (\ref{eq:Kozai window I}), respectively.
}
\label{fig:elim}
\end{centering}
\end{figure}

Figure \ref{fig:elim} shows some examples of the $e_\m(I_0)$ curves. For $\eta\lesssim 1$,
these curves depend mainly on $m_3/a_{\OUT,\eff}^3$ (for given inner binary parameters).
We see that the excitation of eccentricity can only happen within a finite range of $I_0$,
and the achieved maximum $e$ cannot exceed $e_\li$ for any values of $\eta$.

For systems with $m_1\neq m_2$ and $e_\OUT\neq 0$,
$\varepsilon_\oct$ is non-negligible, the octupole effect may become important
\citep[e.g.,][]{Ford,Blaes 2002,Naoz Nature,Katz PRL,Naoz oct,Naoz 2016}.
This tends to widen the inclination window for large eccentricity excitation.
However, the analytic expression for $e_\li$ given by
Equation~(\ref{eq:ELIM}) remains valid even for $\varepsilon_{\rm oct}\neq 0$ \citep{Liu et al 2015,
Anderson 2017b}.
In another word, because of the short-range force effect due to GR, the maximum eccentricity cannot
exceed $e_\li$ even when the octupole potential is significant.
Higher eccentricity may be achieved when the double averaging approximation breaks down (see Section \ref{sec 3 2})

\subsection{Summary of Parameter Regimes}
\label{sec 2 3}
\begin{figure}
\begin{centering}
\begin{tabular}{cc}
\includegraphics[width=8.5cm]{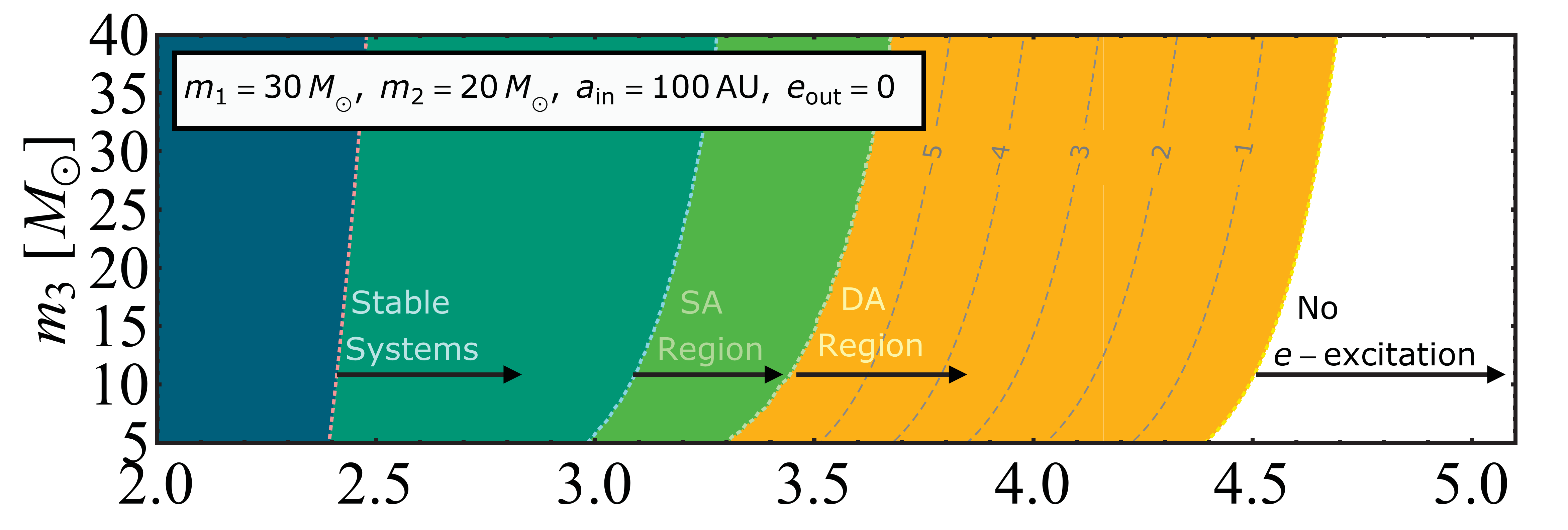}\\
\includegraphics[width=8.5cm]{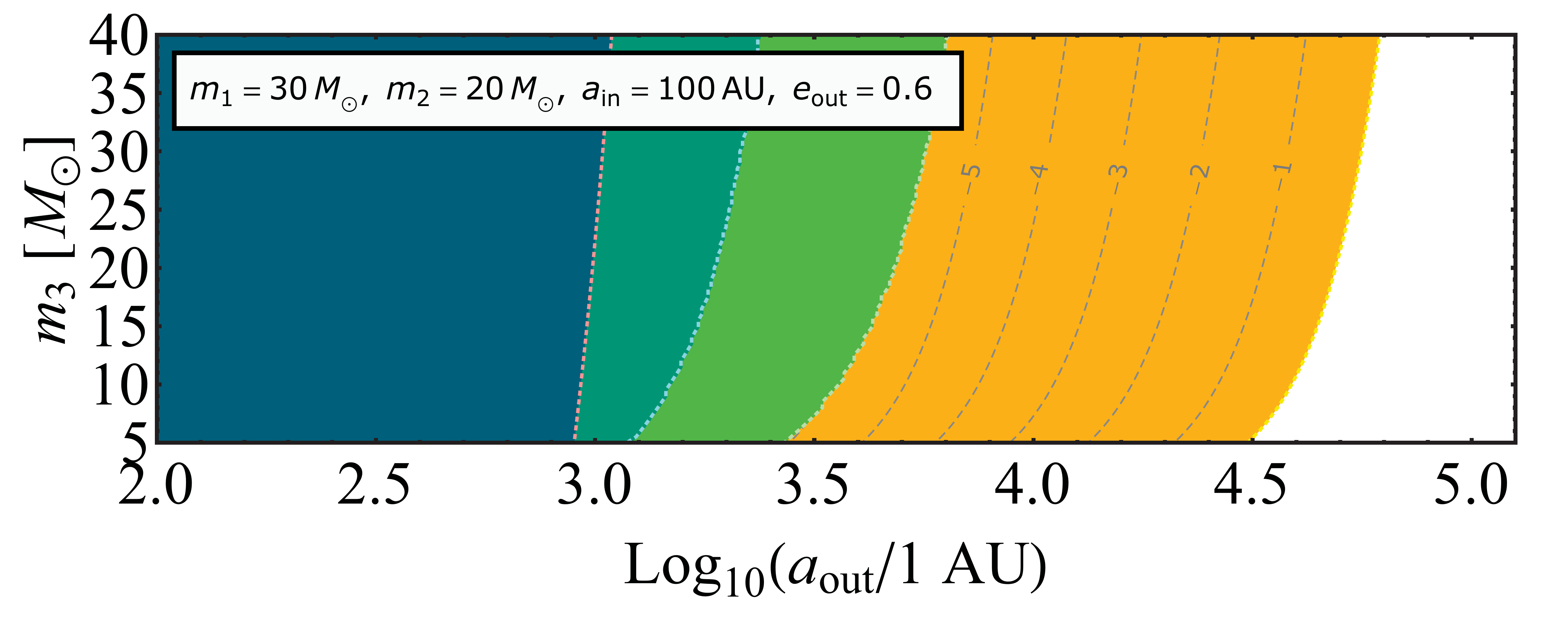}
\end{tabular}
\caption{Parameter space for eccentricity excitation of BH binaries, with $m_3$ and $a_\OUT$
the mass and semi-major axis of the tertiary companion. The parameters for the inner binary are
given in the figure. Five regions are indicated by different colors.
The boundary of ``no $e$-excitation" is given by Equation (\ref{eq:grlim}).
The boundaries of double averaging (DA) and single averaging (SA) approximation
are given by Equations (\ref{eq:DA}) and (\ref{eq:SA}).
The stability condition is given by Equation (\ref{eq:stability}) with $I_0=90^\circ$.
In the yellow region, the dashed curves are contours of constant
$\mathrm{Log}_{10}(1-e_\li)$ (see Equation \ref{eq:ELIM}) with the value indicated.
}
\label{fig:parameterspace}
\end{centering}
\end{figure}

Figure \ref{fig:parameterspace} summarizes the parameter regimes of BH
triples in terms of the mass ($m_3$) and semi-major axis ($a_\OUT$) of the tertiary companion.
For concreteness, we consider a fixed set of inner binary parameters
($m_1=30M_\odot$, $m_2=20M_\odot$ and $a_0=100\au$), with $e_\OUT=0$ (upper panel)
and $e_\OUT=0.6$ (lower panel). The stability of the triple requires \citep[e.g.,][]{Mardling}
\be\label{eq:stability}
\frac{a_\OUT}{a}>2.8\bigg(1+\frac{m_3}{m_{12}}\bigg)^{2/5}\frac{(1+e_\OUT)^{2/5}}{(1-e_\OUT)^{6/5}}\bigg(1-\frac{0.3 I_0}{180^\circ}\bigg).
\ee
In Figure \ref{fig:parameterspace}, several regions have been identified (color coded) and the boundaries are given by the various criteria
(Equations \ref{eq:DA}, \ref{eq:SA}, \ref{eq:grlim} and \ref{eq:stability}; see the dotted curves).
We see that, in the rightmost region, the perturber is so far away that no LK oscillations occur (no $e$-excitation).
In the ``DA region", the dynamics of the system can be well described by the double-averaged (DA)
secular equations. The numbers shown on the dashed curves indicate the values of $\mathrm{Log}_{10}(1-e_\li)$,
suggesting the extent of the excitation of eccentricity (Equation \ref{eq:ELIM}).
In the ``SA region", the outer averaging fails, but the single-averaged (SA)
secular equations are valid.

\section{merger window and merger fraction}
\label{sec 3}
In this section, we use numerical integrations to determine the ``merger window" of BH binaries,
i.e., the range of inclination angles of the tertiary companion such that the inner binary can attain
sufficiently large eccentricities and merge within a critical timescale $T_\mathrm{crit}$
(chosen to be the Hubble time, $10^{10}$ yrs, throughout this paper; but see Section \ref{sec 3 3}).
For a isotropic distribution of the tertiary inclinations, the merger window
then determines the ``merger fraction". Our main goal is to determine how the merger window and merger
fraction depend on the parameters of the triples.

\subsection{Binary Mergers Induced by Quadrupole Lidov-Kozai Effect}
\label{sec 3 1}
\begin{figure}
\begin{centering}
\includegraphics[width=8.5cm]{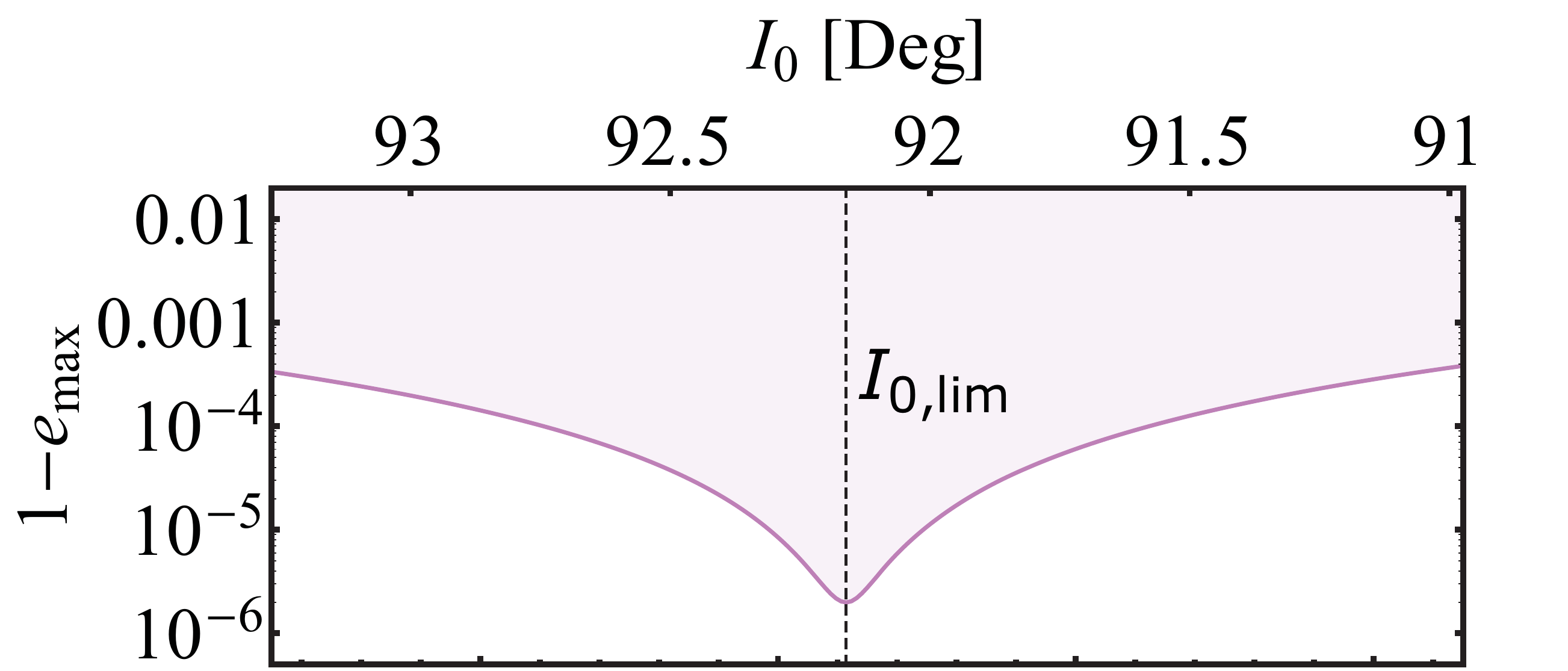}\\
\includegraphics[width=8.5cm]{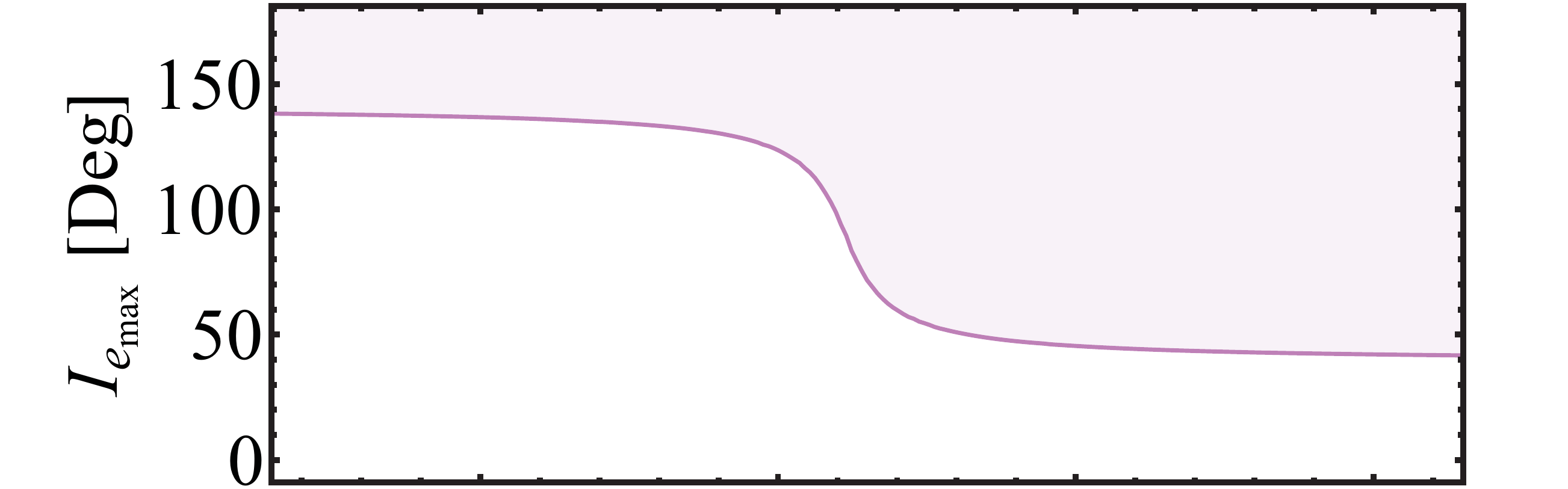}\\
\includegraphics[width=8.5cm]{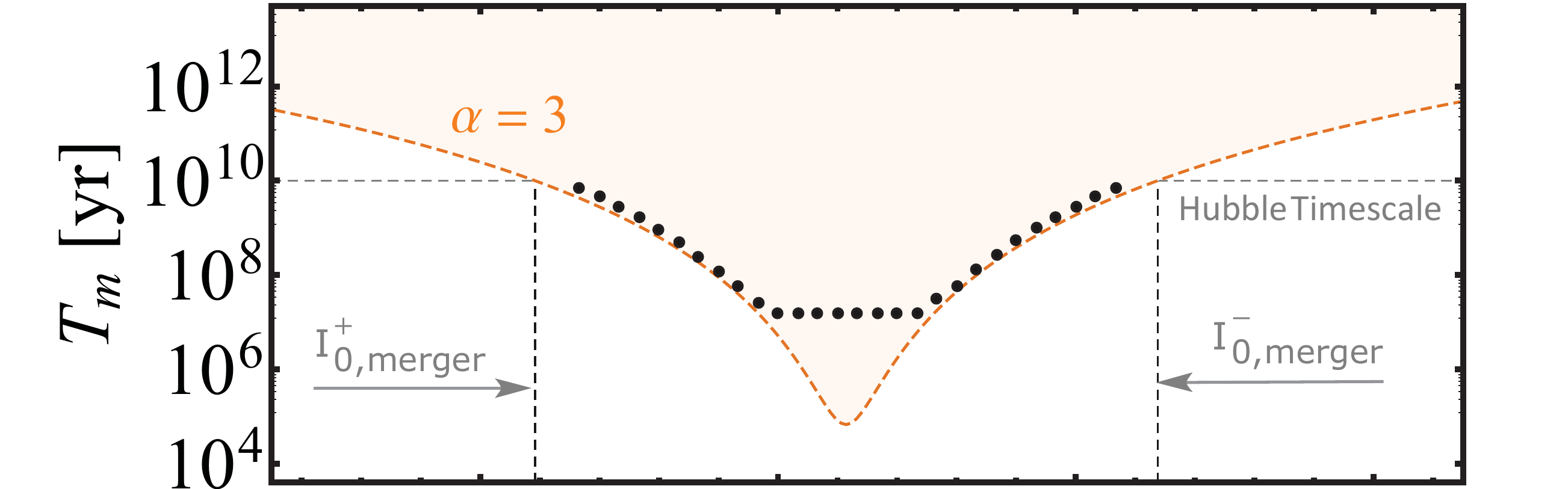}\\
\includegraphics[width=8.5cm]{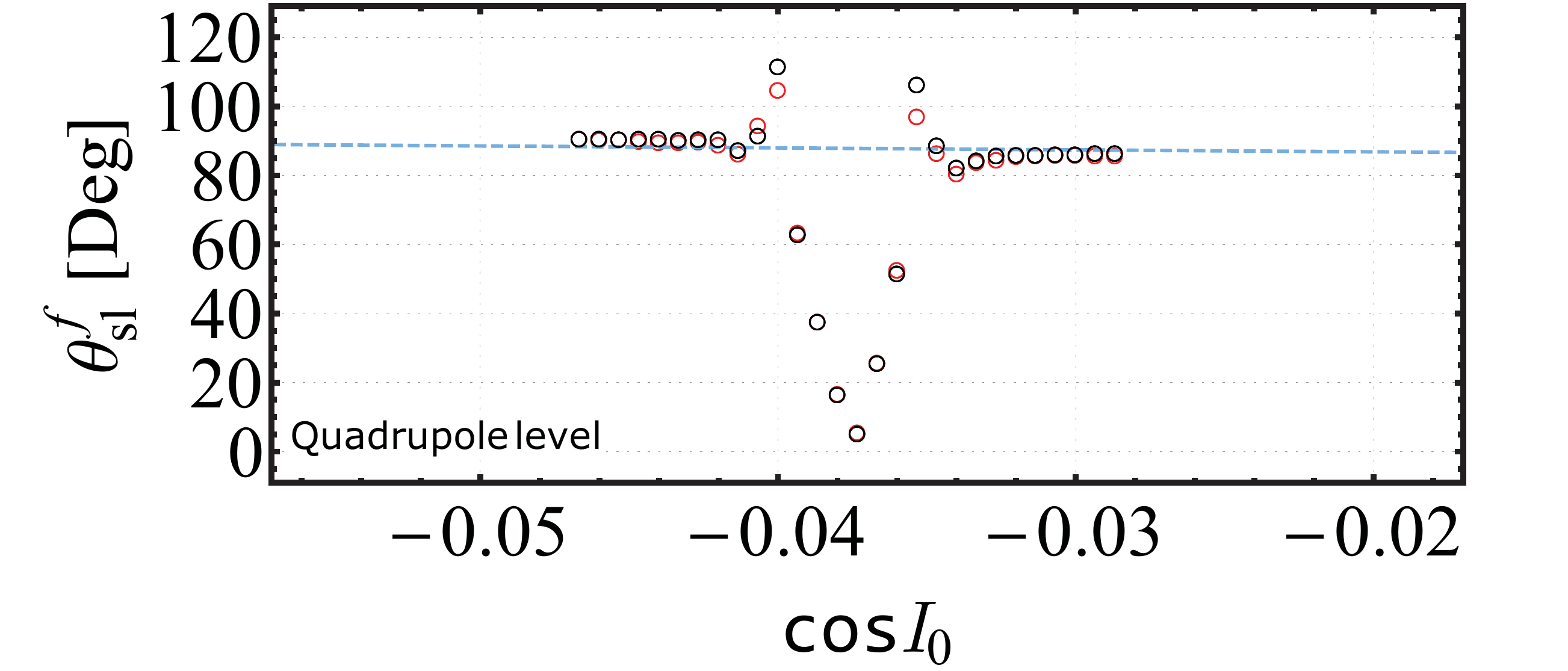}
\caption{BH binary mergers induced by quadrupole LK effect.
From the top to the bottom: the maximum eccentricity $e_\m$, the inclination $I_{e_\m}$
(the value of $I$ at $e=e_\m$; both $e_\m$ and $I_{e_\m}$ are calculated assuming no GW emission),
the inner binary merger time $T_\mathrm{m}$ and the final spin-orbit misalignment angle (with GW emission)
as a function of the initial inclination for the triple system.
The system parameters are $m_1=m_3=30M_\odot$, $m_2=20M_\odot$, $a_0=100\au$
(initial value of $a$), $a_\OUT=4500\au$ and $e_\OUT=0$.
The solid lines in the top two panels are obtained from the analytical
expressions given in Section \ref{sec 2 2}.
The numerical results (dots in the third and bottom panels) are
from the double-averaged secular equations
(each dot represents a successful merger event with the Hubble time, $10^{10}$ yrs).
In the third panel, the dashed curve corresponds to the fitting formula $T_\mathrm{m}\simeq T_\mathrm{m,0}(1-e_\m^2)^3$.
In the bottom panel, the dots show the final spin-orbit misalignment angles for $m_1$ (black)
and $m_2$ (red); note that $\theta_\mathrm{s_1l}^\f$ and $\theta_\mathrm{s_2l}^\f$ nearly overlap.
The dashed curve is given by Equation (\ref{eq:tantheta}).
}
\label{fig:merger window quad}
\end{centering}
\end{figure}

We first consider the cases when the octupole effect is negligible
($\varepsilon_\oct\simeq0$; see Equation \ref{eq:varepsilon oct}).
These apply when the tertiary companion has a circular orbit ($e_\OUT=0$)
or when the inner BHs have equal masses ($m_1=m_2$).
Figure \ref{fig:merger window quad} summarizes our results for a given set of binary and
companion parameters ($m_1=m_3=30M_\odot$, $m_2=20M_\odot$, $a_0=100\au$, and $a_\OUT=4500\au$)
as a function of the initial mutual inclination angle $I_0$.
All initial systems satisfy the criterion of double averaging for triples (Equation \ref{eq:DA}).

The top panel of Figure \ref{fig:merger window quad} shows $e_\m$ for a grid of inclinations
(uniformly distributed in $\cos I_0$) in the absence of GW emission
(this panel is similar to Figure \ref{fig:elim}, but is a zoom-in version).
We see that the eccentricity can be driven to be as large as $e_\m\simeq1-10^{-6}$,
at $I_0=I_{0,\li}\simeq92.16^\circ$ (see Equation \ref{eq:I0lim}).
In the second panel, we plot $I_{e_\m}$, which is the instantaneous inclination at $e=e_\m$,
as a function of $I_0$.
When $e_\m(I_0)$ achieves the maximum, $I_{e_\m}$ become very chose to $I_{0,\li}$,
implying that the range of oscillation in $I$ is small (i.e., $\hat{\bf L}$ exhibits negligible nutation).

\begin{figure*}
\begin{centering}
\begin{tabular}{ccc}
\includegraphics[width=6cm]{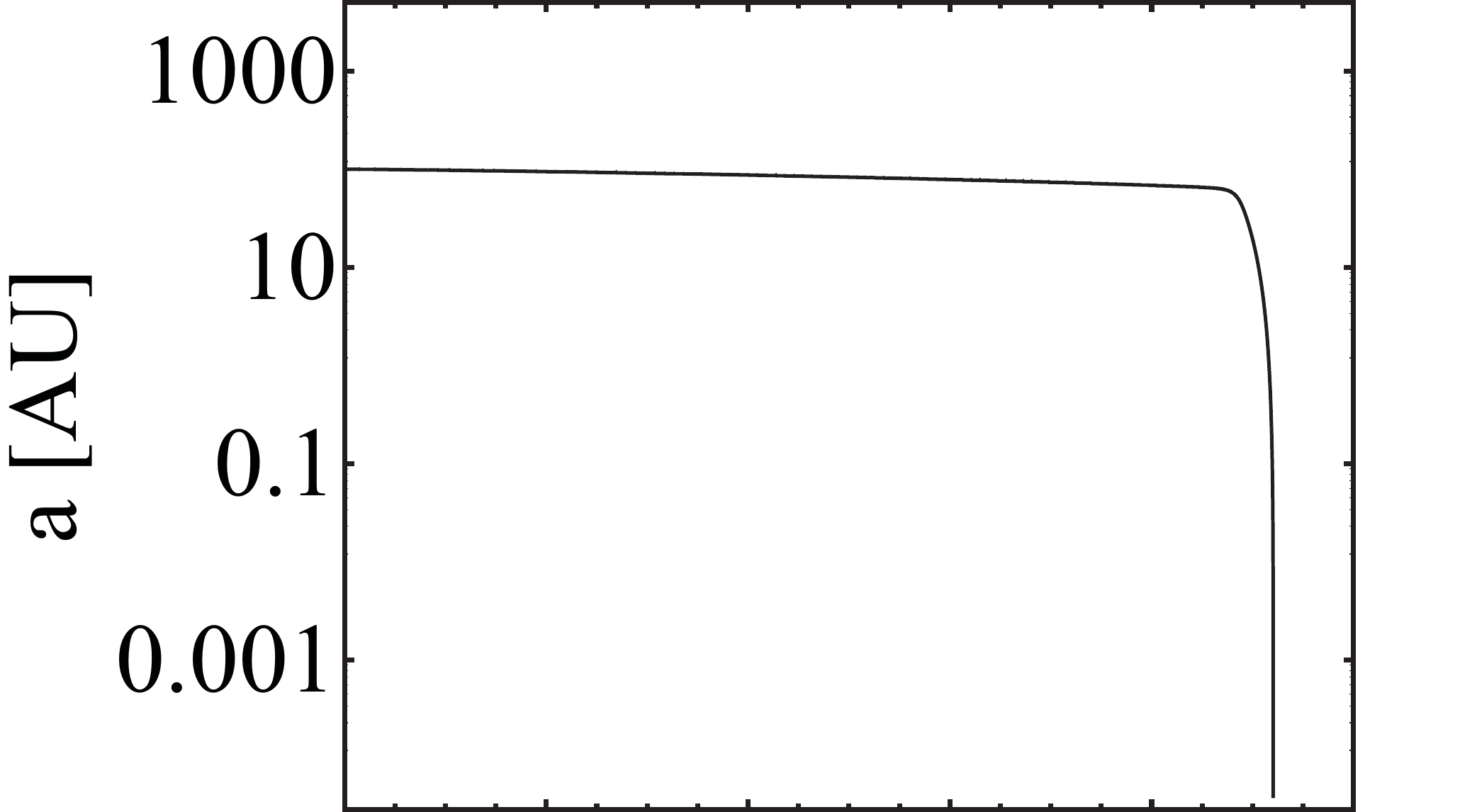}
\includegraphics[width=6cm]{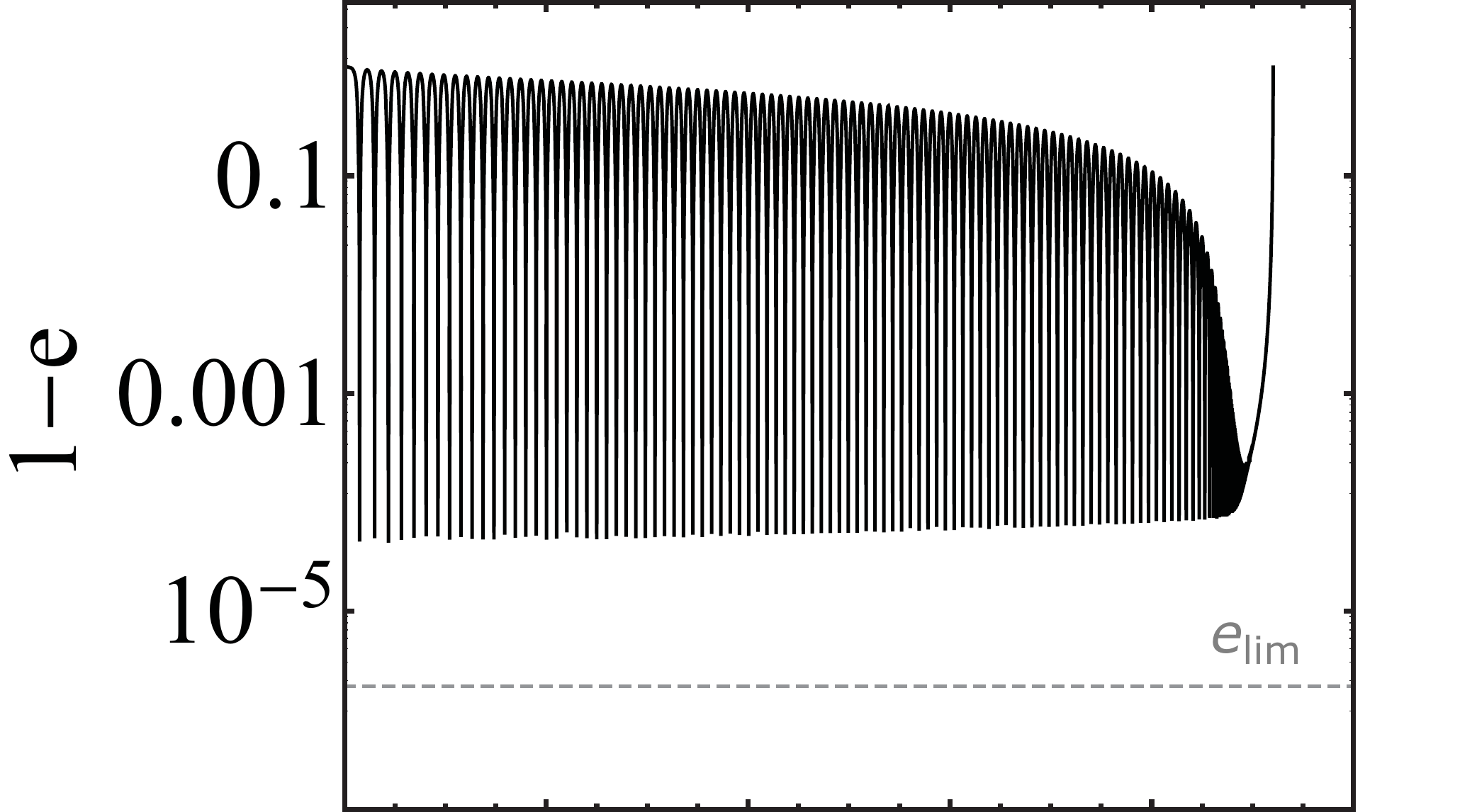}
\includegraphics[width=6cm]{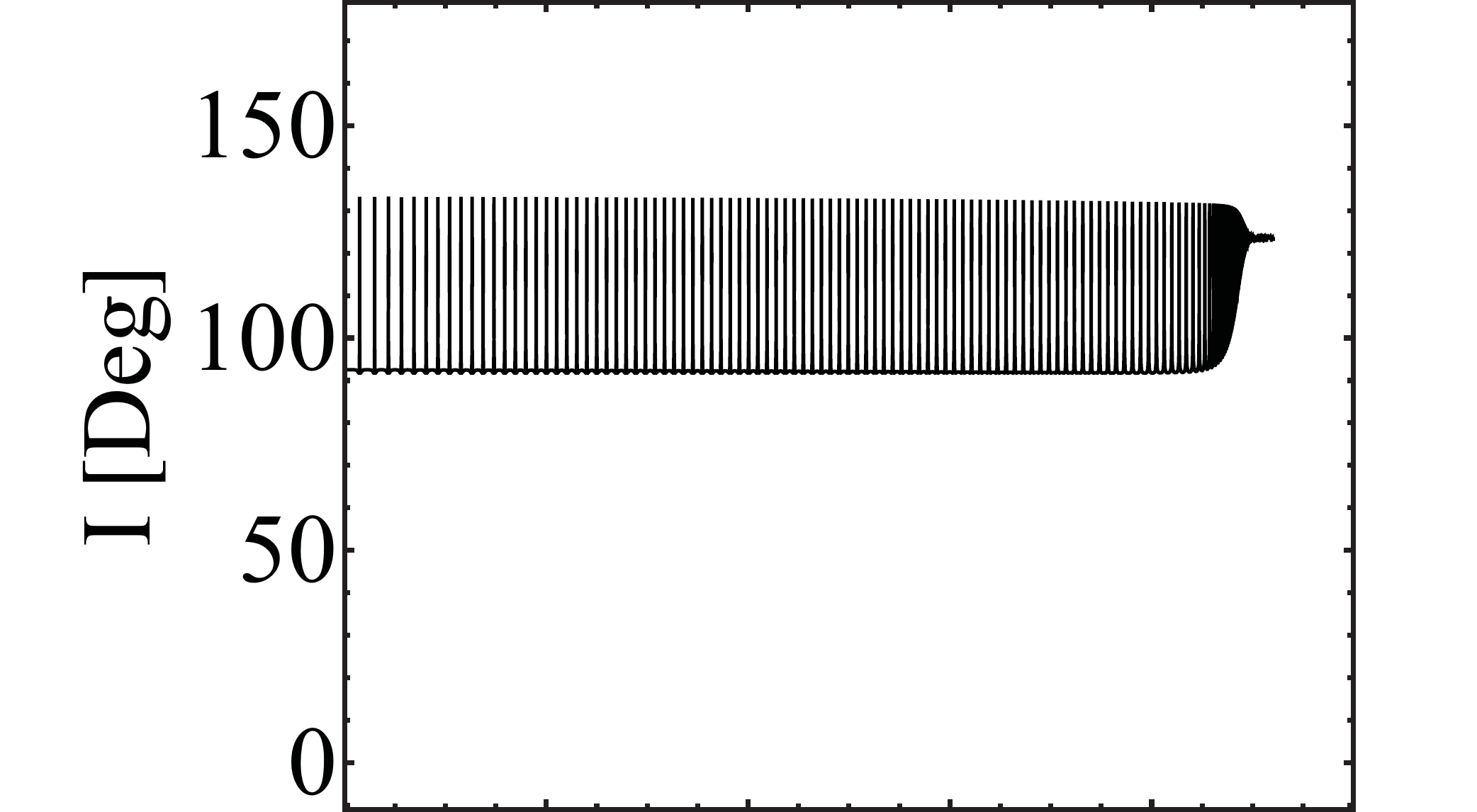}\\
\includegraphics[width=6cm]{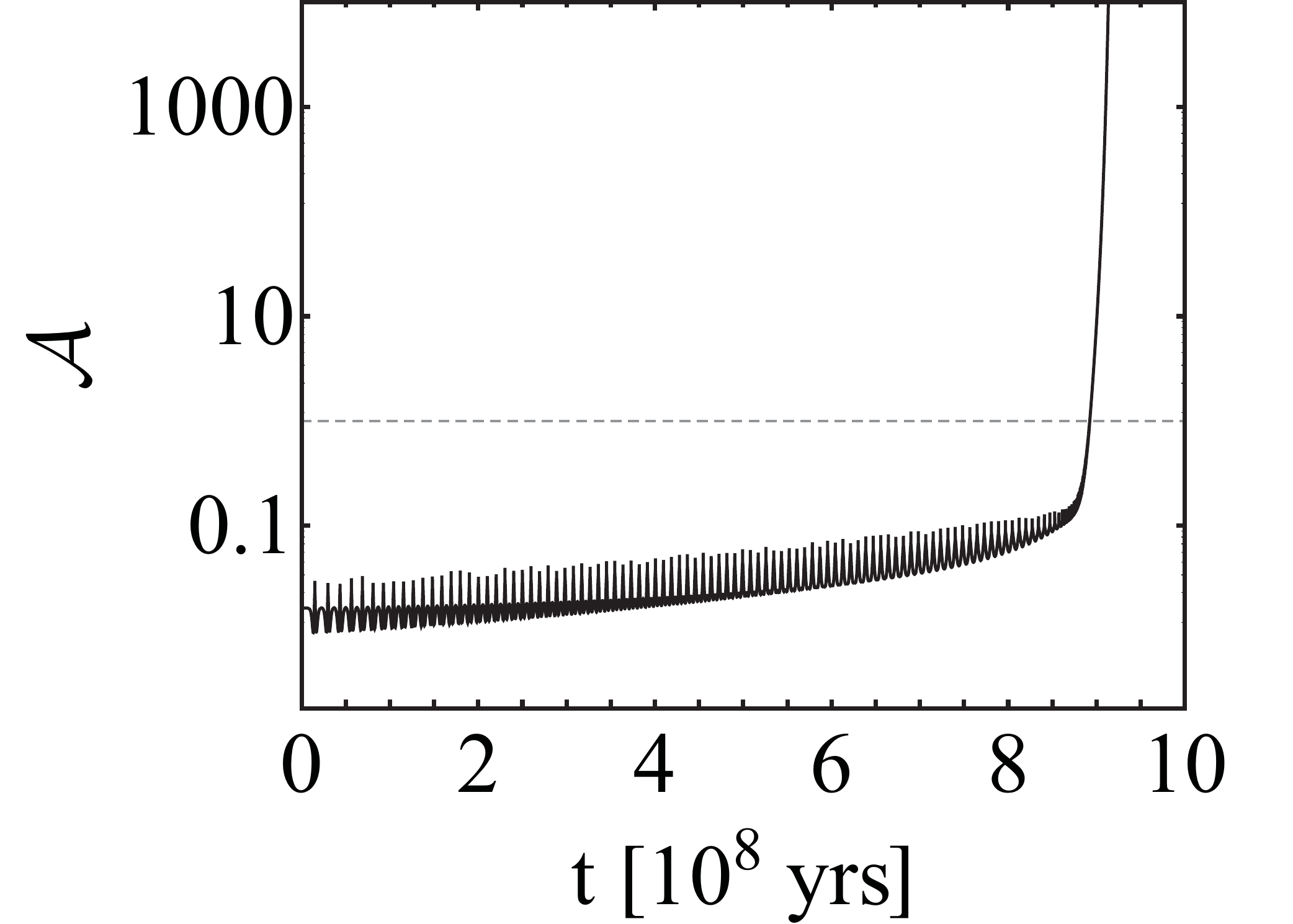}
\includegraphics[width=6cm]{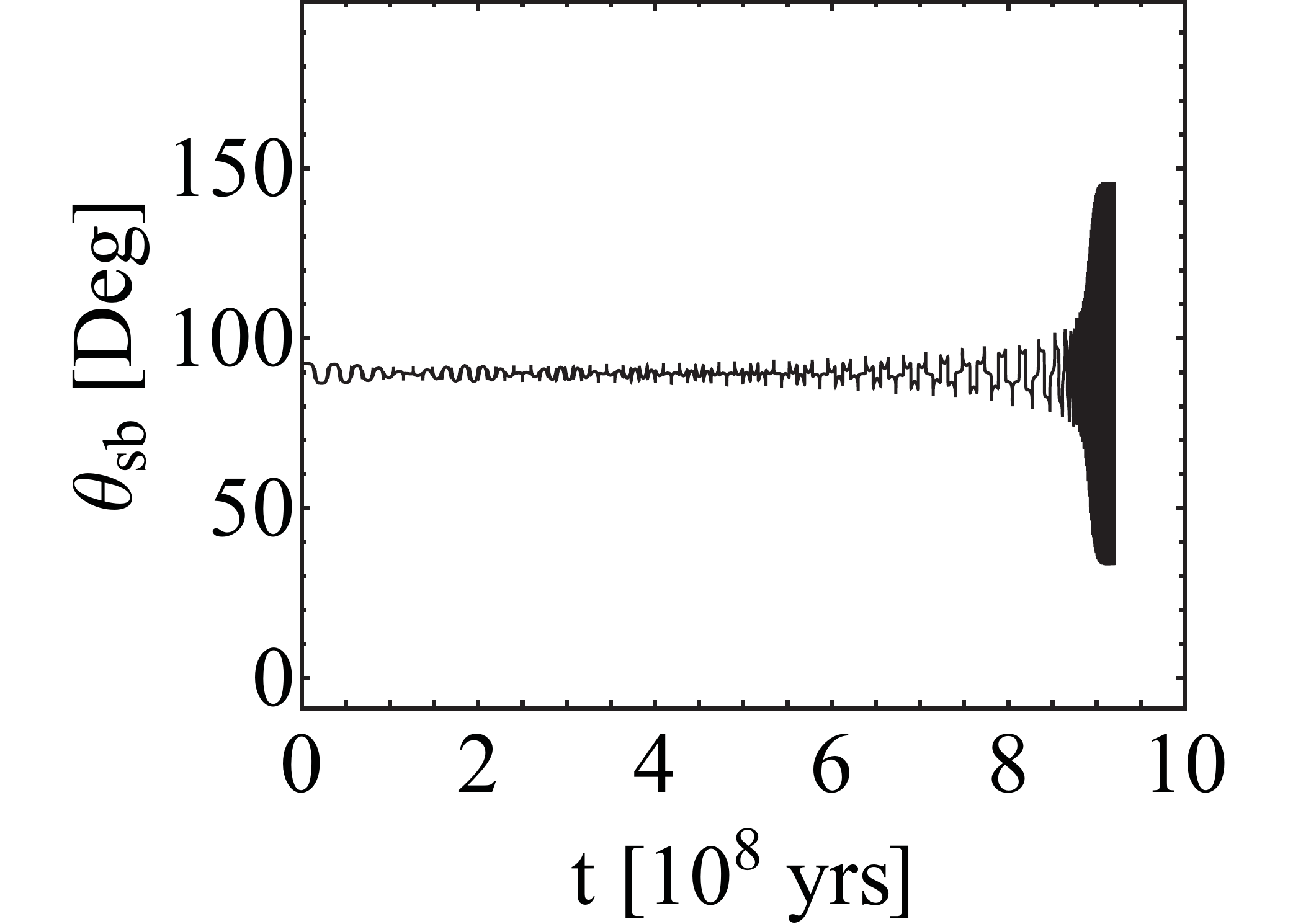}
\includegraphics[width=6cm]{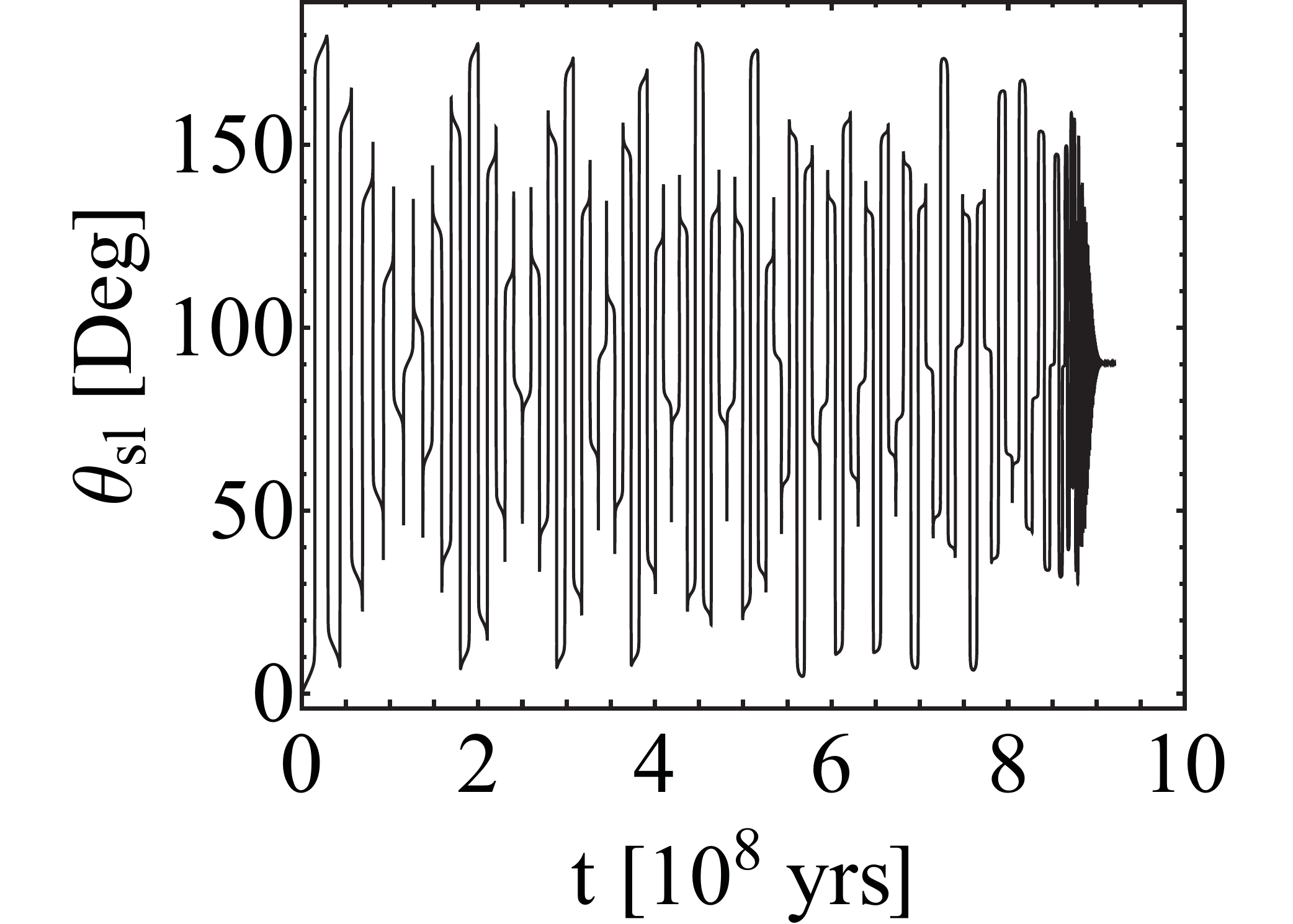}
\end{tabular}
\caption{Sample orbital and spin evolution of a BH binary system with a tertiary companion.
The three top panels show the semi-major axis, eccentricity
and inclination (relative to ${\hat{\bf L}}_\OUT$) of the inner BH binary, and the three bottom panels show
the adiabaticity parameter $\mathcal{A}$ (Equation \ref{eq:adiabaticity parameter}),
the spin-orbit misalignment angle $\theta_\SB$ (the angle between ${\bf S}_1$ and ${\bf L}_\OUT$) and
$\theta_\SL$ (the angle between ${\bf S}_1$ and ${\bf L}$).
The parameters are $m_1=30M_\odot$, $m_2=20M_\odot$, $m_3=30M_\odot$, $a_\OUT=4500\au$, $e_\OUT=0$, and the initial
$a_0=100\au$, $I_0=92.52^\circ$, $e_0=0.001$, and $\theta_\SL^0=0^\circ$.
}
\label{fig:OE quad 1}
\end{centering}
\end{figure*}

\begin{figure*}
\begin{centering}
\begin{tabular}{ccc}
\includegraphics[width=6cm]{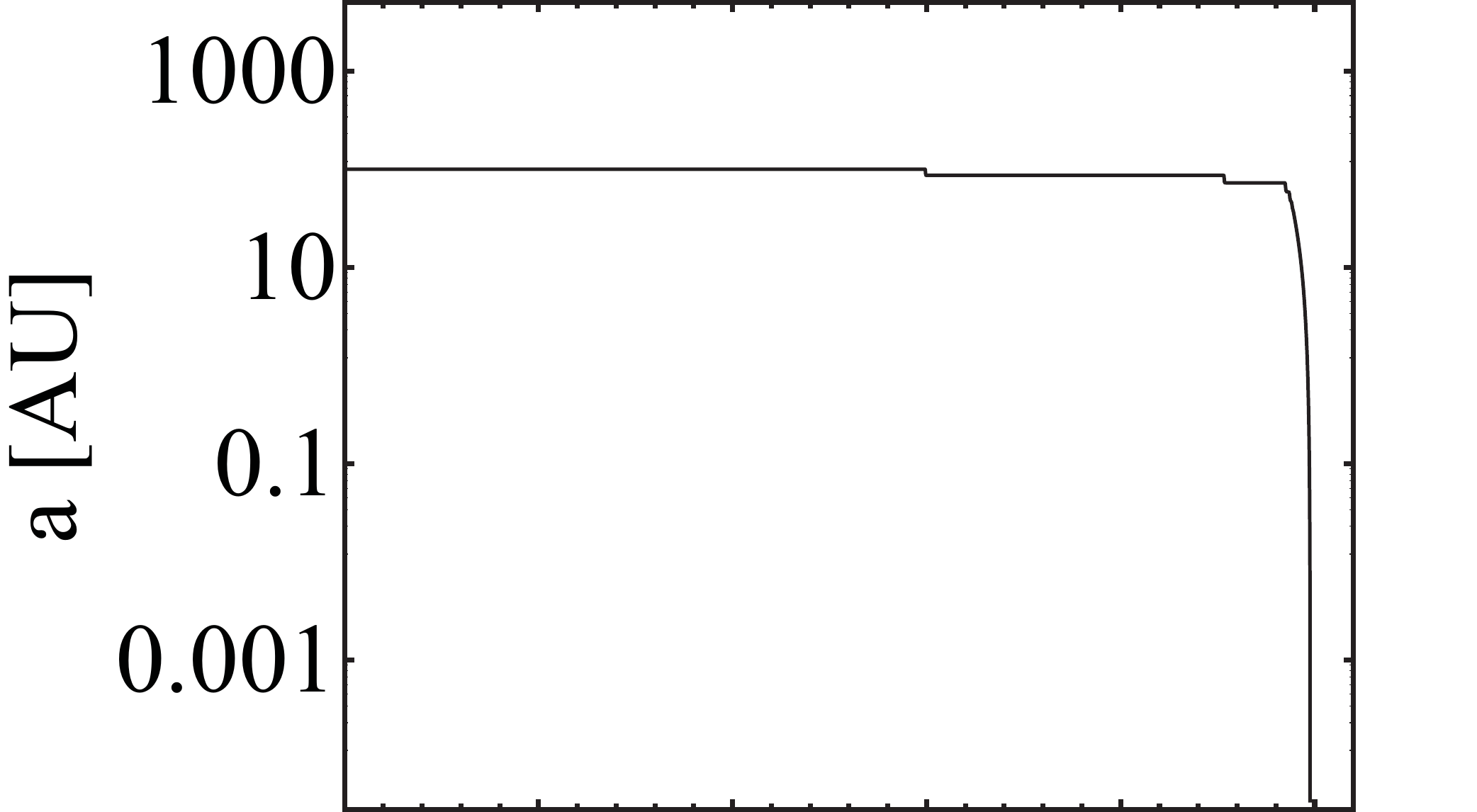}
\includegraphics[width=6cm]{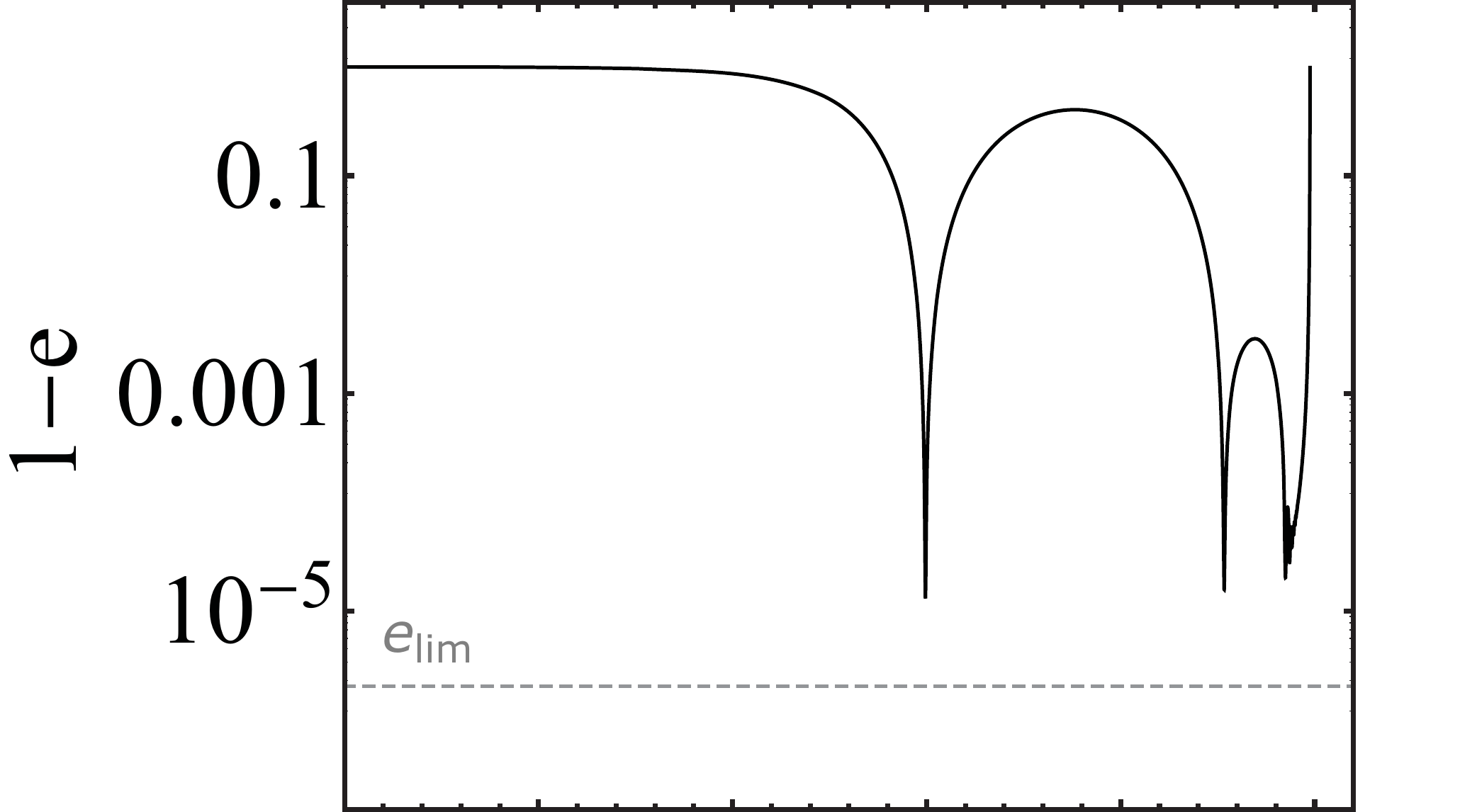}
\includegraphics[width=6cm]{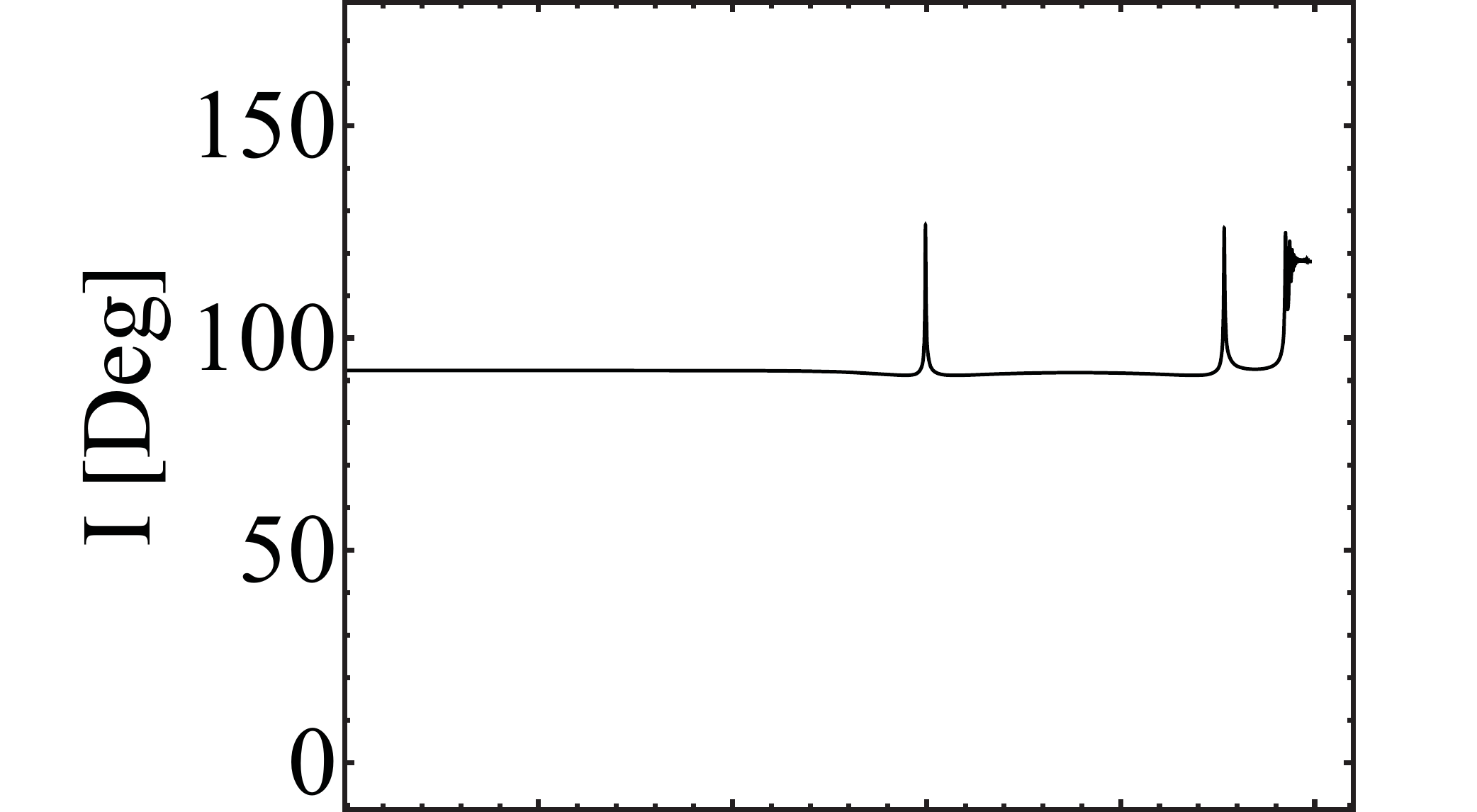}\\
\includegraphics[width=6cm]{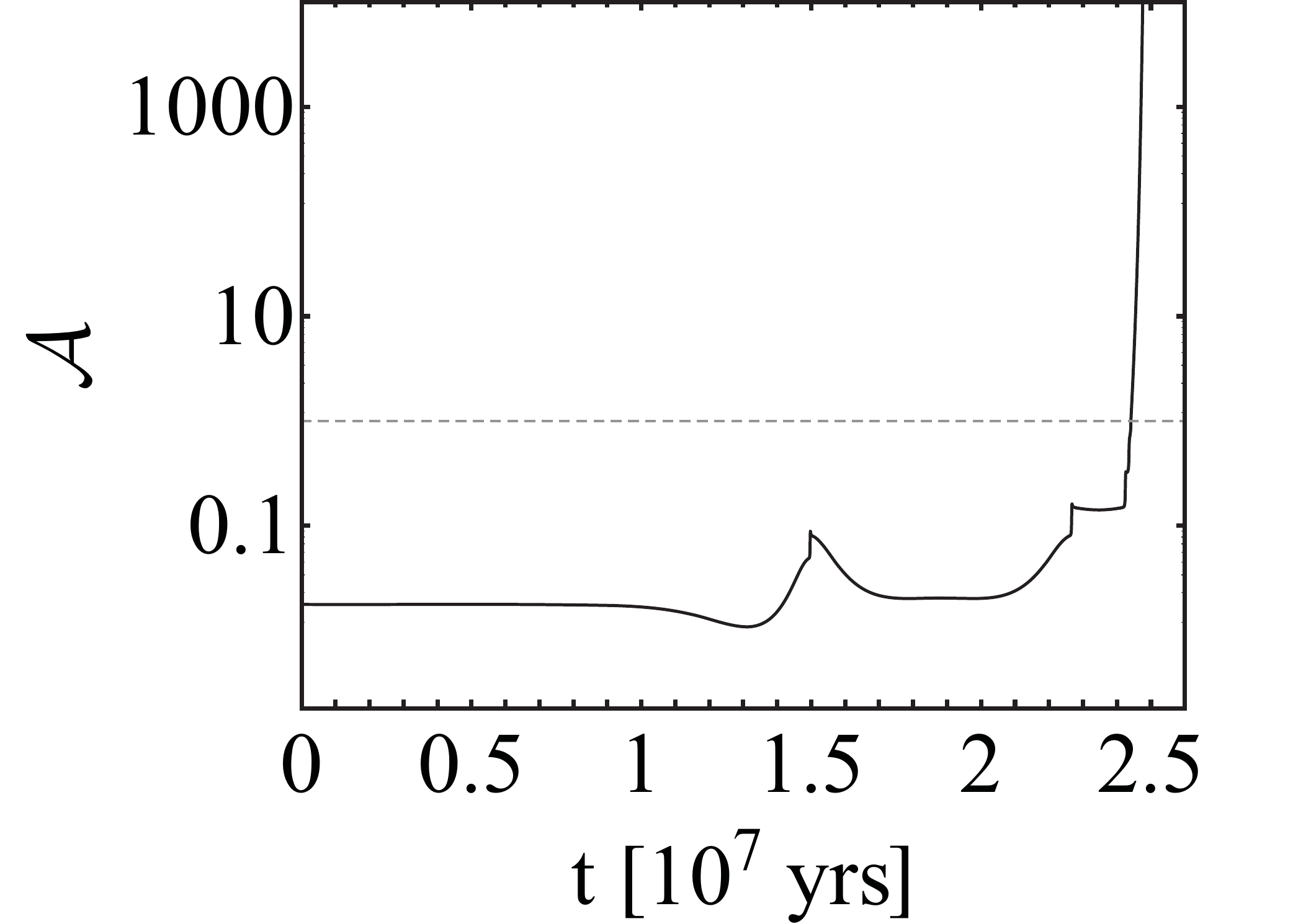}
\includegraphics[width=6cm]{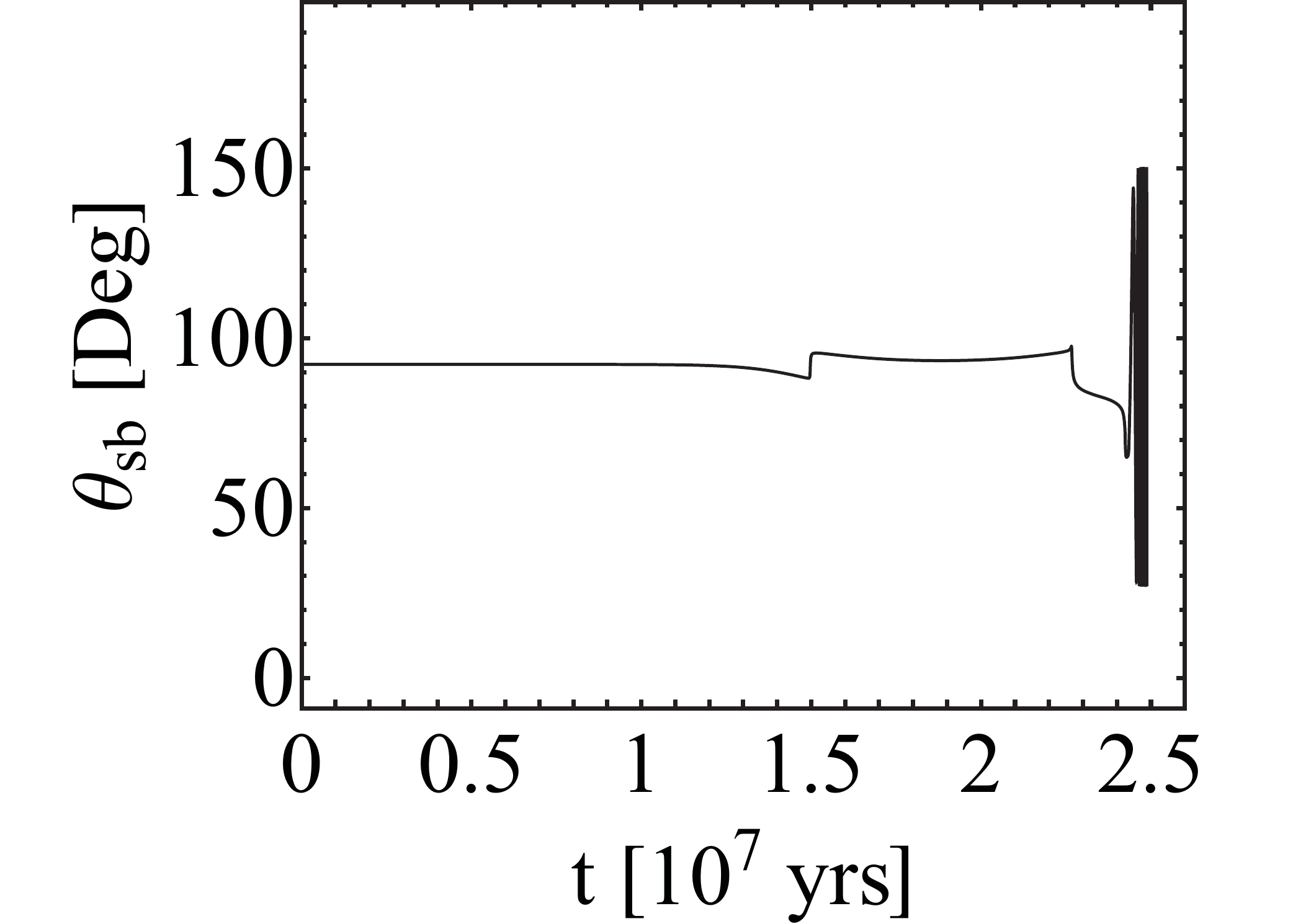}
\includegraphics[width=6cm]{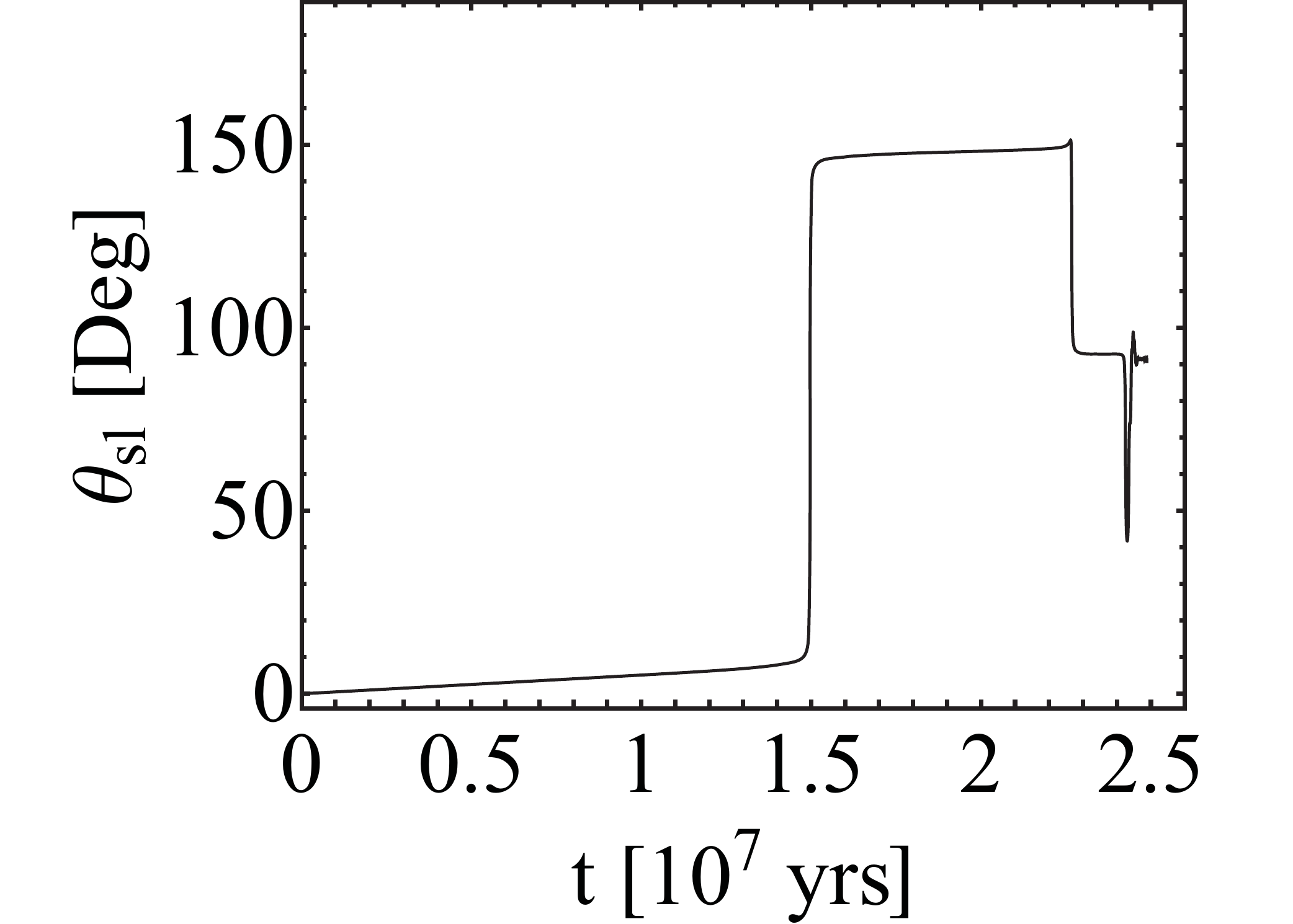}
\end{tabular}
\caption{
Same as Figure \ref{fig:OE quad 1}, except for $I_0=92.33^\circ$.
}
\label{fig:OE quad 2}
\end{centering}
\end{figure*}

\begin{figure*}
\begin{centering}
\begin{tabular}{ccc}
\includegraphics[width=6cm]{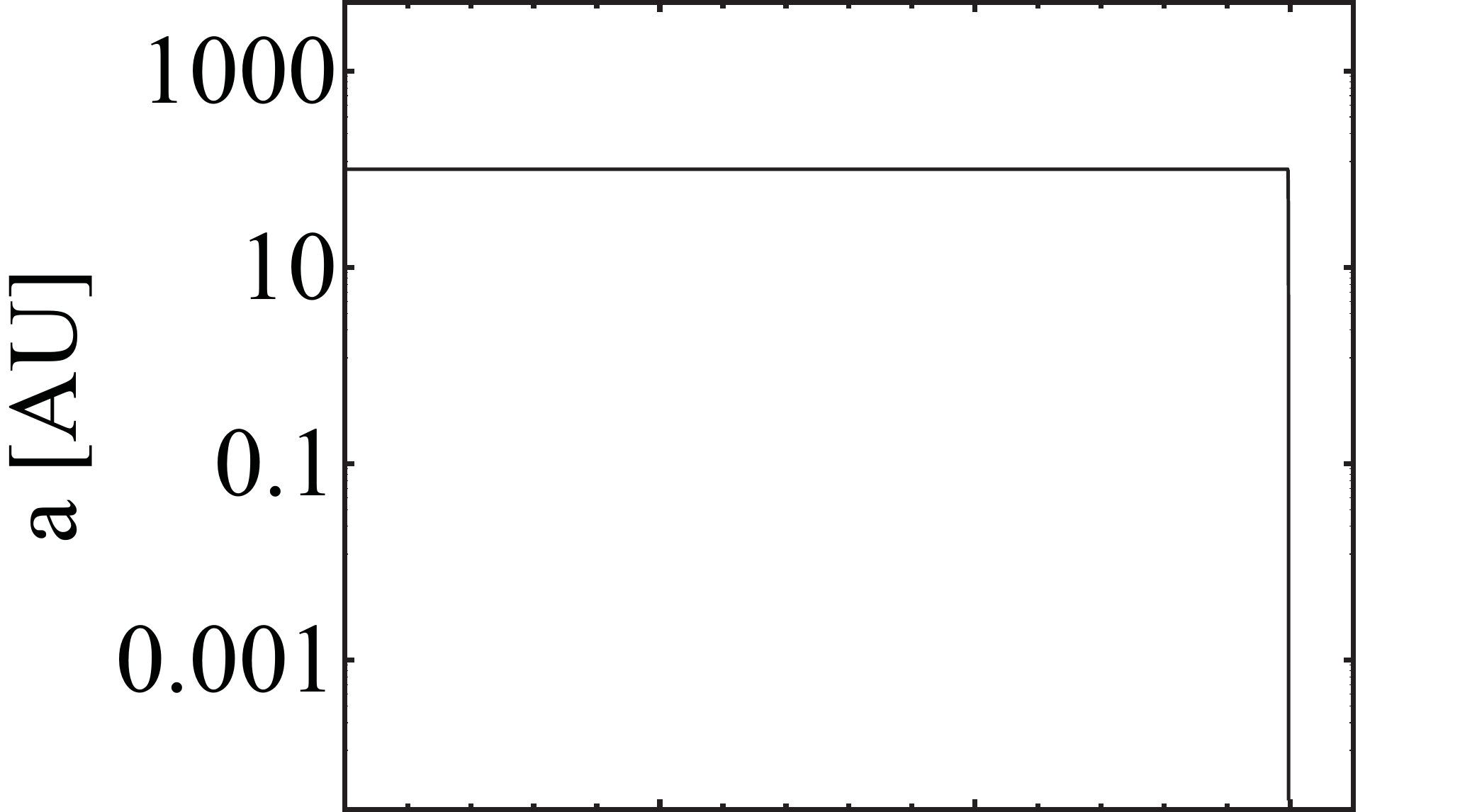}
\includegraphics[width=6cm]{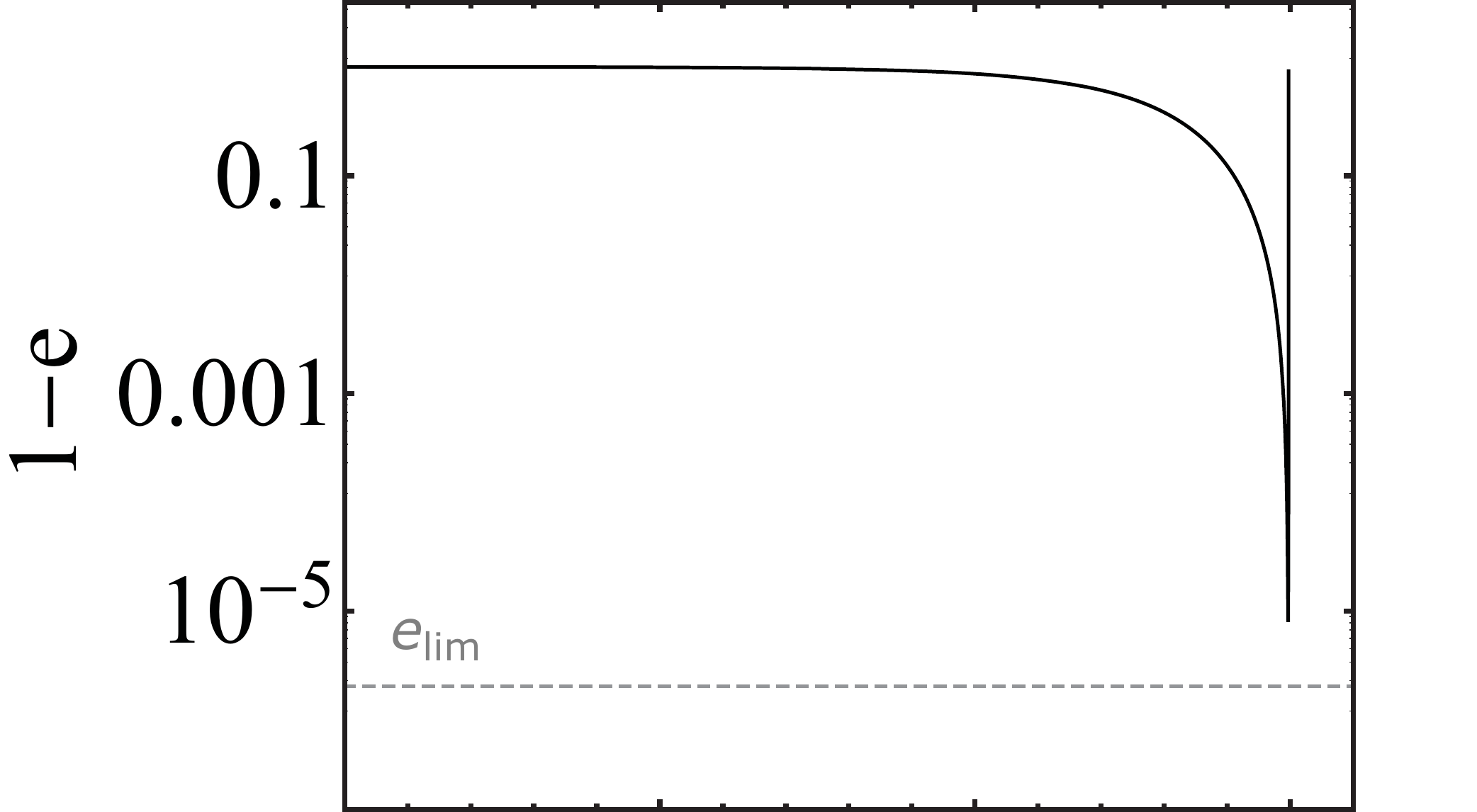}
\includegraphics[width=6cm]{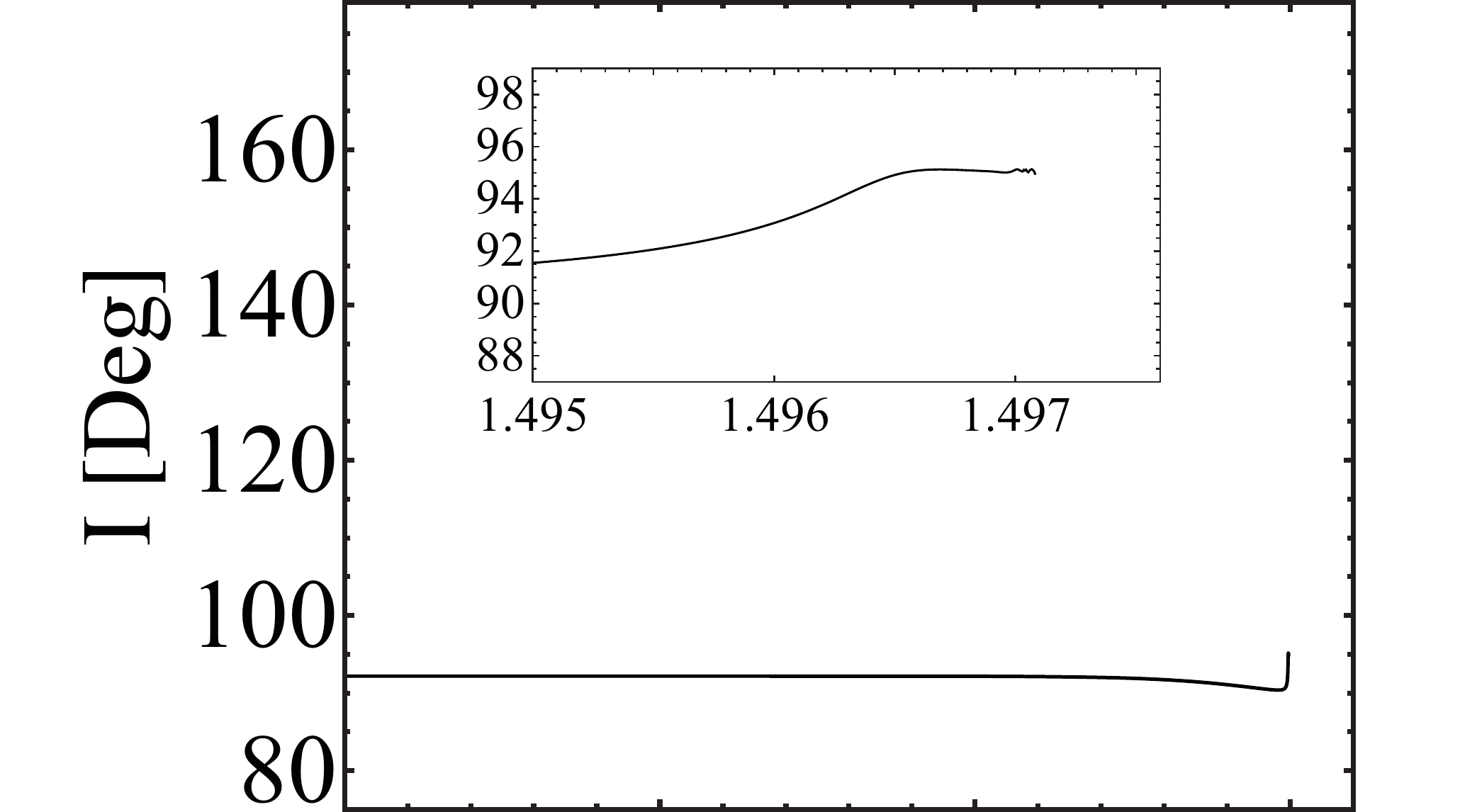}\\
\includegraphics[width=6cm]{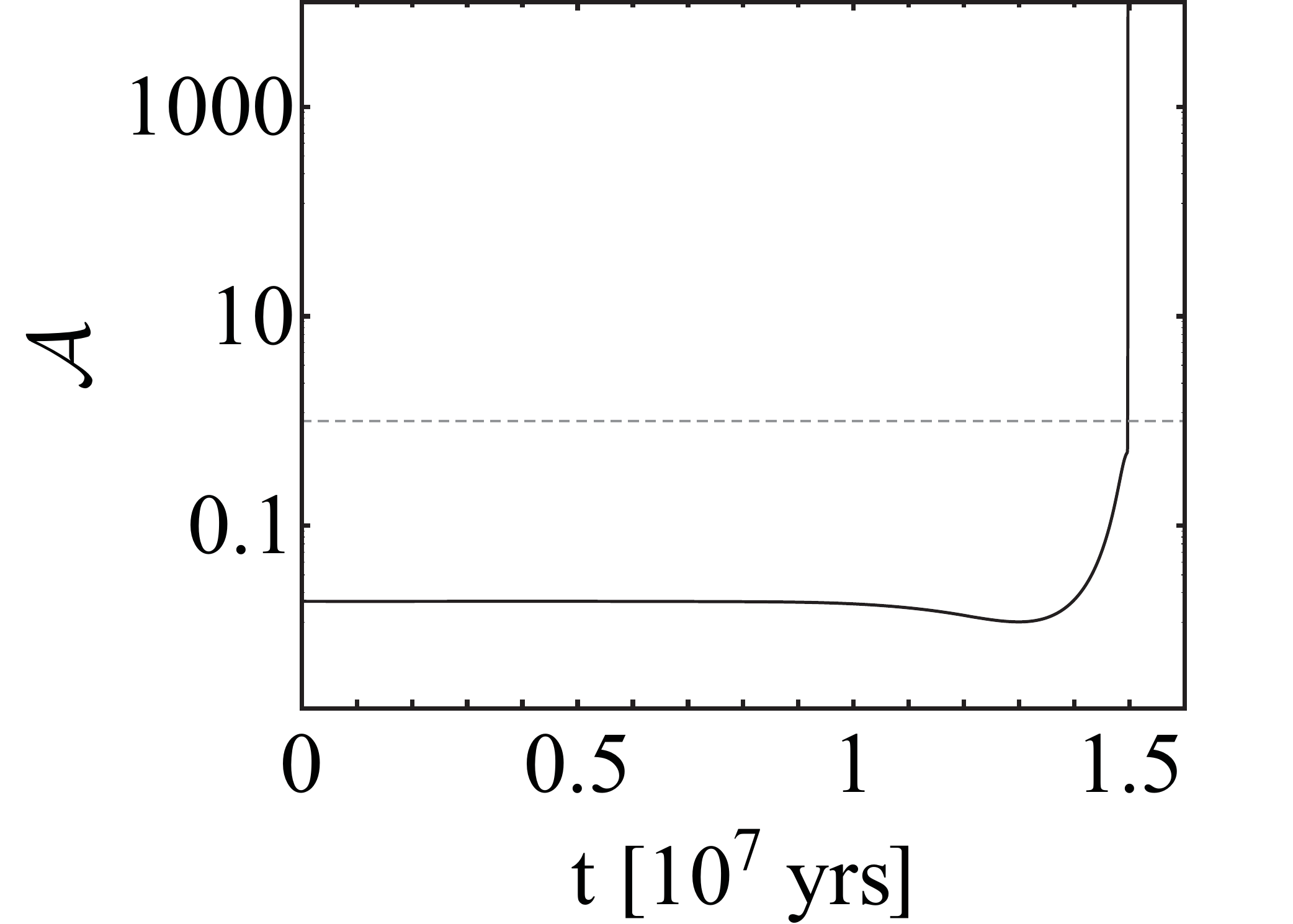}
\includegraphics[width=6cm]{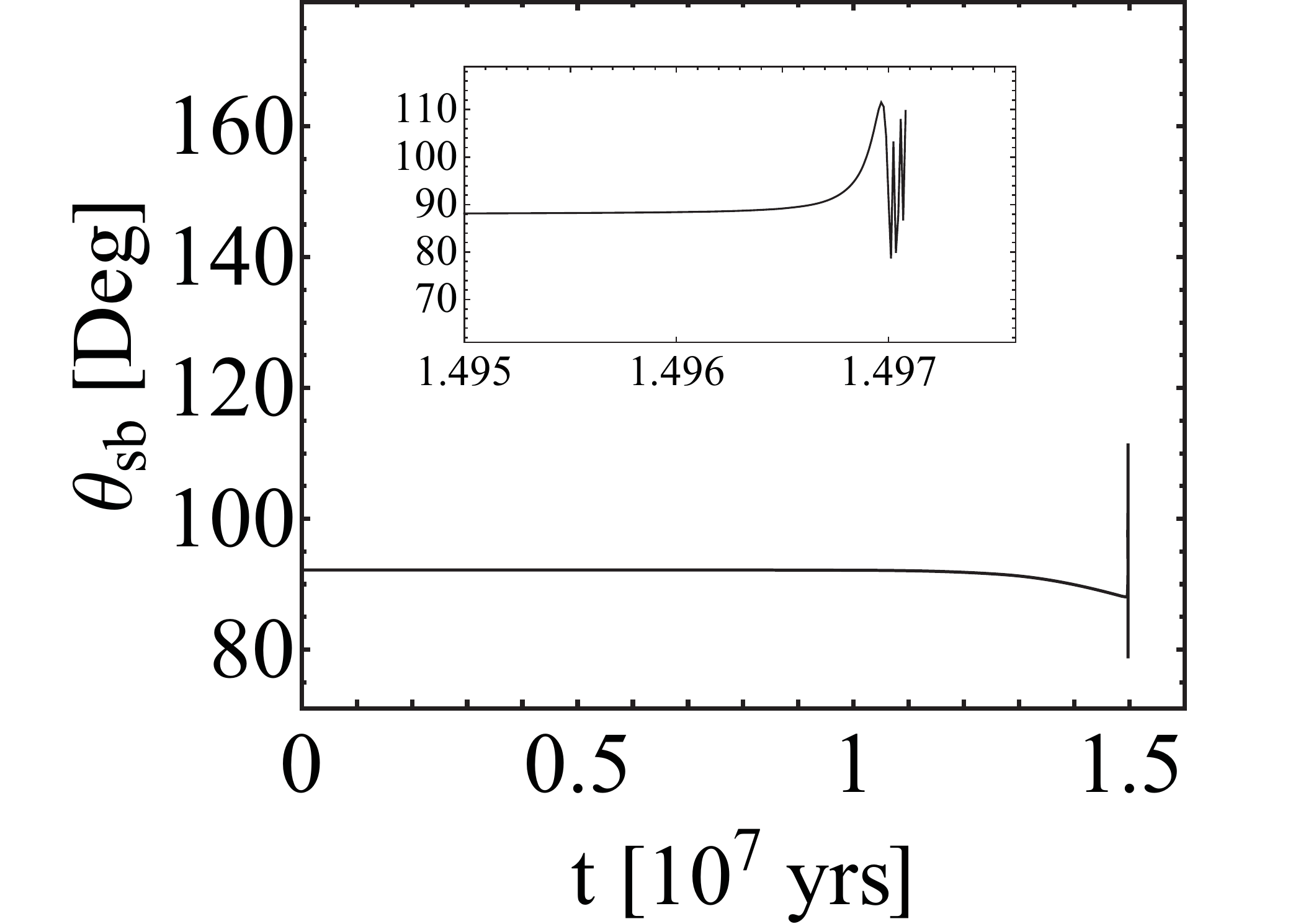}
\includegraphics[width=6cm]{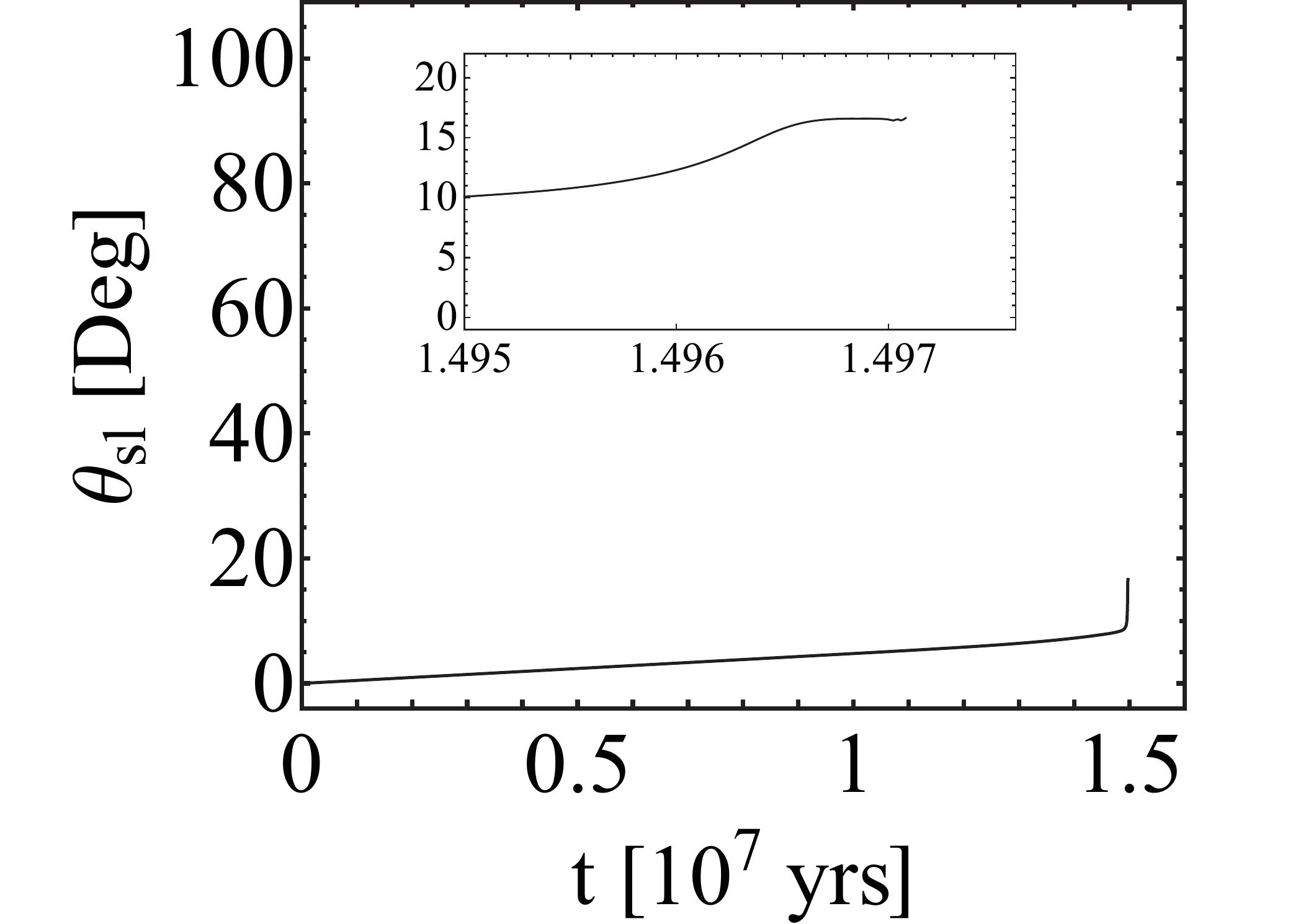}
\end{tabular}
\caption{
Same as Figure \ref{fig:OE quad 1}, except for $I_0=92.18^\circ$.
}
\label{fig:OE quad oneshort}
\end{centering}
\end{figure*}

In the third panel of Figure \ref{fig:merger window quad}, we turn on orbital decay due to gravitational radiation.
The merger of the inner binary is achieved within the Hubble time ($T_\mathrm{m}\lesssim 10^{10}$ yrs)
for a range of inclinations around $I_{0,\li}$.
The eccentricity excitation leads to a shorter binary merger time $T_\mathrm{m}$
compared to the ``circular" merger time $T_{\mathrm{m},0}$ (see Equation \ref{eq:Tmerger}).
In \cite{Liu-ApJL}, we found that the merger timescale in LK-induced mergers
can be described by the fitting formula
$T_\mathrm{m}=T_\mathrm{m,0}(1-e_\m^2)^\alpha$;
the coefficient $\alpha$ depends on
$e_\m$ (from Equation \ref{eq:EMAX}),
with $\alpha\simeq 1.5,~2$ and 2.5 for $e_\m=(0,0.6),~(0.6,0.8)$ and
$(0.8,0.95)$, respectively.
Here we consider the regime where $e_\m$ is much close to unity, and
we find that $T_\mathrm{m}$ can be best fitted by $\alpha=3$, i.e.
\be\label{eq:fitting formula origin}
T_\mathrm{m}\simeq T_\mathrm{m,0}(1-e_\m^2)^3.
\ee
This scaling can be understood as follows: The intrinsic GW-induced orbital decay rate
$|\dot a/a|_\mathrm{GW}$ is proportional to $(1-e^2)^{-7/2}$ (Equation \ref{eq:decay rate}).
In a LK-induced merger, the orbital decay mainly occurs at $e\simeq e_\m$.
During the LK cycle, the binary only spreads a fraction ($\sim \sqrt{1-e_\m^2}$) of the time
near $e\simeq e_\m$. Thus, the LK-averaged orbital decay rate is of order
$T_{\mathrm{m},0}^{-1}(1-e_\m^2)^{-3}$, as indicated by Equation (\ref{eq:fitting formula origin}).
Using Equation (\ref{eq:fitting formula origin}), we
can define the ``merger eccentricity" $e_\mathrm{m}$ via
\be\label{eq:fitting formula}
T_{\mathrm{m},0}(1-e_\mathrm{m}^2)^3=T_\mathrm{crit}.
\ee
Thus, only systems with $e_\m\gtrsim e_\mathrm{m}$ can have the merger time $T_\mathrm{m}$ less than $T_\mathrm{crit}$ -- Throughout
this paper, our numerical results refer to $T_\mathrm{crit}=10^{10}$ yrs (see Section \ref{sec 3 3}).
For the systems shown in Figure \ref{fig:merger window quad}, we find
$1-e_\mathrm{m}\simeq10^{-4}$, and the merger window of initial inclinations is
$I_{0,\merger}^-\leqslant I_0\leqslant I_{0,\merger}^+$,
with $I_{0,\merger}^-=91.56^\circ$ and $I_{0,\merger}^+=92.76^\circ$ (Equation \ref{eq:EMAX}),
in agreement with the direct numerical results. As expected,
the width of the merger window ($I_{0,\merger}^+-I_{0,\merger}^-\simeq 1.2^\circ$) is rather small.
Also note that $T_\mathrm{m}$ shows a constant distribution around $I_0\sim I_{0,\li}$.
This is the result of ``one-shot" merger,
where the system only undergoes the first LK cycle, then ``suddenly" merges during the high-$e$ phase.

Figures \ref{fig:OE quad 1}-\ref{fig:OE quad oneshort} show a few examples of the orbital evolution for the systems
inside the merger window, for which the initial inclination equals to $I_0=92.52^\circ, 92.33^\circ, 92.18^\circ$,
respectively. The evolution of BH spin is also shown, and this will be discussed in Section \ref{sec 4}.
In the three upper panels of Figure~\ref{fig:OE quad 1}, we see that the inner binary undergoes cyclic excursions to the maximum eccentricity $e_\m$, with accompanying oscillations in the inclination $I$.
As the binary decays, the range of eccentricity oscillations becomes smaller,
and the eccentricity ``freezes" to a large value.
In the final phase, GW dissipation
causes the orbit to shrink in the semi-major axis and circularize in the eccentricity.

\begin{figure*}
\begin{centering}
\begin{tabular}{ccc}
\includegraphics[width=6cm]{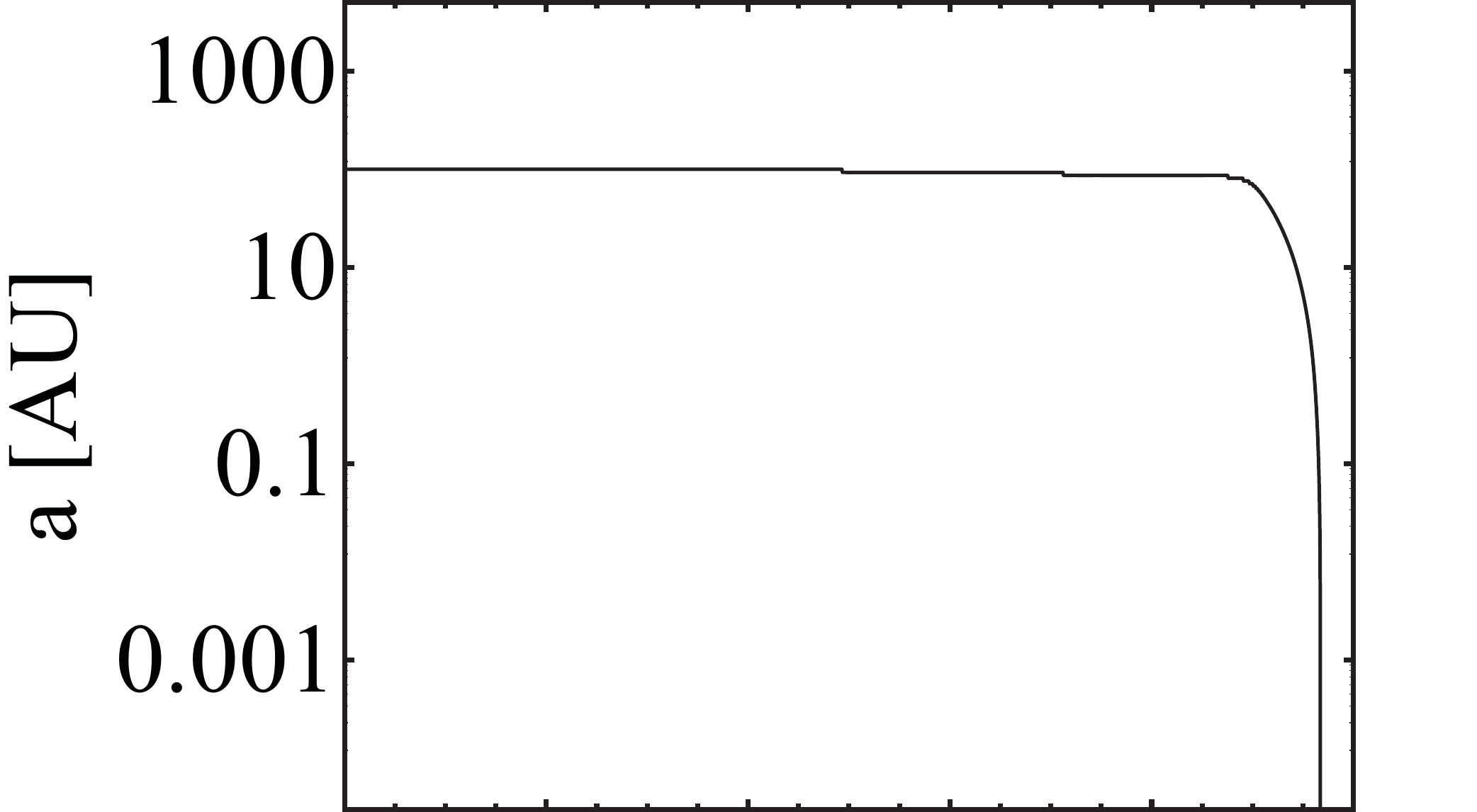}
\includegraphics[width=6cm]{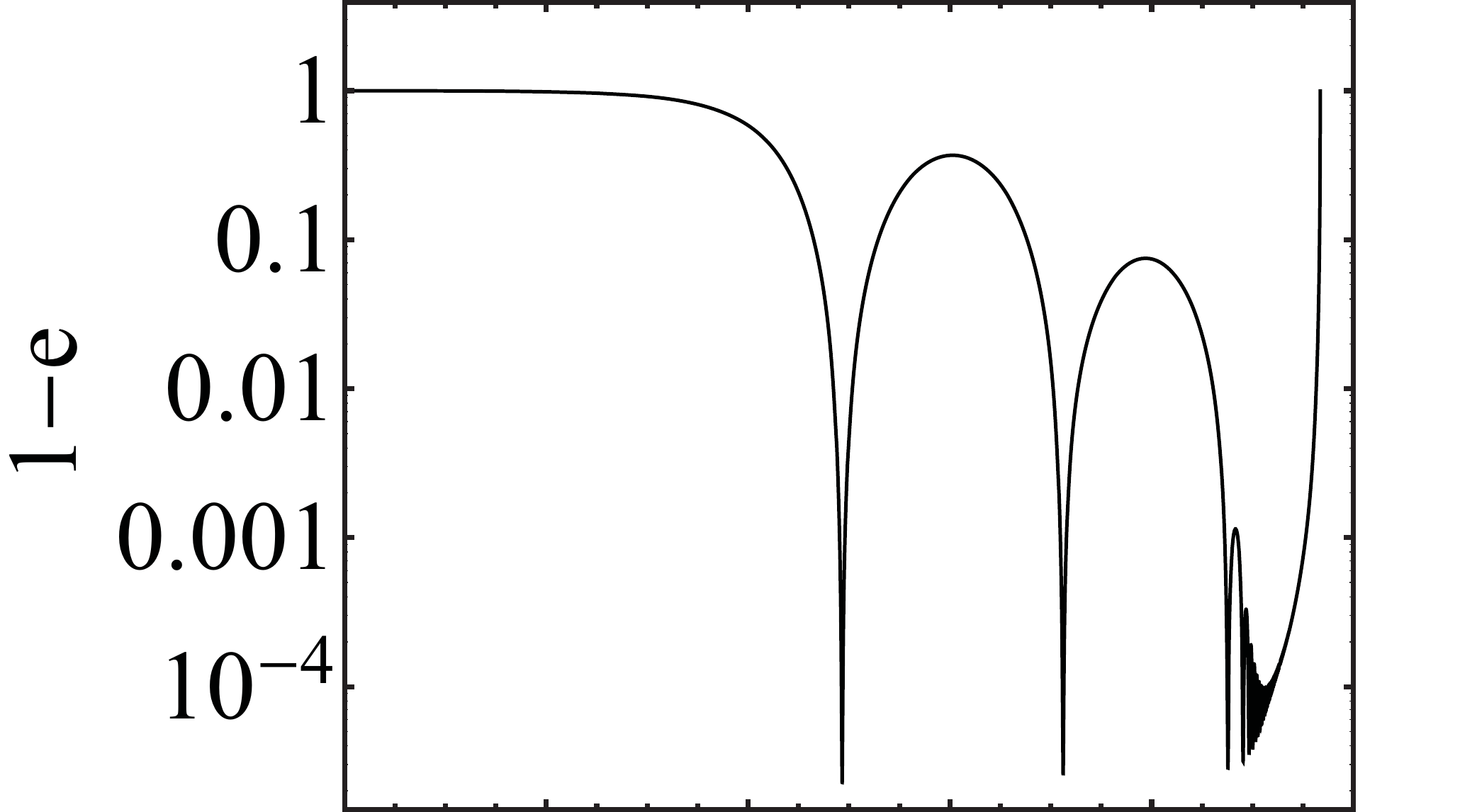}
\includegraphics[width=6cm]{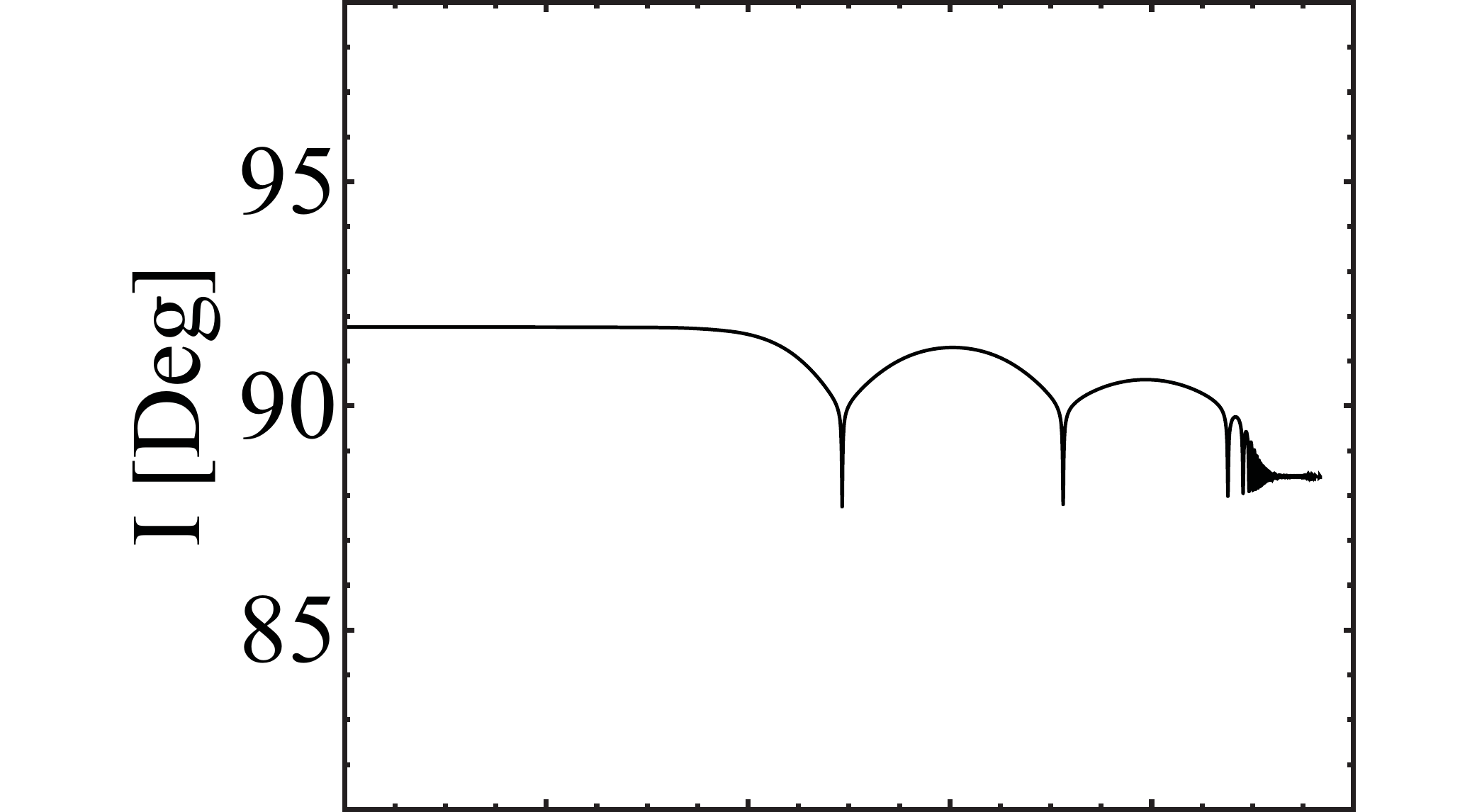}\\
\includegraphics[width=6cm]{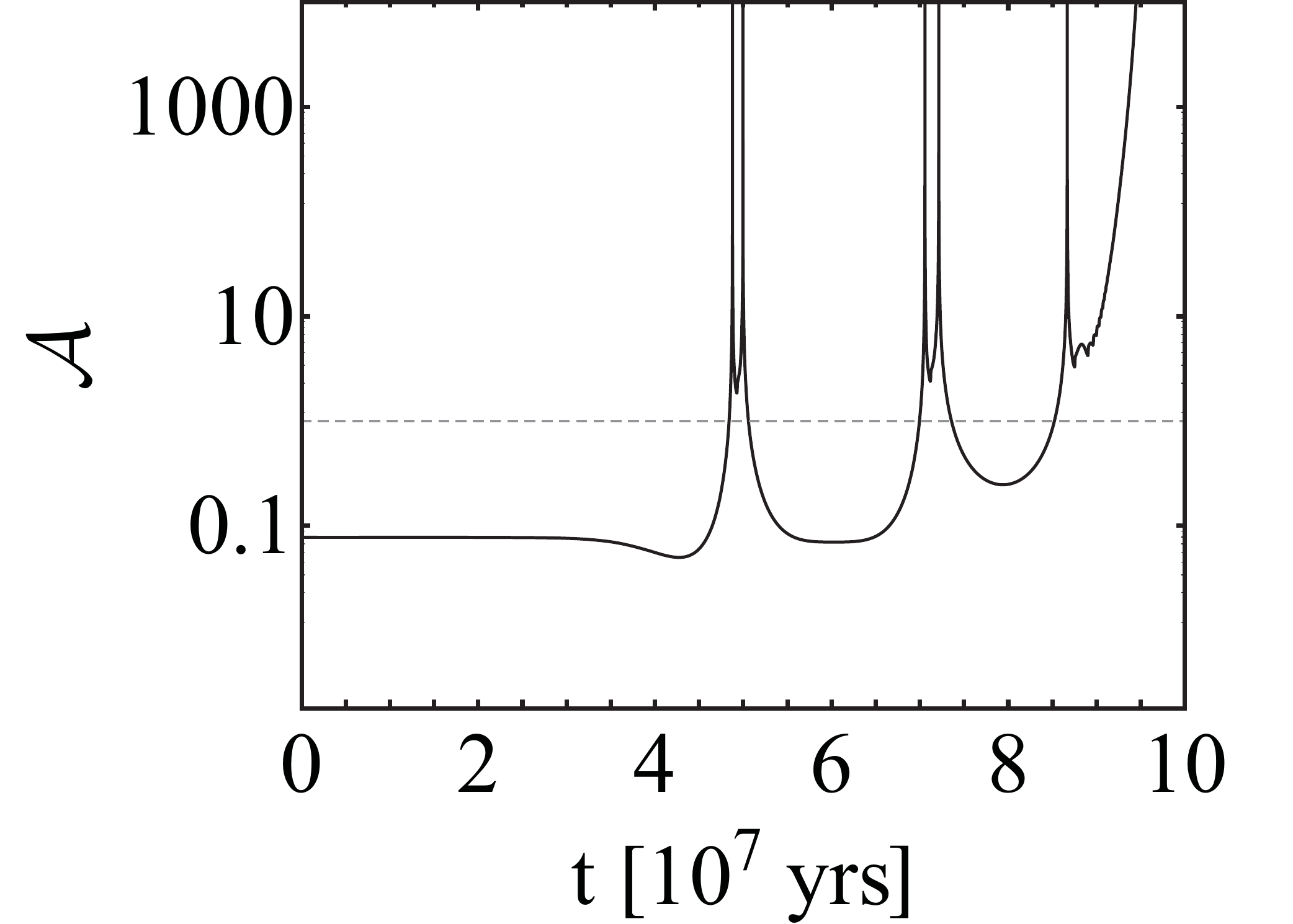}
\includegraphics[width=6cm]{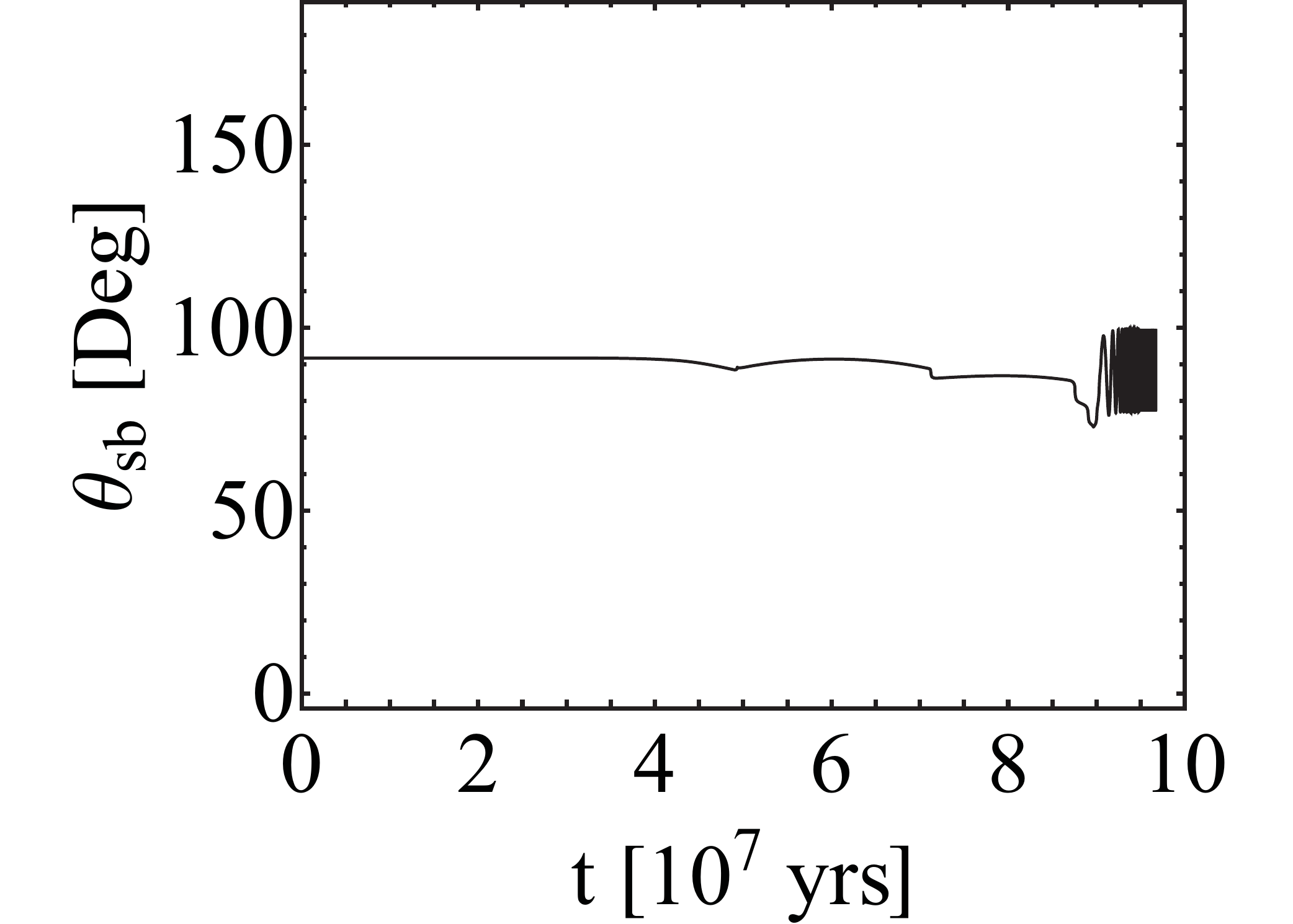}
\includegraphics[width=6cm]{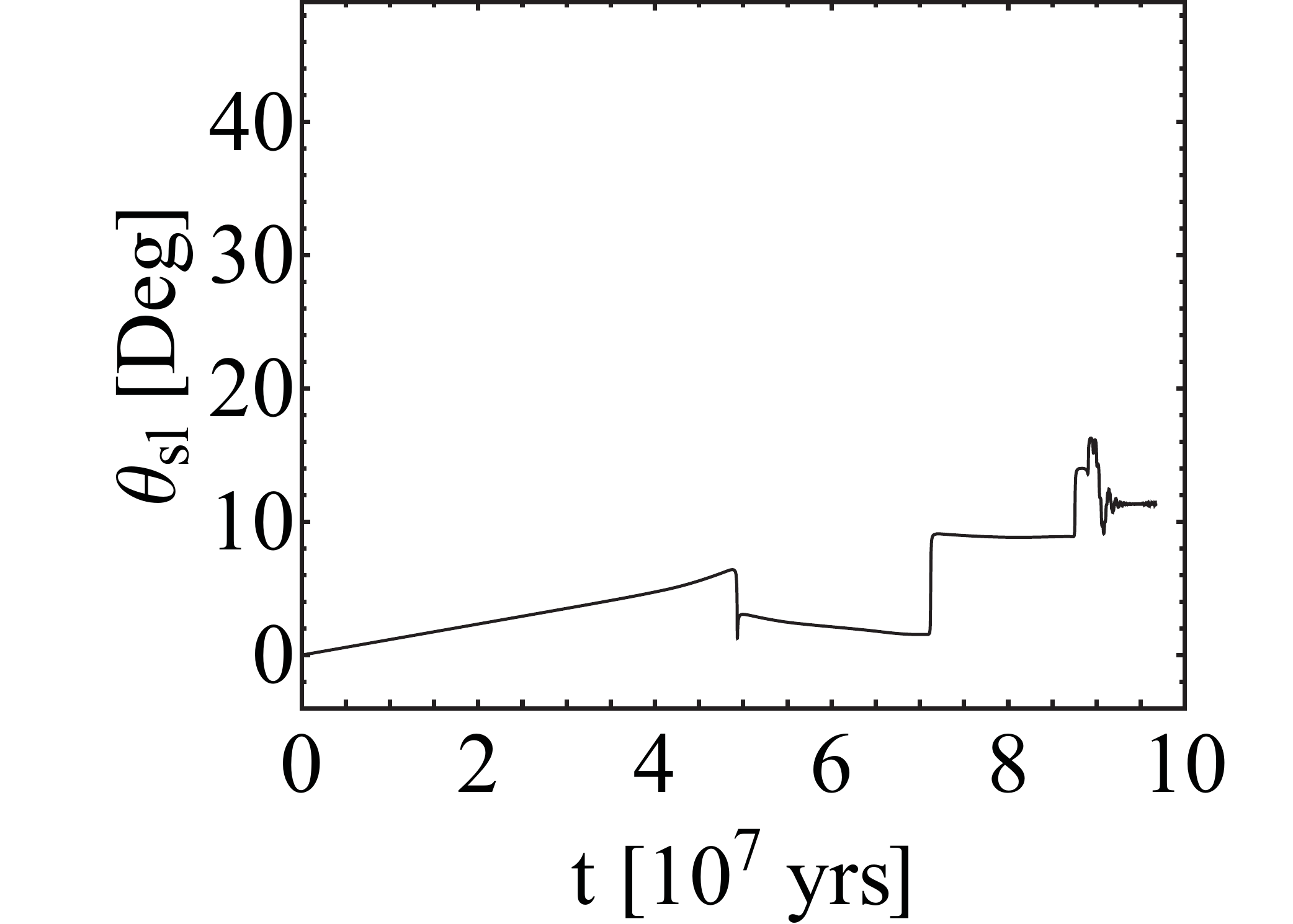}
\end{tabular}
\caption{Same as Figure \ref{fig:OE quad 1}, except for a more distant companion with
$a_\OUT=6700\au$, and $I_0=91.76^\circ$.
}
\label{fig:OE quad 3}
\end{centering}
\end{figure*}

In Figures \ref{fig:OE quad 2}-\ref{fig:OE quad oneshort}, $I_0$ is closer to $I_{0,\li}$,
so that $e_\m$ achieved during LK oscillations is closer to $e_\li$.
The GW-induced orbital decay is more efficient (Equations \ref{eq:GW 1}-\ref{eq:GW 2}),
so the binary only experiences a few or even less than one LK cycles before merging.
In Figure \ref{fig:OE quad 2}, the orbit undergoes the usual freezing of eccentricity oscillations
as in Figure \ref{fig:OE quad 1}.
In Figure \ref{fig:OE quad oneshort}, $a$ decays abruptly,
and the binary merges in the first high-eccentricity episode (``one-shot merger").

Figure \ref{fig:OE quad 3} shows another example for a system with a more distant companion ($a_\OUT=6700\au$).
Even though $I_0\simeq I_{0,\li}$ for this example, the inner BH binary does not attain sufficiently large $e_\m$
to enable ``one-shot" merger.

For all the examples considered in Figures \ref{fig:merger window quad}-\ref{fig:OE quad 3}, we
find that the merging BH binaries have a negligible eccentricity ($e\lesssim0.01$) when
entering the aLIGO band.

\begin{figure}
\begin{centering}
\includegraphics[width=9cm]{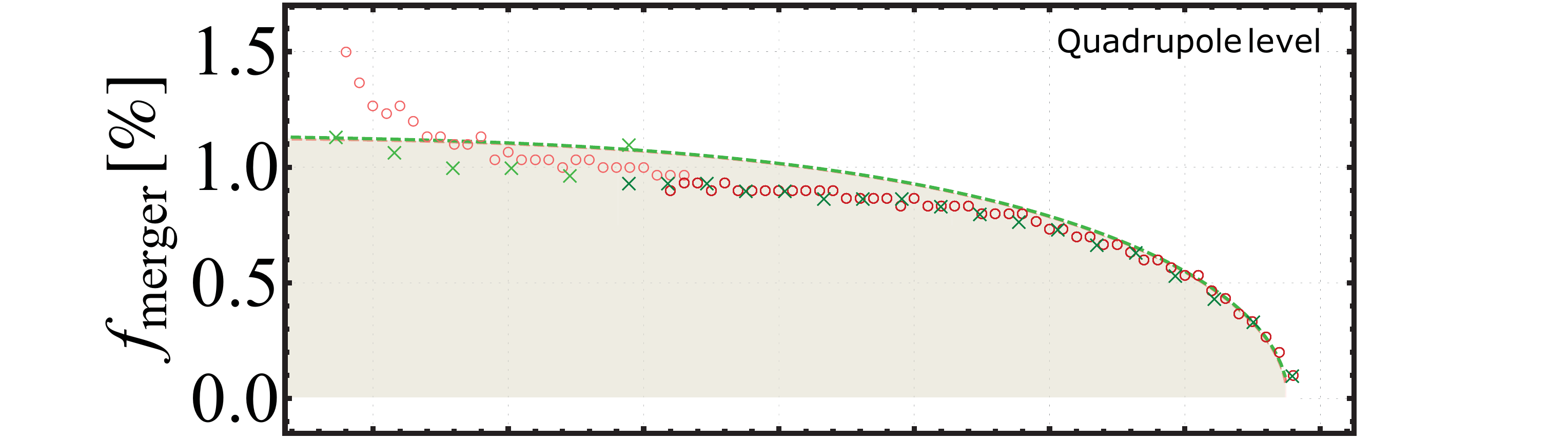}\\
\includegraphics[width=9cm]{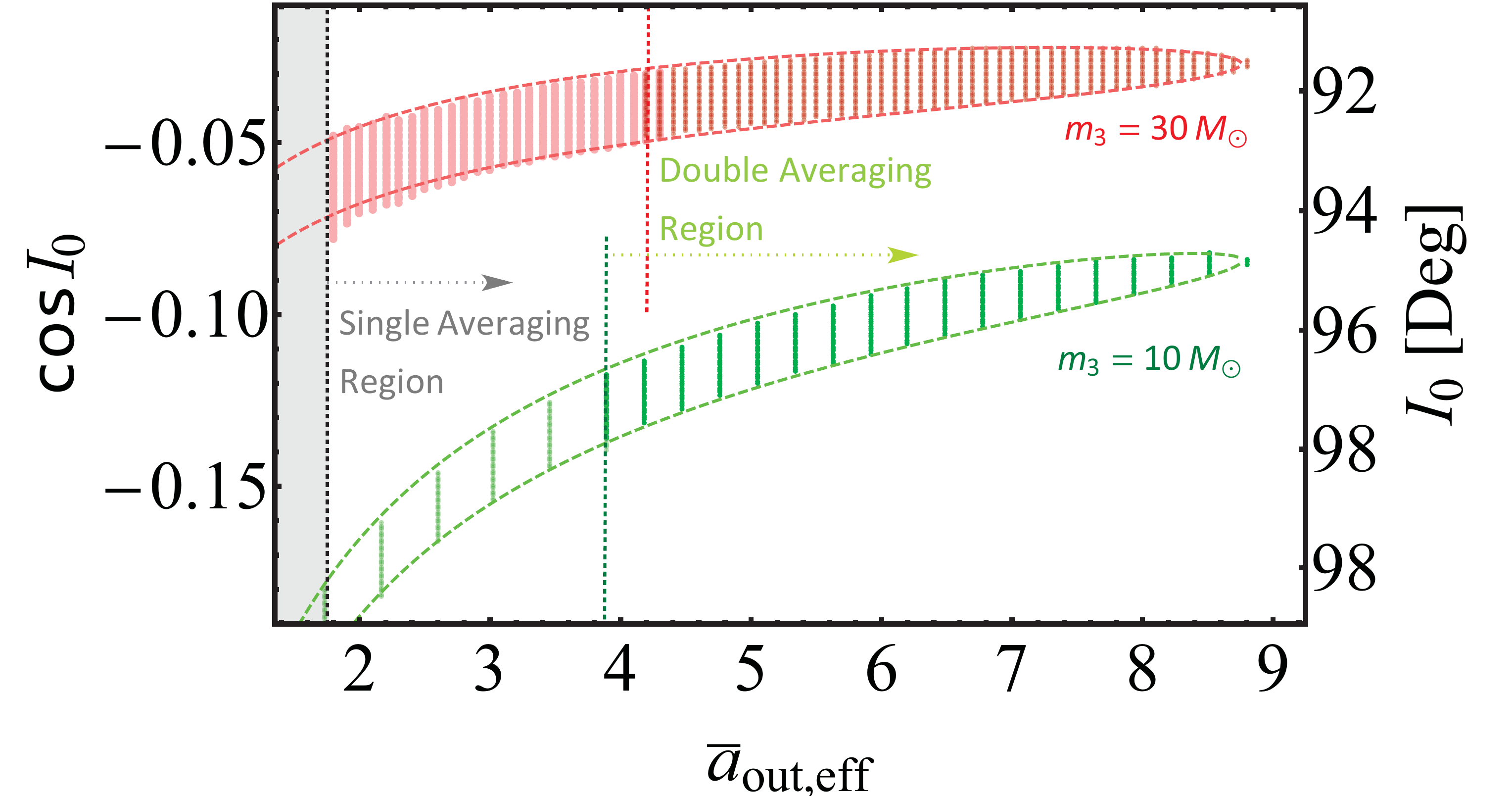}\\
\caption{Merger fraction (upper panel) and merger window (lower panel) as a function of
the effective semi-major axis of tertiary companion $\bar{a}_{\OUT,\eff}$ (Equation \ref{eq:aout bar}).
The system parameters are $m_1=30M_\odot$, $m_2=20M_\odot$, $a_0=100\au$, $e_0=0.001$ and $e_\OUT=0$.
In the lower panel, the color coded dots are obtained by integrating the single/double averaged secular equations
(each dot represents a successful merger within $10^{10}$ yrs),
and the dashed curves (for each $m_3$ value) represent $\cos I_{0,\merger}^+$ and $\cos I_{0,\merger}^-$, and
can be obtained analytically using Equations (\ref{eq:EMAX}) and (\ref{eq:fitting formula}).
In the top panel, the open circles and crosses indicate the merger fraction
from the mergers shown in the lower panel. The dashed curve is the analytical estimate, given by Equation (\ref{eq:merger fraction}).
}
\label{fig:merger fraction quad }
\end{centering}
\end{figure}

The lower panel of Figure \ref{fig:merger fraction quad } shows the merger window (in terms of $\cos I_0$) as a
function of the effective semi-major axis of the tertiary companion.
From Section \ref{sec 2 2} (see Figure \ref{fig:elim}), we have found that in the quadrupole approximation,
the eccentricity excitation depends on $m_3$, $a_\OUT$, $e_\OUT$ through the ratio $m_3/a_{\OUT,\eff}^3$
(where $a_{\OUT,\eff}$ is given by Equation \ref{eq:aout eff}).
We therefore introduce the dimensionless scaled semi-major axis
\be\label{eq:aout bar}
\begin{split}
\bar{a}_{\OUT,\eff}&\equiv \bigg(\frac{a_{\OUT,\eff}}{1000\au}\bigg)\bigg(\frac{m_3}{30M_\odot}\bigg)^{-1/3}\\
&=\bigg(\frac{a_\OUT\sqrt{1-e_\OUT^2}}{1000\au}\bigg)\bigg(\frac{m_3}{30M_\odot}\bigg)^{-1/3}
\end{split}
\ee
to characterize the ``strength" of the outer perturber
(note that Figure \ref{fig:merger fraction quad } neglects the octupole effect, which can complicate the single
dependence of the merger window on $\bar{a}_{\OUT,\eff}$; see Section \ref{sec 3 2}).
For a given BH binary ($m_1=30M_\odot$, $m_2=20M_\odot$, $a_0=100\au$), we fix $e_\OUT=0$ and
$m_3=30M_\odot$ or $10M_\odot$, but vary $a_\OUT$.
For each $\bar{a}_{\OUT,\eff}$, we consider 3000 values of $I_0$ spaced equally in $\cos I_0\in(-1,1)$,
evolve the systems numerically and record every merger event.
The results obtained by the double and single averaged secular equations are marked by dark and light colors, respectively.

The upper panel of Figure \ref{fig:merger fraction quad } shows the merger fraction from the mergers shown in the lower panel,
which can also be characterized by the analyzed expression
\be\label{eq:merger fraction}
f_\merger(a_0,a_\OUT,e_\OUT)=\frac{1}{2}\bigg|\cos I_{0,\merger}^+-\cos I_{0,\merger}^-\bigg|.
\ee
In the upper panel, the merger fraction is around $\sim 1\%$,
and gradually decreases as $\bar{a}_{\OUT,\eff}$ increases.
The merger window is closed when $\bar{a}_{\OUT,\eff}$ exceeds certain critical value.
Figure \ref{fig:merger fraction quad } also shows results for a different value of $m_3$ ($10M_\odot$).
This gives different values of $\eta$ and $I_{0,\li}$ (Equation \ref{eq:I0lim}),
but the merger window is qualitatively similar to the $m_3=30M_\odot$ case, except
shifted to relatively large values of $I_0$.
Both merger windows (for the two values of $m_3$) are well described by
Equations (\ref{eq:EMAX}) and (\ref{eq:fitting formula}).
The merger fractions, $f_\merger$,
are essentially identical (see the dashed curve in the upper panel).
These indicate that the fitting formula (\ref{eq:fitting formula}), together with Equation (\ref{eq:EMAX}),
can be used to predicted what types of systems will undergo merger in less than $10^{10}$ yrs,
at least in the quadrupole order.
For ``pure" quadrupole systems, simple scaling relations for $f_\merger$
(as a function of $a_0$, $m_1$, $m_2$ and $T_\mathrm{crit}$) can be obtained (see Section \ref{sec 3 3}).

\subsection{Eccentric Companions: Mergers Induced by Octupole Lidov-Kozai Effect}
\label{sec 3 2}
For $m_1\neq m_2$ and eccentric companions ($e_\OUT\neq0$),
the octupole effect becomes important when $\varepsilon_\oct$ (Equation \ref{eq:varepsilon oct}) is
appreciable, and some of the analytical expressions given in Section \ref{sec 2 2} break down.
Previous works \citep[]{Liu et al 2015,Anderson et al 2017} have shown that the main effect of the octupole potential
is to broaden the range of the initial $I_0$ for extreme eccentricity excitations ($e_\m=e_\li$),
while the quadrupole expression for limiting eccentricity $e_\li$ (Equation \ref{eq:ELIM}) remains valid.

\begin{figure*}
\centering
\begin{tabular}{cc}
\includegraphics[width=8cm]{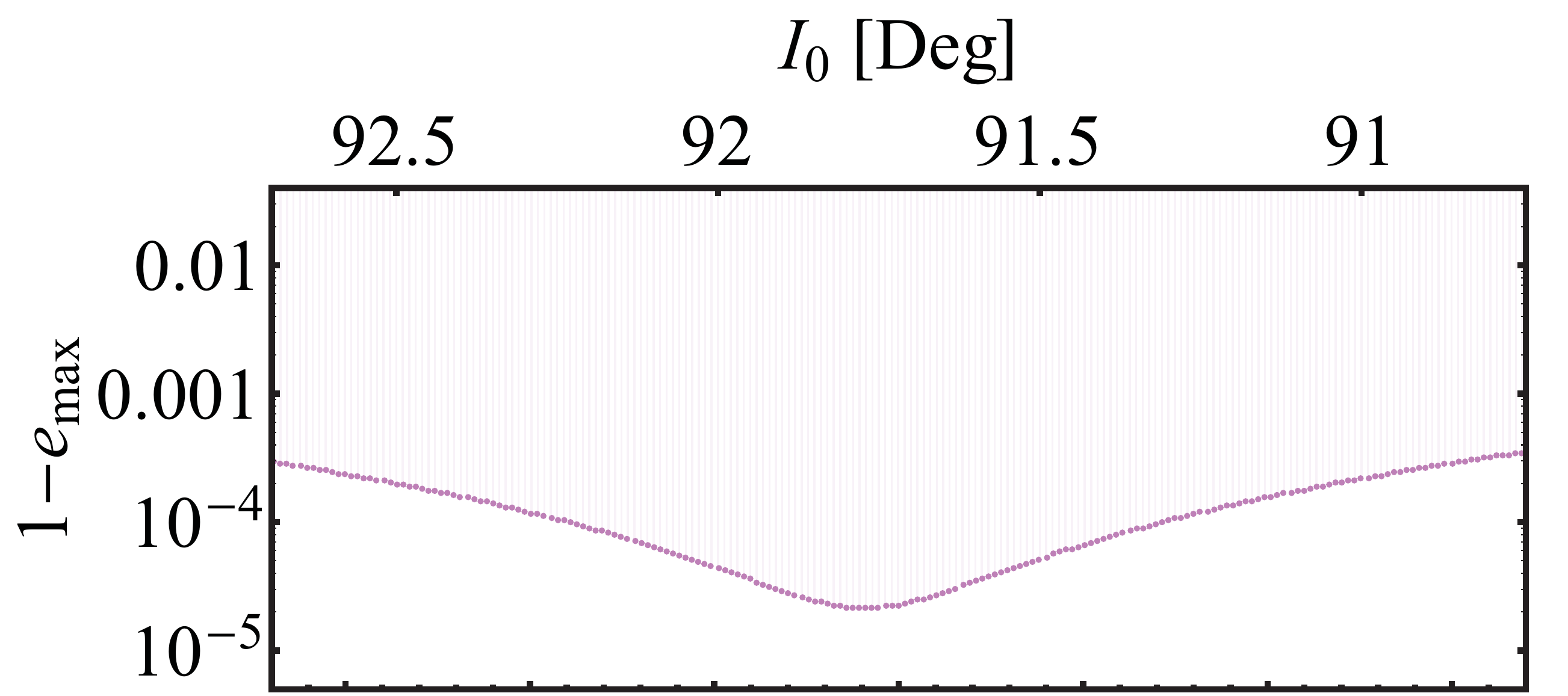}&
\includegraphics[width=8cm]{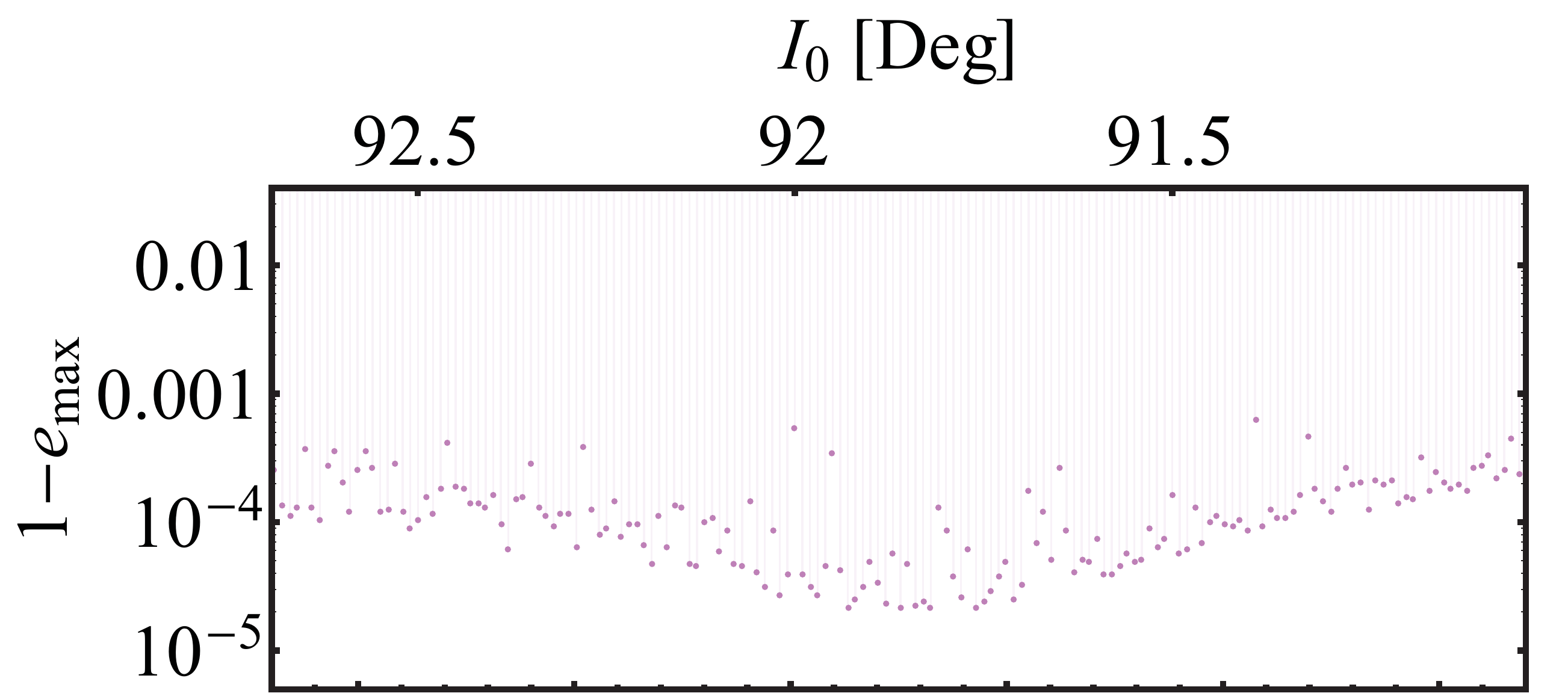}\\
\includegraphics[width=8cm]{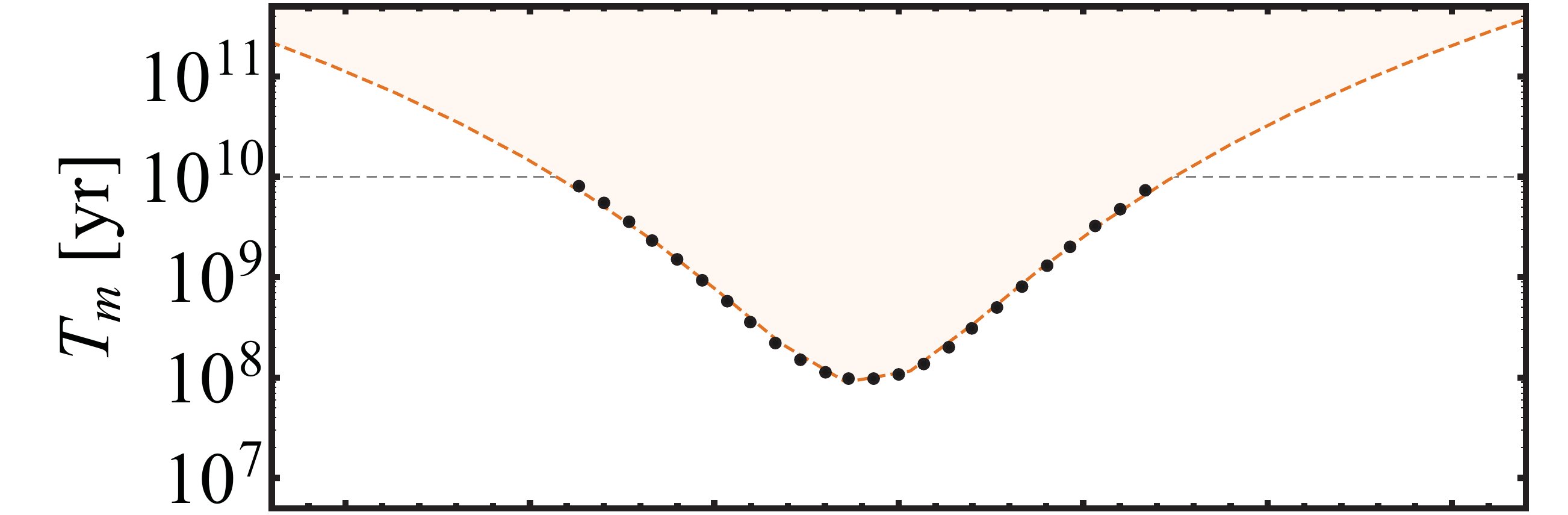}&
\includegraphics[width=8cm]{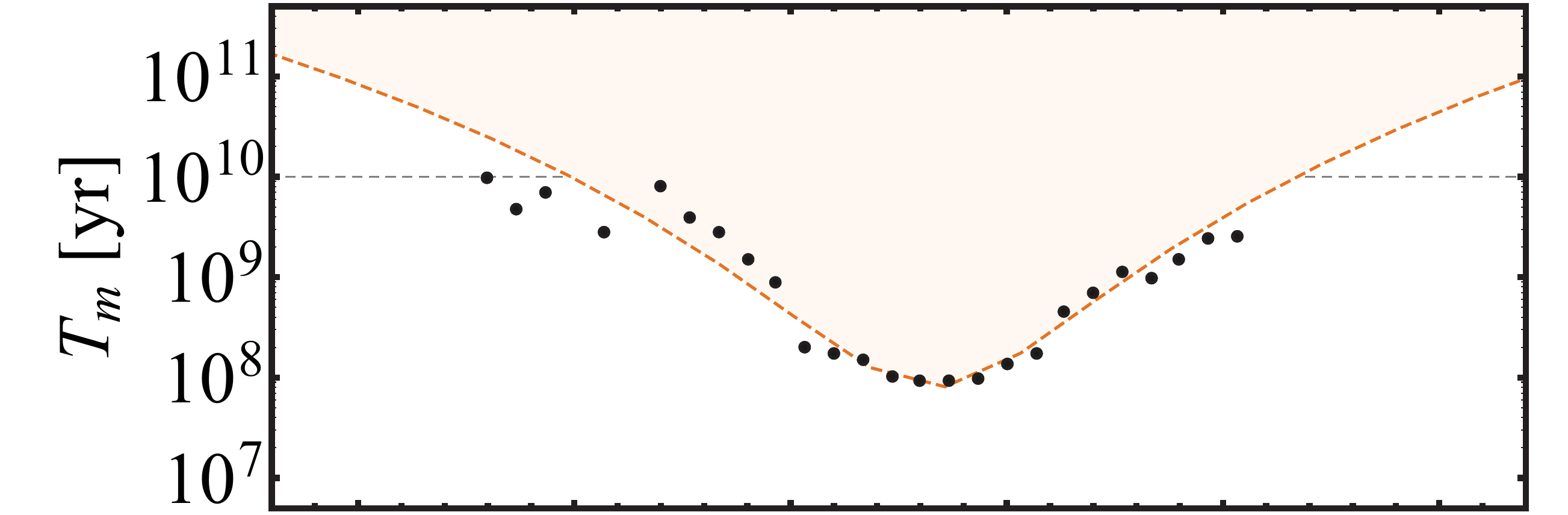}\\
\includegraphics[width=8cm]{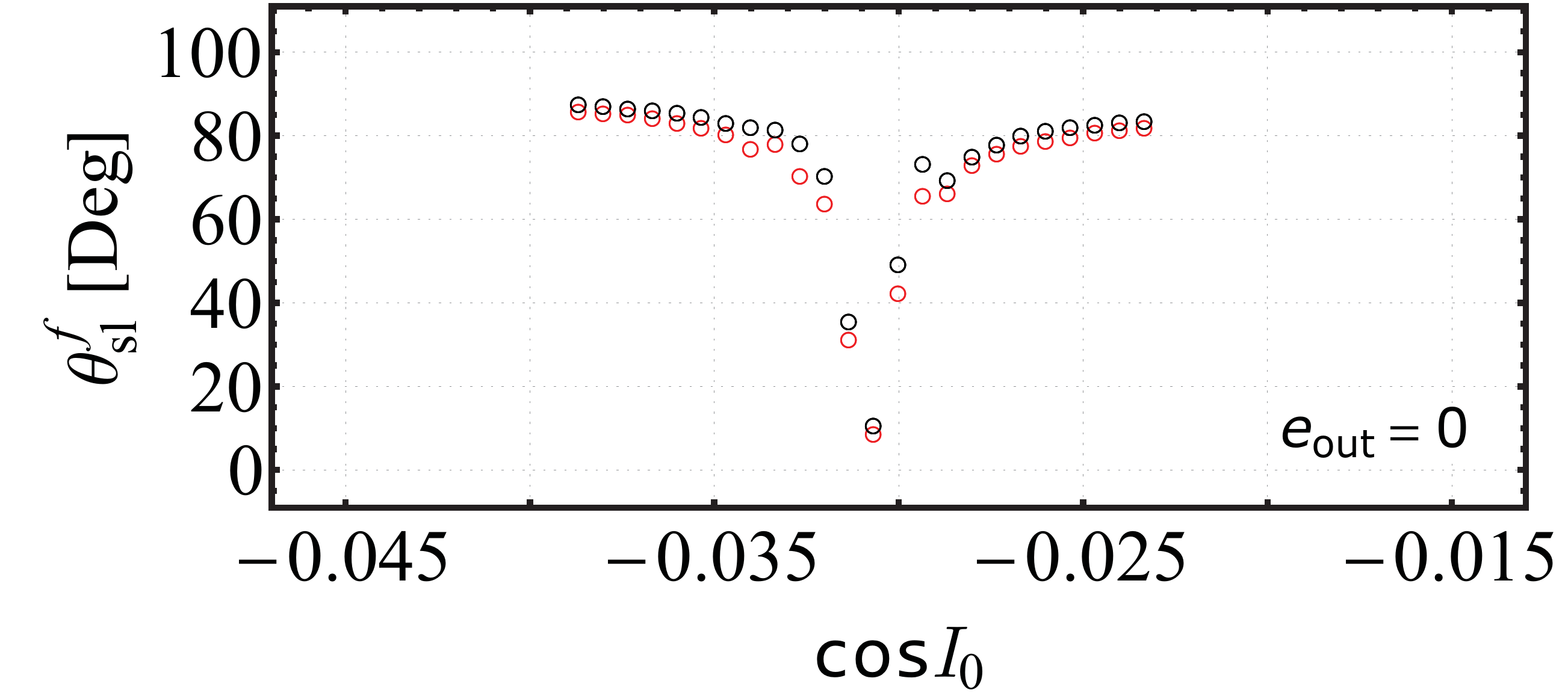}&
\includegraphics[width=8cm]{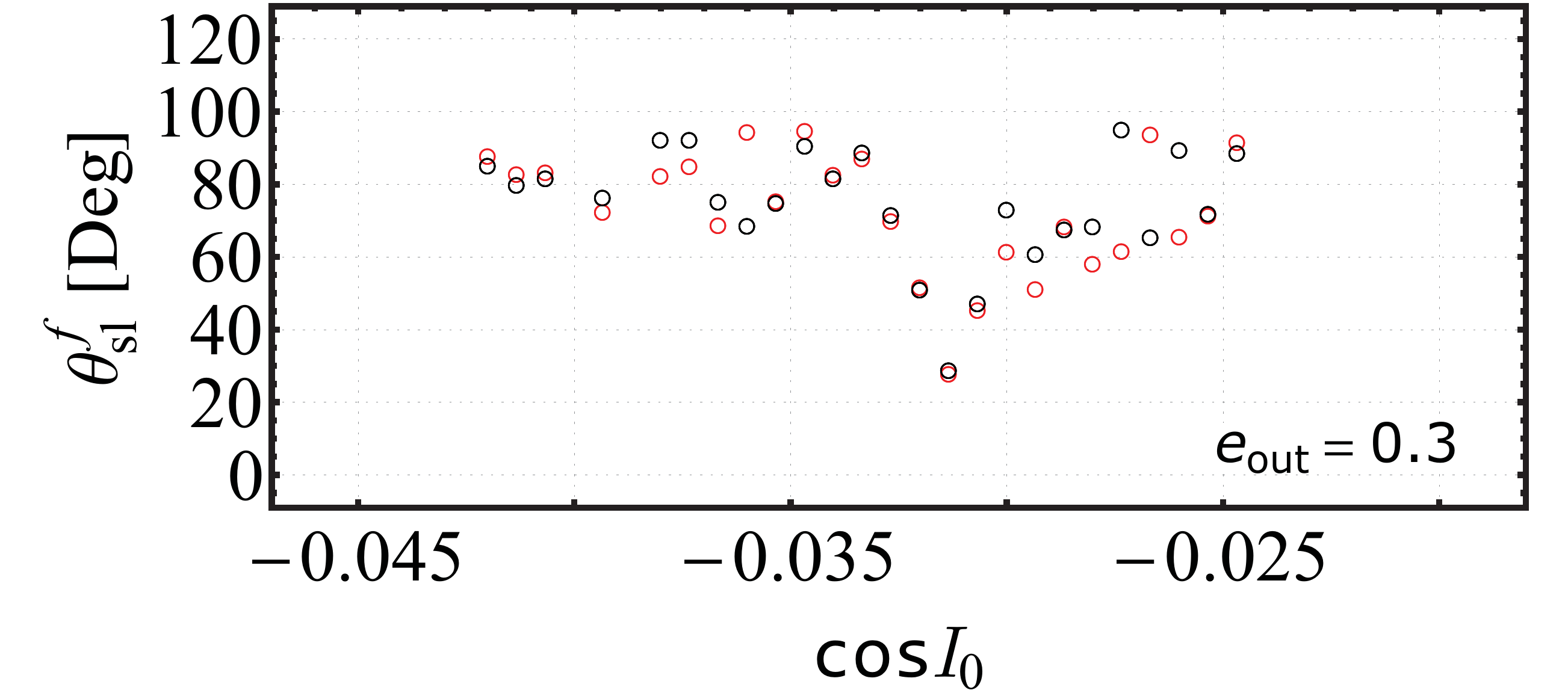}\\
\\
\\
\includegraphics[width=8cm]{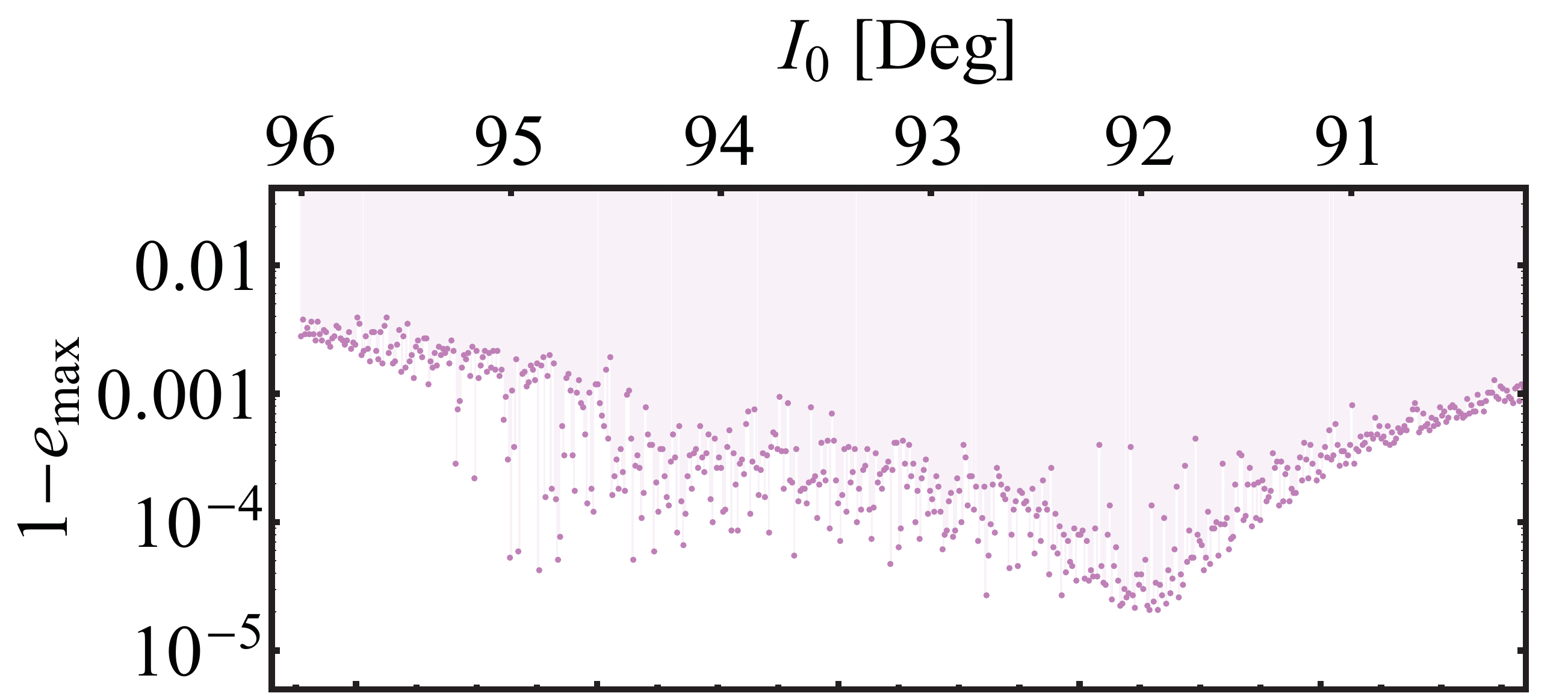}&
\includegraphics[width=8cm]{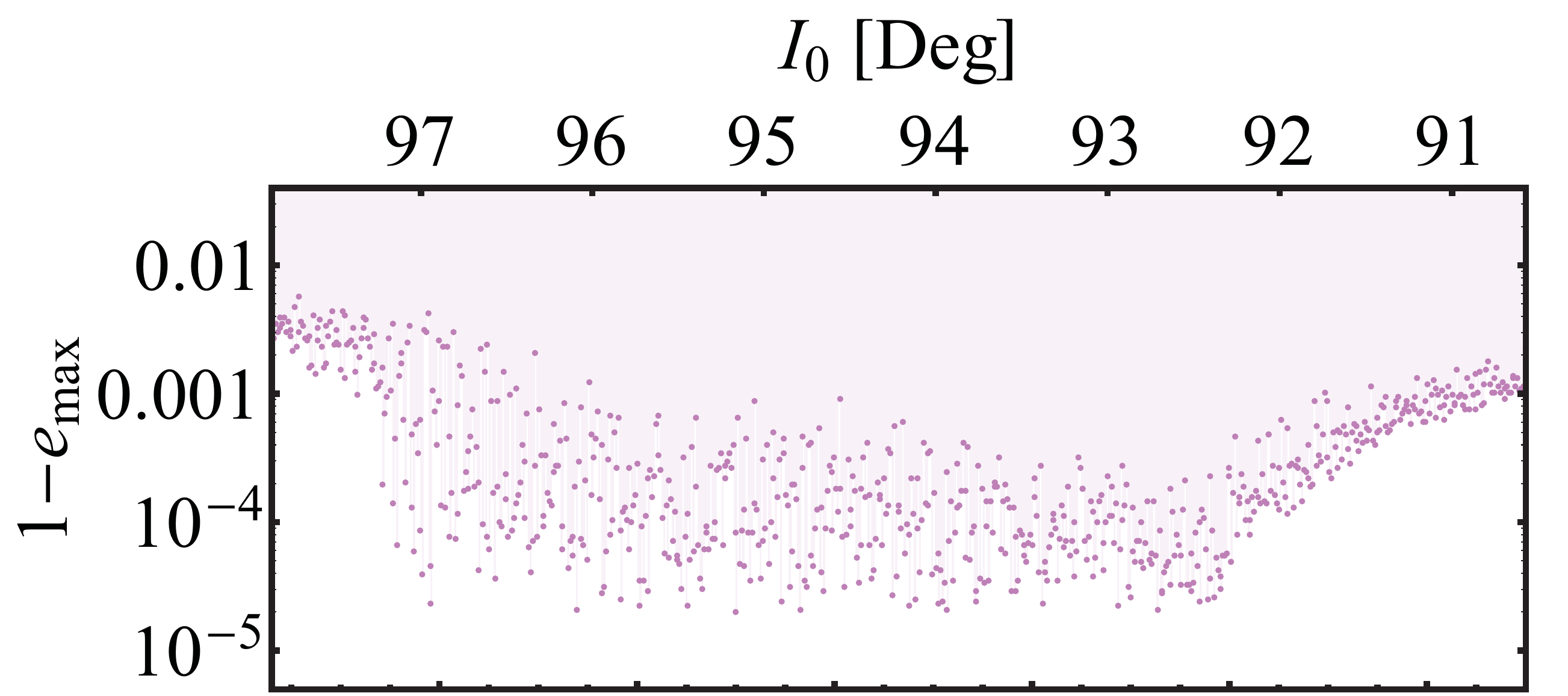}\\
\includegraphics[width=8cm]{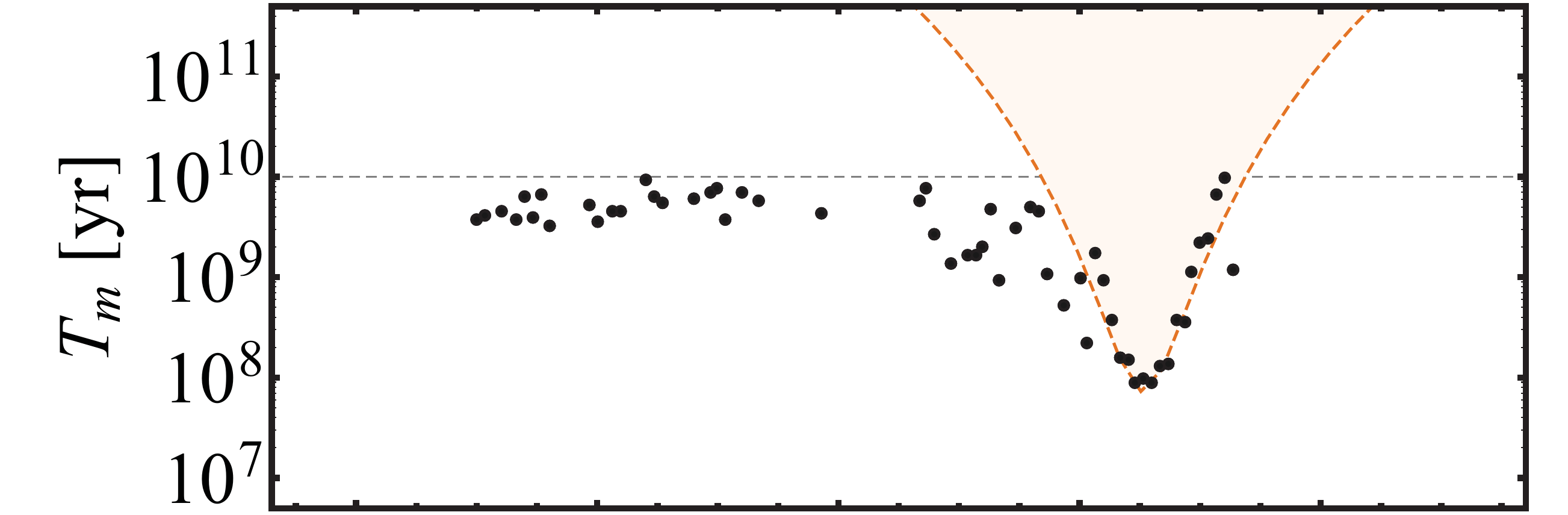}&
\includegraphics[width=8cm]{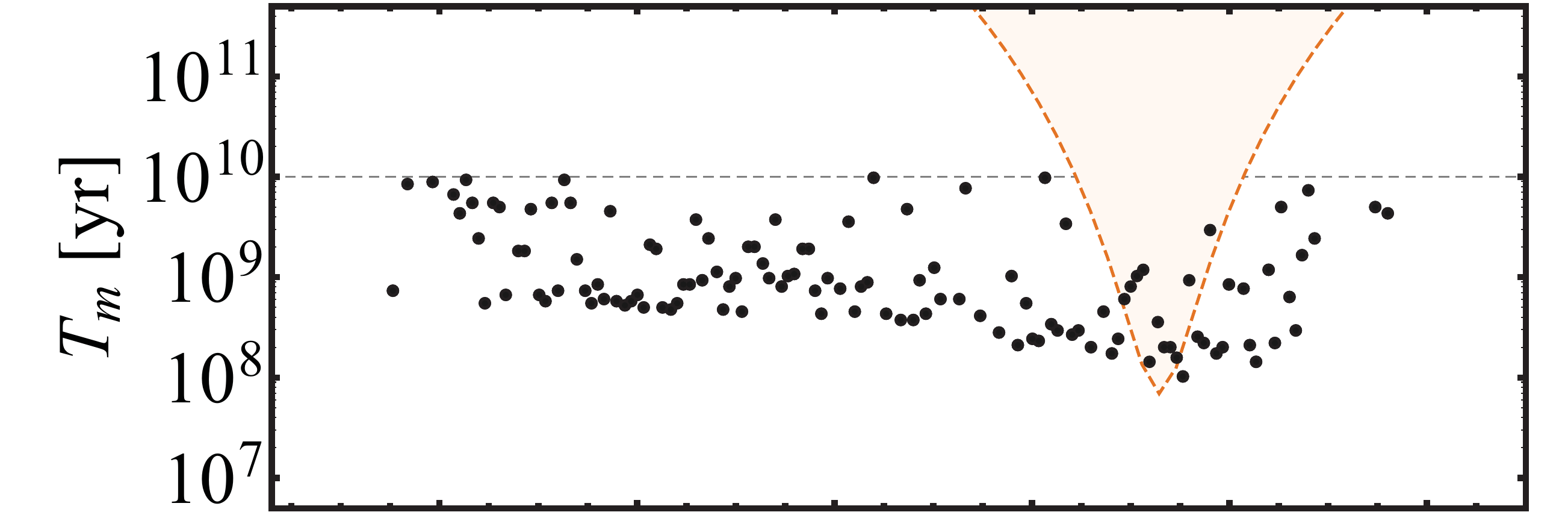}\\
\includegraphics[width=8cm]{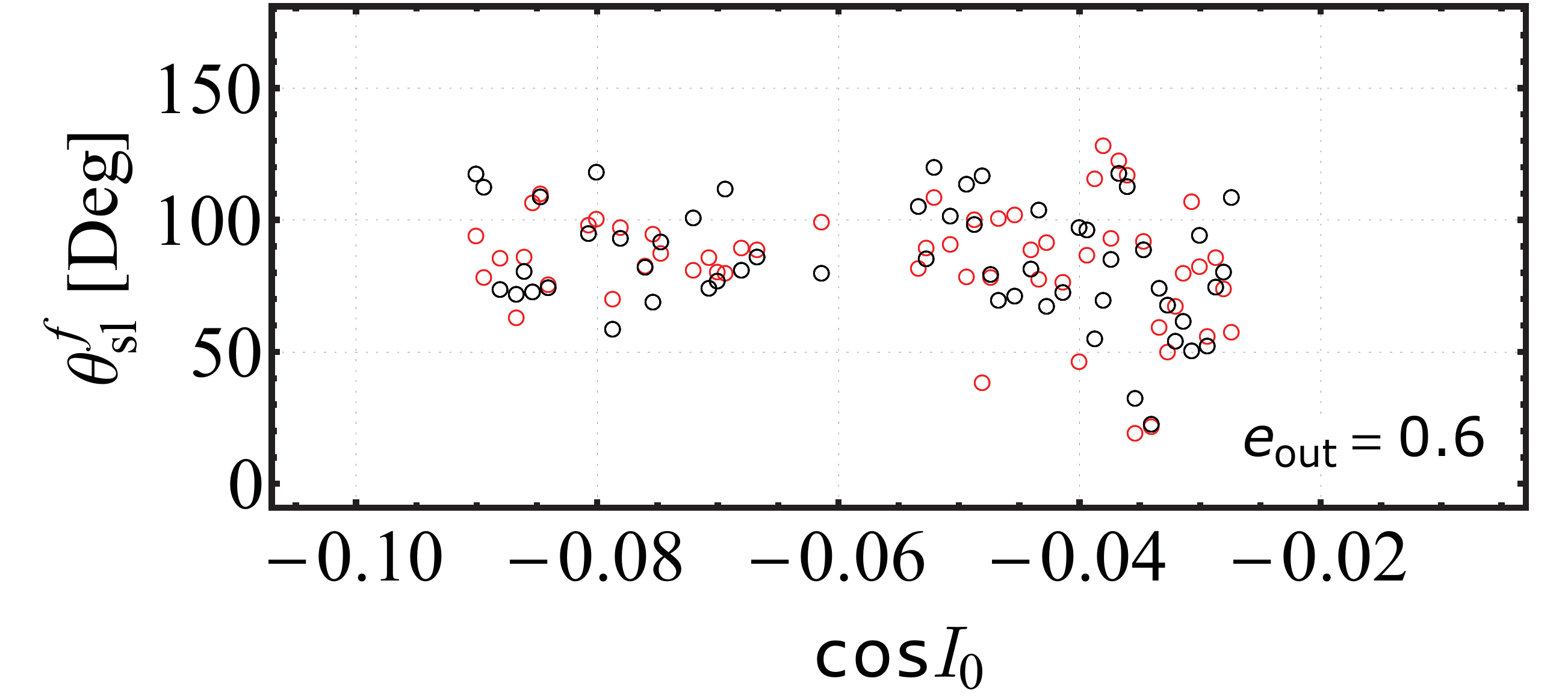}&
\includegraphics[width=8cm]{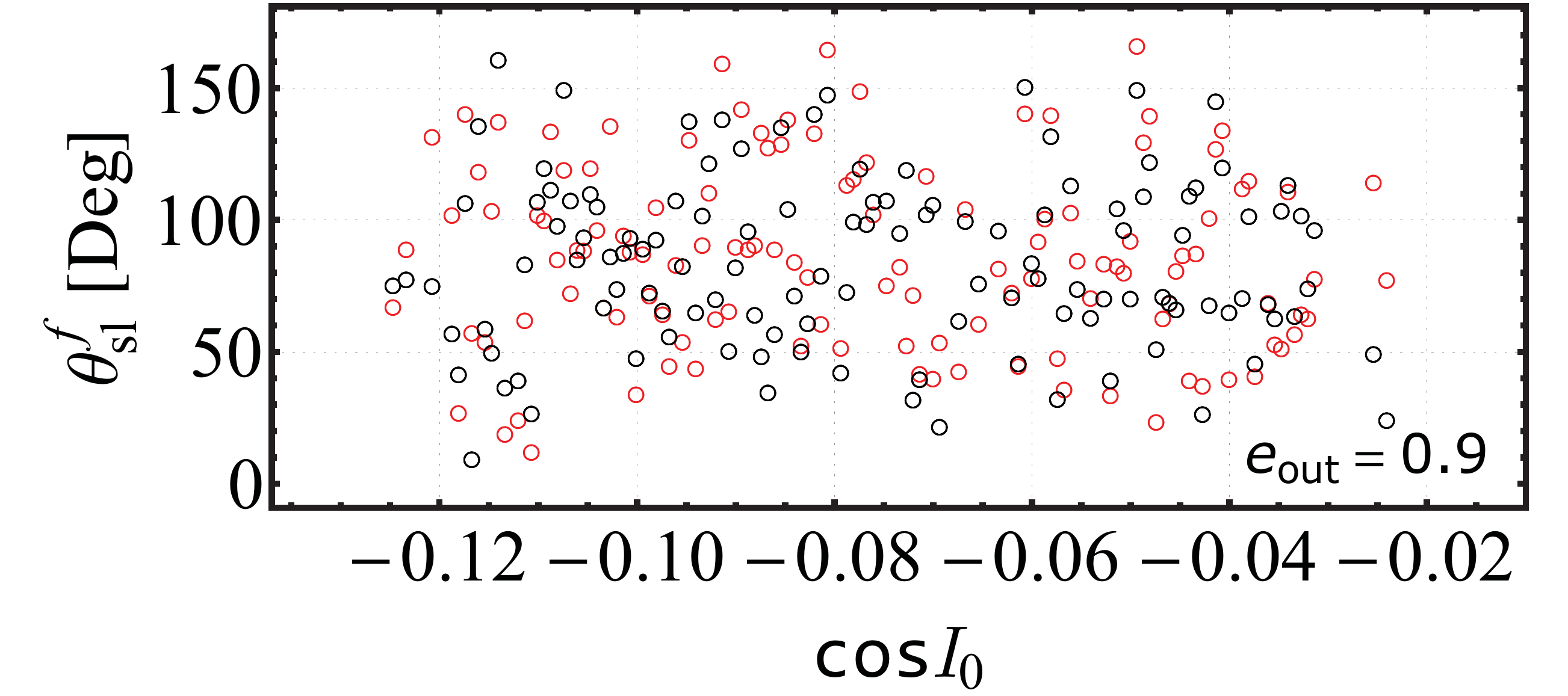}\\
\end{tabular}
\caption{Similar to Figure \ref{fig:merger window quad}, but for four different values of $e_\OUT$.
All four panels have the same $\bar{a}_{\OUT,\eff}\simeq6.65$ (Equation \ref{eq:aout bar}), $m_1=m_3=30M_\odot$, $m_2=20M_\odot$, and $a_0=100\au$.
The semi-major axis of the tertiary companion is $a_\OUT=6700\au$ (top left), $7000\au$ (top right), $8300\au$ (lower left),
and $15200\au$ (lower right), corresponding to $\varepsilon_\oct=0, 0.001, 0.002, 0.006$, respectively.
The orange dashed lines are (quadrupole) analytical expressions (see Equations \ref{eq:EMAX}, \ref{eq:fitting formula origin}).
For each value of $e_\OUT$, the upper panel does not include GW emission, while the middle and bottom panels do
(each dot represents a successful merger event within $10^{10}$ yrs).
}
\label{fig:merger window oct}
\end{figure*}

Figure \ref{fig:merger window oct} shows some examples of the merger windows for
different values of $\varepsilon_\oct$. To illustrated the effect of octupole perturbation,
we consider four cases with the same $\bar{a}_{\OUT,\eff}$ (Equation \ref{eq:aout bar}),
but different $e_\OUT$ ($=0, 0.3, 0.6, 0.9$) and thus different $a_\OUT$.
All parameters in these examples satisfy the double averaging approximation (Equation \ref{eq:DA}).
The initial longitude of the periapse $\omega_\OUT$
is randomly chosen in a range of (0,2$\pi$)
\footnote{
Note that for $e_{\OUT,0}\neq0$ with finite $\varepsilon_\oct$, the orbital evolution
depends not only on $I_0$, but also on the orientation of ${\bf e}_{\OUT,0}$
relative to the initial $\hat{\bf L}$. We can specify this orientation by the
initial longitude of periapse of the outer orbit, $\omega_{\OUT,0}$,
which is angle between ${\bf e}_{\OUT,0}$ and the line of the ascending node of the two (inner and outer) orbits.
When the inner orbit has a finite eccentricity, the orbital evolution will (in general) also depend on
$\omega_{\IN,0}$, the angle between ${\bf e}_{\IN,0}$ and the line of the ascending node.
Recall that in this paper we consider only $e_{\IN,0}\simeq 0$.
}. We find that, when the octupole effect gets stronger,
the eccentricity excitation becomes increasingly erratic as a function of $I_0$,
and more systems have the maximum eccentricity driven to be $1-e_\m\lesssim10^{-4}$.
Consequently, more mergers over the Hubble timescale can be generated, and
the merger window becomes broader noticeably. Because of the erratic variation of $e_\m$,
the merger events are not uniformly spaced in $\cos I_0$.
In this situation, the merger window cannot be described by the fitting formula
(Equation \ref{eq:fitting formula origin}; see the orange dashed curves in Figure \ref{fig:merger window oct}).

\begin{figure*}
\begin{centering}
\begin{tabular}{ccc}
\includegraphics[width=6cm]{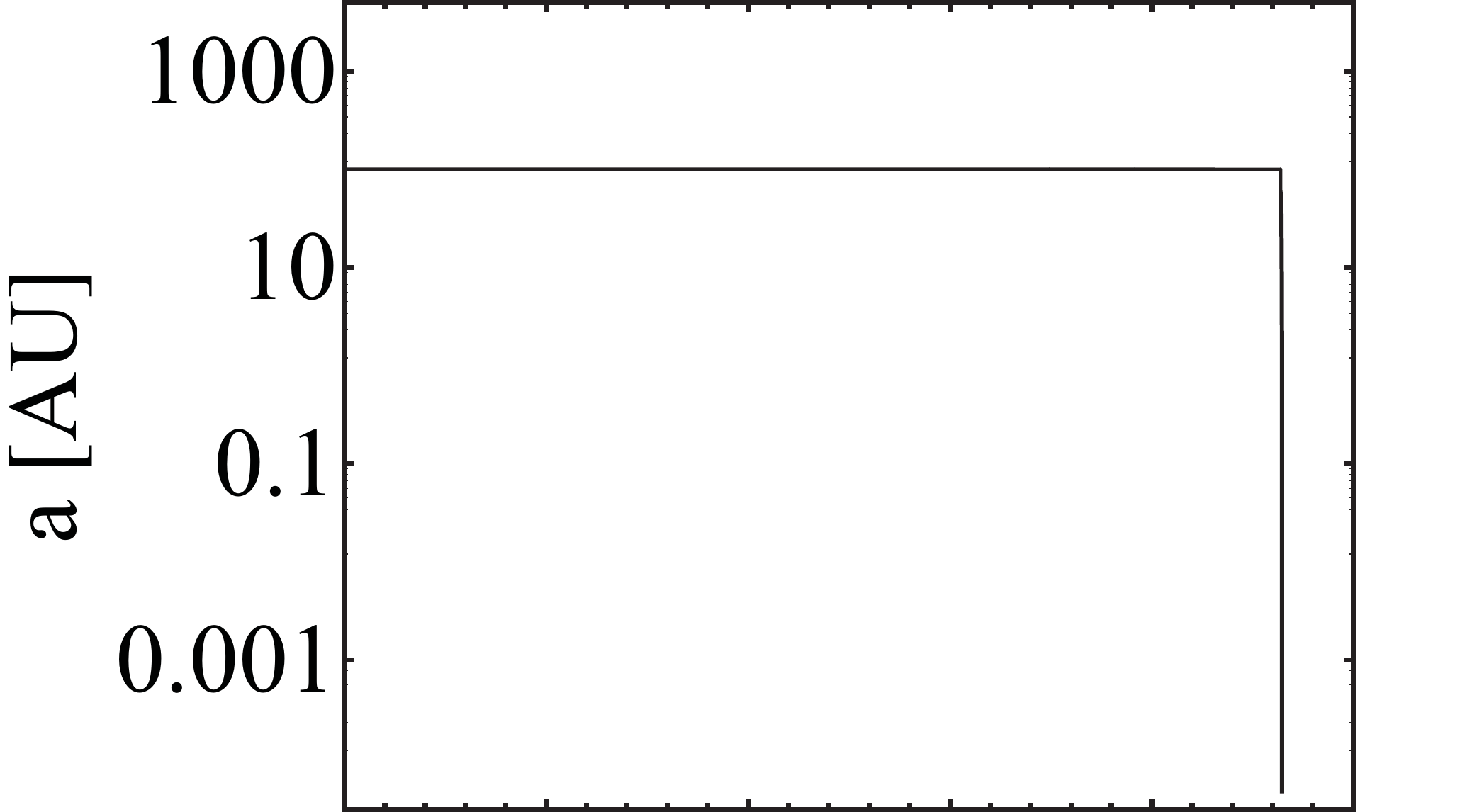}
\includegraphics[width=6cm]{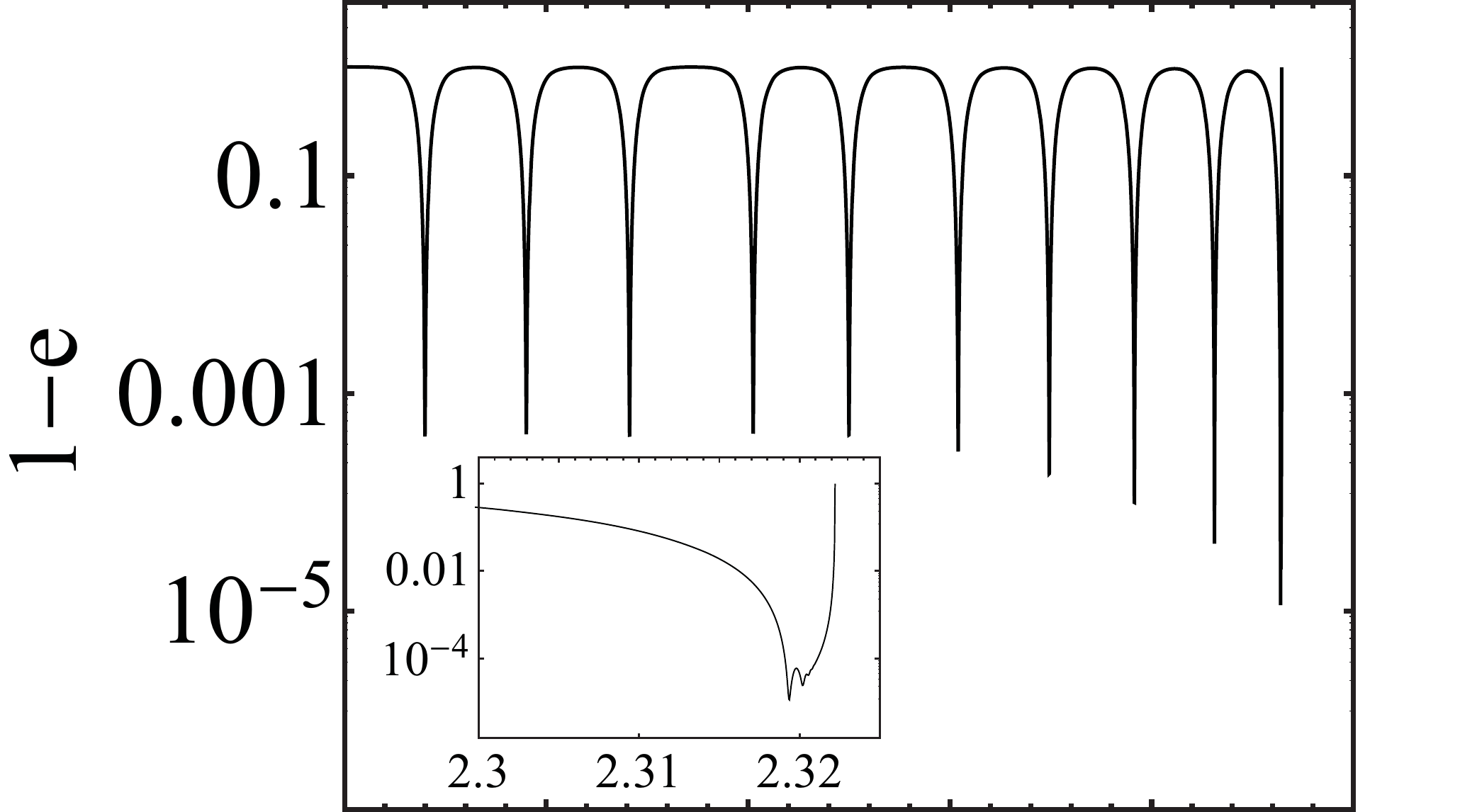}
\includegraphics[width=6cm]{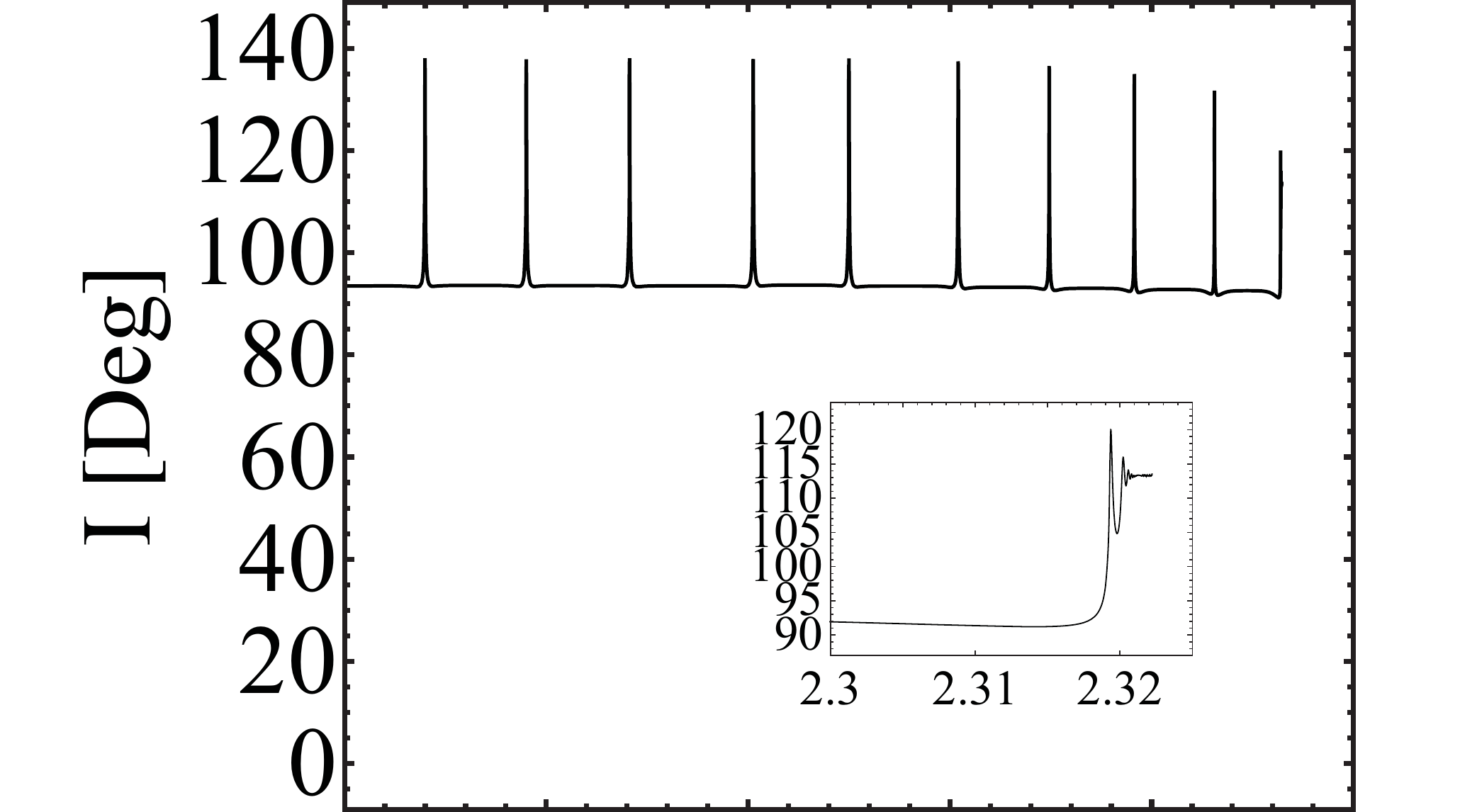}\\
\includegraphics[width=6cm]{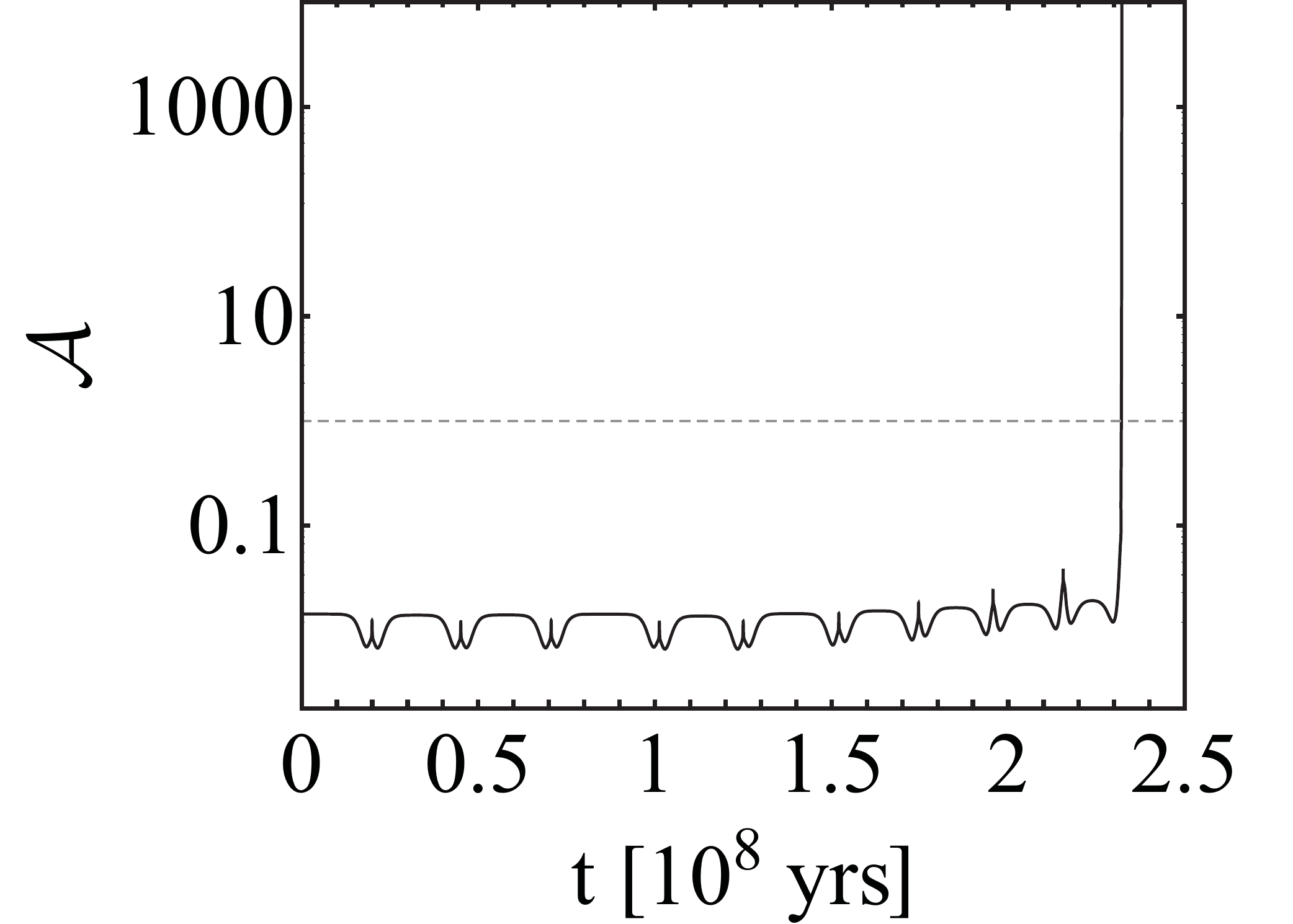}
\includegraphics[width=6cm]{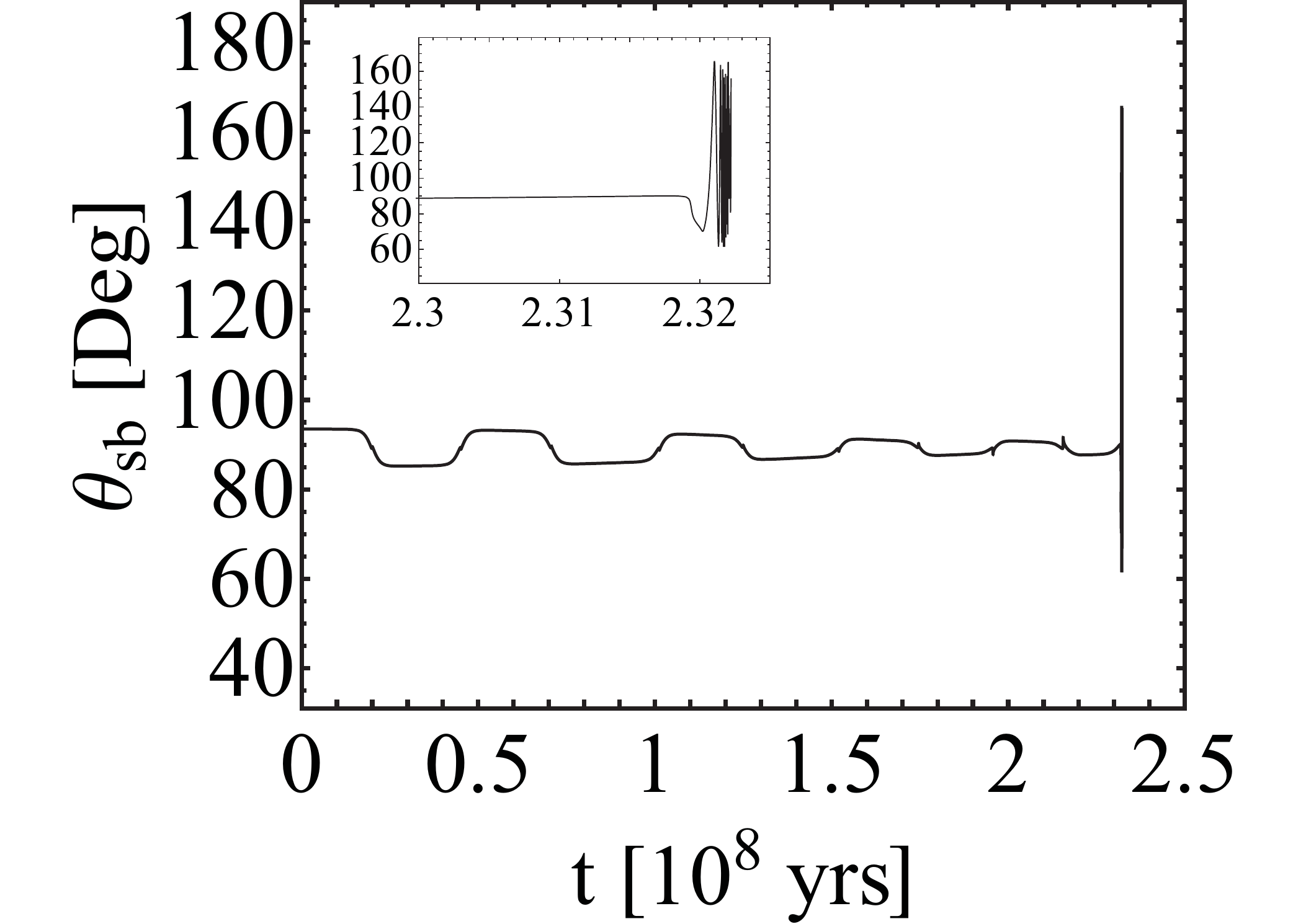}
\includegraphics[width=6cm]{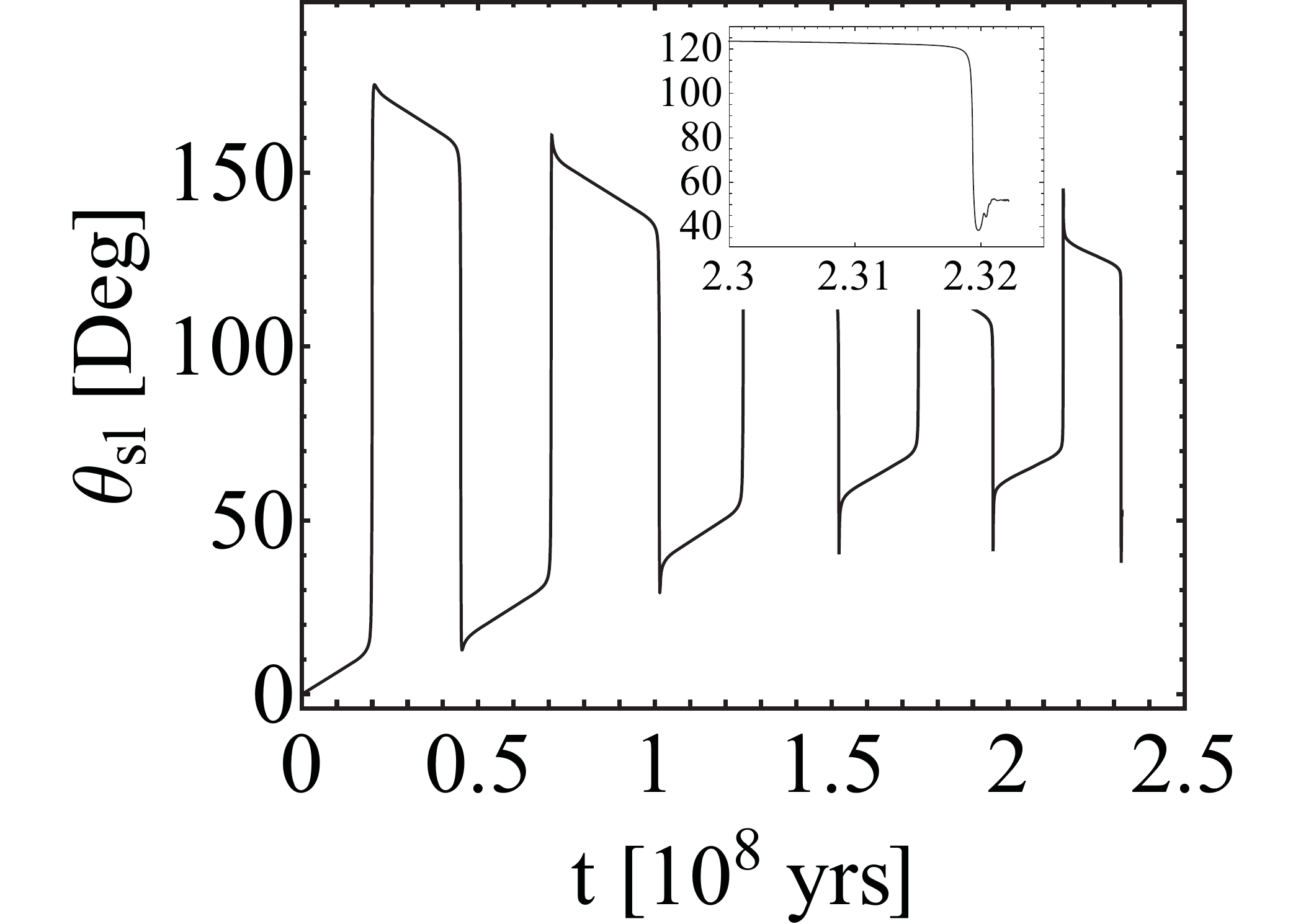}
\end{tabular}
\caption{Sample orbital and spin evolution of a BH binary system with an eccentric tertiary companion.
The top three panels show the time evolution of orbital elements of the inner BH binary and three bottom
panels represent the spin evolution.
Here, the parameters are $m_1=30M_\odot$, $m_2=20M_\odot$, $m_3=30M_\odot$, $a_\OUT=6000\au$, $e_\OUT=0.6$, and the initial
$a_0=100\au$, $e_0=0.001$, $\omega_\OUT=0.7$ rad, $I_0=93.5^\circ$ and $\theta_\SL^0=0^\circ$.
}
\label{fig:OE oct 1}
\end{centering}
\end{figure*}

\begin{figure*}
\begin{centering}
\begin{tabular}{ccc}
\includegraphics[width=6cm]{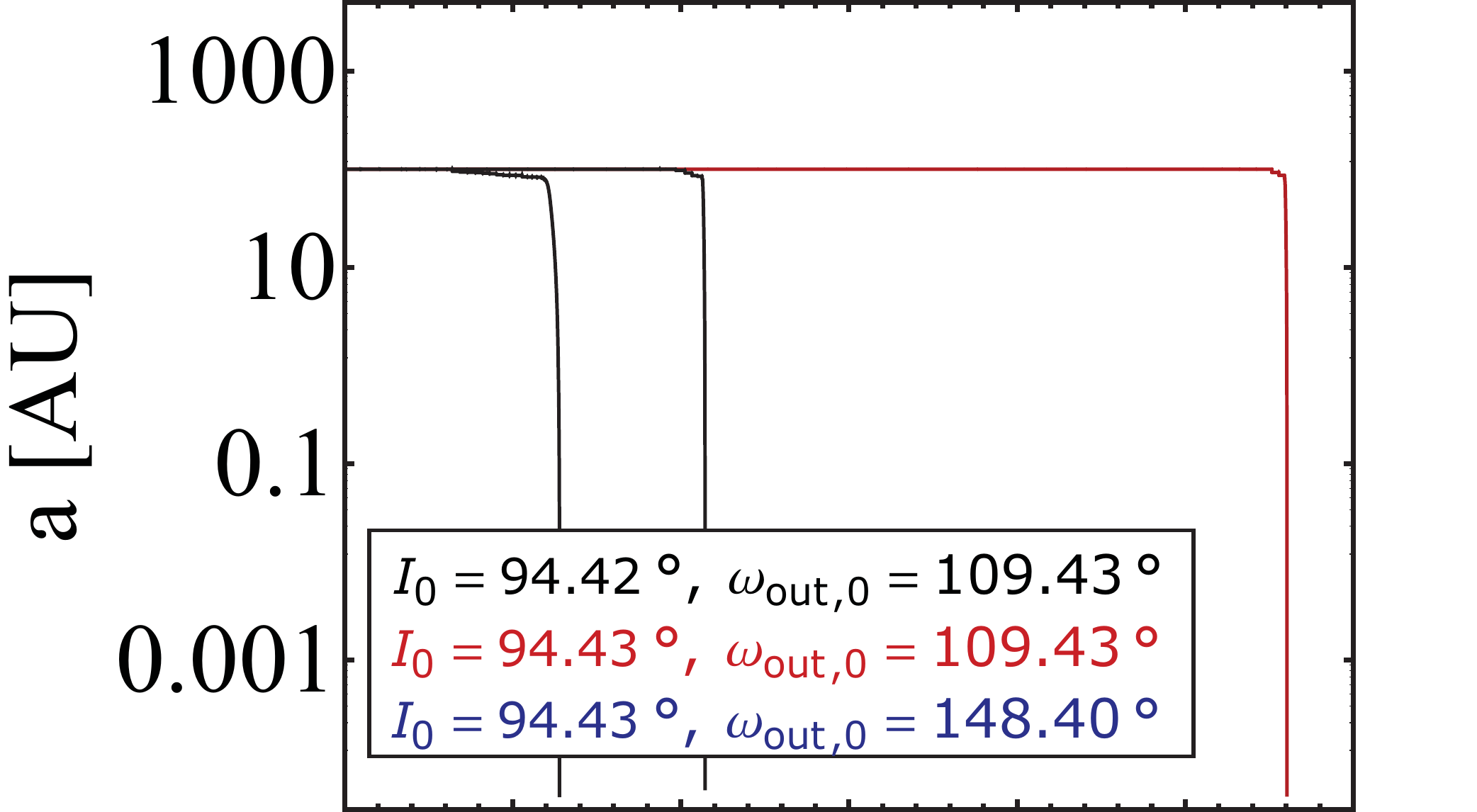}
\includegraphics[width=6cm]{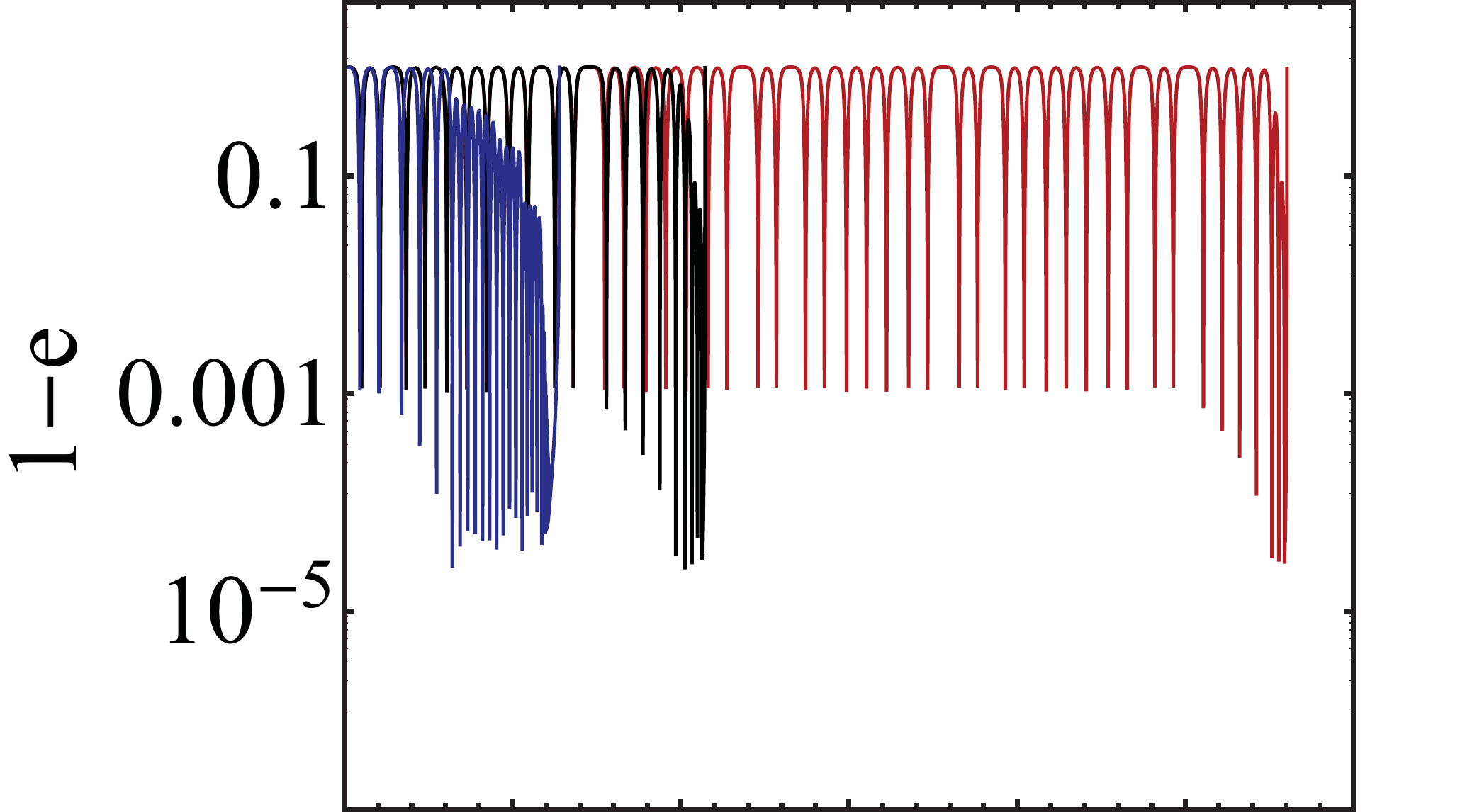}
\includegraphics[width=6cm]{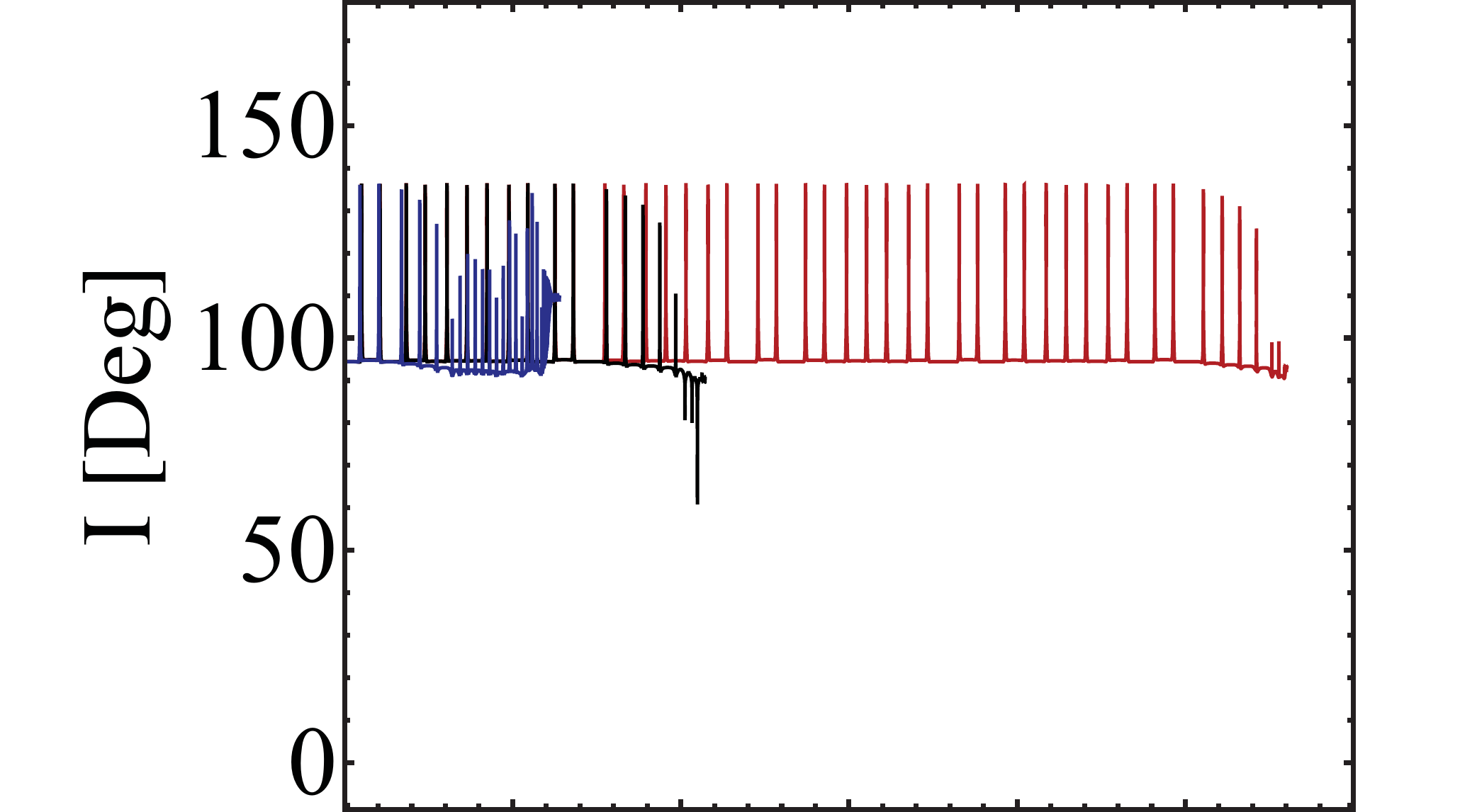}\\
\includegraphics[width=6cm]{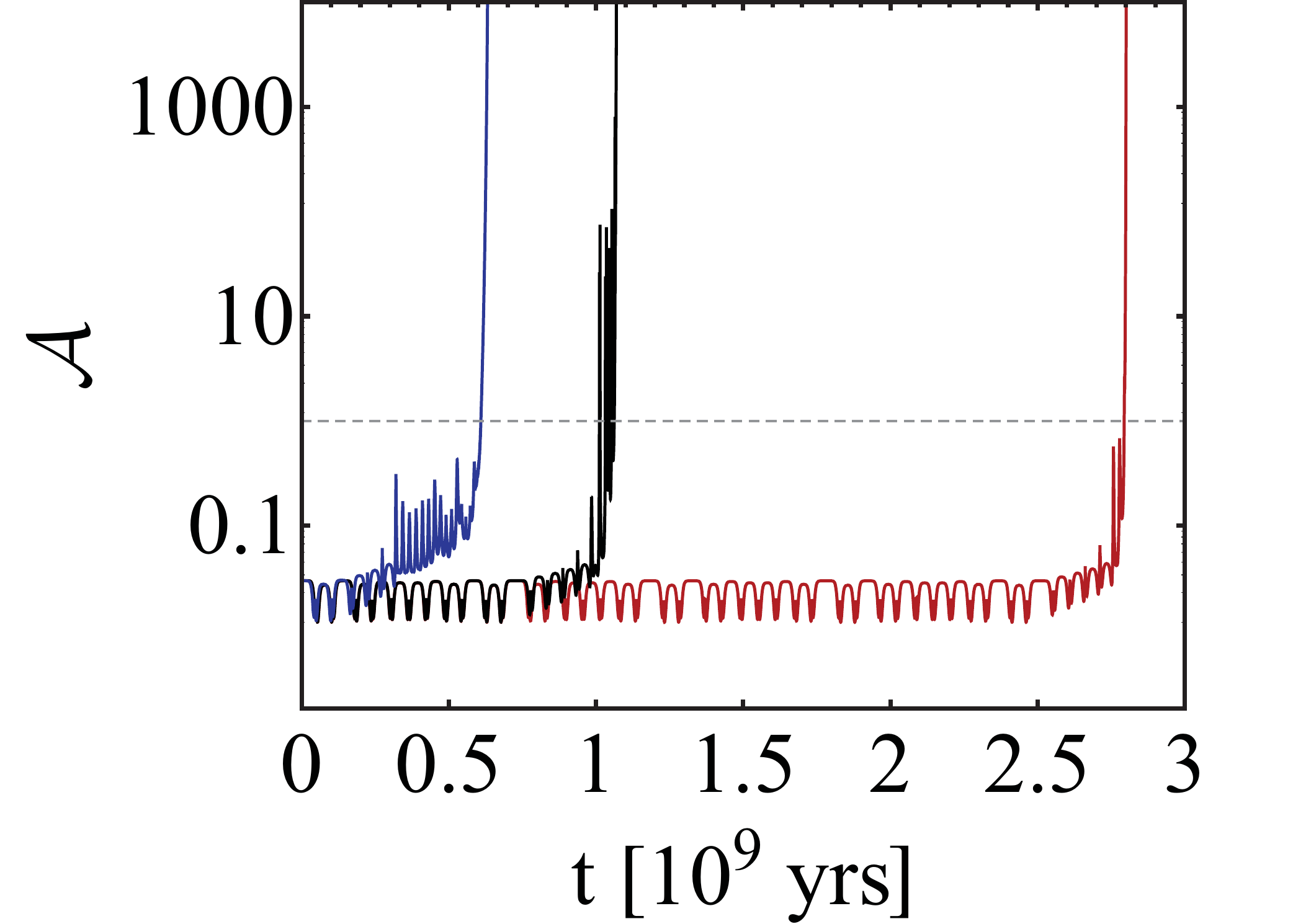}
\includegraphics[width=6cm]{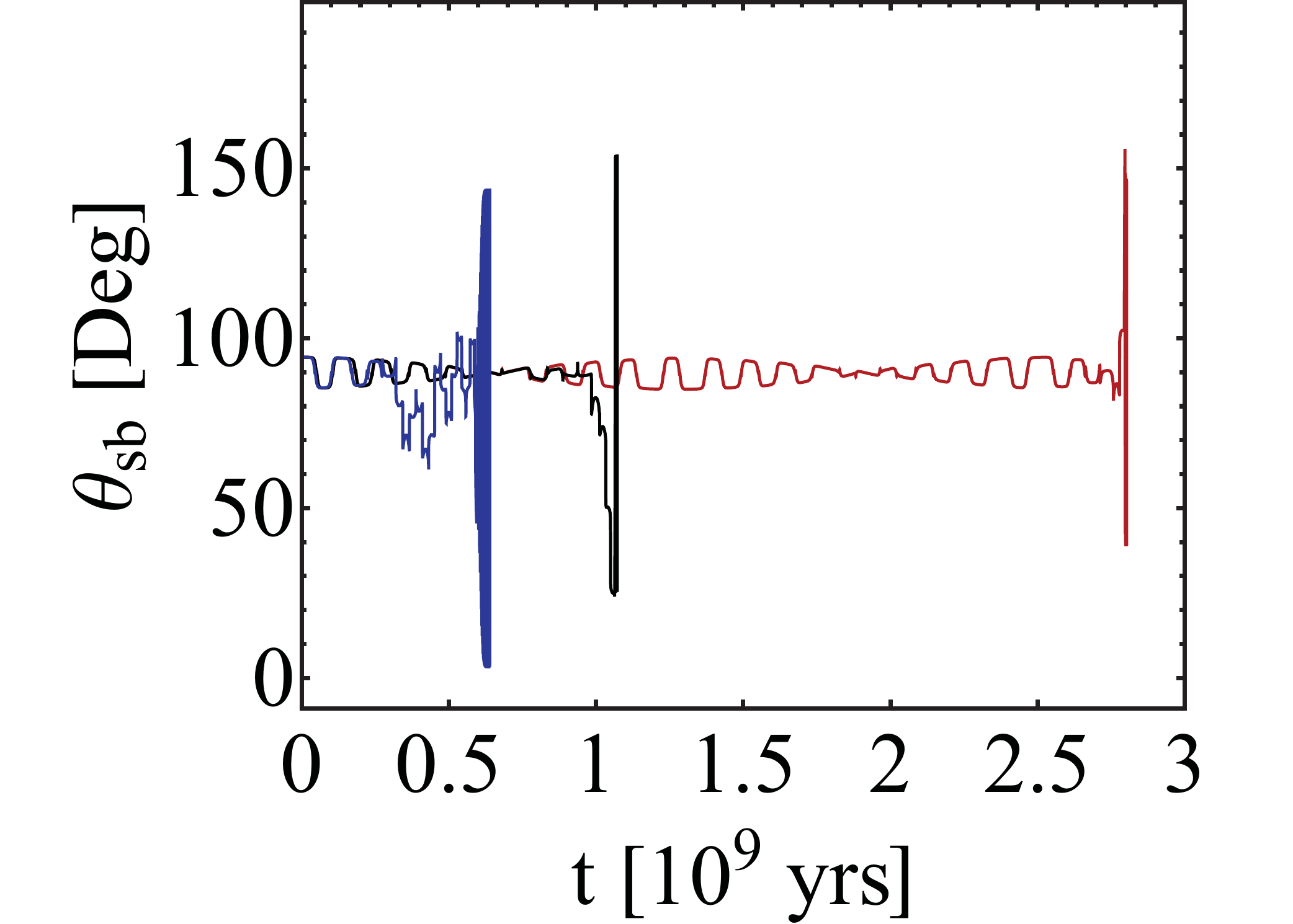}
\includegraphics[width=6cm]{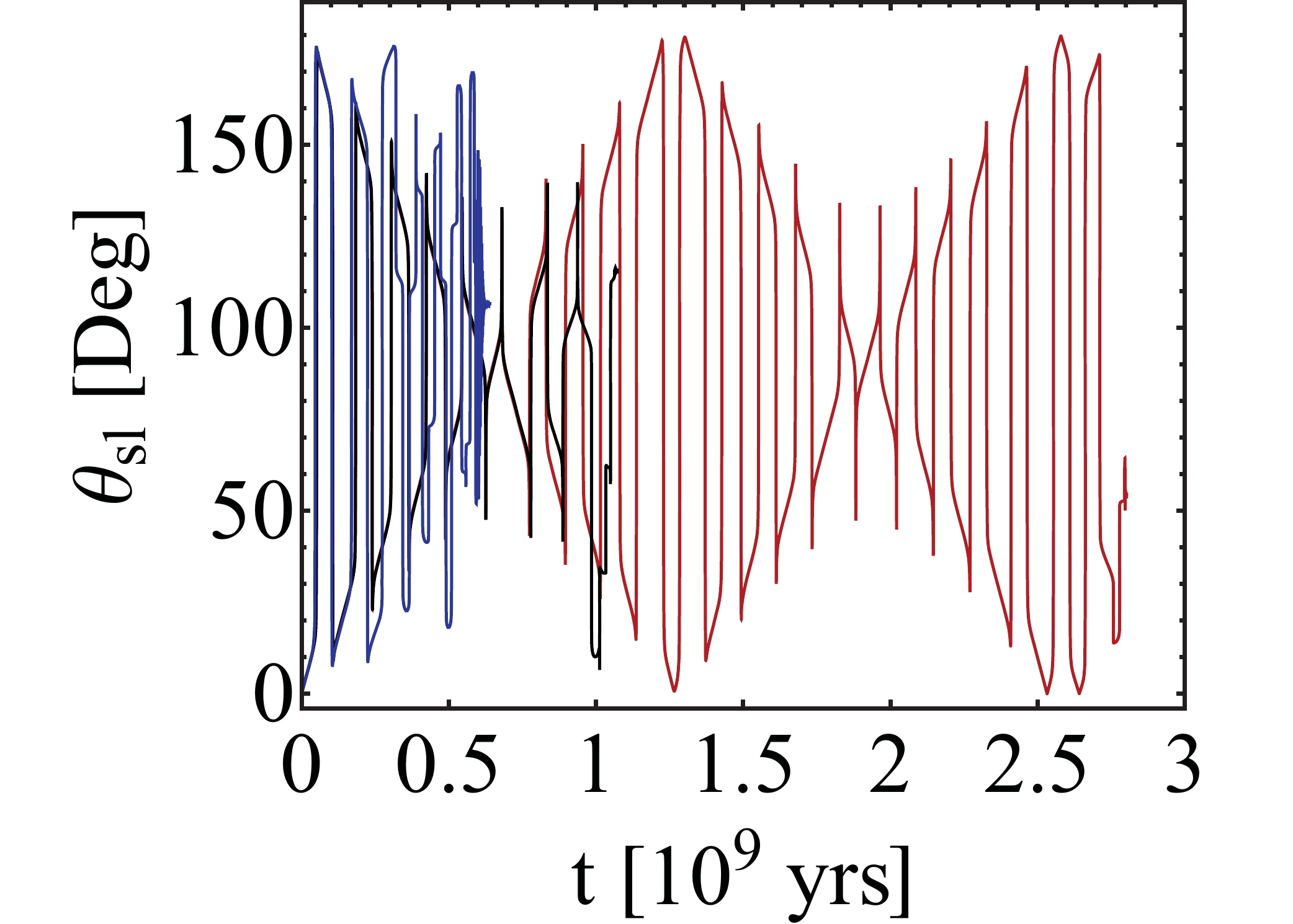}
\end{tabular}
\caption{Same as Figure \ref{fig:OE oct 1}, but for $e_\OUT=0.9$.
The initial parameters $I_0$ and $\omega_{\OUT,0}$ are as labeled.
Because of the chaotic nature of the octupole LK effect, small changes in $I_0$ or $\omega_{\OUT,0}$
lead to very different merger times.
}
\label{fig:OE oct 2}
\end{centering}
\end{figure*}

Figure \ref{fig:OE oct 1} depicts an example of the time evolution of binary merger for $\varepsilon_\oct\neq0$.
Because of the octupole effect, the maximum eccentricities reached in successive (quadrupole) LK cycles increase.
Eventually $e_\m$ becomes sufficiently large and the binary merges quickly.

When $\varepsilon_\oct$ is sufficiently large, the orbital evolution of the inner binary becomes chaotic,
and the evolution shows a strong dependence on the initial conditions \citep[e.g.,][]{Lithwick 2011,Li chaos}.
Figure \ref{fig:OE oct 2} illustrates this chaotic behavior.
We see that the octupole-induced extreme eccentricity excitation occurs in an irregular manner,
shortening or extending the time for mergers.
As a result, $T_\mathrm{m}$ exhibits an irregular dependence on $I_0$, as seem in Figure \ref{fig:merger window oct}.

\begin{figure}
\begin{centering}
\includegraphics[width=8.5cm]{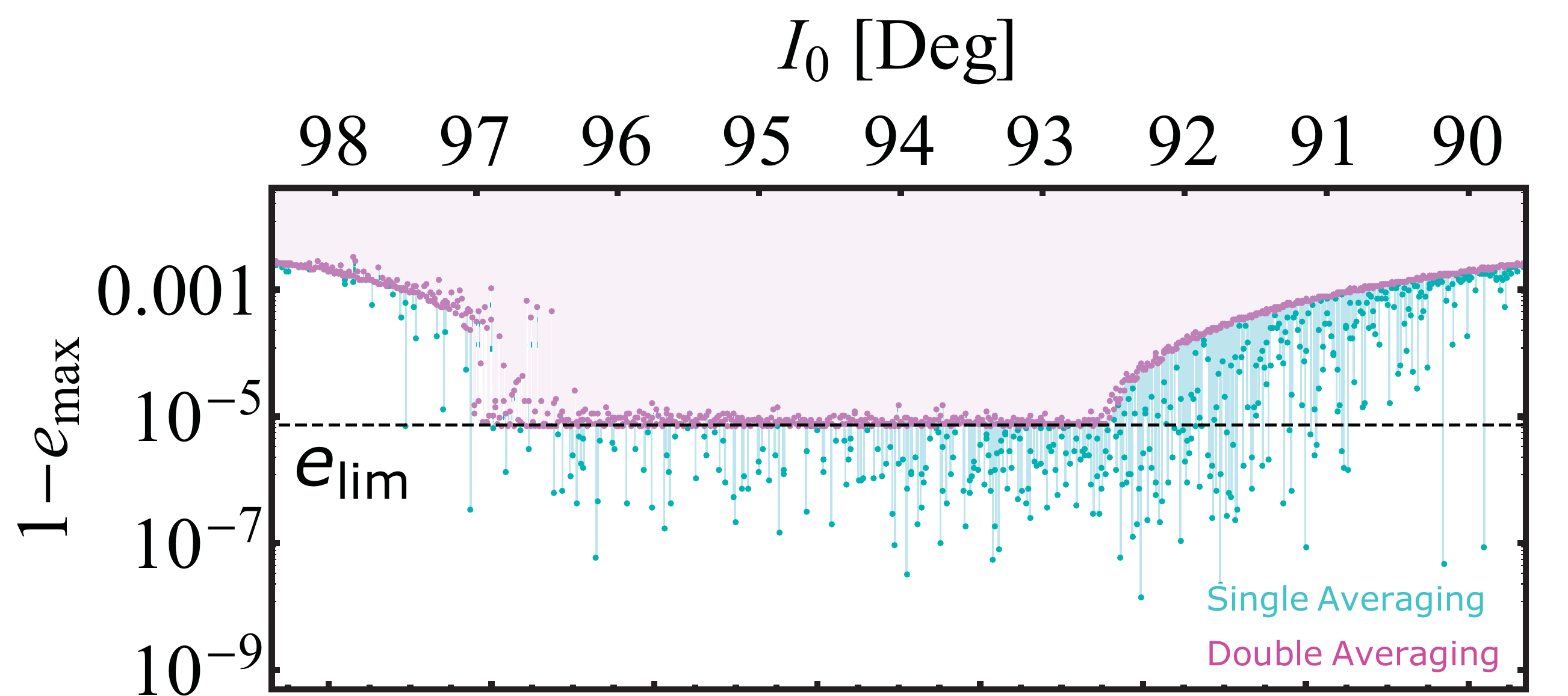}\\
\includegraphics[width=8.5cm]{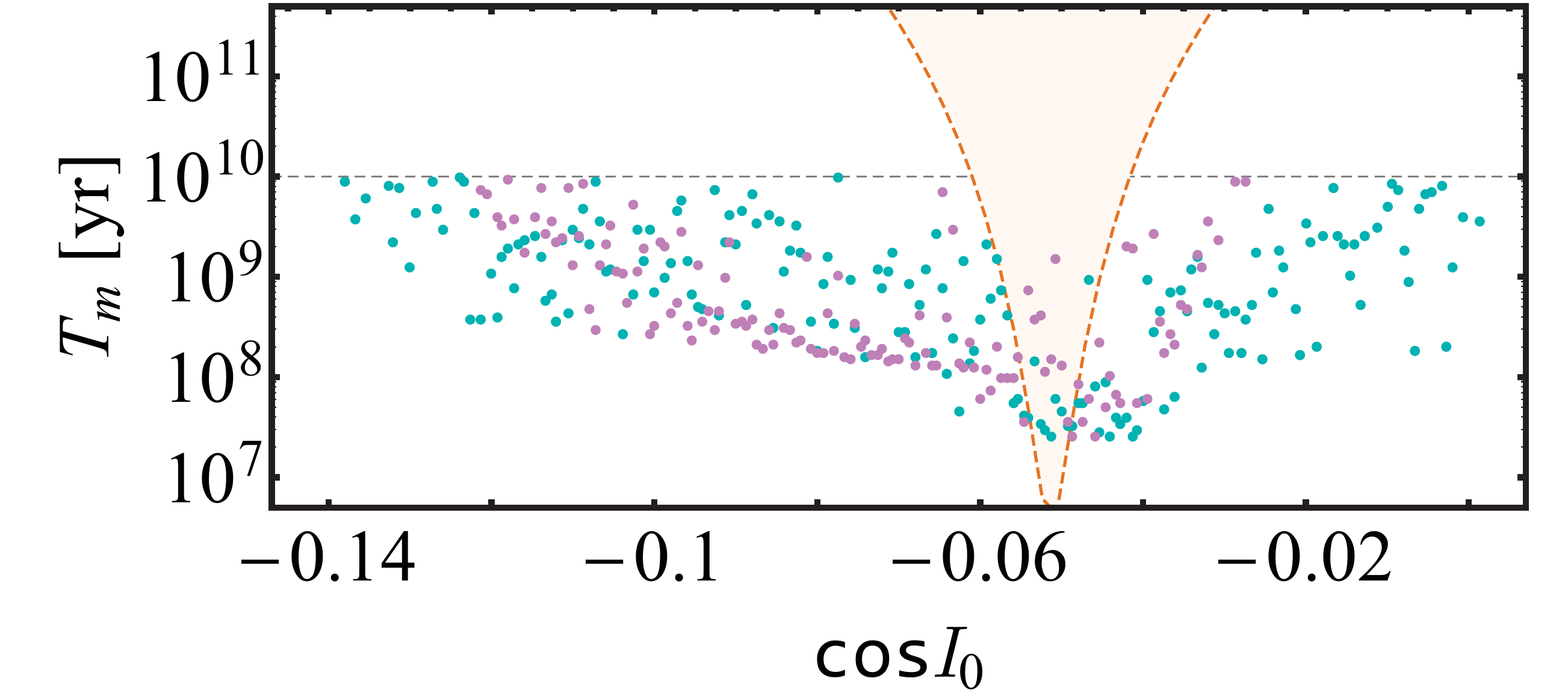}
\caption{Eccentricity excitation (no GW emission; upper panel) and merger time
(with GW emission; lower panel) as a function of $\cos I_0$.
The parameters are the same as the $e_\OUT=0.9$ case of Figure \ref{fig:merger window oct},
except for a closer companion ($a_\OUT=12800\au$).
The cyan and purple dots are obtained from calculations based on the single and double averaged equations, respectively.
}
\label{fig:merger window SA}
\end{centering}
\end{figure}

The merger windows shown in Figure \ref{fig:merger window oct} are based on the double-averaged secular equations.
For close and more eccentric companions, these double-averaged equations break down,
and we can use single-averaged equations (see Section \ref{sec 2 1 2}). Figure \ref{fig:merger window SA}
shows a sample numerical results for the merger windows computed using single-averaged equations and
double-averaged equations. Here, $\bar{a}_{\OUT,\eff}$ is chosen to be relatively small ($\simeq5.6$),
where the system lies near the boundary of parameter regions between single and double averaging.
The upper panel shows the eccentricity excitation for a grid of $I_0$ values when the system is non-dissipative
(i.e. GW emission is turned off).
We find that a portion of the systems computed from single-averaged equations
can reach higher $e_\m$, even beyond $e_\li$ (which is derived from the double averaging approximation).
In particular, when gravitational radiation is included,
a larger number of mergers with $T_\mathrm{m}\lesssim$ (a few)$\times10^8$ yrs occur due to the extreme $e_\m$,
as depicted in the lower panel;
such rapid mergers are relatively rare in the calculations based on the double-averaged equations.

\begin{figure*}
\begin{centering}
\begin{tabular}{cccc}
\includegraphics[width=8.5cm]{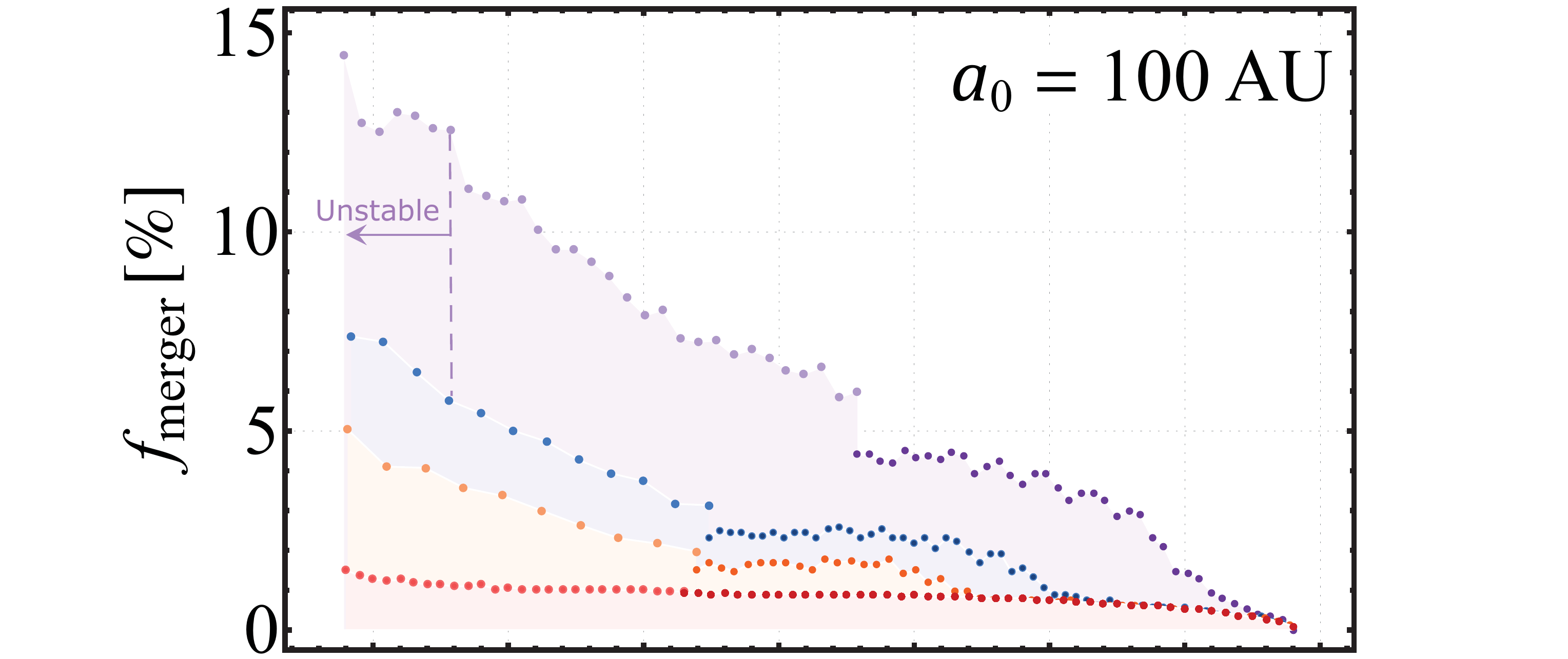}
\includegraphics[width=8.5cm]{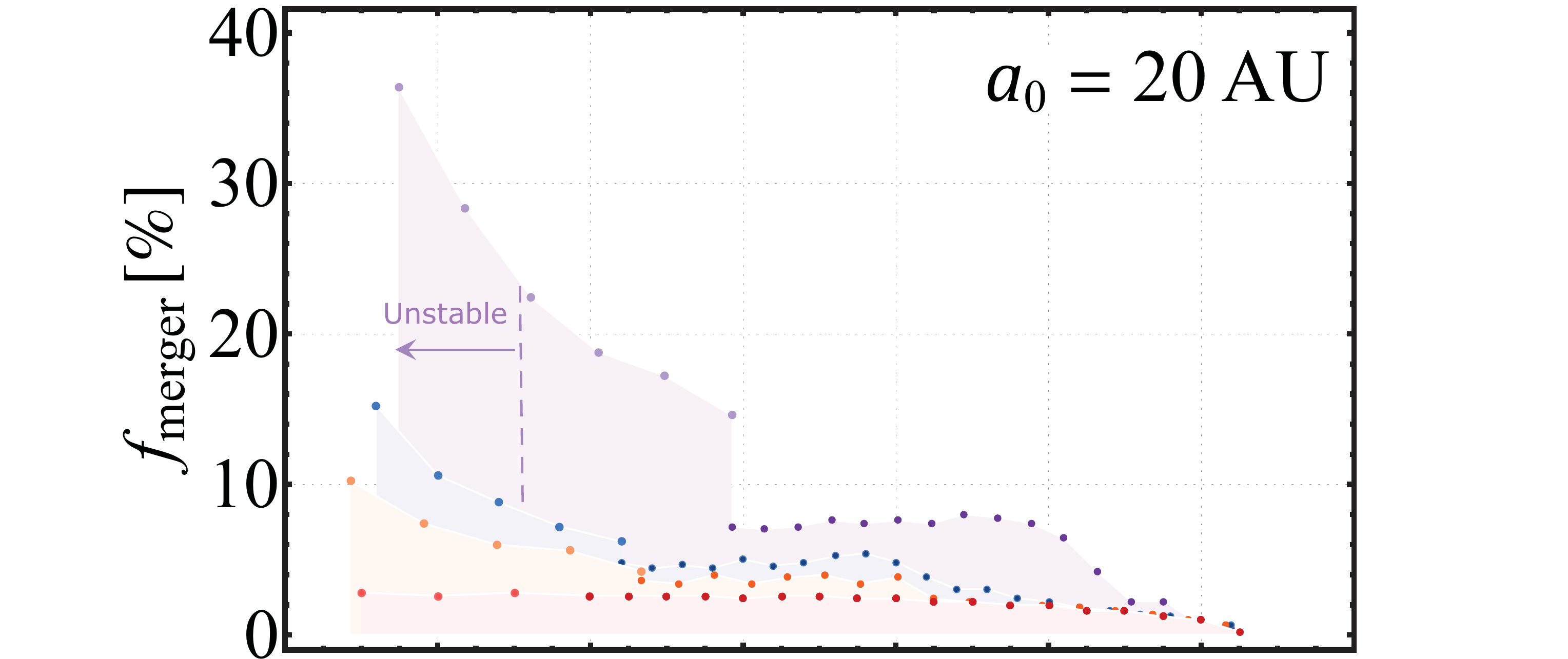}\\
\includegraphics[width=8.5cm]{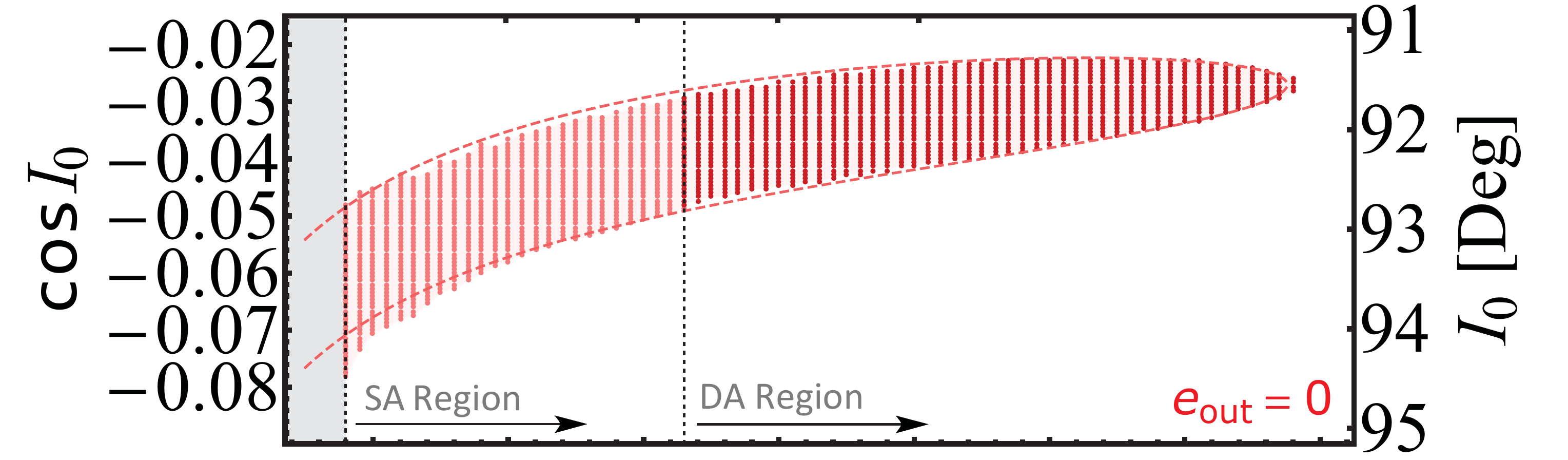}
\includegraphics[width=8.5cm]{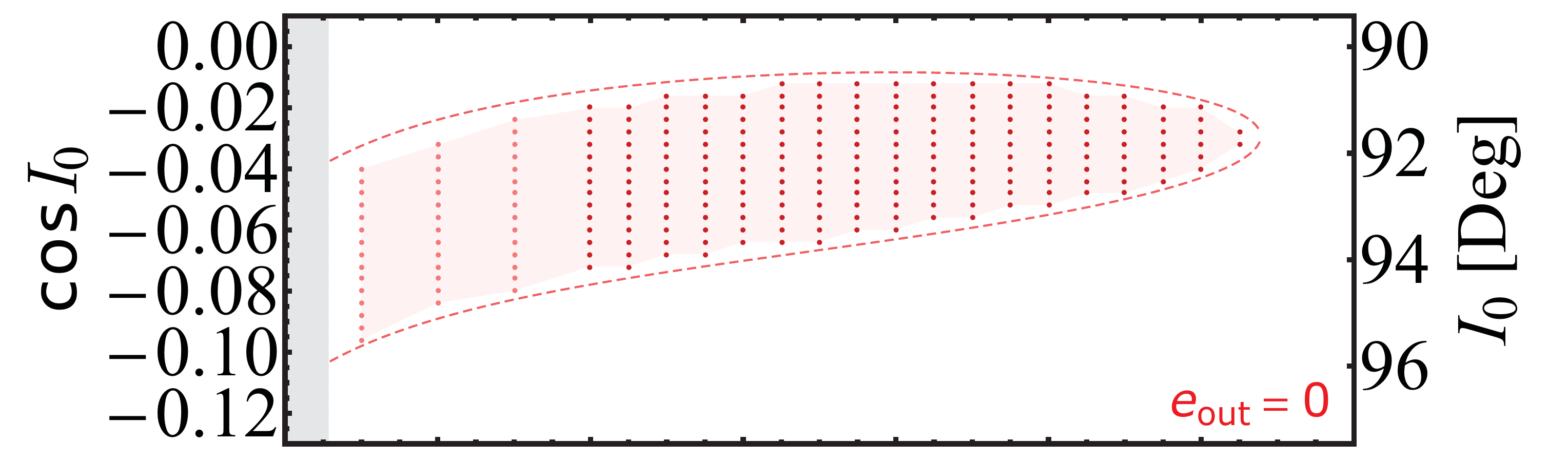}\\
\includegraphics[width=8.5cm]{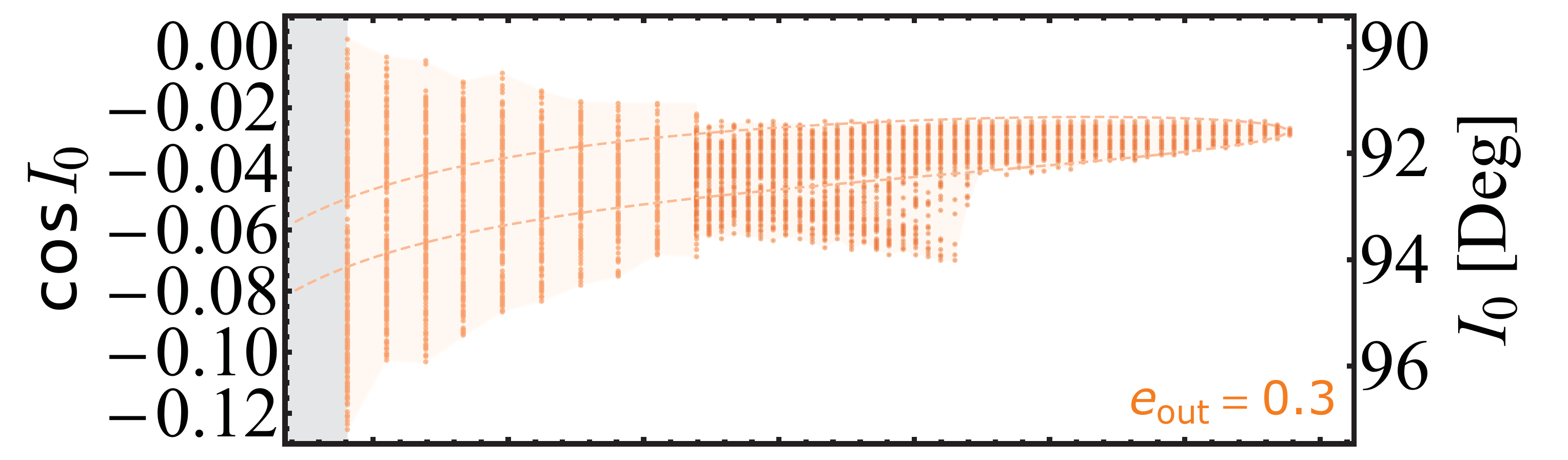}
\includegraphics[width=8.5cm]{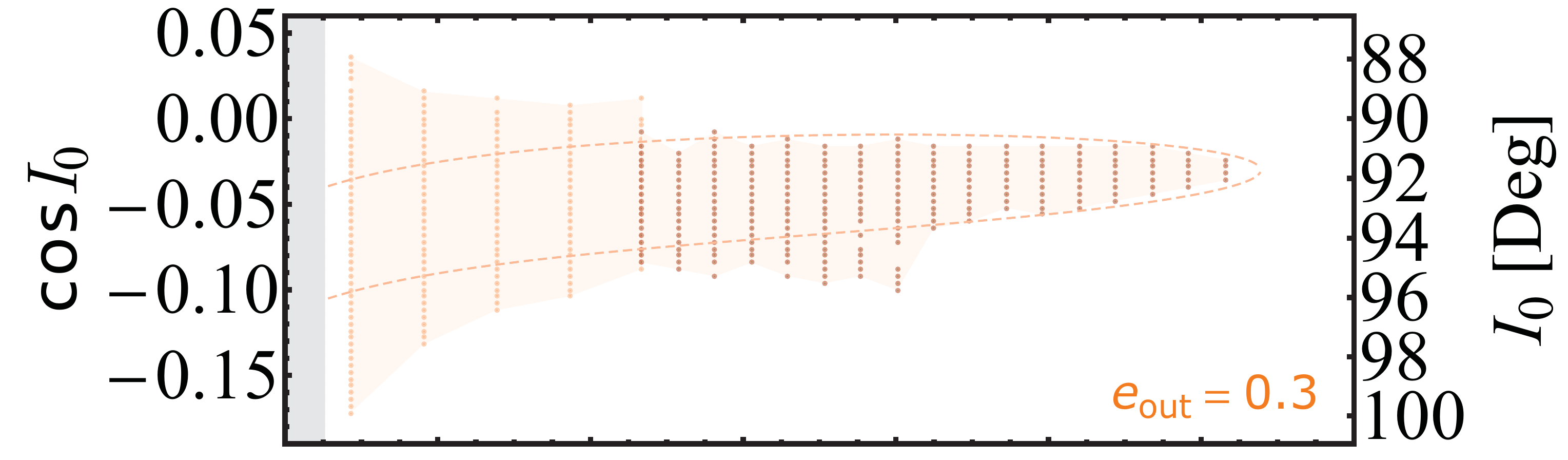}\\
\includegraphics[width=8.5cm]{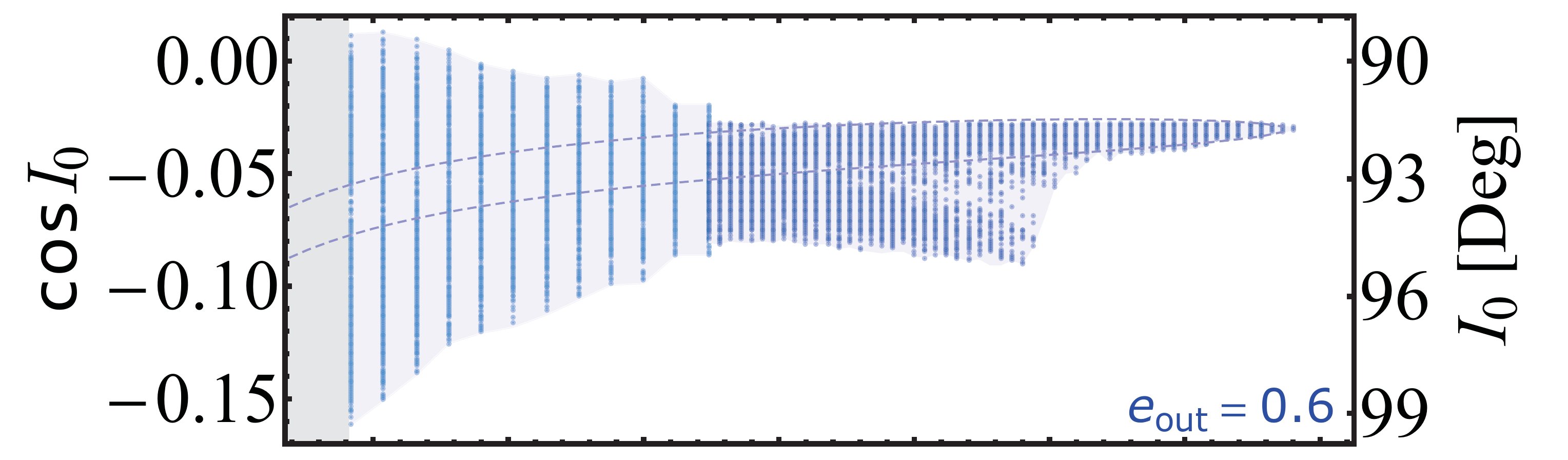}
\includegraphics[width=8.5cm]{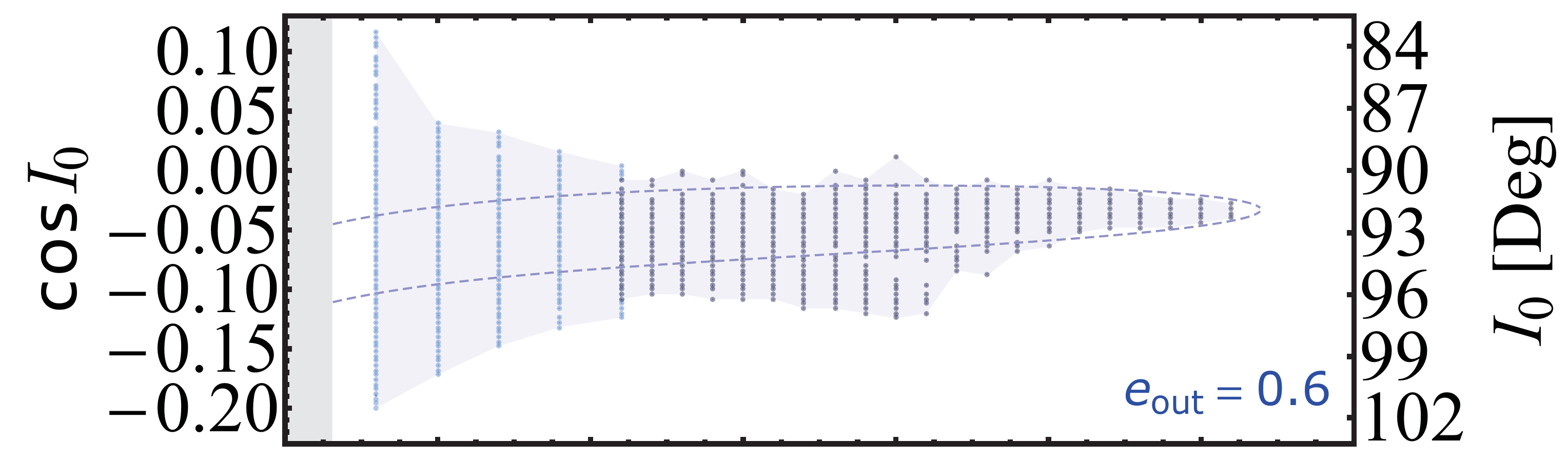}\\
\includegraphics[width=8.5cm]{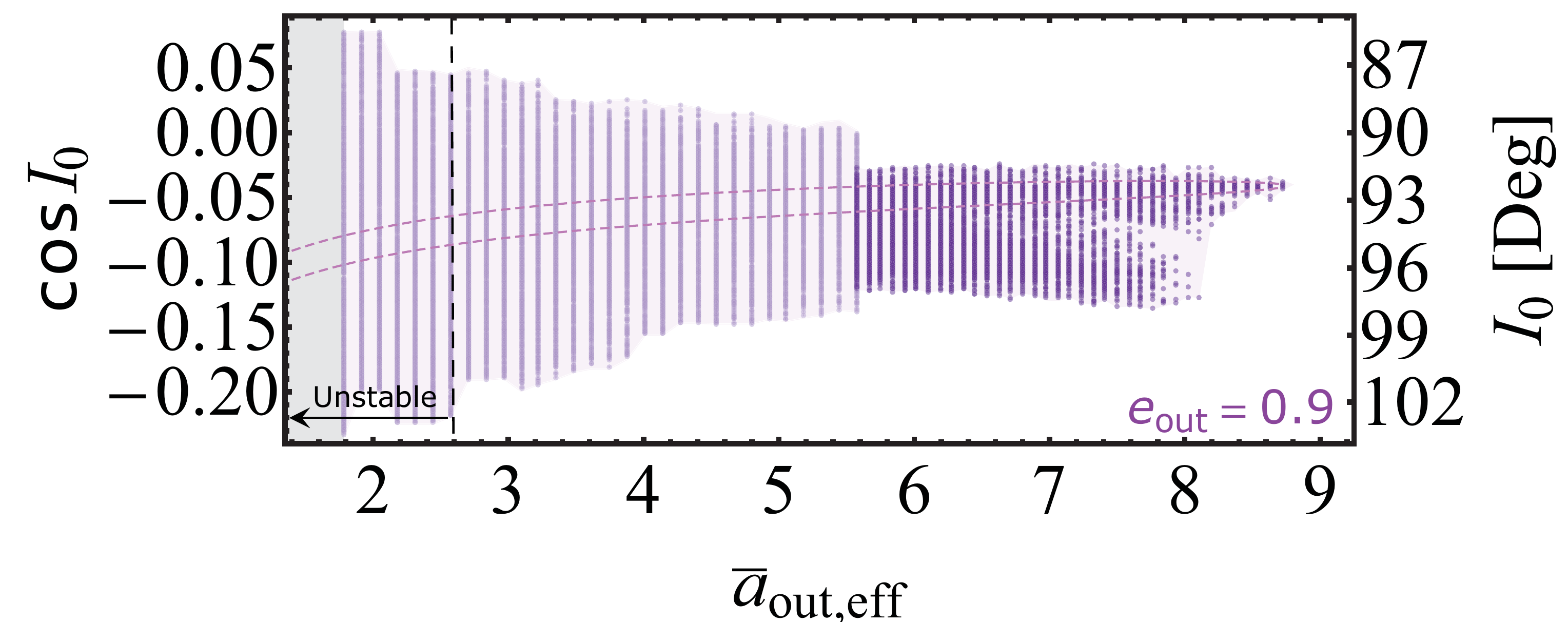}
\includegraphics[width=8.5cm]{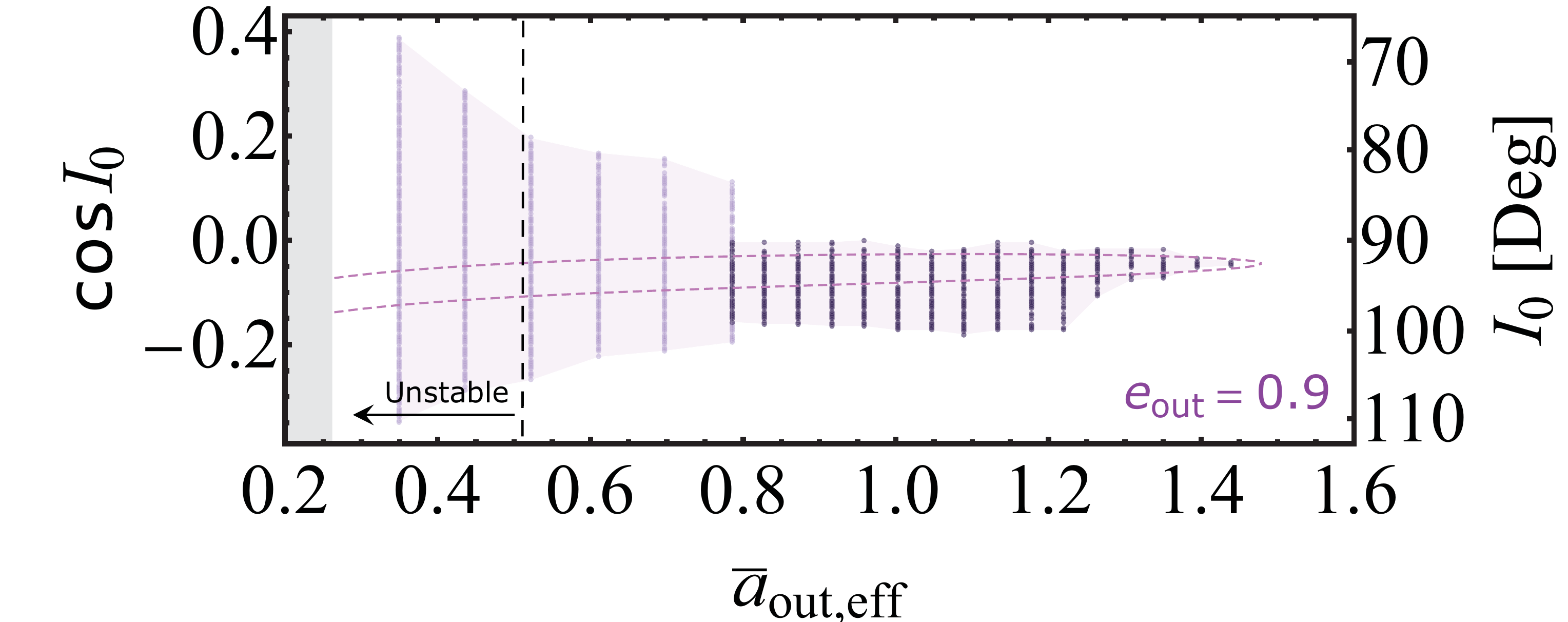}
\end{tabular}
\caption{Similar to Figure \ref{fig:merger fraction quad }, but include the octupole effect.
We fix $m_3=30M_\odot$ but vary $e_\OUT$ as labeled.
The left panels are for $a_0=100\au$, and right panels are for $a_0=20\au$.
In the bottom four panels of each column, each dot represents a successful merger event within the Hubble time ($10^{10}$ yrs).
Note that when $e_\OUT\neq0$, merger events can have an irregular distribution as a function of $\cos I_0$.
}
\label{fig:merger fraction oct}
\end{centering}
\end{figure*}

Figure \ref{fig:merger fraction oct} shows the merger windows and merger fractions
as a function of $\bar{a}_{\OUT,\eff}$ (see Equation \ref{eq:aout bar}) for
different values of $e_\OUT$. In our calculations,
the orientation of the initial $\textbf{e}_\OUT$ (for a given $I_0$) is random
(i.e. $\omega_{\OUT,0}$ is uniformly distributed in $0-2\pi$).
We see that, for a given $e_\OUT$, the merger window shows an general trend of widening as
$\bar{a}_{\OUT,\eff}$ decreases, resulting in an increase of $f_\merger$.
Moreover, for the same value of $\bar{a}_{\OUT,\eff}$ (thus the same quadrupole effect),
the merger window and merger fraction can be very different for different $e_\OUT$.
In general, the larger the eccentricity $e_\OUT$, the stronger the octupole effect, and therefore the wider the window.
Compared to the analytical expressions based on the quadrupole approximation (see Section \ref{sec 3 1}),
$f_\merger$ can be enhanced by a factor of a few.
Note that for some values of $\bar{a}_{\OUT,\eff}$,
the irregular distribution of merger events inside the merger window appears; this results from
the chaotic behaviors of the octupole-order LK oscillations
(see also the examples in Figure \ref{fig:merger window oct},  particulary the $e_\OUT=0.6$ case).

\subsection{Scaling Relations for Quadrupole Systems and Application to Neutron Star Binaries}
\label{sec 3 3}
Although in this paper we focus on BH binaries, similar analysis can be done for neutron star (NS) binary mergers
induced by tertiary companions. A double NS merger event (GW170817) has recently been detected through
gravitational waves and electromagnetic radiation \citep[e.g.,][]{Abbott 2017d}.
We can expect more such detections in the future.

NS binaries differ from BH binaries in that the NS mass is much smaller than the BH mass,
and thus for the same initial $a_\IN=a_0$ ($\gtrsim 1 \au$), a larger eccentricity excitation is required
to induce NS binary merger. Moreover, since the masses of the two members of NS binaries are
typically quite similar, the octupole LK effect is negligible ($\varepsilon_\oct\simeq0$).
Therefore, the mergers of NS binaries in the presence of distant companions
can be well described in the quadrupole approximation (see Section \ref{sec 3 1}).
Thus, the maximum eccentricity required for mergers (within time $T_\mathrm{crit}$) can be obtained form Equation (\ref{eq:fitting formula}),
and the required initially mutual inclination can be calculated using Equation (\ref{eq:EMAX})
by replacing $e_\m$ with $e_\mathrm{m}$.
In another word, the merger window and merger fraction for NS binaries can be calculated analytically
(Equation \ref{eq:merger fraction}),
without the need for numerical integrations of the single or double
averaged equations.

\begin{figure}
\begin{centering}
\includegraphics[width=9cm]{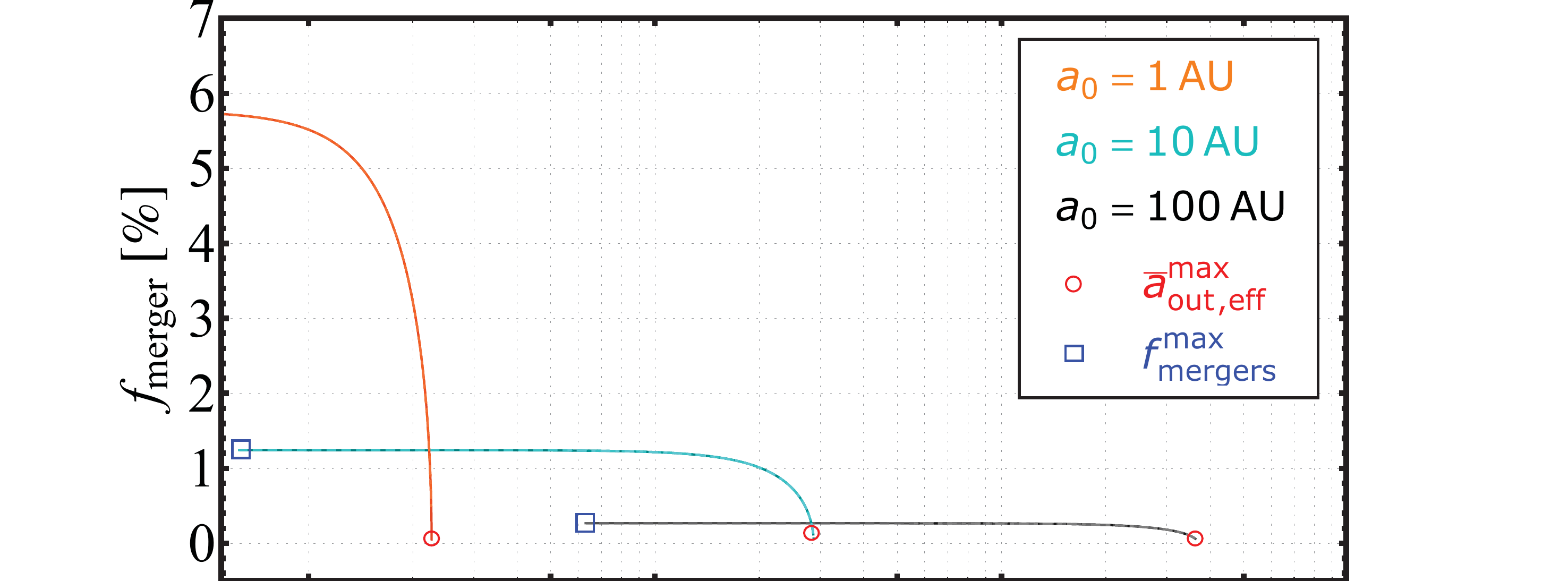}\\
\includegraphics[width=9cm]{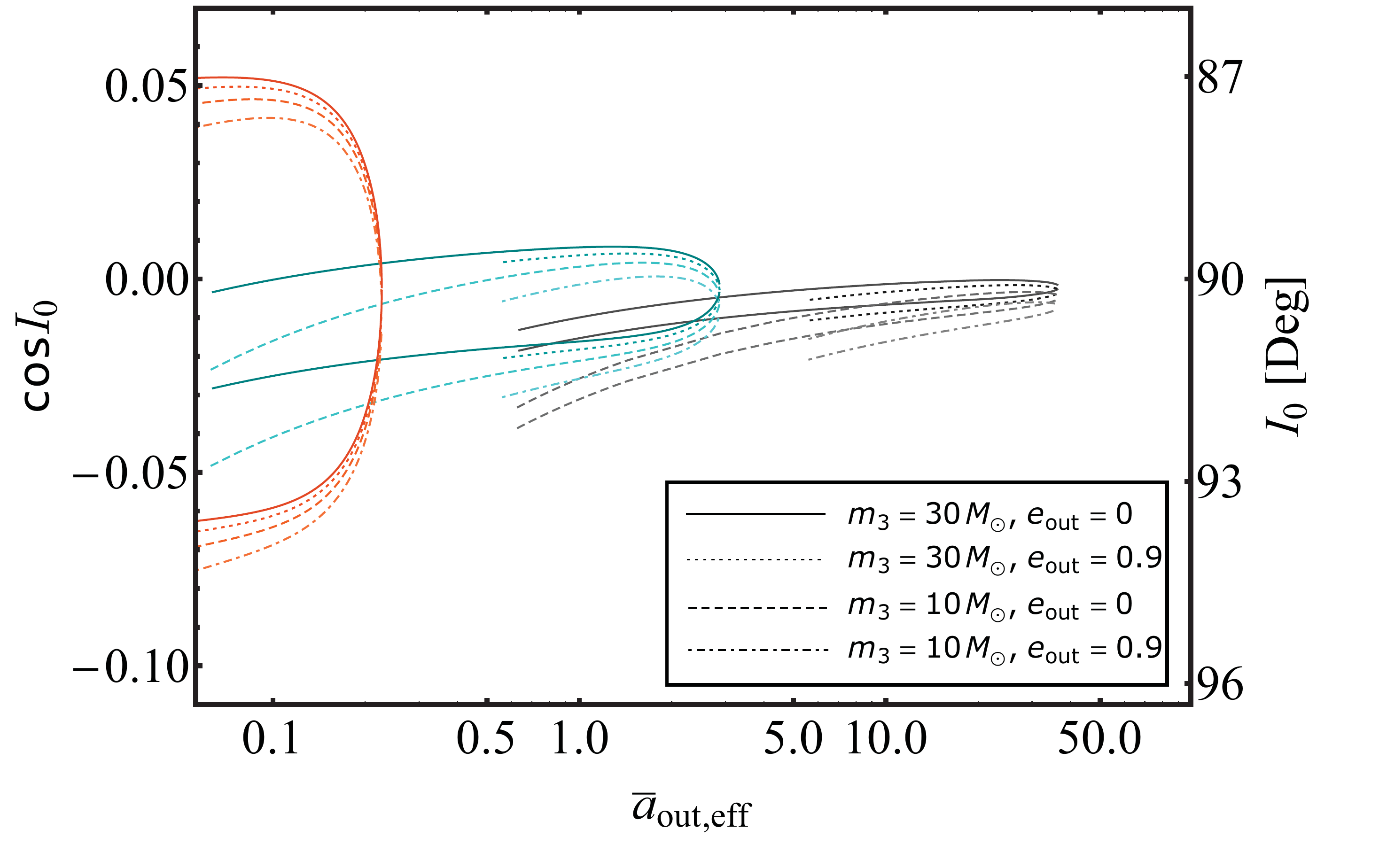}
\caption{
Merger fractions and merger windows as a function of $\bar{a}_{\OUT,\eff}$ (see Equation \ref{eq:aout bar}) for neutron star binaries.
The binary parameters are $m_1=m_2=1.4M_\odot$,
and the tertiary companion parameters are as indicated.
These results are obtained analytically using Equations (\ref{eq:EMAX}), (\ref{eq:fitting formula}) and (\ref{eq:merger fraction}).
Each curve terminates on the left at the instability limit (Equation \ref{eq:stability}).
The maximum value of $\bar{a}_{\OUT,\eff}$ to have merger is denoted by $\bar{a}_{\OUT,\eff}^\m$,
and the maximum value of $f_\merger$ (which occurs at small $\bar{a}_{\OUT,\eff}$)
is denoted by $f_\merger^\m$.
}
\label{fig:merger fraction NS}
\end{centering}
\end{figure}

Figure \ref{fig:merger fraction NS} presents the results of merger window and merger fraction
for equal-mass NS binaries ($m_1=m_2=1.4M_\odot$). All systems shown satisfy the stability criterion.
We choose three different initial semi-major axes ($a_0=1\au$, $10\au$ and $100\au$)
\footnote{
Note that for $a_0\lesssim$ a few AU, binary interactions, such as mass transfer and common envelope
phase may be important. We include the $a_0=1\au$ case to illustrate the dependence of our results
on $a_0$.}.
For each NS binary, we consider a variety of tertiary bodies (different $m_3$ and $e_\OUT$, as labeled).
We find that, for a given $a_0$, different $m_3$ and $e_\OUT$ (with the same $\bar{a}_{\OUT,\eff}$) affect the position of merger window
(i.e. the range of $\cos I_0$) but not the value of $f_\merger$ (cf. Figure \ref{fig:merger fraction quad }).
On the other hand,
the merger windows and fractions have strong dependence on the initial semi-major axis
(e.g. $f_\merger$ for $a_0=1\au$ is about $100$ times larger than that for $a_0=100\au$).
This is because for the small $a_0$, the induced eccentricity in the LK oscillations
does not have to be too large to produce mergers within $10^{10}$ yrs
(e.g. $1-e_\mathrm{m}\simeq 10^{-3}$ for $a_0=1\au$, $10^{-4}$ for $a_0=10\au$ and $10^{-6}$ for $a_0=100\au$, respectively).
In addition, the range of $\bar{a}_{\OUT,\eff}$ producing merger is different for different $a_0$.

\begin{figure}
\begin{centering}
\includegraphics[width=9cm]{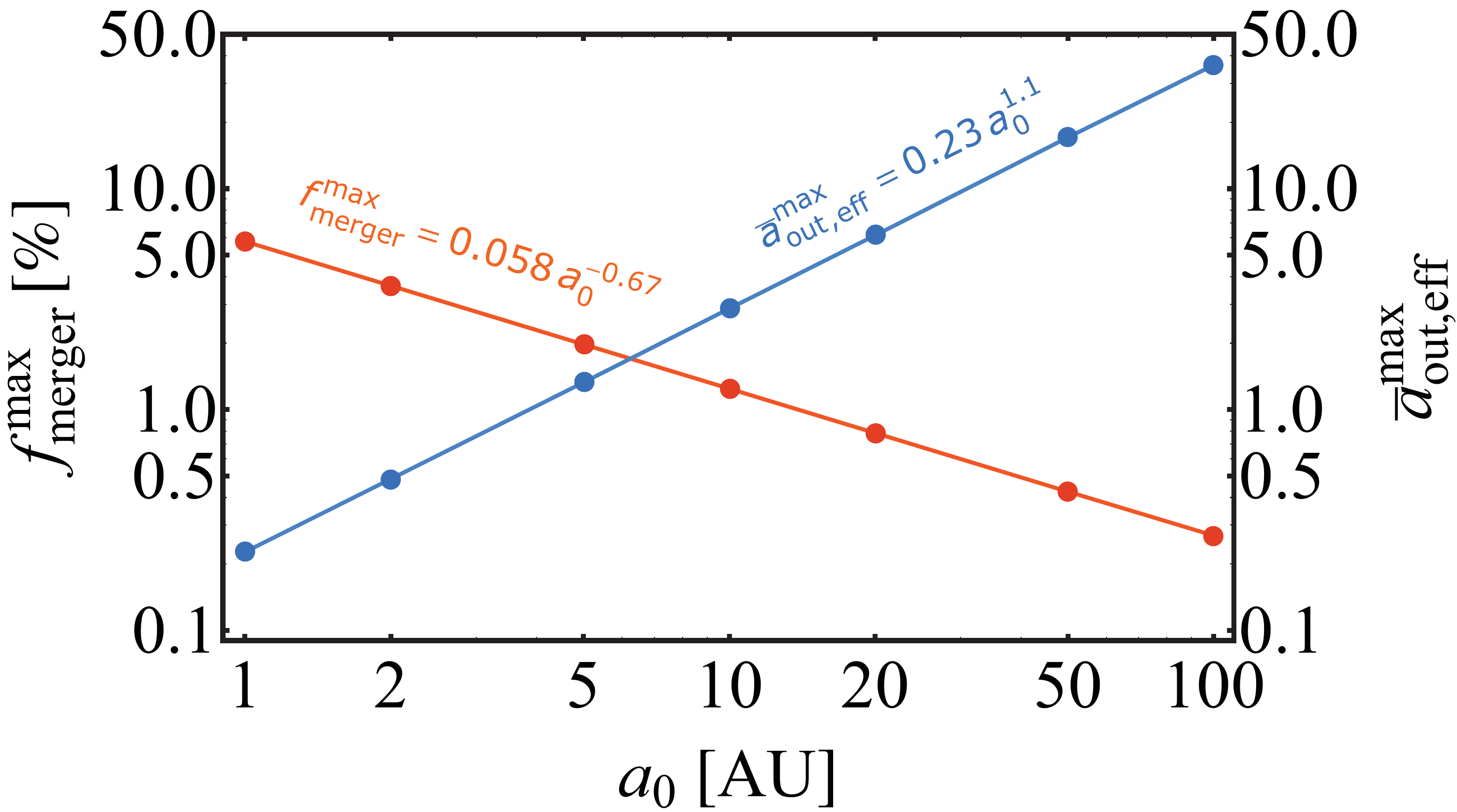}
\caption{
The maximum values of $f_\merger^\m$ and $\bar{a}_{\OUT,\eff}^\m$ as a function of initial semi-major axis $a_0$
of NS binaries.
}
\label{fig:maximum merger fraction NS}
\end{centering}
\end{figure}

\begin{figure}
\begin{centering}
\includegraphics[width=9cm]{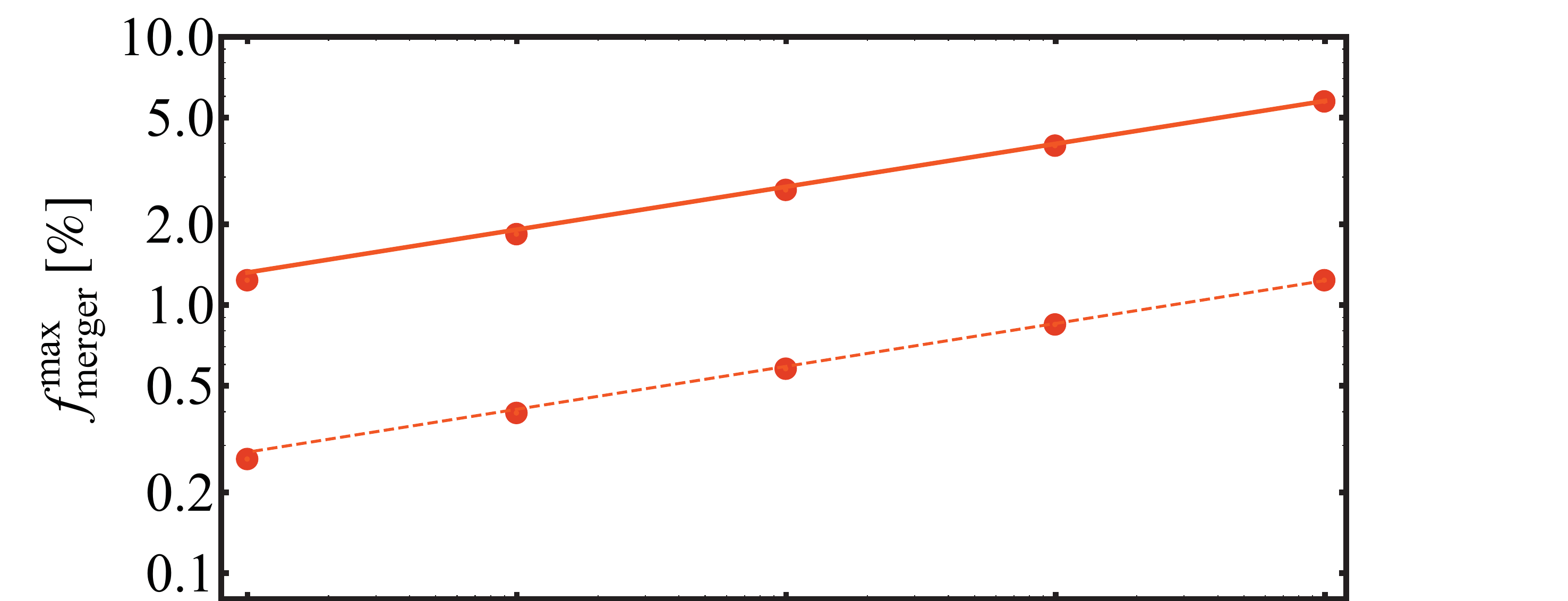}\\
\includegraphics[width=9cm]{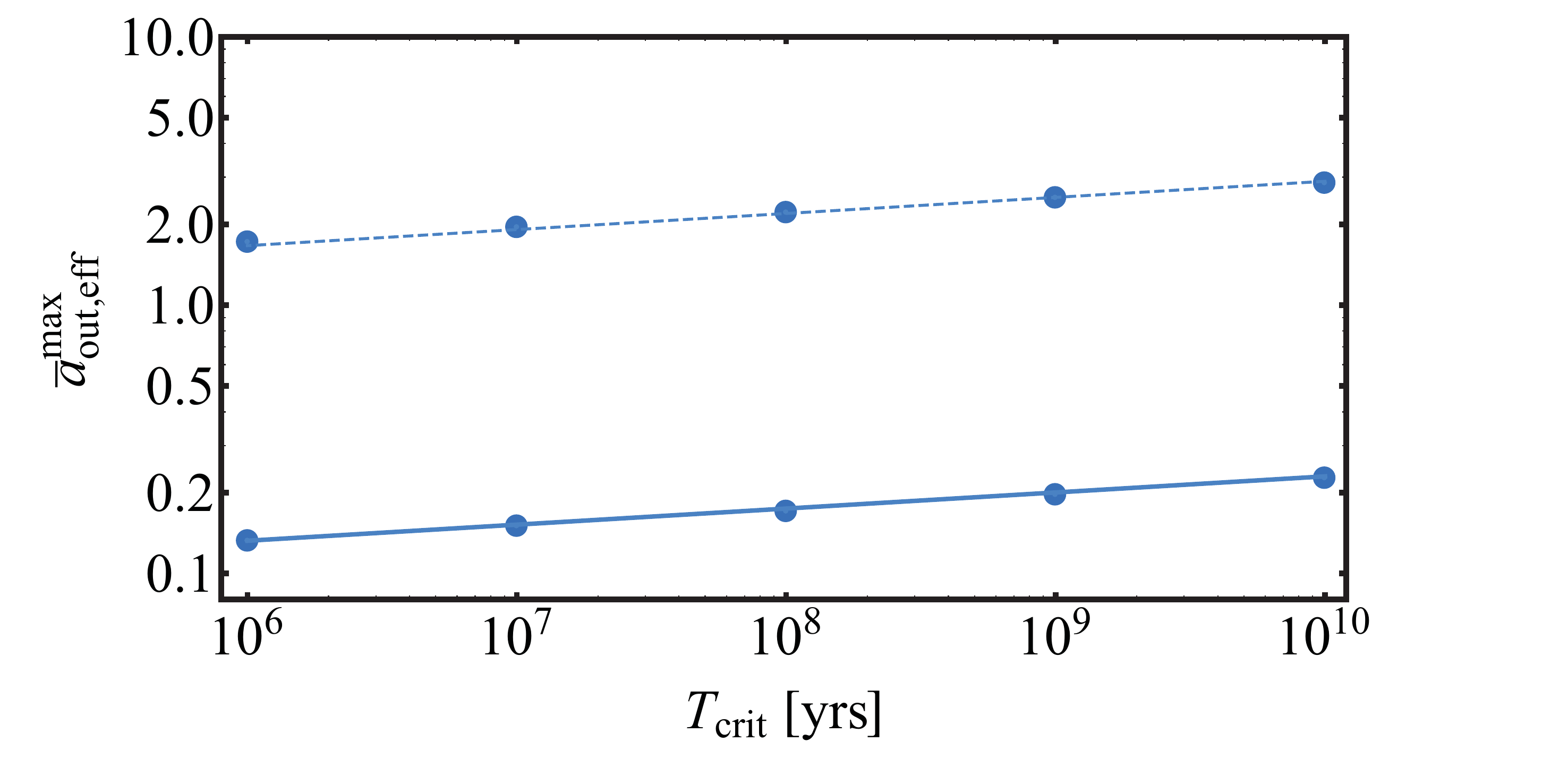}
\caption{
The maximum values of $f_\merger^\m$ and $\bar{a}_{\OUT,\eff}^\m$ as a function of the critical merger time $T_\mathrm{crit}$
of the binaries (for $m_1=m_2=1.4M_\odot$). The points are obtained analytically using Equations
(\ref{eq:EMAX}), (\ref{eq:fitting formula}) and (\ref{eq:merger fraction}).
The solid ($a_0=1\au$) and dashed ($a_0=10\au$) lines are given by the
fitting formulae (\ref{eq:fitting f merger})-(\ref{eq:fitting a bar}).
}
\label{fig:maximum merger fraction NS Tcrit}
\end{centering}
\end{figure}

The result of Figure \ref{fig:merger fraction NS} (upper panel) for the merger fraction can be summarized by
the fitting formula for $f_\merger^\m$, the maximum value of $f_\merger$ (for a given $a_0$),
and $\bar{a}_{\OUT,\eff}^\m$, the maximum value of $\bar{a}_{\OUT,\eff}$ for merger to be possible.
Figure \ref{fig:maximum merger fraction NS} shows that for the parameters of Figure \ref{fig:merger fraction NS}
(with $m_1=m_2=1.4M_\odot$, $T_\mathrm{crit}=10^{10}$ yrs), we have
\be\label{eq:fitting f and a}
f_\merger^\m\simeq5.8\%\bigg(\frac{a_0}{\au}\bigg)^{-0.67},~
\bar{a}_{\OUT,\eff}^\m\simeq0.23\bigg(\frac{a_0}{\au}\bigg)^{1.1}.
\ee

\emph{ Scaling relations for general quadrupole systems}. Equation (\ref{eq:fitting f and a})
can be generalized to other types of systems (with different $m_1$, $m_2$)
and different value of merger time $T_\mathrm{crit}$. From Equations (\ref{eq:Tmerger}) and (\ref{eq:fitting formula}),
we see that the critical eccentricity $e_\mathrm{m}$ required for merger within time $T_\mathrm{crit}$ depends on
$(\mu T_\mathrm{crit})(m_{12}/a_0^2)^2$. From Equation (\ref{eq:EMAX}) we see that
for $\eta\ll 1$ (a good approximation), the critical inclinations
($I_{0,\merger}^\pm$; see Equation \ref{eq:merger fraction}) for a given $e_\m=e_\mathrm{m}$
depend only on $\varepsilon_\gr$, or the combination $(m_{12}/a_0^2)^2(a_{\OUT,\eff}^3/m_3)$.
Thus the merger fraction $f_\merger$ depends on $m_1$, $m_2$, $a_0$ and $T_\mathrm{crit}$ only through
$(m_{12}/a_0^2)$ and $\mu T_\mathrm{crit}$.
We therefore expect from Equation (\ref{eq:fitting f and a}) that
$f_\merger^\m\propto(a_0/m_{12}^{0.5})^{-0.67}(\mu T_\mathrm{crit})^\alpha$ and
$\bar{a}_{\OUT,\eff}^\m\propto(a_0/m_{12}^{0.5})^{1.1}(\mu T_\mathrm{crit})^\beta$,
where $\alpha$, $\beta$ are fitting parameters.
Figure {\ref{fig:maximum merger fraction NS Tcrit}} shows the fitting. We find
\be\label{eq:fitting f merger}
\begin{split}
f_\merger^\m\simeq&5.8\%\bigg[\bigg(\frac{a_0}{\au}\bigg)\bigg(\frac{m_{12}}{2.8M_\odot}\bigg)^{-0.5}\bigg]^{-0.67}\\
&\times\bigg(\frac{\mu}{0.7M_\odot}\frac{T_\mathrm{crit}}{10^{10}\mathrm{yrs}}\bigg)^{0.16},
\end{split}
\ee
and
\be\label{eq:fitting a bar}
\begin{split}
\bar{a}_{\OUT,\eff}^\m\simeq&0.23\bigg[\bigg(\frac{a_0}{\au}\bigg)\bigg(\frac{m_{12}}{2.8M_\odot}\bigg)^{-0.5}\bigg]^{1.1}\\
&\times\bigg(\frac{\mu}{0.7M_\odot}\frac{T_\mathrm{crit}}{10^{10}\mathrm{yrs}}\bigg)^{0.06}.
\end{split}
\ee
These fitting formulae are valid for any type of LK-induced BH/NS mergers in the quadrupole order.

\section{EVOLUTION OF BH SPIN AND SPIN-ORBIT MISALIGNMENT}
\label{sec 4}
\subsection{Spin-Orbit Coupling}
\label{sec 4 1}
We now study how the BH spin evolves
during LK-induced binary mergers.
We present the evolution equation for
$\textbf{S}_1=\mathrm{S}_1 \hat{\textbf{S}}_1$ (where
$\mathrm{S}_1$ is the magnitude of the spin angular momentum of $m_1$ and
$\hat{\textbf{S}}_1$ is the unit vector).
The de-Sitter precession of $\hat{\textbf{S}}_1$ around $\hat{\mathbf{L}}$ (1.5 PN effect)
is govern by \citep[e.g.,][]{Barker}
\be\label{eq:spin}
\frac{d \hat{\textbf{S}}_1}{dt}=\Omega_\mathrm{SL}\hat{\mathbf{L}} \times \hat{\textbf{S}}_1,
\ee
with the orbital-averaged spin precession rate
\be\label{eq:desitter rate}
~~~\Omega_\mathrm{SL}=\frac{3 G n (m_{2}+\mu/3)}{2 c^2 a (1-e^2)}.
\ee
Similar equation applies to the spinning body 2.
Note that $\Omega_\mathrm{SL}$ is of the same order as $\Omega_\gr$ (Equation \ref{eq:GR}) for $m_1\sim m_2$.
There are also back-reaction torques from $\hat{\textbf{S}}_1$ on $\hat{\mathbf{L}}$ and $\hat{\mathbf{e}}$:
\be\label{eq:back reaction}
\begin{split}
\frac{d \mathbf{L}}{dt}\bigg|_{\mathrm{LS}}&=\Omega_\mathrm{LS}\hat{\textbf{S}}_1\times\mathbf{L},\\
\frac{d \mathbf{e}}{dt}\bigg|_{\mathrm{LS}}&=\Omega_\mathrm{LS}\hat{\textbf{S}}_1\times\mathbf{e}-
3\Omega_\mathrm{LS}\big(\hat{\mathbf{L}}\cdot\hat{\textbf{S}}_1\big)\hat{\mathbf{L}}\times\mathbf{e},\\
\end{split}
\ee
where
\be\label{eq:Omega LS}
\Omega_\mathrm{LS}=\Omega_\mathrm{SL}\frac{S_1}{L}=\frac{G S_{1}(4+3m_{2}/m_{1})}{2c^2a^3(1-e^2)^{3/2}}.
\ee
We include this effect in our calculations, although it (Equation \ref{eq:back reaction}) is usually negligible since $S_1\ll L$
\footnote{
With $S_1=\chi_1 G m_1^2/c$, the ratio $S_1/L$ is $\chi_1(m_1/m_2)[Gm_{12}/(c^2a(1-e^2))]^{1/2}$,
which is $\ll1$ for $a(1-e^2)\gg Gm_{12}/c^2$ (i.e., the inner binary pericenter
distance is much larger  than the gravitational radius). The inequity $S_1\ll L$ is well satisfied for our calculations since the
spin-orbit misalignment angle $\theta_\mathrm{s_1l}$ is frozen well before the inner binary reaches the separation
$Gm_{12}/c^2$ (see Equations \ref{eq:adiabaticity parameter}-\ref{eq:adiabaticity parameter initial}).
}.
The spin-spin coupling (2 PN correction) is always
negligible until the final phase of the merger,
and will be ignored in our calculations.
In addition, the de-Sitter precession of $\hat{\mathbf{S}}_1$ induced by the tertiary companion is neglected as well
(since we consider $m_1\sim m_2$, and $m_3$ is not much larger, but $a_\OUT\gg a$).
In all our calculations, we use $\chi_1=\chi_2=0.1$ for concreteness.
But note that the values of $\chi_1$ and $\chi_2$ do not affect the results
of our paper (except Figure \ref{fig:overall spin distribution}, which assumes $\chi_1=\chi_2$). This
is because (i) The de-Sitter precession frequency $\Omega_\mathrm{SL}$ (Equation \ref{eq:desitter rate})
is independent of $\chi_1$, $\chi_2$, (ii) The inequity $S_1, S_2\ll L$ is well satisfied (see footnote 3).

In terms of the inner BH binary axis $\hat{\mathbf{L}}$, the effect of the companion is to induce
precession of $\hat{\mathbf{L}}$ around $\hat{\mathbf{L}}_\OUT$ with nutation (when $e\ne 0$).
In the quadrupole order, we have
\ba\label{eq:L precess around Lout}
\frac{d \mathbf{L}}{dt}\bigg|_{\lk,\qu}=&&-\frac{3 L}{4t_\lk\sqrt{1-e^2}}\\
&&\times \bigg[\Big(\textbf{j}\cdot\hat{\mathbf{L}}_\OUT\Big)\hat{\mathbf{L}}_\OUT\times\textbf{j}+
5\Big(\mathbf{e}\cdot\hat{\mathbf{L}}_\OUT\Big)\mathbf{e}\times\hat{\mathbf{L}}_\OUT\bigg]\nonumber.
\ea
An approximate expression for the rate of change of $\hat{\mathbf{L}}$ is given by \citep{Anderson et al HJ}
\be\label{eq:Omega L}
\Omega_\mathrm{L}=\bigg|\frac{d \hat{\mathbf{L}}}{dt}\bigg|_{\lk,\qu}\simeq
\frac{3(1+4e^2)}{8t_\lk\sqrt{1-e^2}}\Big|\sin2I\Big|.
\ee

The spin evolution is determined by two competing processes:
$\hat{\bf S}_1$ precesses around $\hat{\mathbf{L}}$ at the rate $\Omega_\mathrm{SL}$, and
$\hat{\mathbf{L}}$ varies at the rate $\Omega_\mathrm{L}$.
There are three possible spin behaviors depending on the ratio $\Omega_\mathrm{SL}/\Omega_\mathrm{L}$:

(i) For $\Omega_\mathrm{L}\gg \Omega_\mathrm{SL}$ (``nonadiabatic" regime),
the spin axis $\hat{\bf S}_1$ cannot ``keep up'' with the rapidly changing
$\hat{\mathbf{L}}$, which precesses around a fixed $\hat{\mathbf{L}}_\OUT$ (for $L_\OUT\gg L$).
Thus $\hat{\bf S}_1$ effectively precesses around $\hat{\mathbf{L}}_\OUT$,
keeping the misalignment angle between $\hat{\textbf{S}}_1$ and $\hat{\mathbf{L}}_\OUT$,
$\theta_\SB\equiv\cos^{-1}(\hat{\textbf{S}}_1\cdot\hat{\mathbf{L}}_\OUT)$, approximately
constant.

(ii) For $\Omega_\mathrm{SL}\gg \Omega_\mathrm{L}$ (``adiabatic" regime),
$\hat{\mathbf{S}}_1$ is strongly coupled to $\hat{\mathbf{L}}$.
The spin axis $\hat{\textbf{S}}_1$ closely ``follows'' $\hat{\mathbf{L}}$, maintaining an approximately constant
spin-orbit misalignment angle $\theta_\SL\equiv\cos^{-1}(\hat{\textbf{S}}_1\cdot\hat{\mathbf{L}})$.

(iii) For $\Omega_\mathrm{SL}\sim \Omega_\mathrm{L}$ (``trans-adiabatic" regime), the spin
evolution can be complex, potentially generating large spin-orbit misalignment $\theta_\SL$.
Since both $\Omega_\mathrm{SL}$ and $\Omega_\mathrm{L}$ depend on $e$ during the
LK cycles, the precise transitions between these regimes can be fuzzy.

To help characterize the spin dynamics, we introduce an ``adiabaticity parameter" as
\be\label{eq:adiabaticity parameter}
\mathcal{A}\equiv\bigg|\frac{\Omega_\mathrm{SL}}{\Omega_\mathrm{L}}\bigg|
=\mathcal{A}_0\frac{1}{(1+4e^2)\sqrt{1-e^2}|\sin 2I|},
\ee
where
\be\label{eq:adiabaticity parameter initial}
\begin{split}
\mathcal{A}_0&\equiv \bigg|\frac{\Omega_\mathrm{SL}}{\Omega_\mathrm{L}}\sin2I\bigg|_{e=0}
=\frac{4G(m_2+\mu/3)m_{12}a_{\OUT,\eff}^3}{c^2m_3a^4}\\
&\simeq 2.76\times10^{-5}\bigg[\frac{(m_2+\mu/3)}{35M_\odot}\bigg]\bigg(\frac{m_{12}}{60M_\odot}\bigg)\\
&~~~\times\bigg(\frac{m_3}{30M_\odot}\bigg)^{-1}
\bigg(\frac{a_{\OUT,\eff}}{10^3\au}\bigg)^3\bigg(\frac{a}{10^2\au}\bigg)^{-4}.
\end{split}
\ee
Note that $\mathcal{A}$ has a steep dependence on the eccentricity $e$ and inclination $I$,
and it is time varying, while $\mathcal{A}_0$
is an intrinsic indicator for identifying which system may
undergo potentially complicated spin evolution.
Since $\mathcal{A}_0$ depends sensitively on $a$, during the orbital decay a system may transit from ``non-adiabatic''
at large $a$'s to ``adiabatic'' at small $a$'s, where
the final spin-orbit misalignment angle $\theta_\SL^{\rm f}$ is ``frozen''.
Note that ${\cal A}_0$ is directly related to $\varepsilon_\gr$ (see Equation \ref{eq:epsilonGR}) by
\be
{{\cal A}_0\over\varepsilon_\gr}={4\over 3}{m_{2}+\mu/3\over m_{12}}.
\label{eq:A-eps}\ee
Thus, when the initial value of $\varepsilon_\gr$ (at $a=a_0$)
satisfies $\varepsilon_{\gr,0}\lesssim 9/4$ (a necessary condition
for LK eccentricity excitation; see Equation~\ref{eq:grlim}), we also have the initial
${\cal A}_0\lesssim (3m_{2}+\mu)/m_{12}\sim 1$. This implies that any system that
experiences enhanced orbital decay due to LK oscillations
must go through the ``trans-adiabatic'' regime and therefore possibly complicated spin evolution \citep[]{Liu-ApJL}.

In our previous study \citep[]{Liu-ApJL}, we considered initially compact BH binaries
(with $a_0\sim0.2\au$), which can merge by themselves without the aid of a tertiary companion.
We focused on systems with initial $\mathcal{A}_0$ not much less than unity,
and showed that such systems can experience complex/chaotic spin evolution during the LK-enhanced
mergers. In this paper, we consider the inner BH binaries with large initial semi-major axis
($a_0=20, 100\au$) and initial $\mathcal{A}_0\ll1$.
As we shall see, such systems exhibit a variety of different spin evolutionary behaviors
during the LK-induced mergers.

The bottom panel of Figure~\ref{fig:OE quad 1}
shows a representative example of the spin evolution
during the LK-induced orbital decay.
The system begins with $\mathcal{A}_0\sim10^{-3}\ll1$
(Equation \ref{eq:adiabaticity parameter initial}).
At the early stage of the evolution, $\Omega_\mathrm{SL}\ll \Omega_\mathrm{L}$, leading $\theta_\SB$ to be nearly constant.
Because of the large variation of $\hat{\textbf{L}}$, the spin-orbit angle
$\theta_\SL$ oscillates with a large amplitude.
As the orbit decays and circularizes, $\hat{\textbf{L}}$ becomes frozen relative to $\hat{\textbf{L}}_\OUT$
(with final inclination $I\simeq 125^\circ$),
while $\hat{\textbf{S}}$ precesses rapidly around $\hat{\textbf{L}}$, with $\theta_\SL$ settling down to the
final value ($\simeq90^\circ$).
A non-zero final spin-orbit misalignment has been produced from the originally aligned configuration -- This is
only one example of the complex BH spin evolutionary paths during LK-induced mergers (Section \ref{sec 4 2}).

The problem we study here is similar to the problem of the dynamics of stellar spin driven by a giant
planet undergoing Lidov-Kozai oscillations and migration \citep[]{Dong Science,Storch 2015,Anderson et al HJ,Storch 2017}.
However, there is an important difference: The de-Sitter precession of
the BH spin is always prograde with respect to the orbit
(the precession rate vector is $\Omega_\mathrm{SL}\hat{\mathbf{L}}$), while the Newtonian
precession of the stellar spin driven by the planet arises from the
rotation-induced stellar oblateness and depends on $\cos\theta_\SL$
(the precession rate vector is along the direction of $-\cos\theta_\SL\hat{\mathbf{L}}$).
This difference implies that the (Newtonian) stellar spin axis is prone to resonant
(and potentially chaotic) excitation of spin-orbit misalignment, even
for circular orbit \citep[e.g.,][]{Lai 2014,Lai 2018}, while the BH spin evolution
is more regular: The nodal precession of the inner orbit driven by the external companion
(i.e.the precession of $\hat{\textbf{L}}$ and $\hat{\textbf{L}}_\OUT$) is
retrograde (see Equation \ref{eq:L precess around Lout}), whereas the precession of $\hat{\textbf{S}}$ around
$\hat{\textbf{L}}$ is prograde, so secular resonace does not usually happen when the orbital
evolution is regular.

In the case of NS binaries, Newtonian effect due to the oblateness of the NS
($m_1$) also contributes to the spin precession. Equation (\ref{eq:spin}) is changed to
\be
\frac{d \hat{\textbf{S}}_1}{dt}=\Omega^{(\mathrm{dS})}_\mathrm{SL}\hat{\mathbf{L}} \times
\hat{\textbf{S}}_1+\Omega^{(\mathrm{Newtonian})}_\mathrm{SL}\hat{\mathbf{L}} \times \hat{\textbf{S}}_1,
\ee
where
\be
\Omega^{(\mathrm{Newtonian})}_\mathrm{SL}=-\frac{3Gm_2(I_3-I_1)}{2a^3(1-e^2)^{3/2}}\frac{\cos\theta_\mathrm{s_1l}}{S_1}.
\ee
Here, $I_3$ and $I_1$ are principal moments of inertia of the NS.
For $(I_3-I_1)\equiv k_{q\ast} m_1 R_1^2 \hat{\Omega}_1^2$ and $S_1=I_3\Omega_1=k_\ast m_1 R_1^2\Omega_1$,
where $R_1$ is the NS radius and $\hat{\Omega}_1$ is the rotation rate of the NS in unit of $(Gm_1/R_1^3)^{1/2}$,
we have
\be
\Omega^{(\mathrm{Newtonian})}_\mathrm{SL}=-\frac{3k_{q\ast}}{2k_\ast}\bigg(\frac{m_2}{m_1}\bigg)
\bigg(\frac{R_1}{a}\bigg)^3\frac{\Omega_1}{(1-e^2)^{3/2}}\cos\theta_\mathrm{s_1l}.
\ee
Thus the ratio $\Omega^{(\mathrm{Newtonian})}_\mathrm{SL}/\Omega^{(\mathrm{dS})}_\mathrm{SL}$ is
\ba
\Bigg|\frac{\Omega^{(\mathrm{Newtonian})}_\mathrm{SL}}{\Omega^{(\mathrm{dS})}_\mathrm{SL}}\Bigg|=&&
\bigg(\frac{k_{q\ast}}{k_\ast}\bigg)\frac{m_2}{\sqrt{m_1m_{12}}}\bigg[\frac{R_1}{a(1-e^2)}\bigg]^{1/2}\nonumber\\
&&\times\frac{R_1c^2}{G(m_2+\mu/3)}\hat{\Omega}_1|\cos\theta_\mathrm{s_1l}|.
\ea
For typical NS, $m_1\simeq1.4 M_\odot$, $R_1\simeq10$ km, $k_{q\ast}\simeq0.17$,
$k_\ast\simeq0.26$
\footnote{
For polytropic stellar models (with index n), $k_{q\ast}$
is approximately related to $k_\ast$ via the relations
$k_\ast=2\kappa_n/5$ and $k_{q\ast}\simeq\kappa_n^2/2(1-n/5)$.
For $n=1$, $\kappa_n\simeq0.65$ \citep[see Table 1 of][]{Lai 1993}.
}
and $\hat{\Omega}_1\simeq0.023(P_1/20\mathrm{ms})^{-1}$
(where $P_1$ is the rotation period of the NS).
Since $a(1-e^2)>R_1$, we see that $|\Omega^{(\mathrm{Newtonian})}_\mathrm{SL}/\Omega^{(\mathrm{dS})}_\mathrm{SL}|$
is always $\ll1$.

\subsection{Complex BH Spin Evolution Paths}
\label{sec 4 2}

\begin{figure*}
\centering
\begin{tabular}{cccc}
\includegraphics[width=5cm]{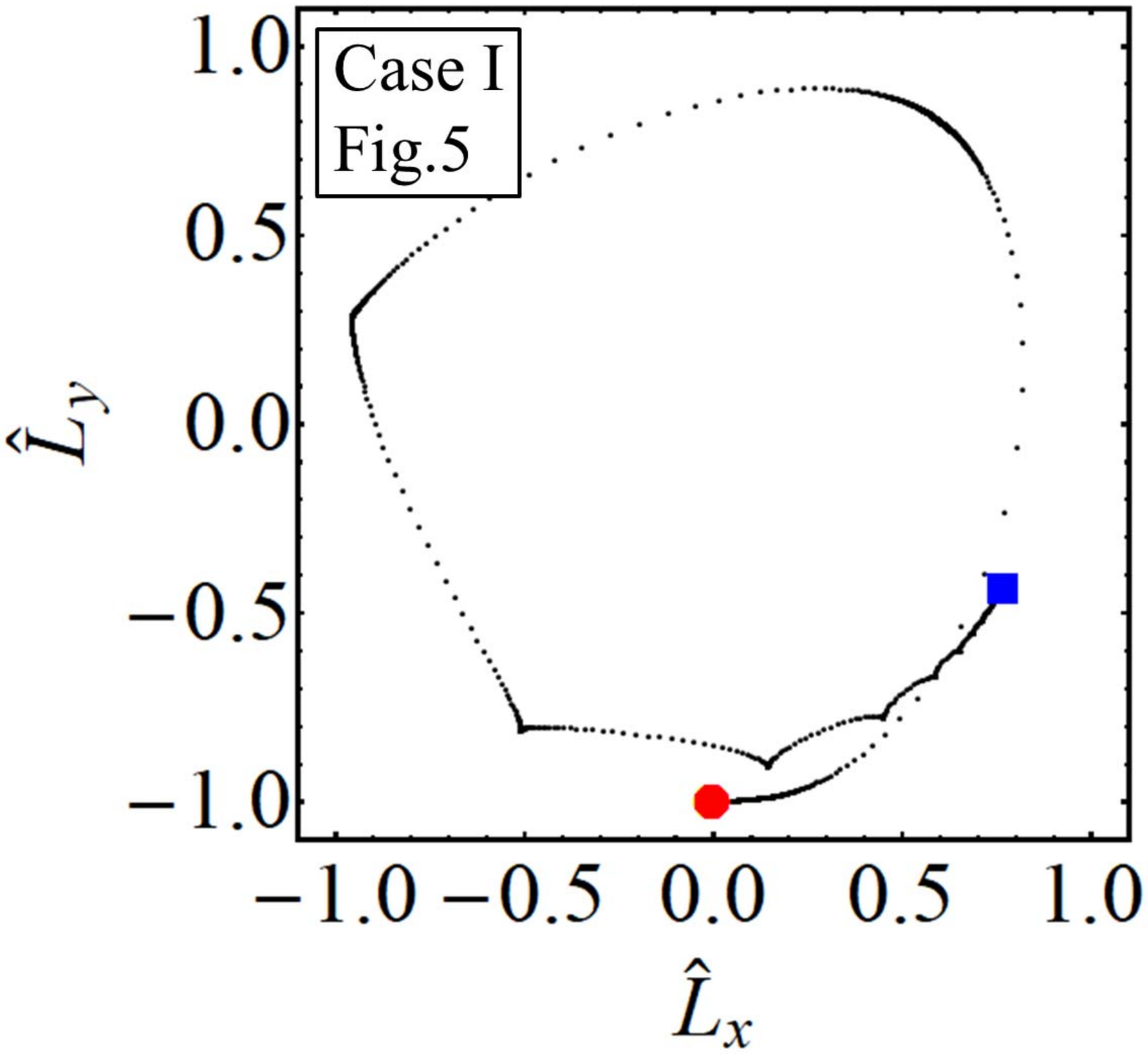}&
\includegraphics[width=5cm]{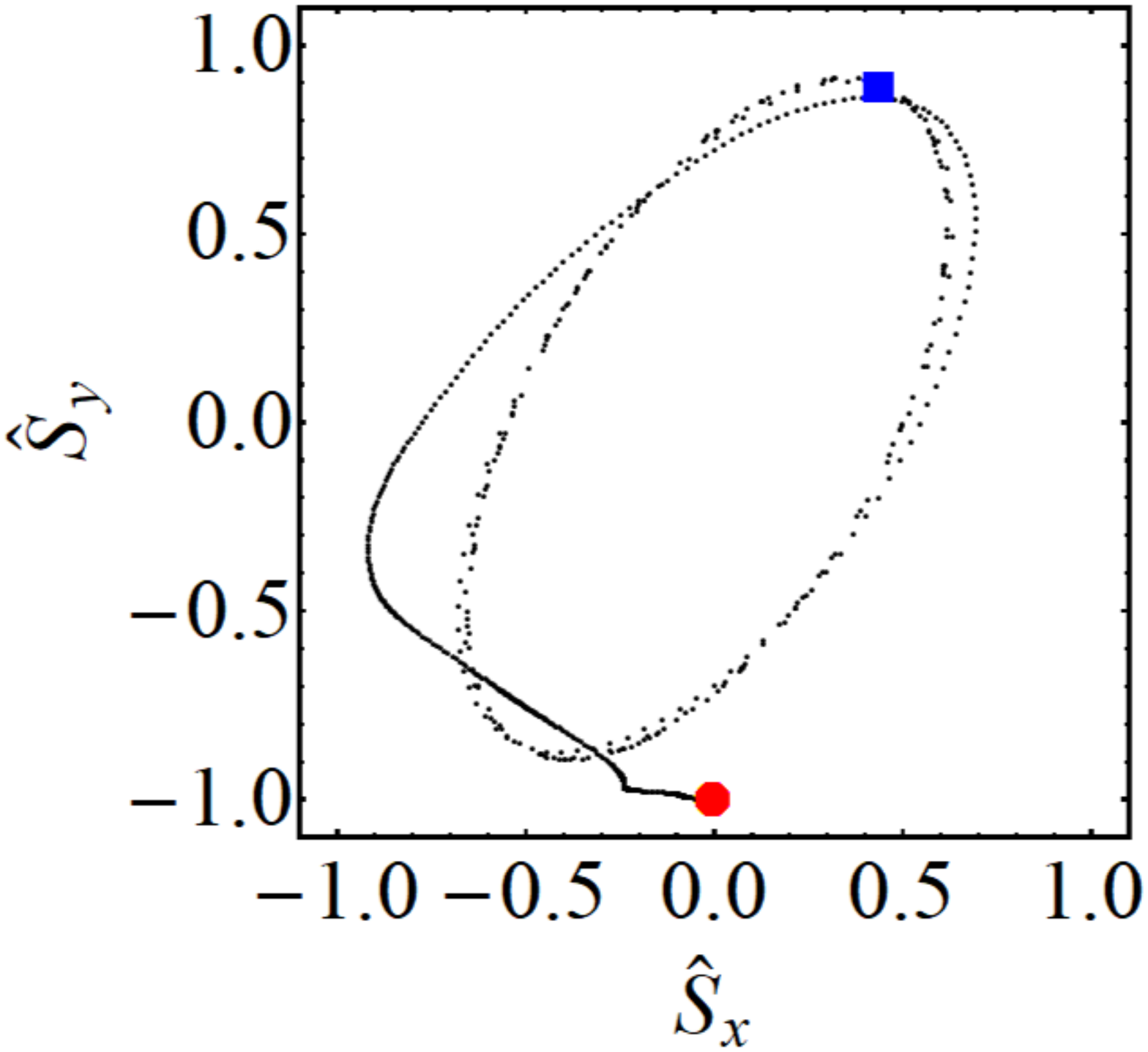}&
\includegraphics[width=5cm]{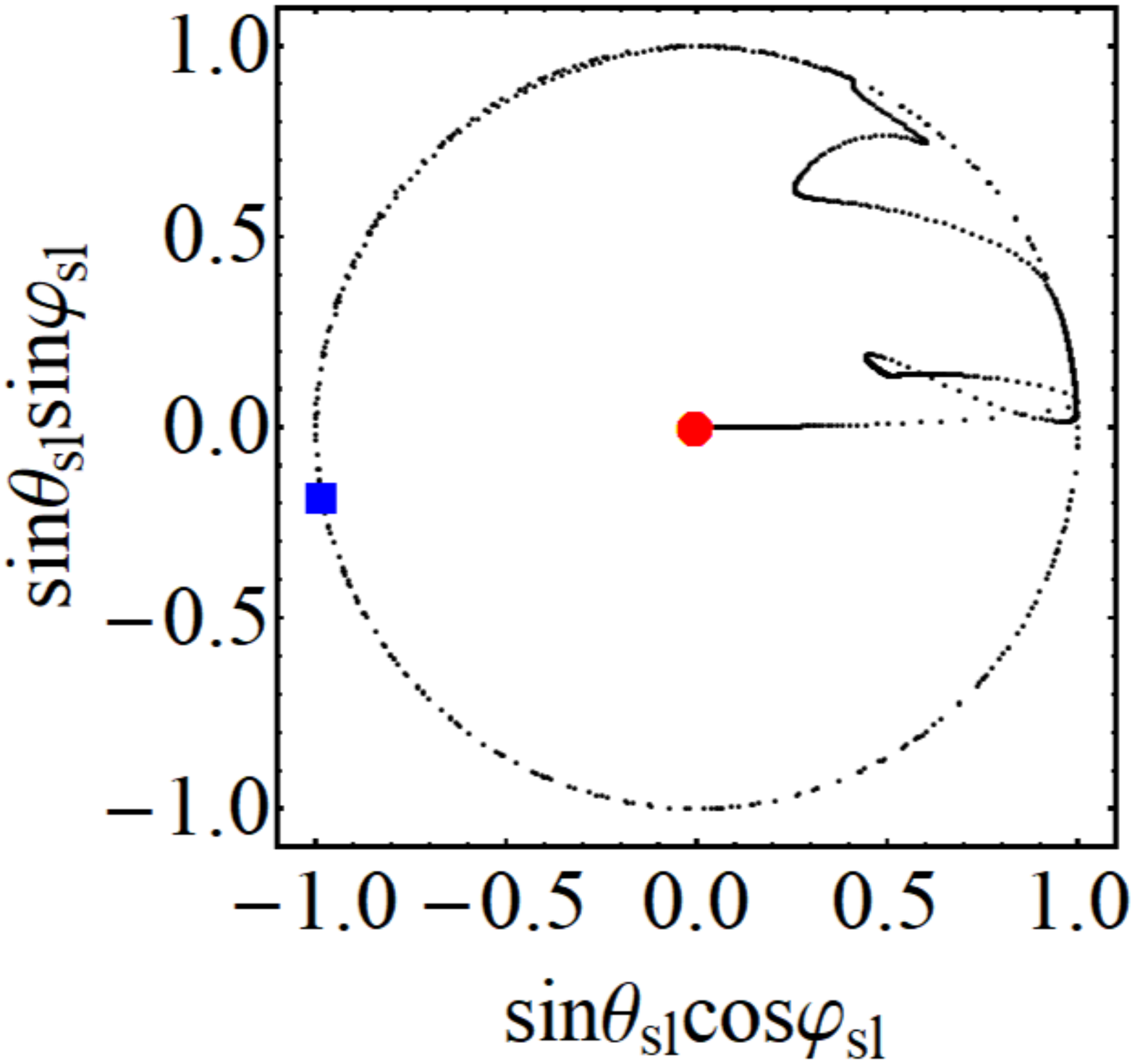}\\
\includegraphics[width=5cm]{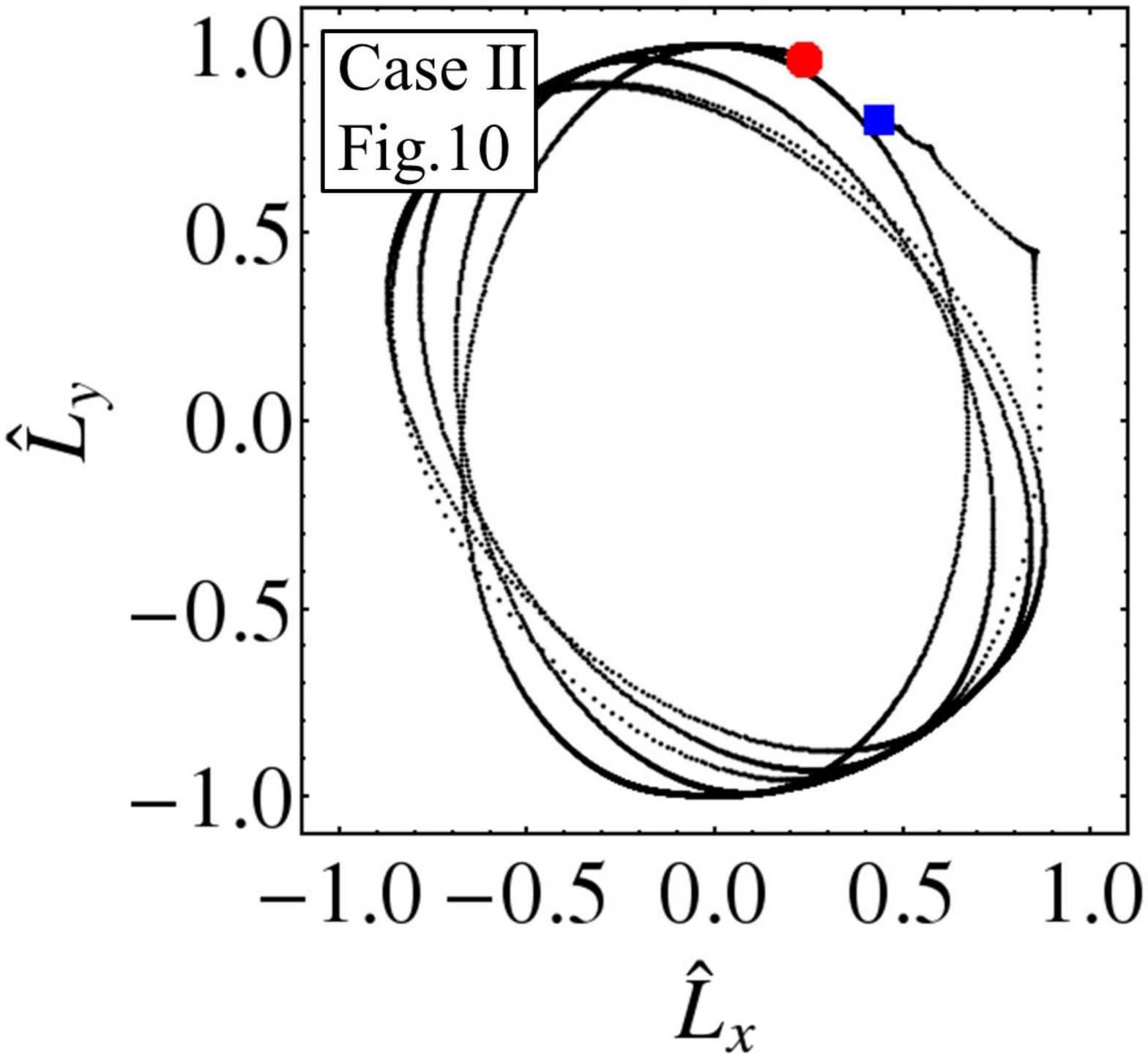}&
\includegraphics[width=5cm]{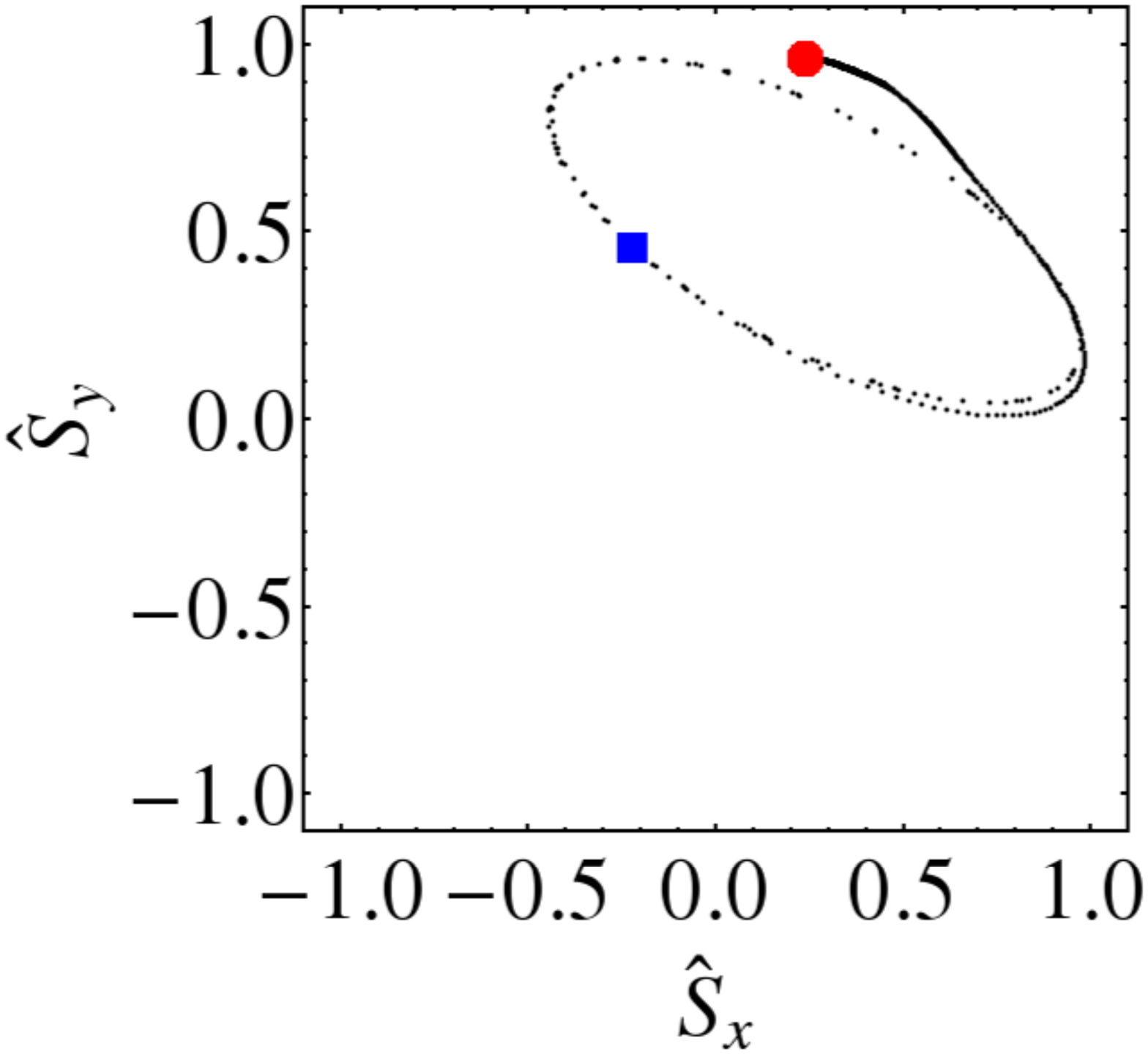}&
\includegraphics[width=5cm]{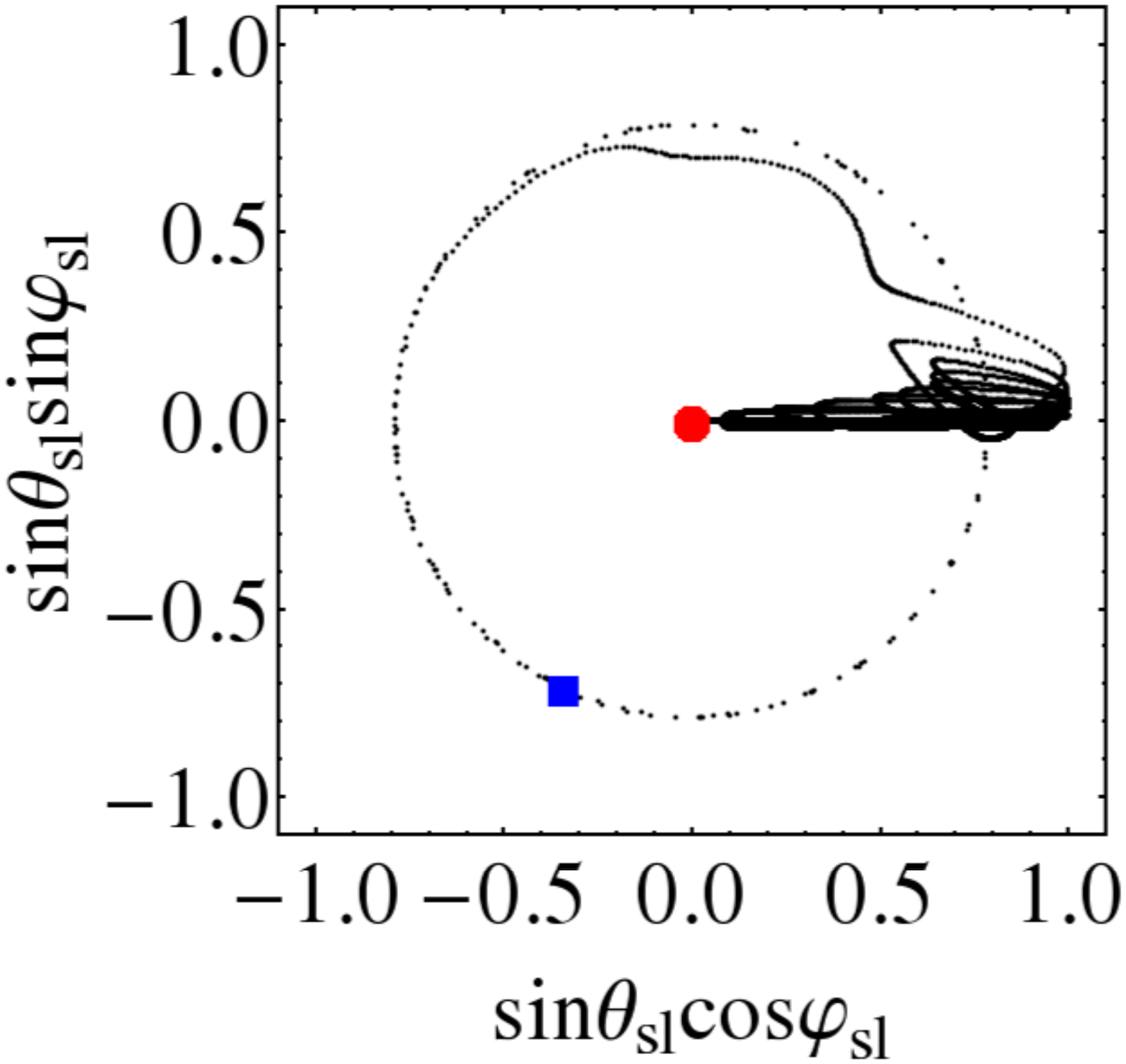}\\
\includegraphics[width=5cm]{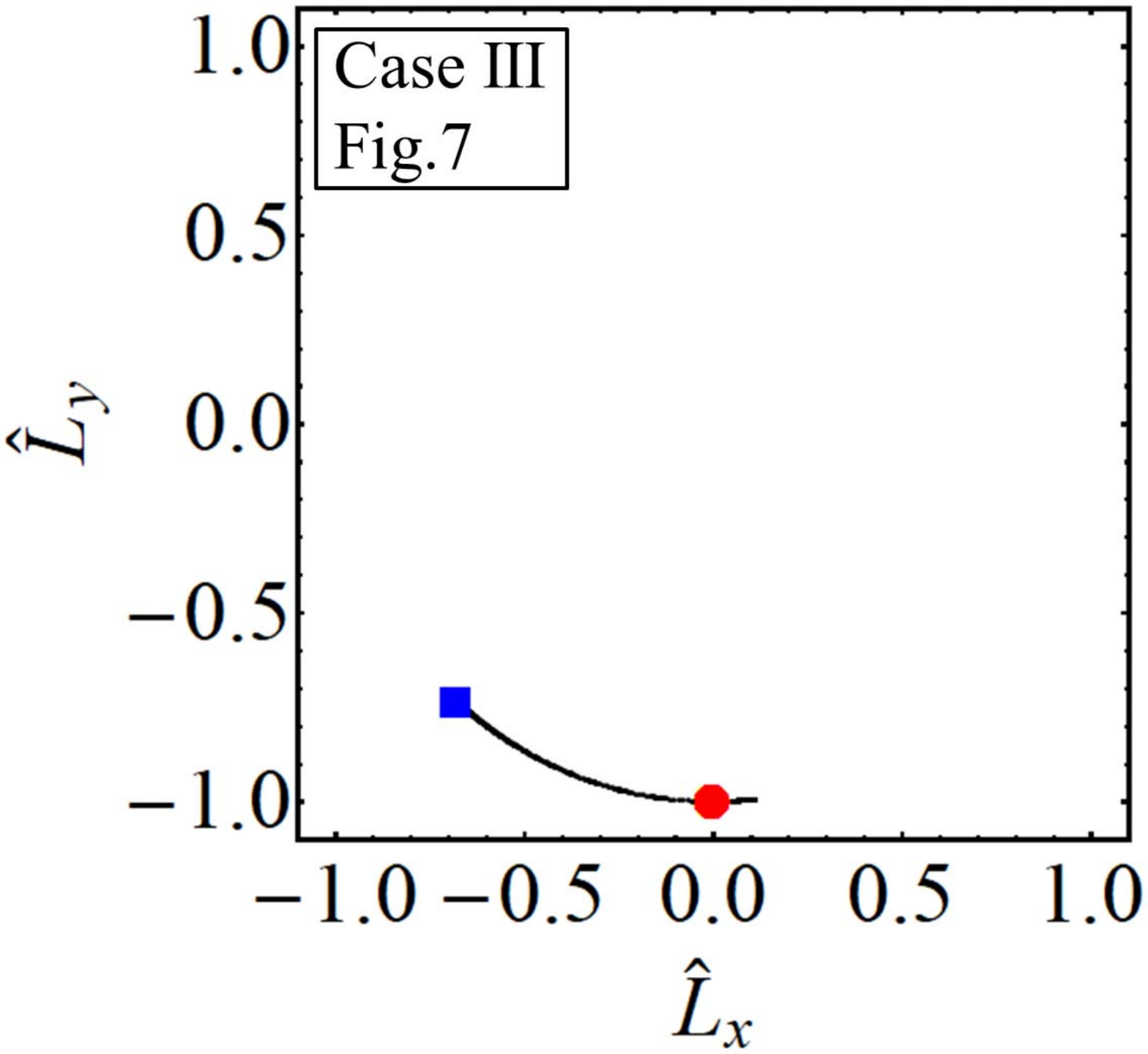}&
\includegraphics[width=5cm]{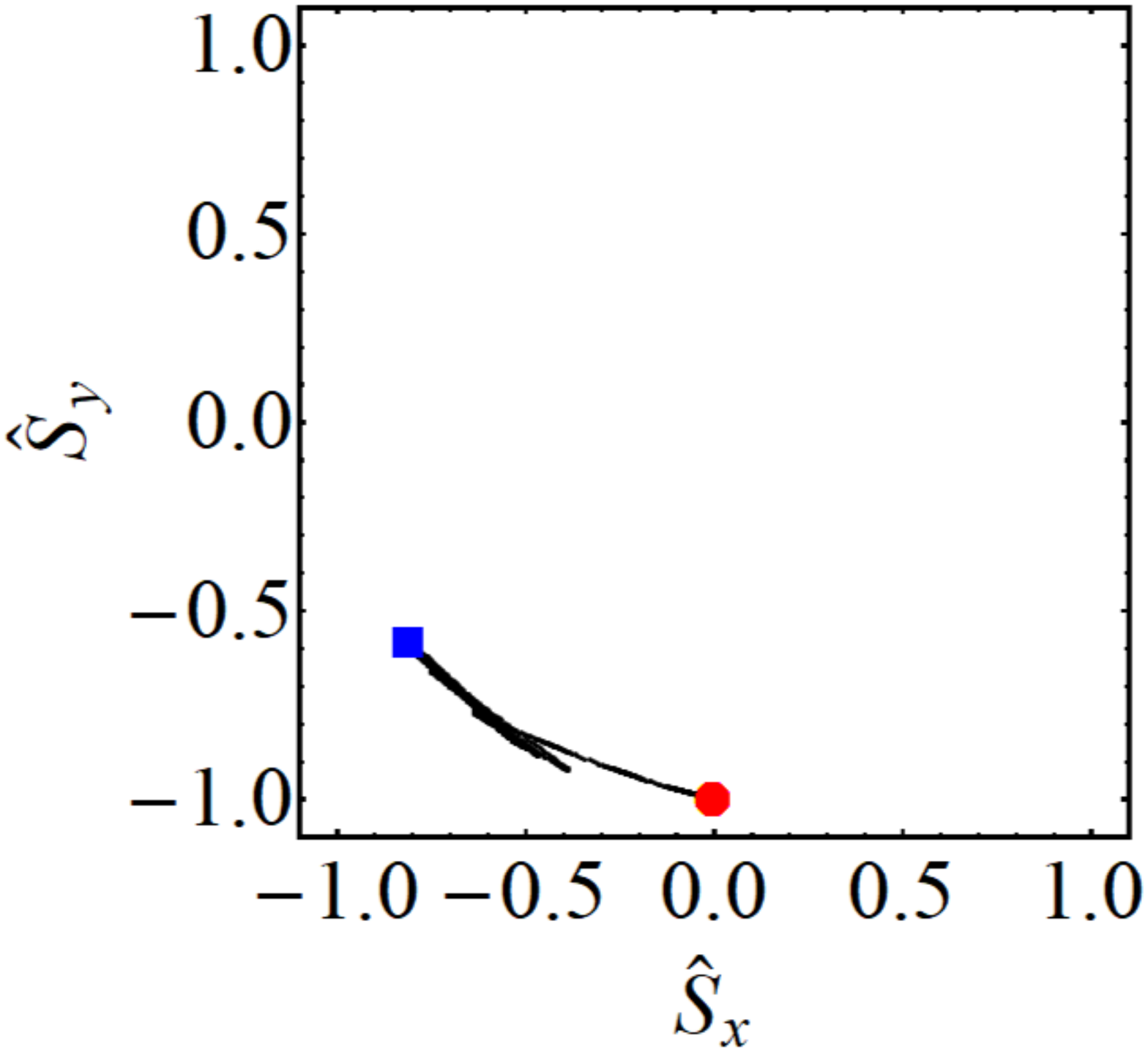}&
\includegraphics[width=5cm]{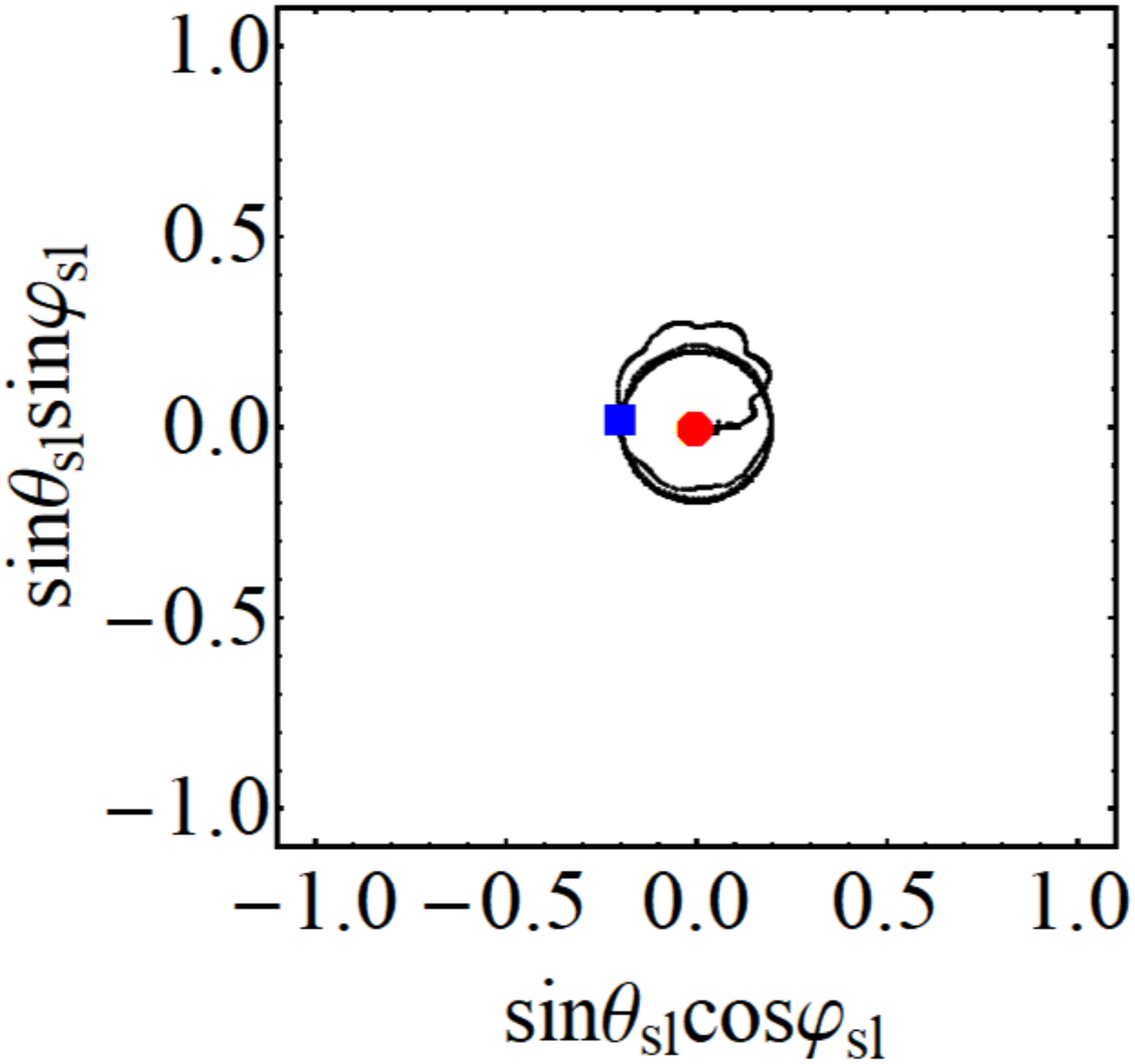}\\
\includegraphics[width=5cm]{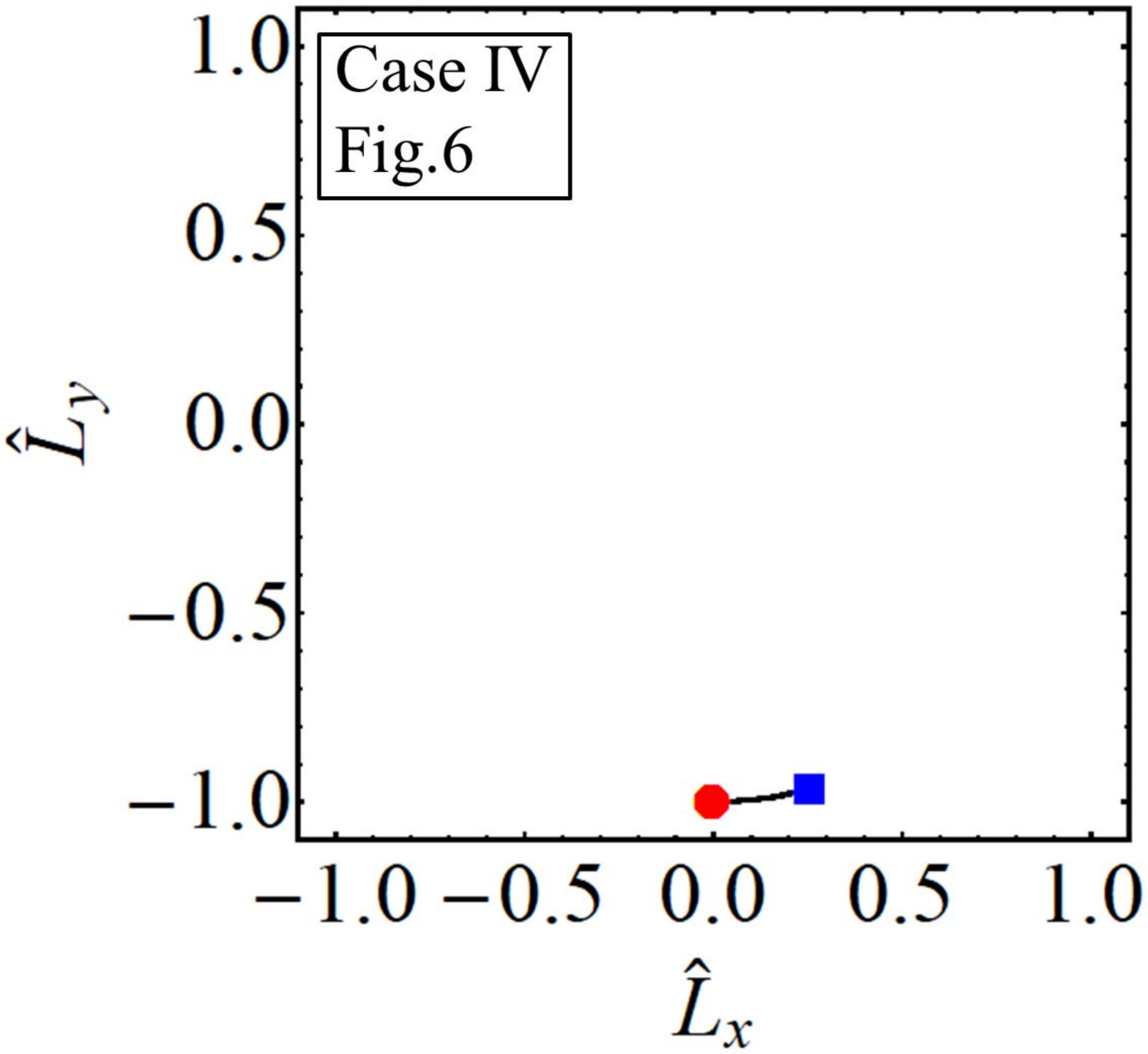}&
\includegraphics[width=5cm]{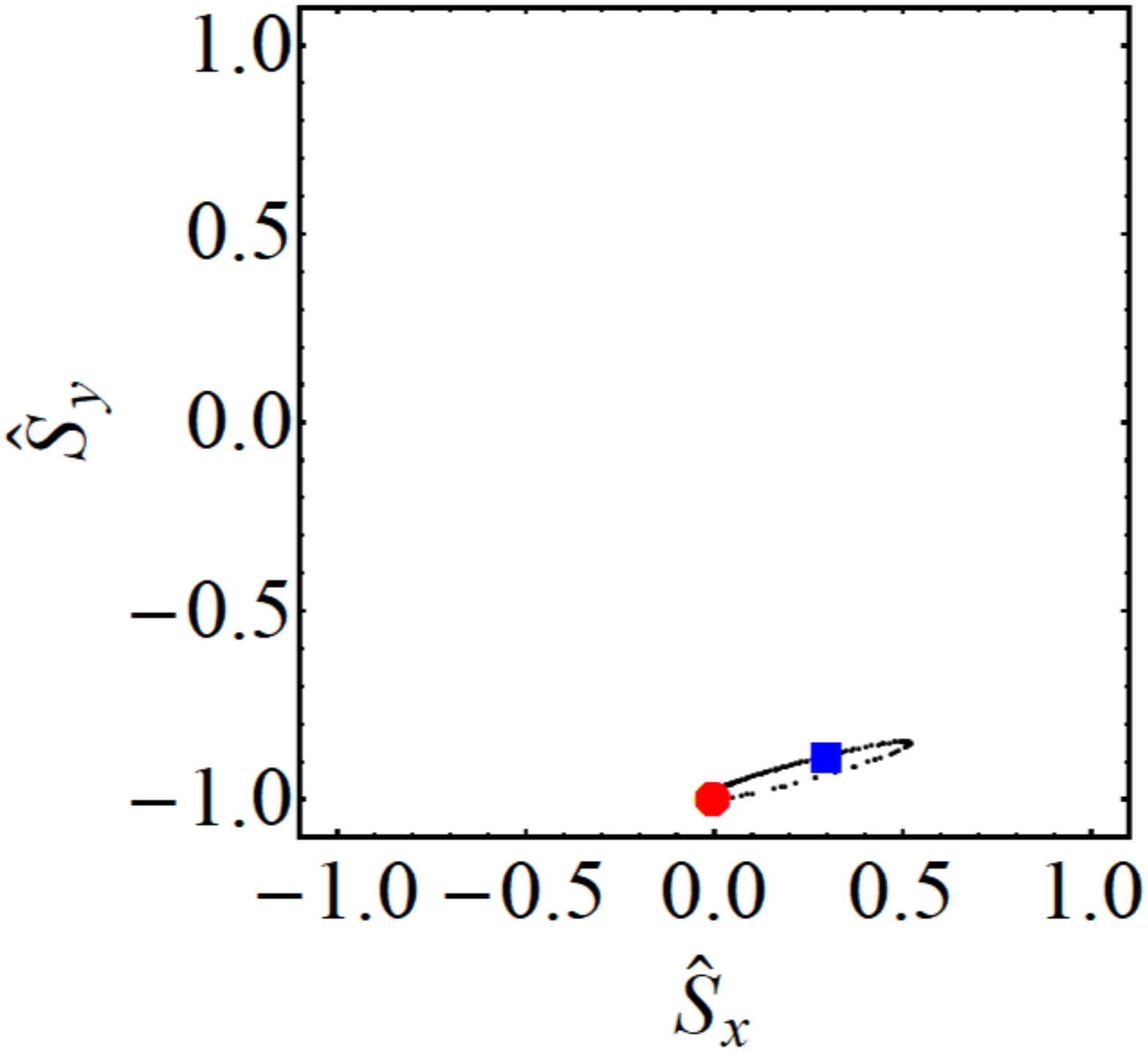}&
\includegraphics[width=5cm]{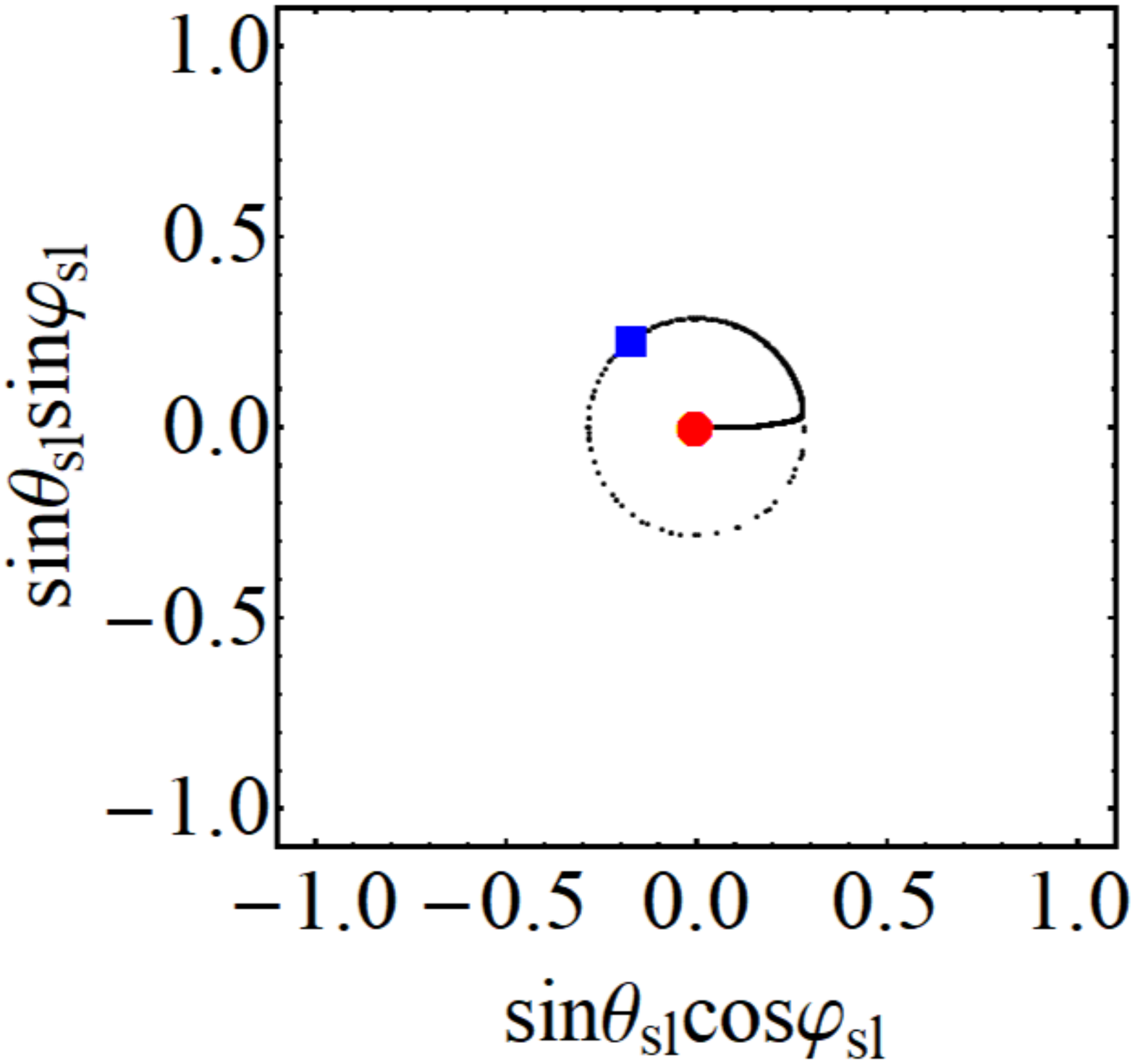}\\
\end{tabular}
\caption{The trajectories of the spin and angular momentum axes of the inner BH binary.
The left panels show the projection of $\hat{\mathbf{L}}$ in the $x$-$y$ plane
(where the $z$-axis is along the initial total angular momentum of the triple,
which is also approximately aligned with $\mathbf{L}_\OUT$).
The middle panels show the similar projection of $\hat{\textbf{S}}$. The right panels show
the projection of $\hat{\textbf{S}}$ in the plane perpendicular to $\hat{\mathbf{L}}$
(with $\theta_\SL$ the angel between $\hat{\textbf{S}}$ and $\hat{\mathbf{L}}$, and $\phi_\SL$
the rotational phase of $\hat{\textbf{S}}$ around $\hat{\mathbf{L}}$). In each panel, the filled circle (square) denotes the initial (final)
position. The four cases shown here correspond to
Figures \ref{fig:OE quad 2}, \ref{fig:OE oct 1}, \ref{fig:OE quad 3} and \ref{fig:OE quad oneshort}, respectively.
}
\label{fig:spin trajectories}
\end{figure*}

We have seem from Section \ref{sec 3} that LK-induced BH binaries can
have a variety of orbital evolution paths toward the final merger.
Correspondingly, the evolution of BH spin in these binaries also exhibit a rich set of
evolutionary behaviors. They can be roughly divided into four cases (see Figure \ref{fig:spin trajectories}).

Case I (see Figures \ref{fig:OE quad 1}-\ref{fig:OE quad 2}): This usually occurs when the initially inclination
$I_0$ is sufficiently different from $I_{0,\li}$ (see Equation \ref{eq:I0lim}), so that
($1-e_{\rm max}$) is much larger than ($1-e_{\rm lim}$). In this case, the inner binary experiences
multiple LK oscillations; the amplitude of the eccentricity oscillations shrinks gradually
as the orbit decays; eventually the binary circularizes and merges quickly.
As shown in the lower panels of Figures \ref{fig:OE quad 1} and \ref{fig:OE quad 2},
during the early stage, the angle $\theta_\SB$ is approximately constant
(since $\mathcal{A}_0\ll 1$), while $\theta_\SL$ exhibits larger variation due to the rapid precession of
$\hat{\mathbf{L}}$ around $\hat{\mathbf{L}}_\OUT$; at the later stage, as the orbit decays, $\hat{\textbf{L}}$ becomes fixed relative to
$\hat{\mathbf{L}}_\OUT$, while $\hat{\textbf{S}}$ precesses rapidly around $\hat{\mathbf{L}}$ with a fixed final
$\theta_\SL$ close to $90^\circ$.

Case II (Figures \ref{fig:OE oct 1}-\ref{fig:OE oct 2}):
This occurs when $I_0$ is not close to $I_{0,\li}$, but
$e_\m$ is driven to a value close to $e_\li$ due to the octupole effect.
As seem from Figures \ref{fig:OE oct 1} and \ref{fig:OE oct 2}, the inner binary
experiences multiple LK cycles, each with increasing $e_\m$ driven by the octupole potential;
eventually $e_\m$ becomes sufficiently large and the orbit decays rapidly.
Unlike case I, the spin evolution transitions from the ``nonadiabatic" regime to the ``adiabatic" regime quickly.
Because of the extremely rapid orbital decay, the oscillation of $\theta_\SL$ continues to the end
(by contrast, In Figures \ref{fig:OE quad 1}-\ref{fig:OE quad 2}, the $\theta_\SL$ oscillation freezes as the orbit decays),
and the final $\theta_\SL$ lies in the range $\theta_\SL^\f \in (0,\pi)$.

Case III (Figure \ref{fig:OE quad 3}): This occurs when $I_0$ is close to $I_{0,\li}$.
Similar to Case I, the orbit goes through eccentricity oscillations, suppression of the oscillations and circularization.
However, since $I_0\approx I_{0,\li}$, the orbital inclination oscillates with a small amplitude and passes through $90^\circ$.
This implies that $\hat{\textbf{S}}_1$ stays fairly close to $\hat{\textbf{L}}$ at the early stage (see Figure \ref{fig:spin trajectories})
and $\theta_\SL$ does not experience large amplitude ($0-\pi$) oscillations.
Eventually, $\theta_\SL$ settles down to a value below $90^\circ$.

Case IV (see Figure \ref{fig:OE quad oneshort}):
This also occurs when $I_0$ is close to $I_{0,\li}$.
Similar to Case II, the inner binary experiences extreme eccentricity excitation,
and merge within one LK cycle (one-shot merger).
Because $\hat{\mathbf{L}}$ basically does not evolve in time (see Figure \ref{fig:spin trajectories}),
a small ($<90^\circ$) $\theta_\SL^\f$ is produced.

It is clear that the spin evolution is complicated and
depends on various parameters and timescales.
Our understanding of the spin behaviors is based largely on the numerical integrations.
The four cases discussed above are representative, and do not
capture the complete sets of spin evolutionary behaviors.

Figures \ref{fig:merger window quad} and \ref{fig:merger window oct} (bottom panels) show the final distribution of $\theta_\SL^\f$
as a function of $\cos I_0$ in the merger window for several different systems.
When $e_\OUT=0$, $\theta_\SL^\f$ has a regular distribution, and most of the values are found around $\lesssim90^\circ$;
the spin evolution follows the examples in Case I ($\theta_\SL^\f\simeq90^\circ$),
Case III and Case IV (one-shot merger) discussed above.
When $e_\OUT\neq0$, $\theta_\SL^\f$ shows a much wider range of values from $0^\circ$ to $180^\circ$
due to the octupole effect (as in Case II discussed above).

Note that for small $e_\OUT$, the final spin-orbit misalignment angles $\theta_\mathrm{s_1l}^\f$ and
$\theta_\mathrm{s_2l}^\f$ are strongly correctly; this correlation is particularly strong for the
$e_\OUT=0$ case (see Figure \ref{fig:merger window oct}). This arises because for small $e_\OUT$, the
orbital evolution is regular. The spin vectors $\hat{\textbf{S}}_1$ and $\hat{\textbf{S}}_2$ evolve independently during
the orbital decay (since spin-spin coupling is negligible). Although the de-Sitter precession rates of $\hat{\textbf{S}}_1$
and $\hat{\textbf{S}}_2$ are different (since $m_1\neq m_2$), the spin evolution is regular as long as ${\cal A}_0\ll1$
(corresponding to Cases I, III and IV discussed in Section \ref{sec 4 2}). In particular, the ``$90^\circ$ attractor"
is a generic feature independent of the precise value of $\Omega_\mathrm{SL}$ (see Section \ref{sec 4 3}).
In contrast, for high $e_\OUT$ (see the $e_\OUT=0.9$ case in Figure \ref{fig:merger window oct}), the octupole
effect makes the orbital evolution chaotic, which also induces chaotic evolution of the spin-orbit misalignment
(see Figure \ref{fig:OE oct 2}). Therefore, $\theta_\mathrm{s_1l}^\f$ and
$\theta_\mathrm{s_2l}^\f$ become largely uncorrelated.

\subsection{Understanding the Spin Evolution: The $90^\circ$ Attractor?}
\label{sec 4 3}

We see from the previous subsections that when the octupole effect is negligible ($\varepsilon_\oct\ll 1$),
the spin-orbit misalignment angle (starting from initial $\theta_\SL^0=0^\circ$) often
evolves toward $\theta_\SL^\f\simeq90^\circ$ as the binary orbit decays.
What is the origin of this $90^\circ$ ``attractor"?

In \citet{Liu-ApJL}, we used the principle of adiabatic invariance to derive an
analytical expression of $\theta_\SL^\f$ for the case where the inner BH
binary remains circular in the presence of an inclined tertiary companion
(i.e. the inner binary merges by itself without eccentricity excitation,
although the binary orbital angular momentum axis $\hat{\mathbf{L}}$ does vary and
precess around $\hat{\mathbf{L}}_\OUT$). We can use similar idea to understand
qualitatively the origin of the $90^\circ$ attractor in quadrupole LK-induced mergers.

In the quadrupole order, the angular momentum axis $\hat{\mathbf{L}}$
varies at the rate given by Equation (\ref{eq:Omega L}).
This variation involves precession around $\hat{\mathbf{L}}_\OUT$ and
nutation (change in $I$). If we neglect nutation, we have
\footnote{Even when nutation is neglected, Equation (\ref{eq:L precess around Lout 2})
is approximate since a fast-varying term in Equation (\ref{eq:Omega pl}) has been neglected.
}
\be\label{eq:L precess around Lout 2}
\frac{d\hat{\mathbf{L}}}{dt}\bigg|_{\lk,\qu}\simeq
-\Omega_\mathrm{pl}\hat{\mathbf{L}}_\OUT\times\hat{\mathbf{L}}=
-\Omega_\mathrm{pl}\frac{\mathbf{L}_\tot}{L_\OUT}\times\hat{\mathbf{L}},
\ee
where $\mathbf{L}_\tot=\mathbf{L}+\mathbf{L}_\OUT$, and
\be\label{eq:Omega pl}
\Omega_\mathrm{pl}=\frac{3\hat{\mathbf{L}}\cdot \hat{\mathbf{L}}_\OUT}{4t_\lk\sqrt{1-e^2}}\big(1+4e^2\big).
\ee
Equation (\ref{eq:L precess around Lout 2}) shows that
$\hat{\bf L}$ rotates around the $\mathbf{L}_\tot$ axis. In this rotating frame, the spin evolution equation
(\ref{eq:spin}) transforms to
\be\label{eq:rotSpin}
\bigg(\frac{d\hat{\textbf{S}}_1}{dt}\bigg)_\mathrm{rot}=
\bm{\Omega}_\eff \times\hat{\textbf{S}}_1,
\ee
where
\be\label{eq:Omega eff}
\bm{\Omega}_\eff=\Omega_\mathrm{SL}\hat{\textbf{L}}+\Omega_\mathrm{pl} \frac{\mathbf{L}_\tot}{L_\OUT}.
\ee
Note that the ratio between $\Omega_\mathrm{SL}$ and $\Omega_\mathrm{pl}$ is
\be
\bigg|\frac{\Omega_\mathrm{SL}}{\Omega_\mathrm{pl}}\bigg|={\cal A}\sin I=
\frac{{\cal A}_0}{2(1+4e^2)\sqrt{1-e^2}|\cos I|},
\ee
where ${\cal A}$, ${\cal A}_0$ are given by Equations (\ref{eq:adiabaticity parameter})-(\ref{eq:adiabaticity parameter initial}).

If we {\it assume} $\hat{\bm{\Omega}}_\eff\equiv\bm{\Omega}_\eff/|\bm{\Omega}_\eff|$ varies slowly
(much slower than $|\bm{\Omega}_\eff|$; see below), then $\theta_{\eff,\mathrm{S}_1}$,
the angle between $\hat{\textbf{S}}_1$ and $\bm{\Omega}_\eff$, is an
adiabatic invariant. Suppose $\hat{\textbf{S}}_1$ and $\hat{\textbf{L}}$ are aligned initially
($\theta_\SL^0=0^\circ$), then the initial $\theta_{\eff,\mathrm{S}_1}^0$ equals the
initial $\theta_{\eff,\mathrm{L}}^0$ (the angle between $\bm{\Omega}_\eff$ and $\hat{\textbf{L}}$),
which is given by
\be
\tan\theta_{\eff,\mathrm{L}}^0=
\frac{\sin I_0}{({\cal A}_0/2\cos I_0) + \eta_0+\cos I_0},
\label{eq:tantheta}\ee
where $\eta_0$ is the initial value of $\eta=L/L_\OUT$ (see Equation \ref{eq:ratio of L})
\footnote{
Note that the definition of ${\cal A}_0$ in this paper is $2$ times that defined in \citet{Liu-ApJL}.
}. On the other hand, after the binary has decayed, $\eta\rightarrow 0$, $|\Omega_\mathrm{SL}|\gg|\Omega_\mathrm{pl}|$,
and thus $\bm{\Omega}_\eff=\Omega_\mathrm{SL}\hat{\textbf{L}}$, which implies
$\theta_\SL^\f\simeq\theta_{\eff,\mathrm{S}_1}^\f$. Therefore, under adiabatic evolution,
we have
\be\label{eq:final misalignment}
\theta_\SL^\f\simeq\theta_{\eff,\mathrm{S}_1}\simeq \theta_{\eff,\mathrm{L}}^0.
\ee
For systems with $\eta_0\ll1$, $|\cos I_0|\ll1$ and ${\cal A}_0/|\cos I_0|\ll1$,
Equation (\ref{eq:tantheta}) gives $\theta_{\eff,\mathrm{L}}^0\approx90^\circ$, and thus adiabatic evolution
predicts $\theta_\SL^\f\approx90^\circ$.

\begin{figure}
\centering
\begin{tabular}{cccc}
\includegraphics[width=9cm]{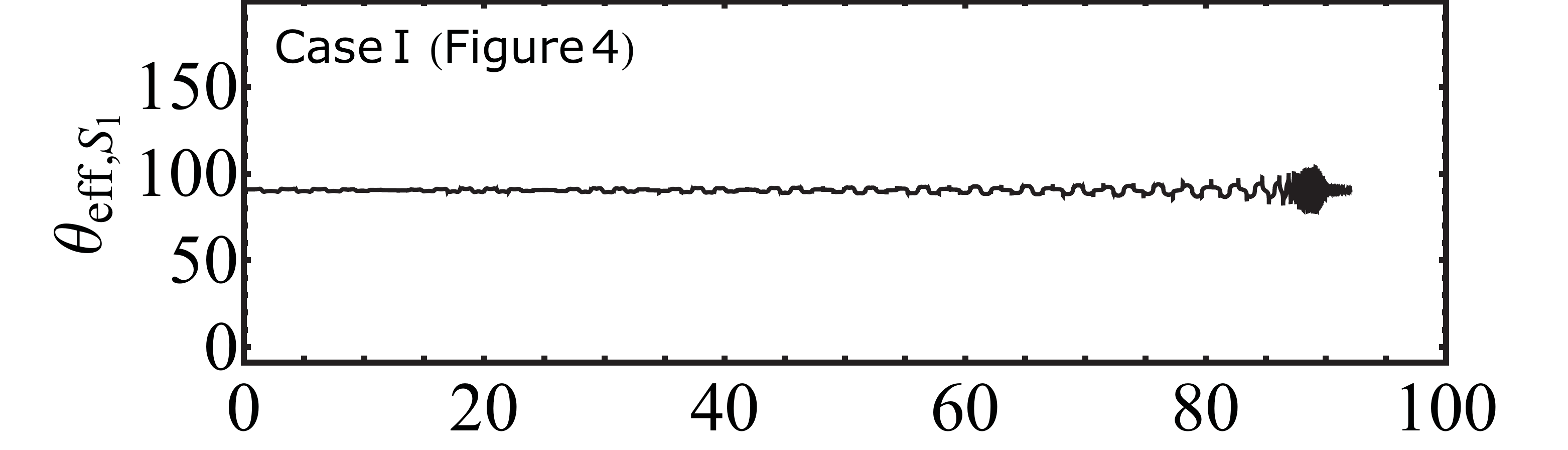}\\
\includegraphics[width=9cm]{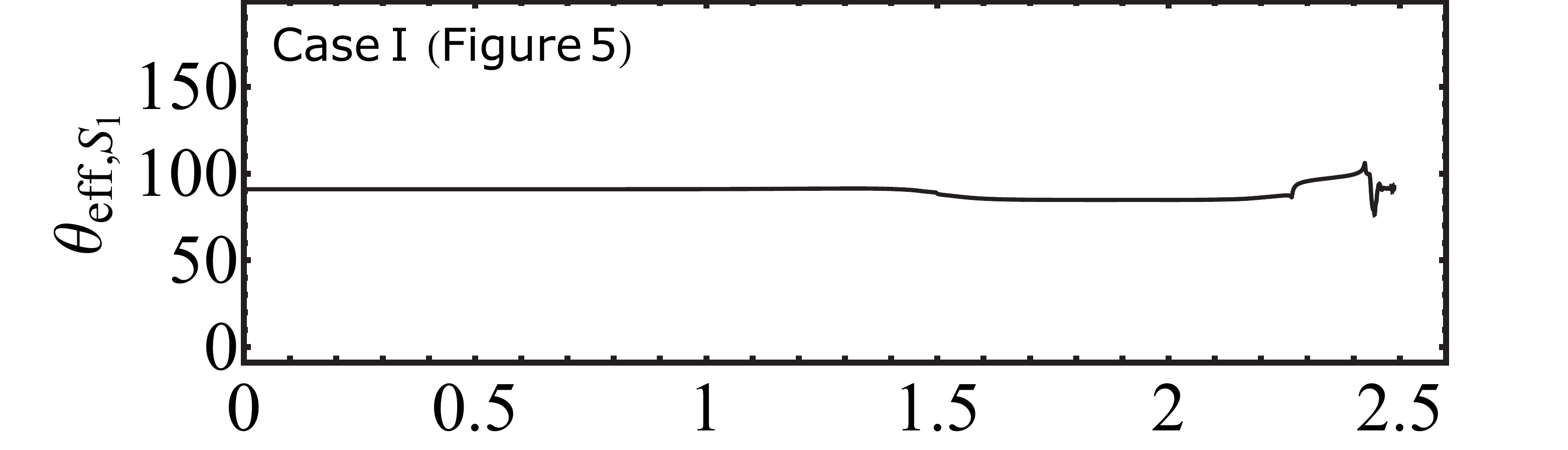}\\
\includegraphics[width=9cm]{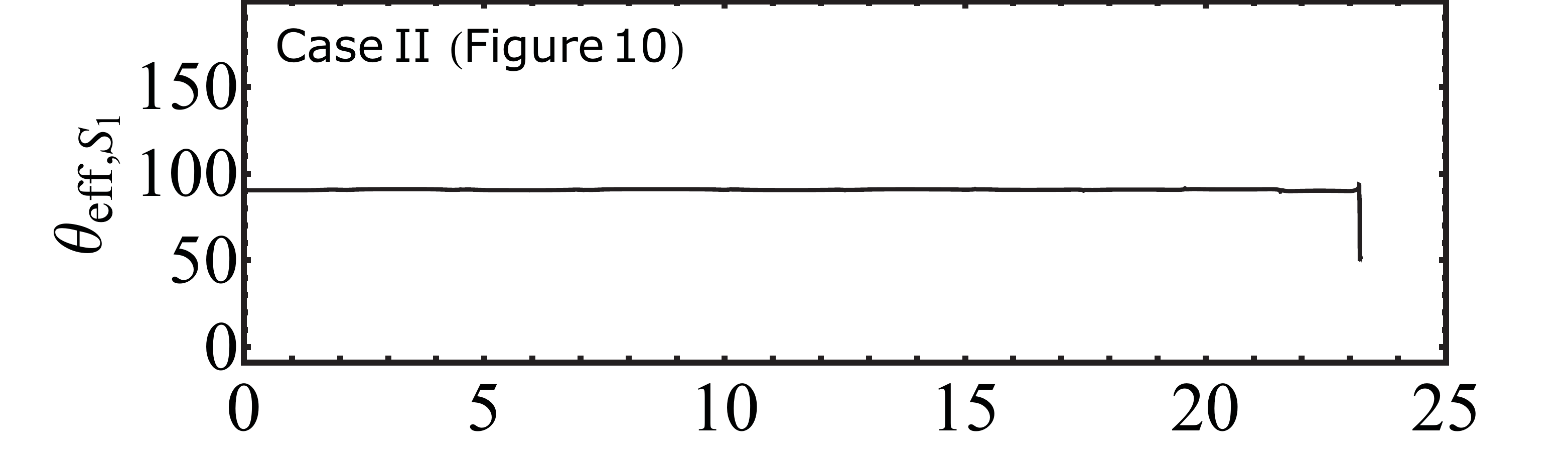}\\
\includegraphics[width=9cm]{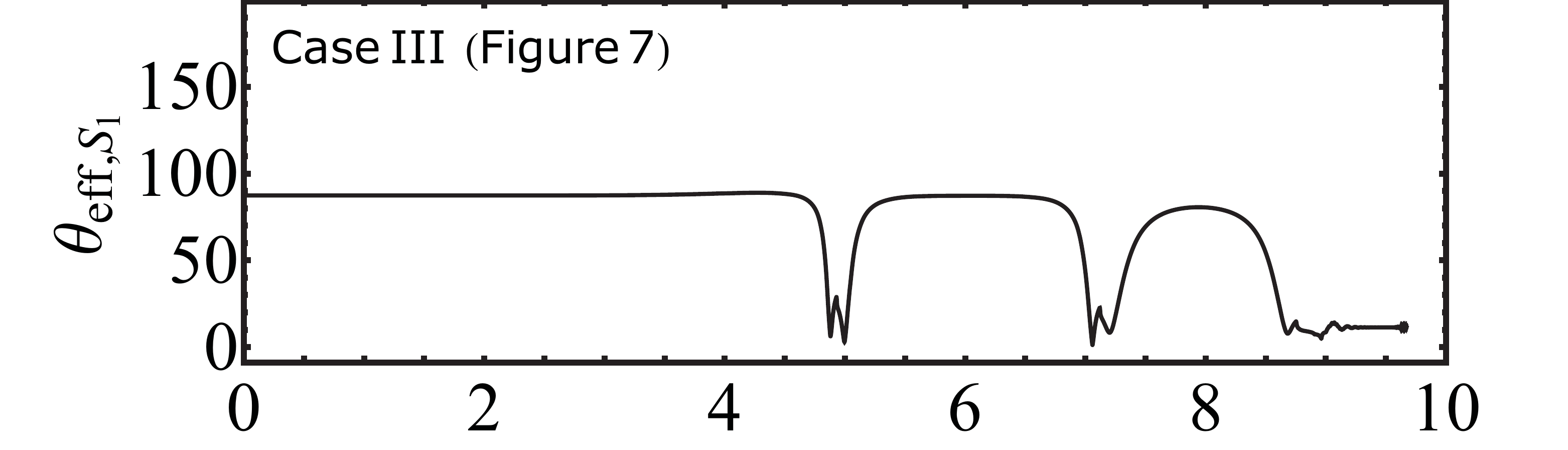}\\
\includegraphics[width=9cm]{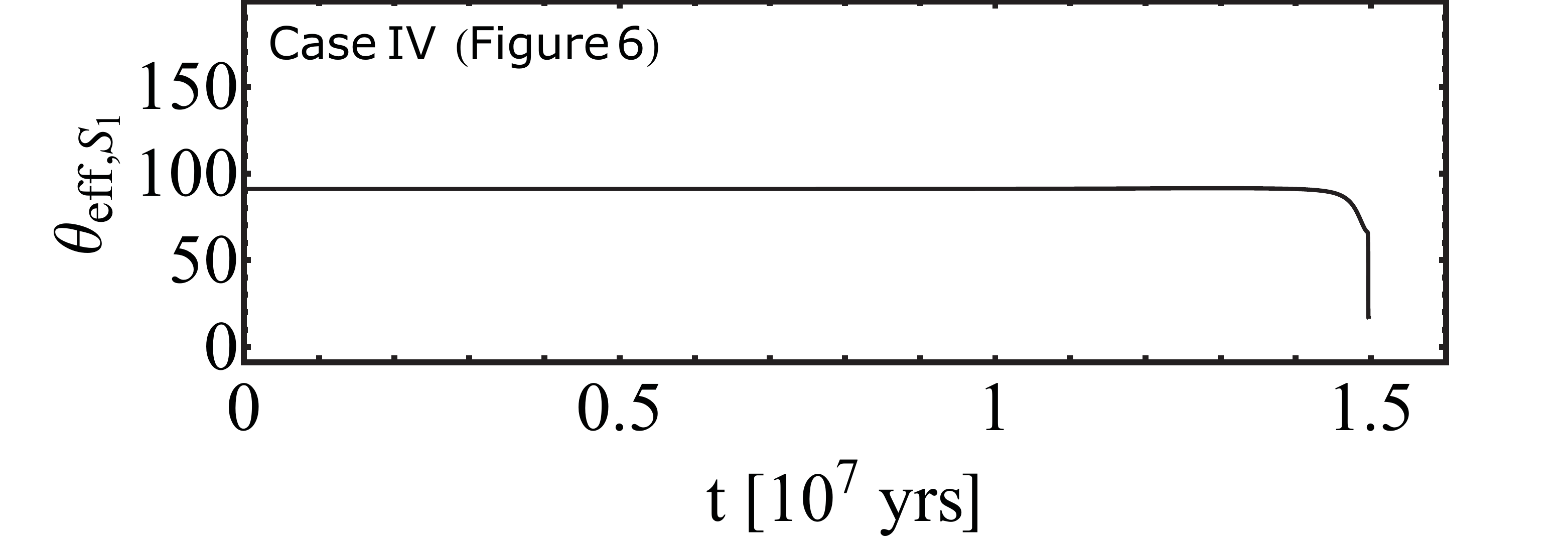}
\end{tabular}
\caption{The time-evolution of $\theta_{\eff,\mathrm{S}_1}$, the angle between $\hat{\mathbf{S}}_1$ and $\bm{\Omega}_\eff$
(Equation \ref{eq:Omega eff}). Each curve ends where the binary enters the aLIGO band,
at which point $\theta_{\eff,\mathrm{S}_1}=\theta_\SL^\f$.
From the top to the bottom, the examples shown correspond to
Figures \ref{fig:OE quad 1}, \ref{fig:OE quad 2}, \ref{fig:OE oct 1}, \ref{fig:OE quad 3} and \ref{fig:OE quad oneshort}, respectively.
}
\label{fig:adiabatic invariant}
\end{figure}

Figure \ref{fig:adiabatic invariant} shows the evolution of $\theta_{\eff,\mathrm{S}_1}$ for the four cases
discussed in Section \ref{sec 4 2}. We see that for Case I (Figures \ref{fig:OE quad 1} and \ref{fig:OE quad 2}),
$\theta_{\eff,\mathrm{S}_1}$ is approximately constant throughout the evolution of the inner binary,
and the adiabatic evolution correctly predicts the $90^\circ$ attractor in the
spin-orbit misalignment. For the other cases (Case II-IV), $\theta_{\eff,\mathrm{S}_1}$ undergoes
significant change during the inner binary's evolution, especially near the final orbital decay phase;
in these cases, Equation (\ref{eq:tantheta}) does not predict the correct $\theta_\SL^\f$.

The validity of adiabatic evolution requires that the rate of change of
$\hat{\bm{\Omega}}_\eff=\bm{\Omega}_\eff/|\bm{\Omega}_\eff|$ be much slower than $|\bm{\Omega}_\eff|$,
i.e.$|d\hat{\bm{\Omega}}_\eff/dt|\ll|\bm{\Omega}_\eff|$. In order of magnitude, we have
$|d\hat{\bm{\Omega}}_\eff/dt|\sim T_\gw^{-1}$ (see Equation \ref{eq:decay rate})
\footnote{
Since $\Omega_\mathrm{SL}$ and $\Omega_\mathrm{pl}$ both depend on $e$, the vector $\bm{\Omega}_\eff$
also varies on the timescale $t_\lk\sqrt{1-e^2}$, which can be comparable to
$|\Omega_\mathrm{pl}|^{-1}$. However, in the early phase, $|\Omega_\mathrm{pl}|\gg \Omega_\mathrm{SL}$
(this breaks down when $I$ crosses $90^\circ$, as in Case III; see Figure \ref{fig:OE quad 3}),
we have $\hat{\bm{\Omega}}_\eff\simeq\hat{\mathbf{L}}_\tot\simeq\hat{\mathbf{L}}_\OUT$, which is
nearly constant. As the orbit decays, $\Omega_\mathrm{SL}$ becomes large relative to $\Omega_\mathrm{pl}$,
and $\hat{\bm{\Omega}}_\eff$ transitions to $\hat{\mathbf{L}}$.
}. In Case I, $e_\m$ induced by the tertiary companion is not too extreme.
So the orbital decay is ``gentle" and the adiabatic condition is satisfied.
In Cases II, III and IV, the rapid orbital decay at high eccentricity implies $T_\gw^{-1}\gtrsim|\bm{\Omega}_\eff|$,
so the adiabatic evolution breaks down.

We reiterate that the above analysis cannot be considered rigorous, since the precession rate in
Equation (\ref{eq:L precess around Lout 2}) is approximate and the nutation of $\hat{\mathbf{L}}$ has been
neglected. Nevertheless, this analysis (especially Figure \ref{fig:adiabatic invariant}) provides a
qualitative understanding as to why $\theta_\SL$ evolves towards $90^\circ$ under some conditions.

\subsection{Final Distribution of Spin-Orbit Misalignment Angles}
\label{sec 4 4}

Having studied the various spin evolutionary paths in the previous subsections, we now
calculate the distribution of final spin-orbit
misalignment angle for the merging systems studied in Figure \ref{fig:merger fraction oct}.
We consider the spins of both BHs, and assume that both $\mathbf{S}_1$ and $\mathbf{S}_2$ are initially aligned
with respect to the binary orbital angular momentum axis.

\begin{figure*}
\centering
\begin{tabular}{cc}
\includegraphics[width=8cm]{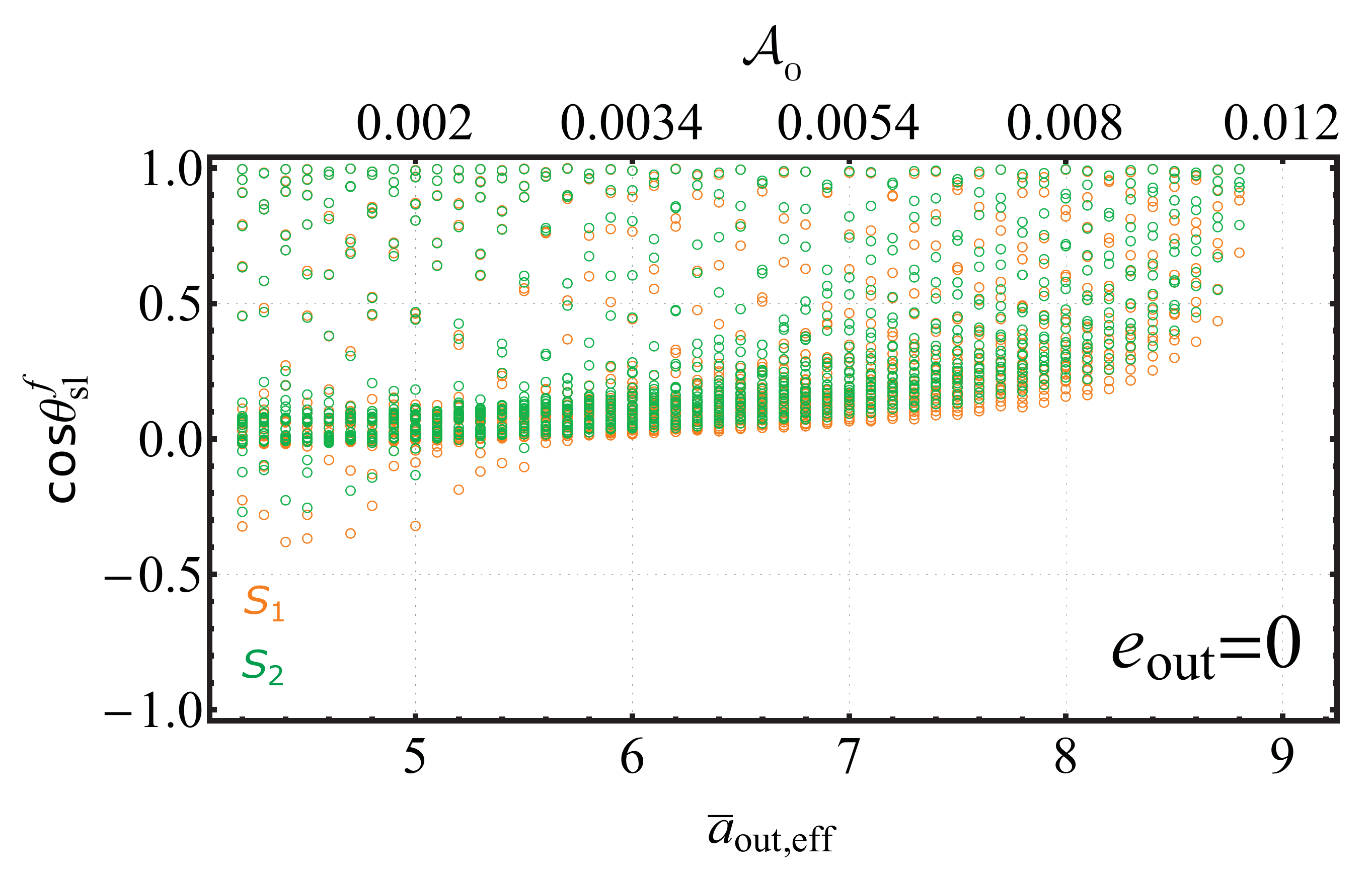}&
\includegraphics[width=8cm]{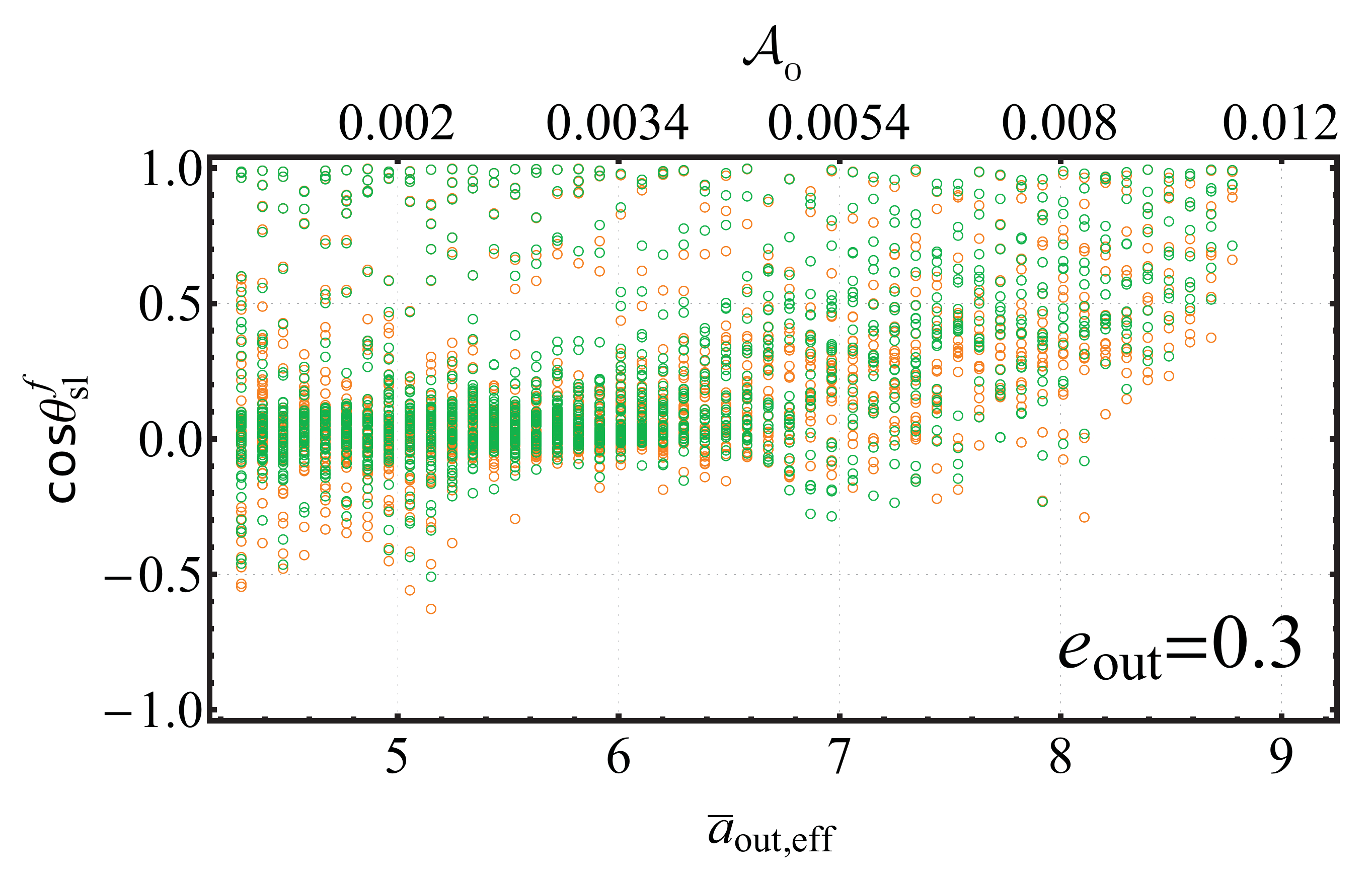}\\
\includegraphics[width=4.6cm]{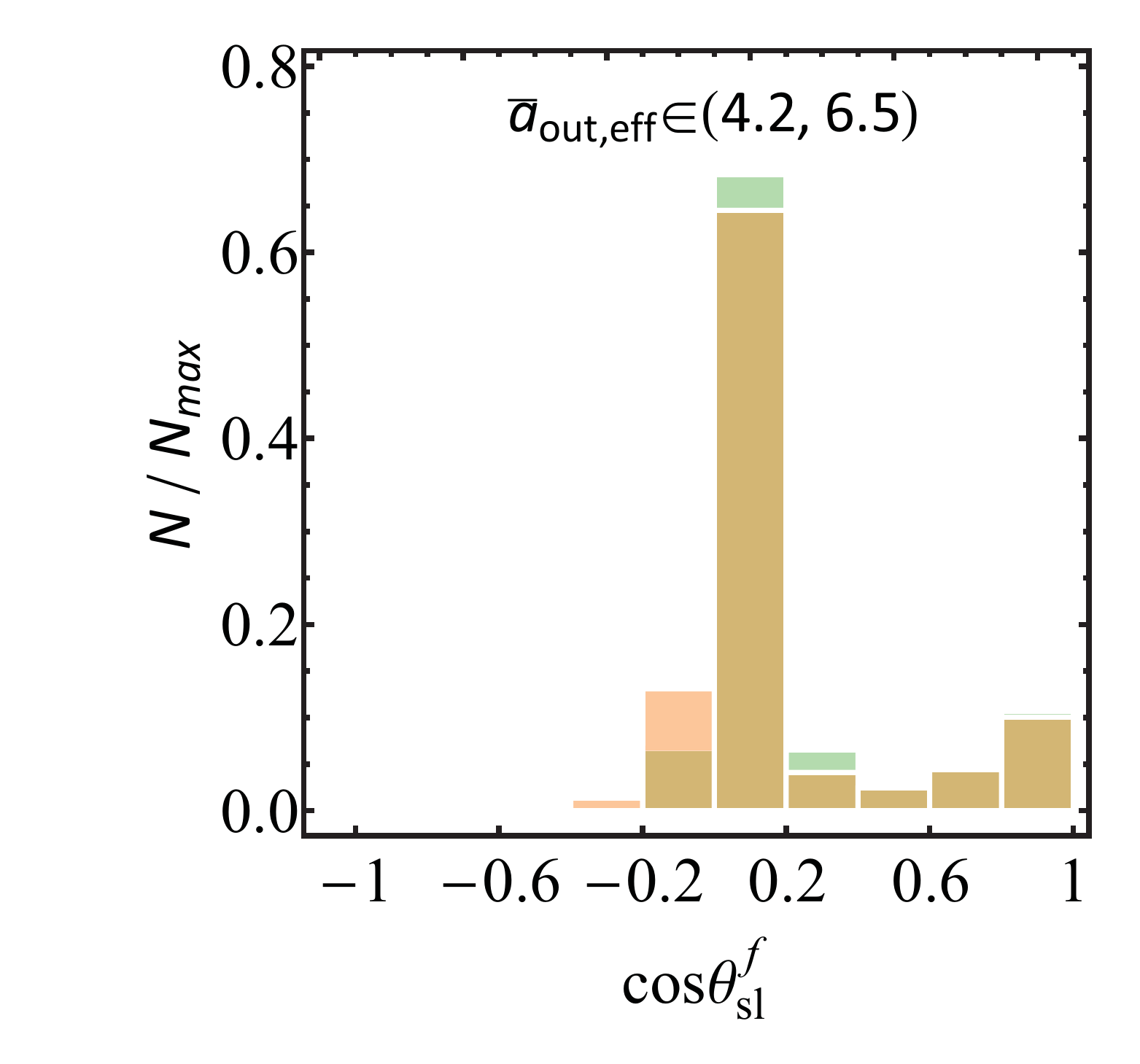}
\includegraphics[width=3.4cm]{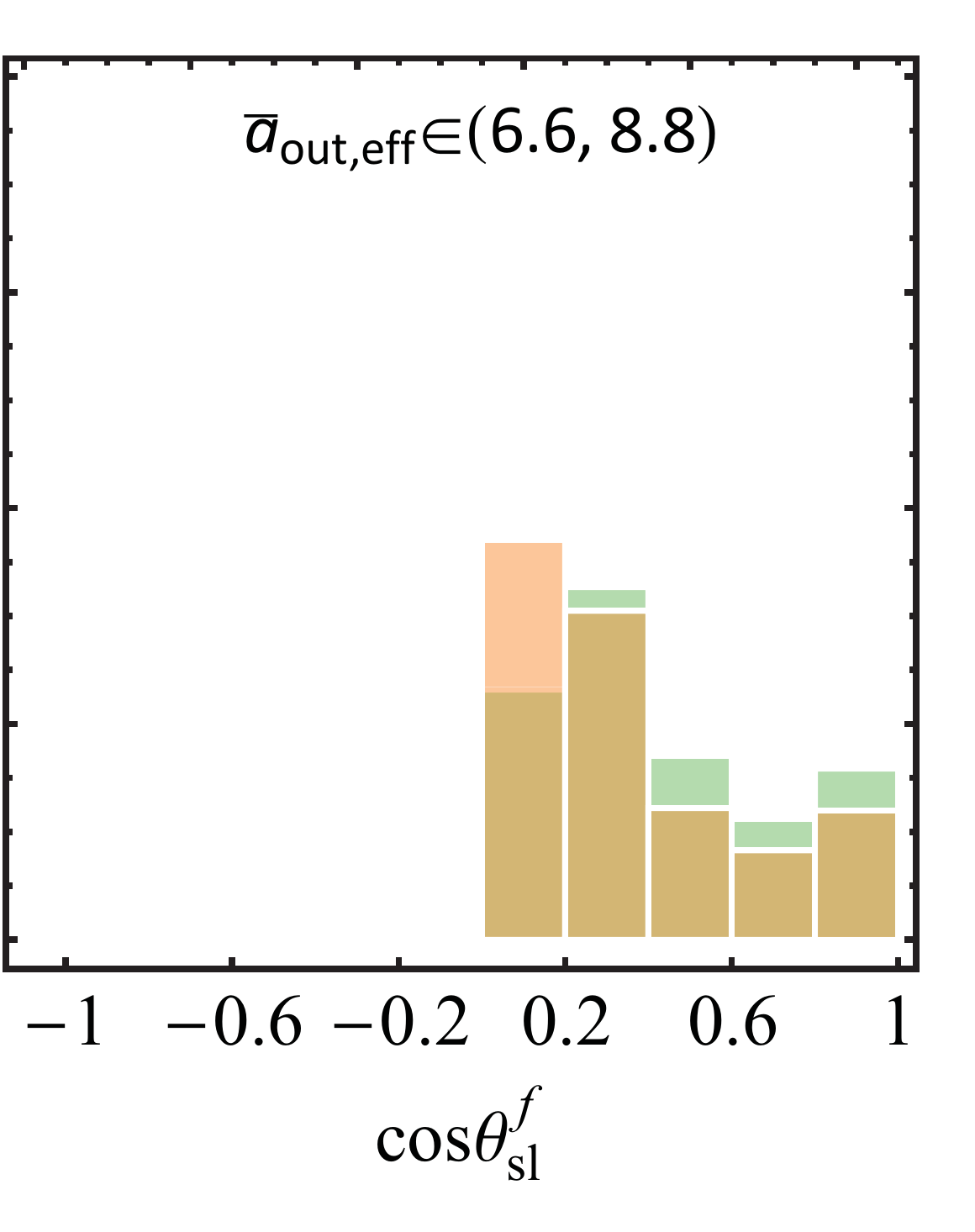}&
\includegraphics[width=4.6cm]{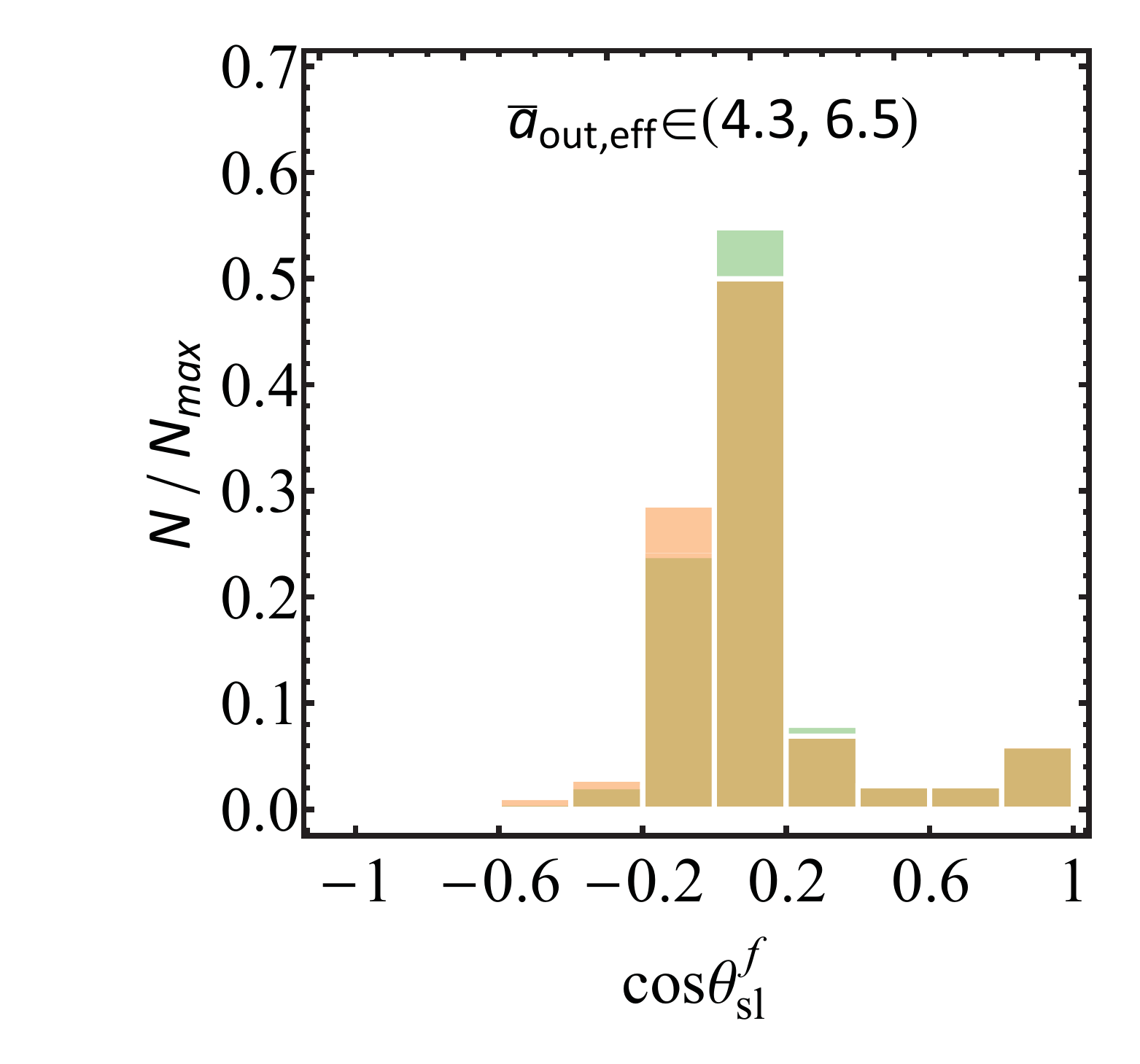}
\includegraphics[width=3.4cm]{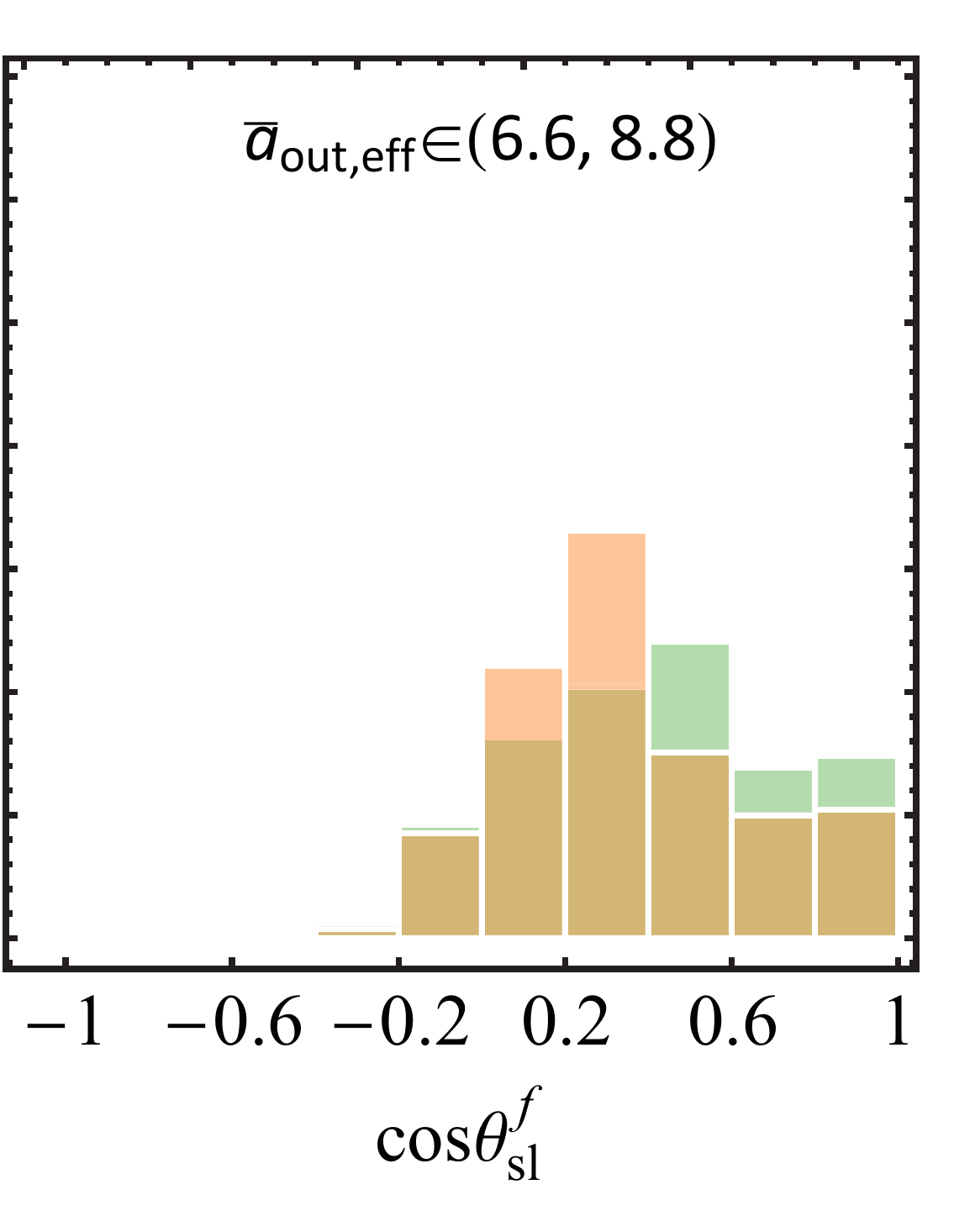}\\
\\
\\
\includegraphics[width=8cm]{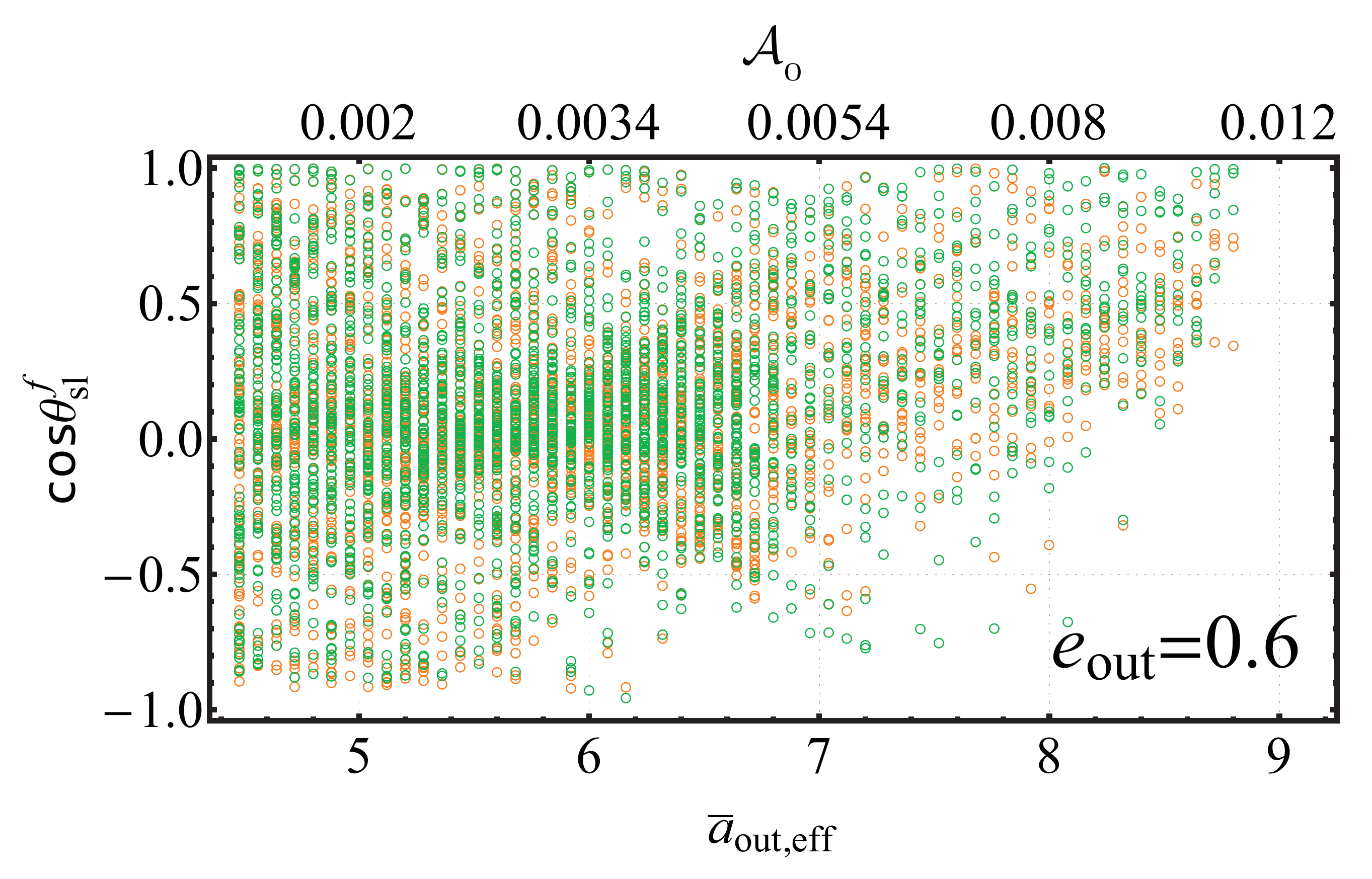}&
\includegraphics[width=8cm]{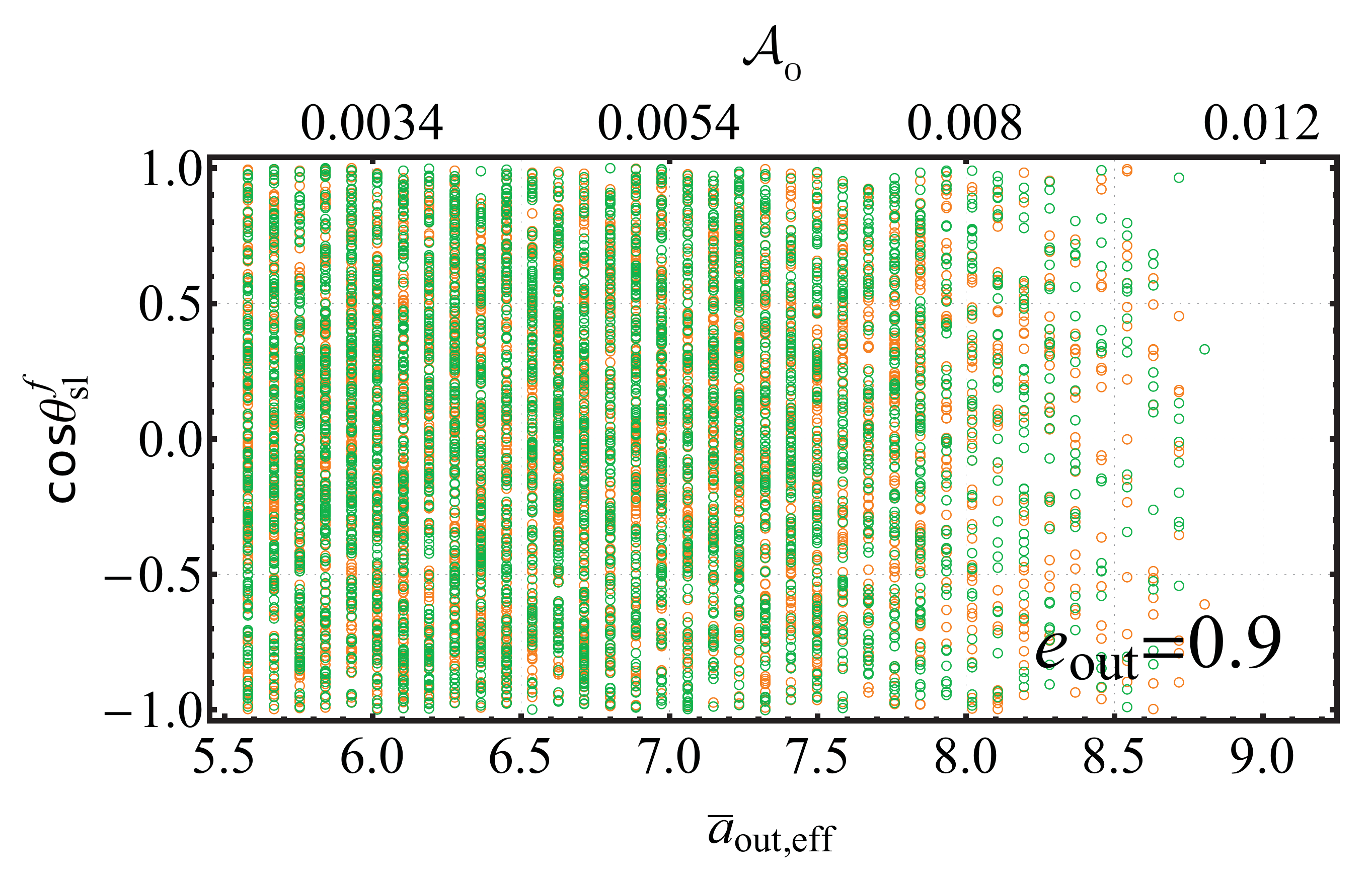}\\
\includegraphics[width=4.6cm]{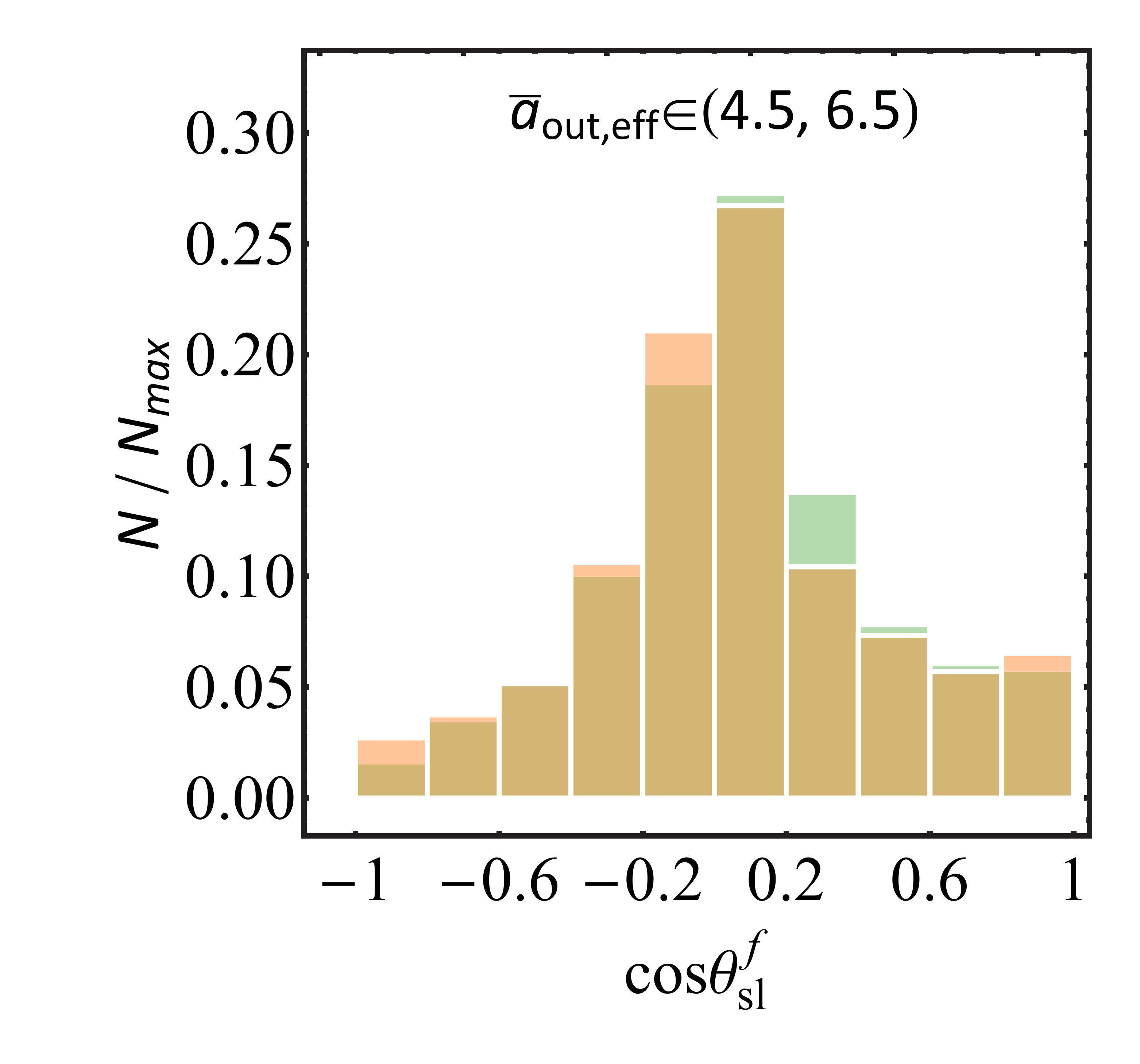}
\includegraphics[width=3.4cm]{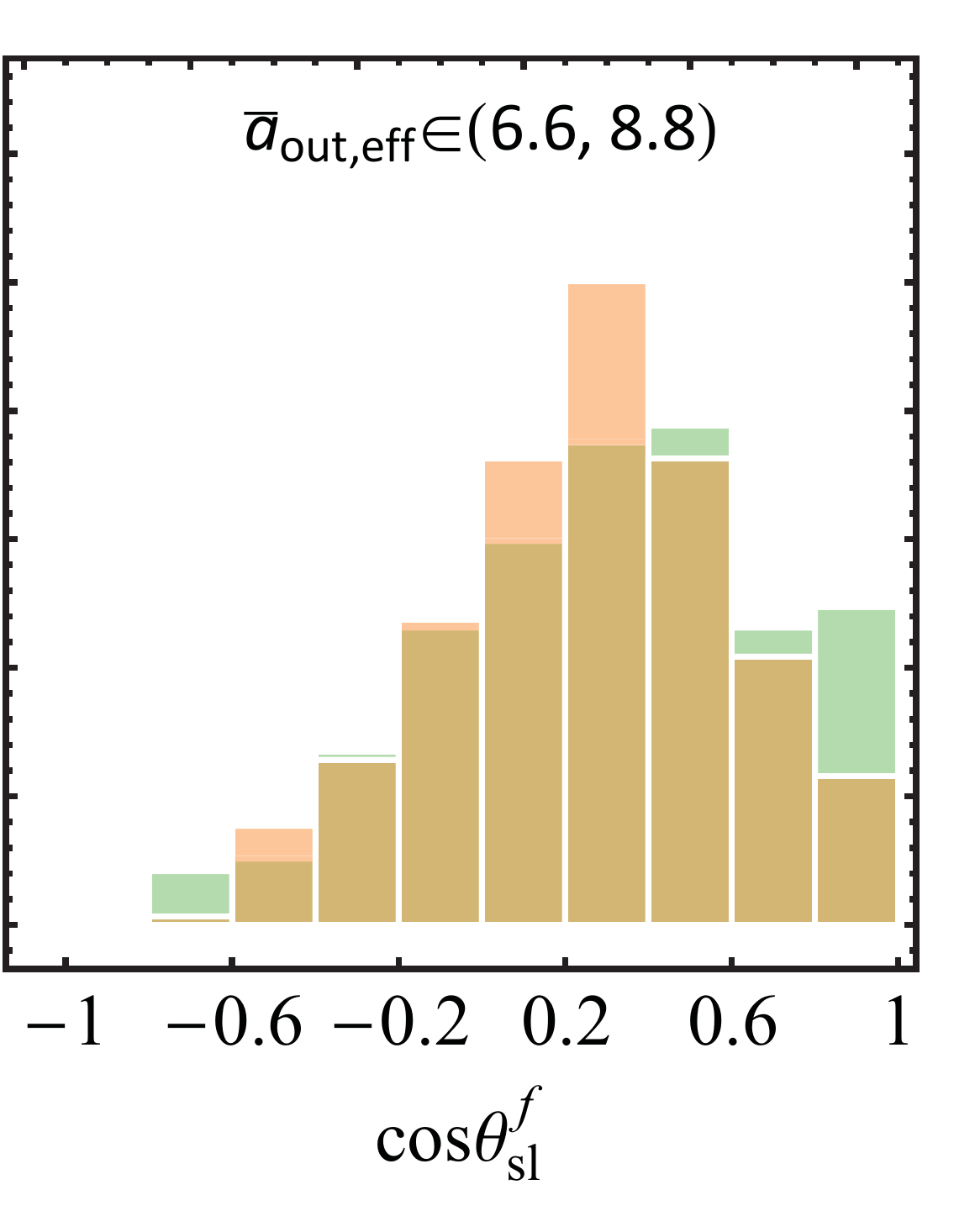}&
\includegraphics[width=8cm]{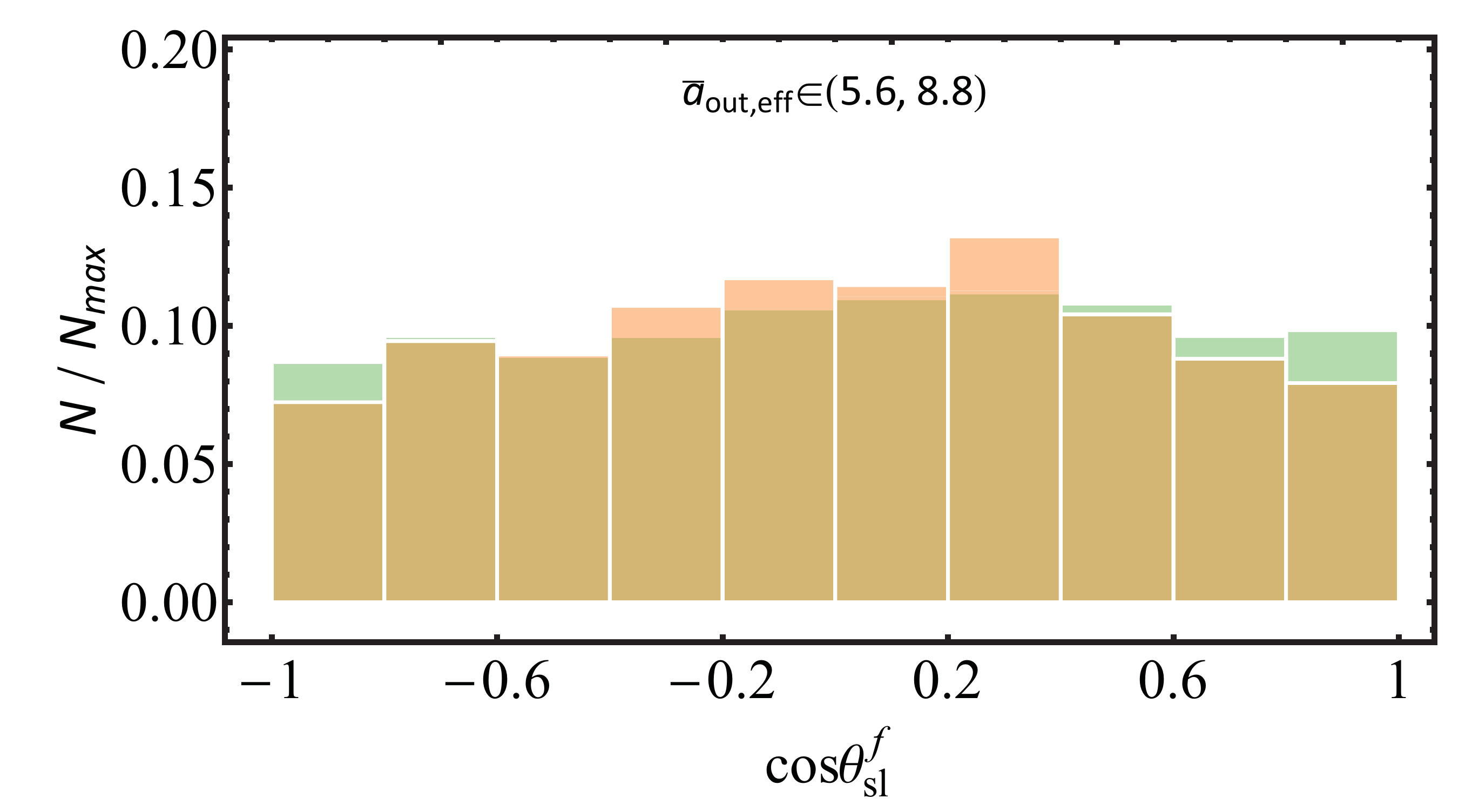}\\
\end{tabular}
\caption{The final spin-orbit misalignment angles for both $\mathbf{S}_1$ and $\mathbf{S}_2$ as a function of $\bar{a}_{\OUT,\eff}$
(see Equation \ref{eq:aout bar}), and the associated distribution.
The four cases ($e_\OUT=0, 0.3, 0.6, 0.9$)
shown here are from the mergers achieved by the double-averaged secular equations as depicted
in the left panels of Figure \ref{fig:merger fraction oct}.
The system parameters are $m_1=30M_\odot$, $m_2=20M_\odot$, $m_3=30M_\odot$, $a_0=100\au$.
The parameter $\mathcal{A}_0$ (Equation \ref{eq:adiabaticity parameter initial}) here corresponds to the spinning body $m_1$.
In the distribution ($N/N_\m$ versus $\cos \theta_\SL^\f$), the range of $\bar{a}_{\OUT,\eff}$ is
specified, and $N_\m$ is the number of the merger events for the corresponding range of $\bar{a}_{\OUT,\eff}$.
In all panels, orange corresponds to $\mathbf{S}_1$, green corresponds to $\mathbf{S}_2$,
and brown corresponds to the overlapped region.
}
\label{fig:spin distribution}
\end{figure*}

Figure \ref{fig:spin distribution} summarizes our results
for $a_0=100\au$ (the results for $a_0=20\au$ are very similar).
The range of $\bar{a}_{\OUT,\eff}$ values considered in this figure all lie in the
regime where the double averaging approximation is valid (see Section \ref{sec 2 3}).
As in Figure \ref{fig:merger fraction oct}, four different values of $e_\OUT$ are considered.
When $e_\OUT=0$ (the top left panels), the spin evolution is regular, following the examples of
Case I, Case III and Case IV (see Section \ref{sec 4 2}).
For $\bar{a}_{\OUT,\eff}\in(4.2, 6.5)$, the tertiary companion is relatively close,
many BH binaries inside the merger window pass through successive stages of LK oscillations, LK suppression and orbital circularization
(Case I), producing a large number of systems with $\theta_\SL^\f$ around $90^\circ$.
For $\bar{a}_{\OUT,\eff}\in(6.6, 8.8)$,
the tertiary companion is relatively distant; the eccentricity cannot grow to
be as in the case of small $\bar{a}_{\OUT,\eff}$, even
when $I_0\simeq I_{0,\li}$. Thus, the spin mainly evolves as described in Case III,
and $\theta_\SL^\f$ lies in the range of $0^\circ-90^\circ$.

When the companions are eccentric ($e_\OUT\neq0$),
the octupole effect comes into play and
the spin may follow the dynamics of Case II.
For a given $e_\OUT$, when the companion is relatively close ($\bar{a}_{\OUT,\eff}$ is small),
the octupole effect becomes more prominent.
In the case of $e_\OUT=0.9$, the orbital evolution is dominated by the octupole effect,
and the distribution of $\theta_\SL^\f$ is close to being isotropic
(i.e. uniform distribution in $\cos\theta_\SL^\f$),
as shown in the bottom-right panel of Figure \ref{fig:spin distribution}.

Since the two components of the BH binary have comparable masses, the de-Sitter precession rates are
similar. Thus it is not surprising that the distributions of $\theta_\SL^\f$ for both spins are similar.
Note that $\theta_\mathrm{s_1l}^\f$ and $\theta_\mathrm{s_2l}^\f$ are strongly correlated for
$e_\OUT=0$, and this correlation becomes much weaker as the octupole effect becomes stronger (see Figure \ref{fig:merger window oct}).

\begin{figure}
\begin{centering}
\begin{tabular}{cc}
\includegraphics[width=8cm]{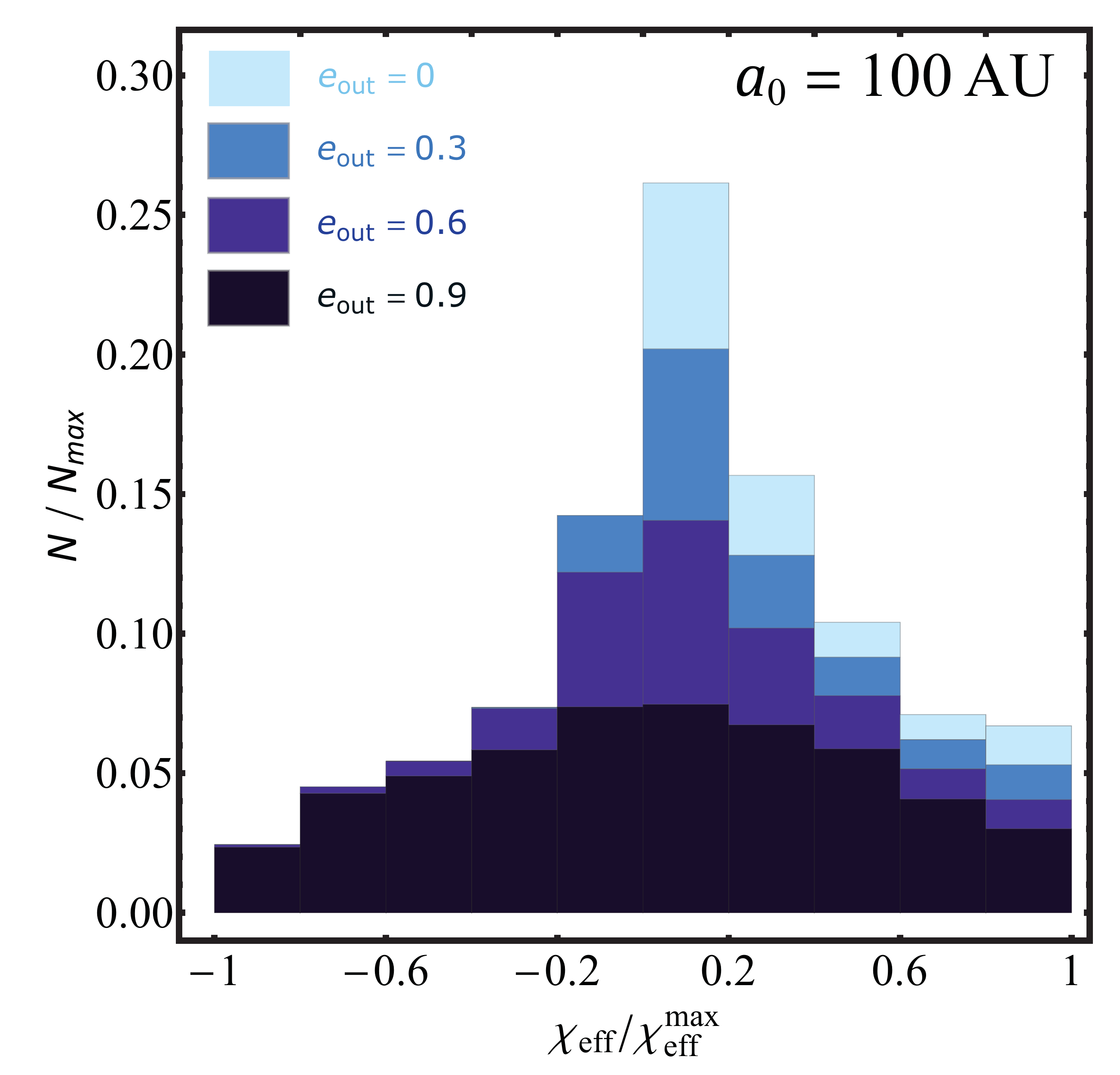}\\
\includegraphics[width=8cm]{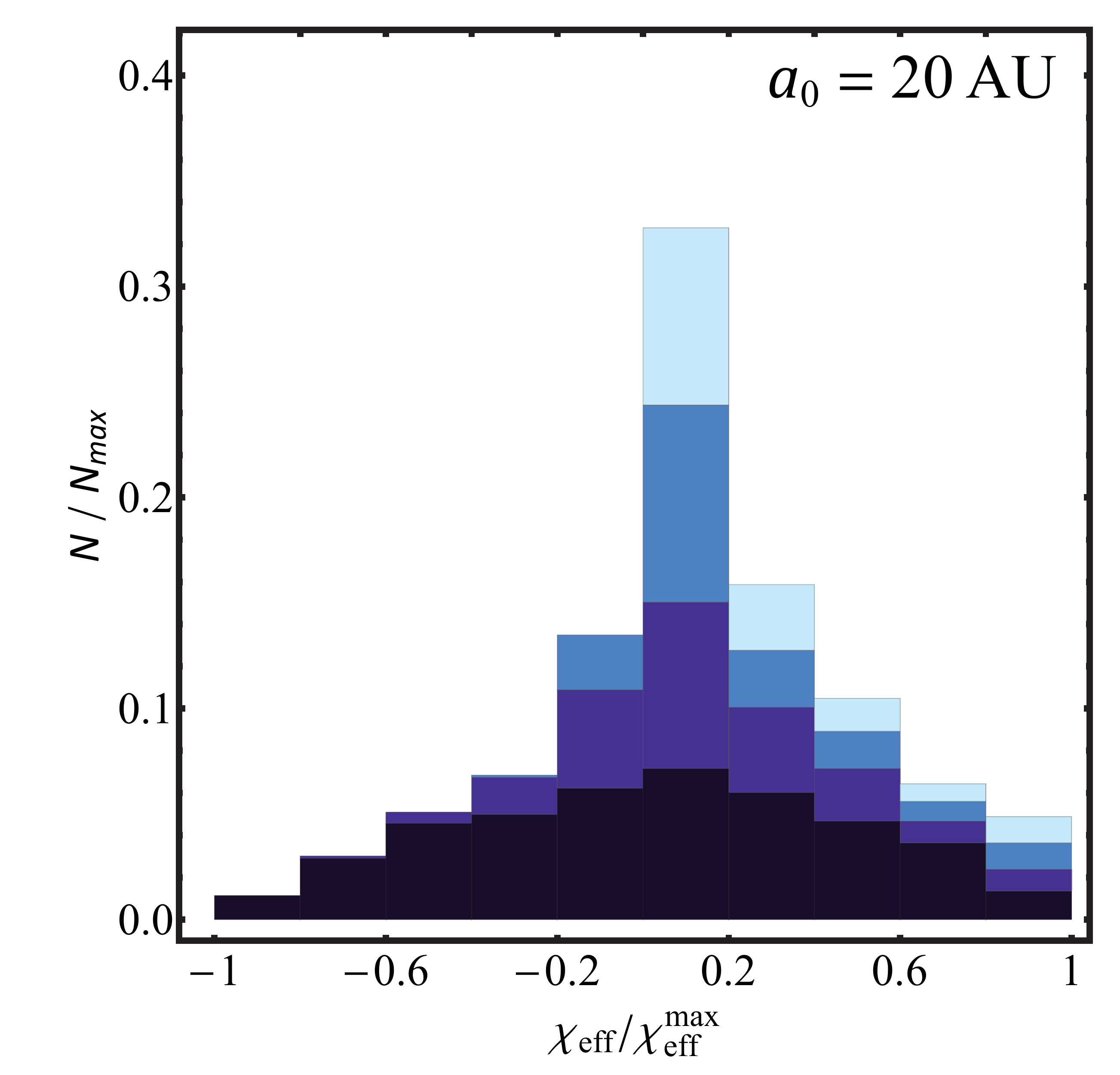}
\end{tabular}
\caption{The overall distribution of the rescaled binary spin parameter
$\chi_\eff$ (Equations \ref{eq:chi eff} and \ref{eq:bar chi eff})
normalized by the total number of mergers.
In these two examples, we set $\chi_1=\chi_2$.
The top panel is for $a_0=100\au$ (see Figure \ref{fig:spin distribution}) and we include merging systems with
$\bar{a}_{\OUT,\eff}\in(5.6, 8.8)$; for each $e_\OUT$, the
number of mergers is 673 ($e_\OUT=0$), 790 ($e_\OUT=0.3$), 1159 ($e_\OUT=0.6$), and 2828 ($e_\OUT=0.9$), respectively,
so that $N_\m=5450$. The lower panel is for $a_0=20\au$, and we include systems with
$\bar{a}_{\OUT,\eff}\in(0.8, 1.4)$; the
number of mergers is 146 ($e_\OUT=0$), 180 ($e_\OUT=0.3$), 227 ($e_\OUT=0.6$), and 411 ($e_\OUT=0.9$), respectively,
so that $N_\m=964$.
The other parameters are the same as in Figure \ref{fig:spin distribution}.
}
\label{fig:overall spin distribution}
\end{centering}
\end{figure}

Having obtained the distributions of $\cos\theta_\mathrm{s_1l}^\f$ and $\cos\theta_\mathrm{s_2l}^\f$
for a range of systems with different parameters,
we can compute the distribution of the effective spin parameter for the merging binaries
(see Equation \ref{eq:effective spin parameter})
\be\label{eq:chi eff}
\chi_{\rm eff}={m_1 \chi_1\cos\theta_\mathrm{s_1l}^\f + m_2 \chi_2\cos\theta_\mathrm{s_2l}^\f\over m_{12}},
\ee
where $\chi_{1,2}$ are the dimensionless BH spins (we set $\chi_1=\chi_2=0.1$ in our calculations,
although our results for $\theta_\mathrm{s_1l}^\f$ and $\theta_\mathrm{s_2l}^\f$
are not affected by this choice since $S_1, S_2\ll L$ for all the systems
considered in this paper).
Figure \ref{fig:overall spin distribution} shows two examples
(for $a_0=100\au$ and $20\au$; see Figures \ref{fig:merger fraction oct} and \ref{fig:spin distribution}),
assuming $\chi_1=\chi_2$
\footnote{
Note that although the distribution of $\cos\theta_\mathrm{s_1l}^\f$ and $\cos\theta_\mathrm{s_2l}^\f$
are independent of the values of $\chi_1$ and $\chi_2$
(see the discussion in the paragraph following Equation \ref{eq:Omega LS}),
the distribution of $\chi_\eff/\chi_\eff^\m$ obviously depends on $\chi_1$ and $\chi_2$.
}.
To obtain the $\chi_\eff$ distribution,
we consider systems with $\bar{a}_{\OUT,\eff}\in(5.6, 8.8)$ for the $a_0=100\au$ case
and $\bar{a}_{\OUT,\eff}\in(0.8, 1.4)$ for the $a_0=20\au$ case, and assume that the eccentricity of the
tertiary companion has a uniform distribution in $e_\OUT$
(i.e. $e_\OUT=0, 0.3, 0.6, 0.9$ are equally probable),
and the initial mutual inclination is randomly distributed (uniform in $\cos I_0$).
We see that although the systems with the most eccentric companion ($e_\OUT=0.9$) contribute a substantial fraction of mergers,
the overall distribution of $\chi_\eff$
has a peak around $0$.
Importantly, our result indicates that LK-induced BH binary mergers can easily have $\chi_\eff<0$ \citep[see also][]{Liu-ApJL}.
This is quite different from the standard isolated binary evolution channel,
where we typically expect spin-orbit alignment and $\chi_\eff>0$.

If the distributions of $\cos\theta_\mathrm{s_1l}^\f$ and $\cos\theta_\mathrm{s_2l}^\f$ are uncorrelated,
as we may expect to be the case for $m_1\neq m_2$, when the octupole effect is significant (see Figure \ref{fig:merger window oct}
and the discussion in the last paragraph of Section \ref{sec 4 2}),
the distribution of $\chi_\eff$ can be derived directly from
$P_1 (\cos\theta_\mathrm{s_1l}^\f)$ and $P_2 (\cos\theta_\mathrm{s_2l}^\f)$,
the distribution functions of $\cos\theta_\mathrm{s_1l}^\f$ and $\cos\theta_\mathrm{s_2l}^\f$.
Define
\ba
&&\bar{\chi}_1\equiv\frac{m_1\chi_1}{m_1\chi_1+m_2\chi_2},\\
&&\bar{\chi}_2\equiv\frac{m_2\chi_2}{m_1\chi_1+m_2\chi_2},\\
&&\bar{\chi}_\eff\equiv\frac{\chi_\eff}{\chi_\eff^\m}\equiv\frac{m_{12}\chi_\eff}{m_1\chi_1+m_2\chi_2},\label{eq:bar chi eff}
\ea
where $\chi_\eff^\m=(m_1\chi_1+m_2\chi_2)/m_{12}$ is the
maximum possible value of $\chi_\eff$ for given $m_1\chi_1$ and $m_2\chi_2$
(this maximum is achieved at $\cos\theta_\mathrm{s_1l}^\f=\cos\theta_\mathrm{s_2l}^\f=1$).
Then Equation (\ref{eq:chi eff}) becomes
\be
\bar{\chi}_\eff=\bar{\chi}_1\mu_1+\bar{\chi}_2\mu_2,
\ee
where $\mu_1\equiv\cos\theta_\mathrm{s_1l}^\f$ and $\mu_2\equiv\cos\theta_\mathrm{s_2l}^\f$.
Note that $\bar{\chi}_1+\bar{\chi}_2=1$ and $\bar{\chi}_\eff\in[-1,1]$.
Given $P_1 (\mu_1)$ and $P_2 (\mu_2)$, the distribution function of $\bar{\chi}_\eff$ is
\be\label{eq:probability}
\begin{split}
P(\bar{\chi}_\eff)=&\int^1_{-1}d\mu_1P_1(\mu_1)\int^1_{-1}d\mu_2P_2(\mu_2)\\
&\times\delta(\bar{\chi}_\eff-\bar{\chi}_1\mu_1-\bar{\chi}_2\mu_2).
\end{split}
\ee

In the special case when $\mu_1$ and $\mu_2$ are uniformly distributed, we
have $P_1=P_2=1/2$, and Equation (\ref{eq:probability}) gives
\be\label{eq:p chi eff}
P(\bar{\chi}_\eff)=\left\{
\begin{array}{ccc}
\begin{split}
&(1-\bar{\chi}_\eff)/(4\bar{\chi}_1\bar{\chi}_2),~~~~~\bar{\chi}_\eff\geq\bar{\chi}_1-\bar{\chi}_2\\
&1/(2\bar{\chi}_1),~~~~~\bar{\chi}_2-\bar{\chi}_1\leq\bar{\chi}_\eff\leq\bar{\chi}_1-\bar{\chi}_2\\
&(1+\bar{\chi}_\eff)/(4\bar{\chi}_1\bar{\chi}_2),~~~~~\bar{\chi}_\eff\leq\bar{\chi}_2-\bar{\chi}_1,
\end{split}
\end{array}
\right.
\ee
where we have assumed $\bar{\chi}_1\geq\bar{\chi}_2$ without loss of generality.
Thus, even for uniform distributions of $\cos\theta_\mathrm{s_1l}^\f$ and $\cos\theta_\mathrm{s_2l}^\f$
(see the case of $e_\OUT=0.9$ in Figure \ref{fig:spin distribution}),
the effective spin parameter $\chi_\eff$ is preferentially distributed around $\chi_\eff=0$
(see Figure \ref{fig:overall spin distribution}).

Note that the spin-orbit misalignment distribution and $\chi_\eff$ distribution obtained above refer to
relatively wide BH binary systems ($a_0\gtrsim10-100\au$) that experience merger due to large LK eccentricity
excitation. Such systems necessarily have ${\cal A}_0\ll1$.
For BH binaries with smaller separations ($a_0\lesssim1\au$) and
${\cal A}_0$ not much less than unity, the
spin-orbit misalignment distribution can be quite different \citep[see][]{Liu-ApJL}.

\citet{Antonini spin} \citep[see also][]{Rodriguez Spin} have carried population studies of BH
mergers in triple systems (based on double-averaged secular equations) and
have found a similar peak around $\chi_\eff=0$ in the $\chi_\eff$ distribution.
\citet{Rodriguez Spin} also showed an example of the final spin-orbit misalignment distribution with a peak
around $90^\circ$, in qualitative agreement with our result.
They did not distinguish the difference in the spin-orbit misalignment distributions between largely
quadrupole systems (small $\varepsilon_\oct$) and strong octupole systems (large $\varepsilon_\oct$).
We do not agree with the reason(s) they gave for the peak in the
$\chi_\eff$ distribution. In particular, it is important to recognize that
during the orbital decay, the adiabaticity parameter $\mathcal{A}$ (Equation \ref{eq:adiabaticity parameter})
transitions from $\ll1$ to $\gg1$, and this transition determines $\theta_\SL^\f$.

\section{Summary and Discussion}
\label{sec 5}

In this paper we have studied black hole (BH) binary mergers in triple
systems: A sufficiently inclined tertiary companion excites large
eccentricity in the BH binary orbit through gravitational perturbations
(the Lidov-Kozai mechanism), significantly shortening its merger
timescale due to gravitational wave emission. We focus on binaries
with initial separations sufficiently large ($\gtrsim 10$~AU) so that
merger is not possible without large eccentricity excitations. While
this problem has been studied before in various contexts (see
references in Section \ref{sec 1}), we make progress by (1) systematically
determining the merger fractions for various system parameters
(e.g. the masses and orbital properties of the binary and perturber) and deriving the
relevant scaling relations, (2) examining the spin evolution of the
BHs to predict the final the spin-orbit misalignments of the merging
binaries. Although our numerical examples focus on BH binaries with
stellar mass companions, our results (with appropriate re-scalings)
can be applied to neutron star binaries (see Section \ref{sec 3 3}) and other
types of perturbers (e.g. supermassive BHs).

\subsection{Summary of Key Results}

1. For BH binaries with a given initial separation ($a_0\gtrsim 10$~AU),
the merger window (i.e., the range of initial inclination angles $I_0$ between the
inner binary and the outer companion that induces binary merger within
$\sim10^{10}$~years) and merger fraction depend on the effective semi-major axis
${\bar a}_{\rm out,eff}\propto a_{\rm out}\sqrt{1-e_{\rm out}^2}/m_3^{1/3}$
(Equation \ref{eq:aout bar}) and eccentricity $e_{\rm out}$ of the companion.
The results are summarized in Figure \ref{fig:merger fraction oct}.
Assuming that the inclination of the companion is randomly distributed,
we find that the merger fraction (for typical BH masses $m_1=30M_\odot$,
$m_2=20M_\odot$) increases rapidly with increasing $e_{\rm out}$,
from $\sim 1\%$ at $e_{\rm out}=0$ to $10-20\%$ at
$e_\OUT=0.9$. This is because as the octupole potential ($\propto
\varepsilon_\oct\propto e_\OUT$; see Equation \ref{eq:varepsilon oct}) of the tertiary companion increases,
extreme eccentricity excitation of the inner binaries becomes possible for
a wide range of $I_0$ (see Figure \ref{fig:merger window oct}). Regardless of the importance of the
octupole effect, the maximum ${\bar a}_{\rm out,eff}$ value for which the inner binary
has a chance to merge within $10^{10}$~years (or any other values) can be determined analytically
(using Equations \ref{eq:ELIM} and \ref{eq:fitting formula}, setting $e_\mathrm{m}$ to $e_\li$;
see also Equation \ref{eq:fitting a bar}).

2. For systems where the octupole effect is negligible (such as those with
$m_1=m_2$ or $e_{\rm out}=0$), the merger window and merger fraction can be
determined analytically (see Figure \ref{fig:merger fraction quad }).
In particular, these analytical
results can be applied to NS-NS binaries with external companions
(see Section \ref{sec 3 3}, Figure \ref{fig:merger fraction NS}). We have also obtained fitting formulae
relevant to the merger fractions of various systems (Equations \ref{eq:fitting f merger}-\ref{eq:fitting a bar}).

3. On the technical side, we have developed new dynamical
equations for the evolution of triples (Section \ref{sec 2 1 2}) in the
single averaging approximation (i.e., the equations of motion are only
averaged over the inner orbit). These single-averaged equations have a
wider regime of validity in the parameter space than the usual
double-averaged secular equations (see Section \ref{sec 2 3} and Figure \ref{fig:parameterspace}). For
systems where the octupole effect is negligible, we find that the
double-averaged equations accurately predict the merger window and
merger fractions even in the regime where the equations formally break
down (see Figure \ref{fig:merger fraction quad }). However, when the octupole effect is strong (large
$\varepsilon_\oct$), using the single-averaged equations leads to wider merger windows
and larger merger fractions (see Figures \ref{fig:merger window SA}-\ref{fig:merger fraction oct}).

4. During the tertiary-induced binary decay, the spin axes of the BHs
exhibit a variety of evolutionary behaviors due to the combined effects of
spin-orbit coupling (de-Sitter precession), Lidov-Kozai orbital precession/nutation
and gravitational wave emission. These spin behaviors are correlated with the
orbital evolution of the BH binary (Section \ref{sec 4 2}).
Starting from aligned spin axes (relative to the orbital angular momentum axis),
a wide range of spin-orbit misalignments can be generated when the binary enters
the LIGO/VIRGO band:
\begin{itemize}
\item For systems where the octupole effect is negligible
(such as those with $m_1\simeq m_2$ or $e_{\rm out}\sim 0$), the BH spin axis evolves
regularly, with the final spin-orbit misalignment angle $\theta_{\rm sl}^\f$
depending on the initial companion inclination angle $I_0$ in a well-defined
manner (see Figure \ref{fig:merger window quad} and the top left panels of Figure \ref{fig:merger window oct})
\footnote{This conclusion applies to the parameter regime studied in
  this paper, where the inner binary has a large initial separation
  $a_0$ and thus is capable of merging only because of the extreme
  eccentricity excitation induced by the companion; this requires that
  the initial $\varepsilon_\gr\ll 1$ (Equation \ref{eq:epsilonGR}) or the initial
  adiabaticity parameter ${\cal A}_0 \ll 1$ (see Equations \ref{eq:adiabaticity parameter}-\ref{eq:A-eps}). By
  contrast, for systems that have smaller $a_0$ and experience only
  modest eccentricity excitations, ${\cal A}_0$ is not much smaller
  than unity, the BH spin may evolve chaotically even when
  $\varepsilon_\oct=0$ \citep[]{Liu-ApJL}.}.  We find that
when $I_0$ is not too close to $I_{\rm lim}$ (the initial
inclination angle for maximum/limiting eccentricity excitation; see
Equations \ref{eq:I0lim}-\ref{eq:ELIM}), the spin-orbit misalignment evolves into a 90 degree ``attractor''
(Figures \ref{fig:OE quad 1}-\ref{fig:OE quad 2}), a feature that can be qualitatively understood using
adiabatic invariance (see Section \ref{sec 4 3}). When $I_0$ is close to $I_{\rm lim}$,
a qualitatively different spin evolution leads to smaller $\theta_{\rm sl}^\f$
(Figures \ref{fig:OE quad oneshort}-\ref{fig:OE quad 3}).
\item For systems with stronger octupole effect (larger $\varepsilon_\oct$),
the BH spin evolution becomes increasingly chaotic, with the final
spin-orbit misalignment angle depending sensitively on the initial conditions
(see Figures \ref{fig:OE oct 1}-\ref{fig:OE oct 2}). As a result, a wide range of $\theta_{\rm sl}^\f$ values are
produced, including retrograde configurations (see Figure \ref{fig:merger window oct}). The final
spin-orbit misalignment distribution typically peaks around $90^\circ$,
but becomes isotropic (uniform in $\cos\theta_{\rm sl}^\f$) for systems with sufficiently
large $\varepsilon_\oct$ (Figure \ref{fig:spin distribution}).
\end{itemize}

5. We have computed the distribution of the mass-weighted spin parameter
$\chi_{\rm eff}$ (Equation \ref{eq:chi eff}) of merging BH binaries in triples (Figure \ref{fig:overall spin distribution}). While
details of this distribution depend on various parameters
(e.g. distribution of the companion eccentricities), it has a characteristic shape
with peak around $\chi_\eff\simeq0$, extending to the maximum possible positive and negative values (see Equation \ref{eq:bar chi eff}).

\subsection{Discussion}

The merger fraction $f_\merger$ computed in this paper (and particularly the
dependence of $f_{\rm merger}$ on various parameters) can be used to
obtain an estimate of the rate of Lidov-Kozai-induced BH binary
mergers in the galactic field, provided that one makes certain assumption
about the BH populations in triples and their properties. We do not
present such an estimate here since such a calculation
necessarily contains large uncertainties \citep[see][]{Silsbee and Tremaine 2017,Antonini 2017},
like all other scenarios of producing merging BH binaries.
Suffice it to say that with our computed $f_{\rm merger}$ of a few to 10 percent,
it is possible to produce (with large error bars) the observed BH binary merger
rate (10-200 Gpc$^{-3}$yr$^{-1}$).

As noted in Section \ref{sec 1}, the mass-weighted spin parameter $\chi_{\rm eff}$
may serve as useful indicator of binary BH formation mechanism.
The five discovered BH binaries all have low values of $\chi_\eff$,
which could be either the result of slowly-spinning
BHs \citep[e.g.,][]{Zaldarriaga} or large spin-orbit misalignments.
The event GW170104 has $\chi_\eff=-0.12^{+0.21}_{-0.3}$, which may require
retrograde spinning BHs, especially if low individual spins ($\chi_{1,2}\lesssim 0.2$)
can be ruled out.
Such a retrograde spin-orbit misalignment
would challenge the isolated binary BH formation channel, and
point to the importance of some flavors of dynamical formation mechanisms.
We note that the Lidov-Kozai-induced BH mergers lead to a unique
{\it shape} of $\chi_{\rm eff}$ distribution (Figure \ref{fig:overall spin distribution}) that may be used
to distinguish it from other types of dynamical interactions.
For example, a completely random distribution of spin-orbit misalignments, as expected
from the mechanisms involving multiple closer encounters and exchange interactions
in dense clusters \citep[e.g.,][]{Rodriguez 2015,Chatterjee 2017}, would lead
to a specific distribution given by Equation (\ref{eq:p chi eff}).  As the number of
detected BH merger events increases in the coming years, the
distribution of $\chi_{\rm eff}$ will be measured experimentally,
therefore providing valuable constraints on the binary BH formation
mechanisms.

Although we have focused on isolated BH triples in this paper, many
aspects of our results (with proper rescalings) can be applied
to triples that dynamically form in globular clusters
or BH binaries moving around a supermassive BH
\citep[e.g.,][]{Miller 2002,Wen 2003,Thompson 2011,Antonini 2012,Antonini 2014,Petrovich 2017,Hoang 2017}.
In a dense cluster, the orbits of a
triple system can be perturbed or even disrupted by close fly-bys of
other objects.  Therefore the survival timescale of the triple may not
be as long as $10^{10}$~years, depending on the mean density of the
surroundings. In this case, the merger window and merger fraction may be
reduced (see Equations \ref{eq:fitting f merger}-\ref{eq:fitting a bar}), and the remaining systems can lead to extremely large
eccentricities and shorter merger times.
Also, our conclusion on the distribution of near-merger spin-orbit misalignments
depends on the initial BH spin orientations.
We have assumed initial spin-orbit alignment throughout this paper,
but this may not be valid for dynamically formed binaries and triples in dense
clusters. We plan to address some of these issues in a future paper.

\section{Acknowledgments}

This work is supported in part by grants from the National
Postdoctoral Program and NSFC (No. BX201600179, No. 2016M601673,
No. 11703068, No. 11661161012 and No. OP201705). DL is supported by
the NSF grant AST-1715246 and NASA grant NNX14AP31G. This work made use of the High
Performance Computing Resource in the Core Facility for Advanced
Research Computing at Shanghai Astronomical Observatory.


\begin{thebibliography}{}
\bibitem[Abbott et al.(2016a)]{Abbott 2016a} Abbott, B. P., Abbott, R., Abbott, T. D., et al.
(LIGO Scientific and Virgo Collaboration) \ 2016a, PhRvL, 116, 061102

\bibitem[Abbott et al.(2016b)]{Abbott 2016b} Abbott, B. P., Abbott, R., Abbott, T. D., et al.
(LIGO Scientific and Virgo Collaboration) \ 2016b, PhRvL, 116, 241103

\bibitem[Abbott et al.(2017a)]{Abbott 2017a} Abbott, B. P., Abbott, R., Abbott, T. D., et al.
(LIGO Scientific and Virgo Collaboration) \ 2017a, PhRvL, 118, 221101

\bibitem[Abbott et al.(2017b)]{Abbott 2017b} Abbott, B. P., Abbott, R., Abbott, T. D., et al.
(LIGO Scientific and Virgo Collaboration) \ 2017b, ApJL, 851, L35

\bibitem[Abbott et al.(2017c)]{Abbott 2017c} Abbott, B. P., Abbott, R., Abbott, T. D., et al.
(LIGO Scientific and Virgo Collaboration) \ 2017c, PhRvL, 119, 141101

\bibitem[Abbott et al.(2017d)]{Abbott 2017d} Abbott, B. P., Abbott, R., Abbott, T. D., et al.
(LIGO Scientific and Virgo Collaboration) \ 2017d, PhRvL, 119, 161101

\bibitem[Anderson et al.(2016)]{Anderson et al HJ} Anderson, K. R., Storch, N. I., \& Lai, D. \ 2016, MNRAS, 456, 3671

\bibitem[Anderson et al.(2017a)]{Anderson et al 2017} Anderson, K.~R., Lai, D., \& Storch, N. I. \ 2017a, MNRAS, 467, 3066

\bibitem[Anderson \& Lai(2017b)]{Anderson 2017b} Anderson, K.~R., \& Lai, D. \ 2017b, MNRAS, 472, 3692

\bibitem[Antonini \& Perets(2012)]{Antonini 2012} Antonini, F., \& Perets, H. B. \ 2012, ApJ, 757, 27

\bibitem[Antonini et al.(2014)]{Antonini 2014} Antonini, F., Murray, N., \& Mikkola, S. \ 2014, ApJ, 781, 45

\bibitem[Antonini \& Rasio(2016)]{Antonini and Rasio 2016} Antonini, F., \& Rasio, F. A. \ 2016, ApJ, 831, 187

\bibitem[Antonini et al.(2017)]{Antonini 2017} Antonini, F., Toonen, S., \& Hamers, A. S. \ 2017a, ApJ, 841, 77

\bibitem[Antonini et al.(2018)]{Antonini spin} Antonini, F., Rodriguez, C.~L., Petrovich, C., \& Fischer C.~L. \ 2017, MNRAS, 480, L58

\bibitem[Barker \& O'Connell(1975)]{Barker} Barker, B.~M., \& O'Connell, R.~F. \ 1975, PhRvD, 12, 329

\bibitem[Banerjee et al.(2010)]{Banerjee 2010} Banerjee, S., Baumgardt, H., \& Kroupa, P. \ 2010, MNRAS, 402, 371

\bibitem[Belczynski et al.(2010)]{Belczynski 2010} Belczynski, K., Dominik, M., Bulik, T., O'Shaughnessy, R., Fryer, C., \& Holz, D.~E. \ 2010, ApJ, 715, L138

\bibitem[Belczynski et al.(2016)]{Belczynski 2016} Belczynski, K., Holz, D.~E., Bulik, T., \& O'Shaughnessy, R. \ 2016, Nature, 534, 512

\bibitem[Belczynski et al.(2017)]{Belczynski 2017} Belczynski, K., Klencki, J., Meynet, G., et al. \ 2017, arXiv:1706.07053

\bibitem[Blaes et al.(2002)]{Blaes 2002} Blaes, O., Lee, M.~H., \& Socrates, A. \ 2002, ApJ, 578, 775

\bibitem[Chatterjee et al.(2017)]{Chatterjee 2017} Chatterjee, S., Rodriguez, C.~L., Kalogera, V., \& Rasio, F.~A. \ 2017, ApJL, 836, L26

\bibitem[Chen \& Amaro-Seoane(2017)]{Chen xian} Chen, X., \& Amaro-Seoane, P. \ 2017, ApJ, 842, L2

\bibitem[Cholis et al.(2016)]{Cholis 2016} Cholis, I., Kovetz, E.~D., Ali-Ha{\"i}moud, Y., Bird, S., Kamionkowski, M., Mu{\~n}oz, J.~B., \& Raccanelli,
A. \ 2016, PhRvD, 94, 084013

\bibitem[Dominik et al.(2012)]{Dominik 2012} Dominik, M., Belczynski, K., Fryer, C., Holz, D.~E., Berti, E., Bulik, T., Mandel, I., \& O'Shaughnessy, R. \ 2012, ApJ, 759, 52

\bibitem[Dominik et al.(2013)]{Dominik 2013} Dominik, M., Belczynski, K., Fryer, C., Holz, D.~E., Berti, E., Bulik, T., Mandel, I., \& O'Shaughnessy, R. \ 2013, ApJ, 779, 72

\bibitem[Dominik et al.(2015)]{Dominik 2015} Dominik, M., Berti,E., O'Shaughnessy, R., et al. \ 2015, ApJ, 806, 263

\bibitem[Downing et al.(2010)]{Downing 2010} Downing, J.~M.~B., Benacquista, M.~J., Giersz, M., \& Spurzem, R. \ 2010, MNRAS, 407, 1946

\bibitem[Fabrycky \& Tremaine(2007)]{Fabrycky and Tremaine 2007} Fabrycky, D., \& Tremaine, S. \ 2007, ApJ, 669, 1298

\bibitem[Farr et al.(2017)]{Will Nature} Farr, W.~M., Stevenson, S., Miller, M.~C., Mandel, I., Farr, B., \& Vecchio, A. \ 2017, Natur, 548, 426

\bibitem[Ford et al.(2000)]{Ford} Ford, E. B., Kozinsky, B., \& Rasio, F. A. \ 2000, ApJ, 535, 385

\bibitem[G{\"u}ltekin et al.(2006)]{Gultekin 2006} G{\"u}ltekin, K., Miller, M.~C., \& Hamilton, D.~P. \ 2006, ApJ, 640, 156

\bibitem[Harrington(1968)]{Harrington} Harrington, R. S. \ 1968, AJ, 73, 190

\bibitem[Hoang et al.(2018)]{Hoang 2017} Hoang, B.-M., Naoz, S., Kocsis, B., Rasio, F.~A., \& Dosopoulou, F. \ 2018, ApJ, 856, 140

\bibitem[Holman et al.(1997)]{Holman} Holman, M., Touma, J., \& Tremaine, S. \ 1997, Nature, 386, 254

\bibitem[Katz et al.(2011)]{Katz PRL} Katz, B., Dong, S., \& Malhotra, R. \ 2011, PhRvL, 107, 181101

\bibitem[Kozai(1962)]{Kozai} Kozai, Y. \ 1962, AJ, 67, 591

\bibitem[Lai et al.(1993)]{Lai 1993} Lai, D., Rasio, F.A., \& Shapiro, S.L. \ 1993, ApJS, 88, 205

\bibitem[Lai(2014)]{Lai 2014} Lai, D. \ 2014, MNRAS, 440, 3532

\bibitem[Lai et al.(2018)]{Lai 2018} Lai, D., Anderson, K. R., \& Pu, B. \ 2018, MNRAS, 475, 5231

\bibitem[Leigh et al.(2018)]{Leigh 2018} Leigh, N.~W.~C., Geller, A.~M., McKernan, B., et al. \ 2018, MNRAS, 474, 5672

\bibitem[Li et al.(2014)]{Li chaos} Li, G., Naoz, S., Holman, M., \& Loeb, A. \ 2014, ApJ, 791, 86

\bibitem[Lidov(1962)]{Lidov} Lidov, M. L. \ 1962, Planetary and Space Science, 9, 719

\bibitem[Lithwick \& Naoz(2011)]{Lithwick 2011} Lithwick, Y., \& Naoz, S. \ 2011, ApJ, 742, 94

\bibitem[Lipunov et al.(1997)]{Lipunov 1997} Lipunov, V. M., Postnov, K. A., \& Prokhorov, M. E \ 1997, AstL, 23, 492

\bibitem[Lipunov et al.(2017)]{Lipunov 2017} Lipunov, V. M., Kornilov, V., Gorbovskoy, E., et al. \ 2017, MNRAS, 465, 3656

\bibitem[Liu et al.(2015a)]{Liu et al 2015} Liu, B., Mu{\~n}oz, D.~J., \& Lai, D. \ 2015, MNRAS, 447, 747

\bibitem[Liu \& Lai(2017)]{Liu-ApJL} Liu, B., \& Lai, D. \ 2017, ApJL, 846, L11

\bibitem[Mandel \& de Mink(2016)]{Mandel and de Mink 2016} Mandel, I., \& de Mink, S.~E. \ 2016, MNRAS, 458, 2634

\bibitem[Marchant et al.(2016)]{Marchant 2016} Marchant, P., Langer, N., Podsiadlowski, P., Tauris, T.~M., \& Moriya, T.~J. \ 2016, A\&A, 588, A50

\bibitem[Mardling \& Aarseth(2001)]{Mardling} Mardling, R. A., \& Aarseth, S. J. \ 2001, MNRAS, 321, 398

\bibitem[Miller \& Hamilton(2002)]{Miller 2002} Miller, M. C., \& Hamilton, D. P. \ 2002, ApJ, 576, 894

\bibitem[Miller \& Lauburg(2009)]{Miller 2009} Miller, M.~C., \& Lauburg, V.~M. \ 2009, ApJ, 692, 917

\bibitem[Naoz et al.(2011)]{Naoz Nature} Naoz, S., Farr, W.~M., Lithwick, Y., Rasio, F.~A., \& Teyssandier, J. \ 2011, Nature, 473, 187

\bibitem[Naoz et al.(2013a)]{Naoz oct} Naoz, S., Farr, W.~M., Lithwick, Y., Rasio, F.~A., \& Teyssandier, J. \ 2013, MNRAS, 431, 2155

\bibitem[Naoz(2016)]{Naoz 2016} Naoz, S. \ 2016, ARA\&A, 54, 441

\bibitem[O'Leary et al.(2006)]{O'Leary 2006} O'Leary, R.~M., Rasio, F.~A., Fregeau, J.~M., Ivanova, N., \& O'Shaughnessy, R. \ 2006, ApJ, 637, 937

\bibitem[O'Leary et al.(2009)]{O'Leary 2009} O'Leary, R.~M., Kocsis, B., \& Loeb, A. \ 2009, MNRAS, 395, 2127

\bibitem[Peters(1964)]{Peters 1964} Peters, P. C. 1964, PhRv., 136, B1224

\bibitem[Petrovich \& Antonini(2017)]{Petrovich 2017} Petrovich, C., \& Antonini, F. \ 2017, ApJ, 846, 146

\bibitem[Podsiadlowski et al.(2003)]{Podsiadlowski 2003} Podsiadlowski, P., Rappaport, S., \& Han, Z. \ 2003, MNRAS, 341, 385

\bibitem[Portegies Zwart \& McMillan(2000)]{Portegies 2000} Portegies, Zwart, S.~F., McMillan, \& S.~L.~W. \ 2000, ApJ, 528, L17

\bibitem[Postnov \& Kuranov(2017)]{Postnov 2017} Postnov, K., \& Kuranov, A. \ 2017, arXiv:1706.00369

\bibitem[Rodriguez et al.(2015)]{Rodriguez 2015} Rodriguez, C.~L., Morscher, M., Pattabiraman, B, et al. \ 2015, PhRvL, 115, 051101

\bibitem[Rodriguez et al.(2016)]{Rodriguez ApJL} Rodriguez, C.~L., Zevin, M., Pankow, C., Kalogera, V., \& Rasio, F.~A. \ 2016, ApJ, 832, L2

\bibitem[Rodriguez \& Antonini(2018)]{Rodriguez Spin} Rodriguez, C.~L., \& Antonini, F. \ 2018, arXiv:1805.08212

\bibitem[Samsing \& Ramirez-Ruiz(2017)]{Samsing 2017} Samsing, J., \& Ramirez-Ruiz, E. \ 2017, ApJ, 840, L14

\bibitem[Samsing et al.(2018)]{Samsing 2018} Samsing, J., D'Orazio, D.~J., Askar, A., \& Giersz, M. \ 2018, arXiv:1802.08654

\bibitem[Seto (2013)]{Seto PRL} Seto, N. \ 2013, PhRvL, 111, 061106

\bibitem[Silsbee \& Tremaine(2017)]{Silsbee and Tremaine 2017} Silsbee, K., \& Tremaine, S. \ 2017, ApJ, 836, 39

\bibitem[Storch et al.(2014)]{Dong Science} Storch, N. I., Anderson, K. R., \& Lai, D. \ 2014, Science 345, 1317

\bibitem[Storch \& Lai(2015)]{Storch 2015} Storch, N. I., \& Lai, D. \ 2015, MNRAS, 448, 1821

\bibitem[Storch et al.(2017)]{Storch 2017} Storch, N. I., Lai, D., \& Anderson, K. R. \ 2017, MNRAS, 465, 3927

\bibitem[Thompson(2011)]{Thompson 2011} Thompson, T.~A. \ 2011, ApJ, 741, 82

\bibitem[Tremaine et al.(2009)]{Tremaine 2009} Tremaine, S., Touma, J., \& Namouni, F. \ 2009, AJ, 137, 3706

\bibitem[VanLandingham et al.(2016)]{VanLandingham 2016} VanLandingham, J.~H., Miller, M.~C., Hamilton, D.~P., \& Richardson, D.~C. \ 2016, ApJ, 828, 77

\bibitem[Wen(2003)]{Wen 2003} Wen, L. \ 2003, ApJ, 598, 419

\bibitem[Zaldarriaga et al.(2017)]{Zaldarriaga} Zaldarriaga, M., Kushnir, D., \& Kollmeier, J. A. \ 2017, MNRAS, 473, 4174

\end{thebibliography}
\end{document}